\documentclass{aa}
\usepackage[varg]{txfonts}
\usepackage{graphicx}
\usepackage{amsmath}
\usepackage{aas_macros}
\usepackage[caption=false]{subfig}
\usepackage{graphics}
\usepackage{times}
\usepackage{newtxmath}
\usepackage{color}

\newcommand{\msun}{\ensuremath{\textup{ M}_{\odot}}}
\newcommand{\MSUN}{\ensuremath{\textup{M}_{\odot}}}

\newcommand{\rsun}{\ensuremath{\textup{R}_{\odot}}}
\newcommand{\RSUN}{\ensuremath{\textup{R}_{\odot}}}
\newcommand{\mjup}{\ensuremath{\textup{ M}_{\textsc{j}}}}
\newcommand{\MJUP}{\ensuremath{\textup{M}_{\textsc{j}}}}

\newcommand{\dif}{\mathrm{d}}
\renewcommand{\vec}[1]{\boldsymbol{#1}} 

\title{Planet formation around M dwarfs via disc instability}
\subtitle{Fragmentation conditions and protoplanet properties}
\author{Anthony Mercer \and Dimitris Stamatellos\thanks{dstamatellos@uclan.ac.uk}} 
\institute{Jeremiah Horrocks Institute for Mathematics,
Physics \& Astronomy, University of Central Lancashire, Preston, PR1 2HE, UK}
\authorrunning{A. Mercer \and D. Stamatellos}
\date{Received 2019 / Accepted 2019}
 
\begin{document}

\abstract
{ 
  { Around 30 per cent of the observed exoplanets} that orbit M dwarf stars are gas giants that are  more massive than Jupiter. These planets are prime candidates for formation by  disc instability.
}
{
  We want to determine the conditions for disc fragmentation around M dwarfs and the properties of the planets that are formed by disc instability.
}
{ 
  We performed hydrodynamic simulations of M dwarf protostellar discs in order to determine the minimum disc mass required for gravitational fragmentation to occur. Different stellar masses, disc radii, and metallicities were considered. The mass of each protostellar disc was steadily increased until the disc fragmented and a protoplanet was formed.
}
{
  We find that a disc-to-star mass ratio between $\sim 0.3$ and $\sim 0.6$ is required for fragmentation to happen. The minimum mass at which a disc fragments increases with the stellar mass and the disc size. Metallicity does not significantly affect the minimum disc fragmentation mass  but high metallicity may suppress fragmentation. Protoplanets form quickly (within a few thousand years) at distances around $\sim50$~AU from the host star, and they are initially very hot; their centres have temperatures similar to the ones expected at the accretion shocks around planets formed by core accretion (up to 12,000~K). The final properties of these planets (e.g. mass and orbital radius) are determined through long-term disc-planet or planet-planet interactions. 
}
{ 
    Disc instability is a plausible way to form  gas giant planets around M dwarfs provided that discs have at least 30\% the mass of their host stars during the initial stages of their formation. Future observations of massive M dwarf discs or planets around very young M dwarfs are required to establish the importance of disc instability for planet formation around low-mass stars.  
}

\keywords{Accretion, accretion disks - protoplanetary disks -stars:low-mass - planets and satellites: formation - hydrodynamics}

\maketitle

\section{Introduction}
\label{sec:introduction}

M dwarfs are the most common stars in the Galaxy \citep{Kroupa:2001a, Chabrier:2003a} and so
their study is important, especially in the context of planet formation. Among
the few thousand planets that have been observed since the discovery of 51~Pegasi~b, the first exoplanet around a main-sequence star \citep{Mayor:1995a},
many planets have been observed orbiting around M dwarfs. These planets have been discovered
either indirectly with the radial velocity and transit methods
\citep[e.g][]{Bonfils:2013a, Reiners:2018b} or directly by imaging
\citep[e.g.][see \citealt{Bowler:2016a} for a review]{Marois:2008a, Bowler:2015b}.

\begin{figure*}
  \begin{center}
  \subfloat{\resizebox{0.5\hsize}{!}
  {\includegraphics{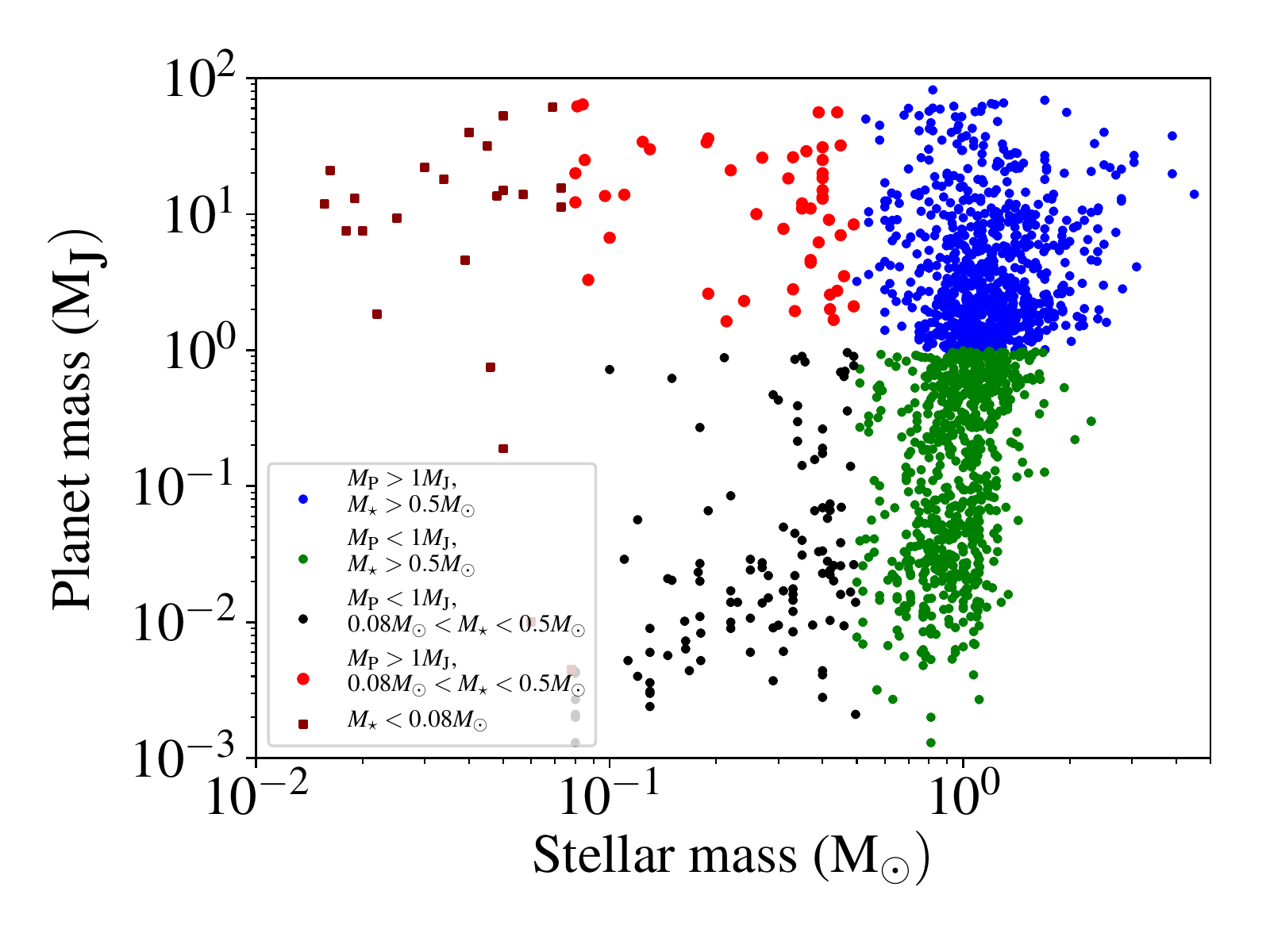}}}
  \subfloat{\resizebox{0.5\hsize}{!}
  {\includegraphics{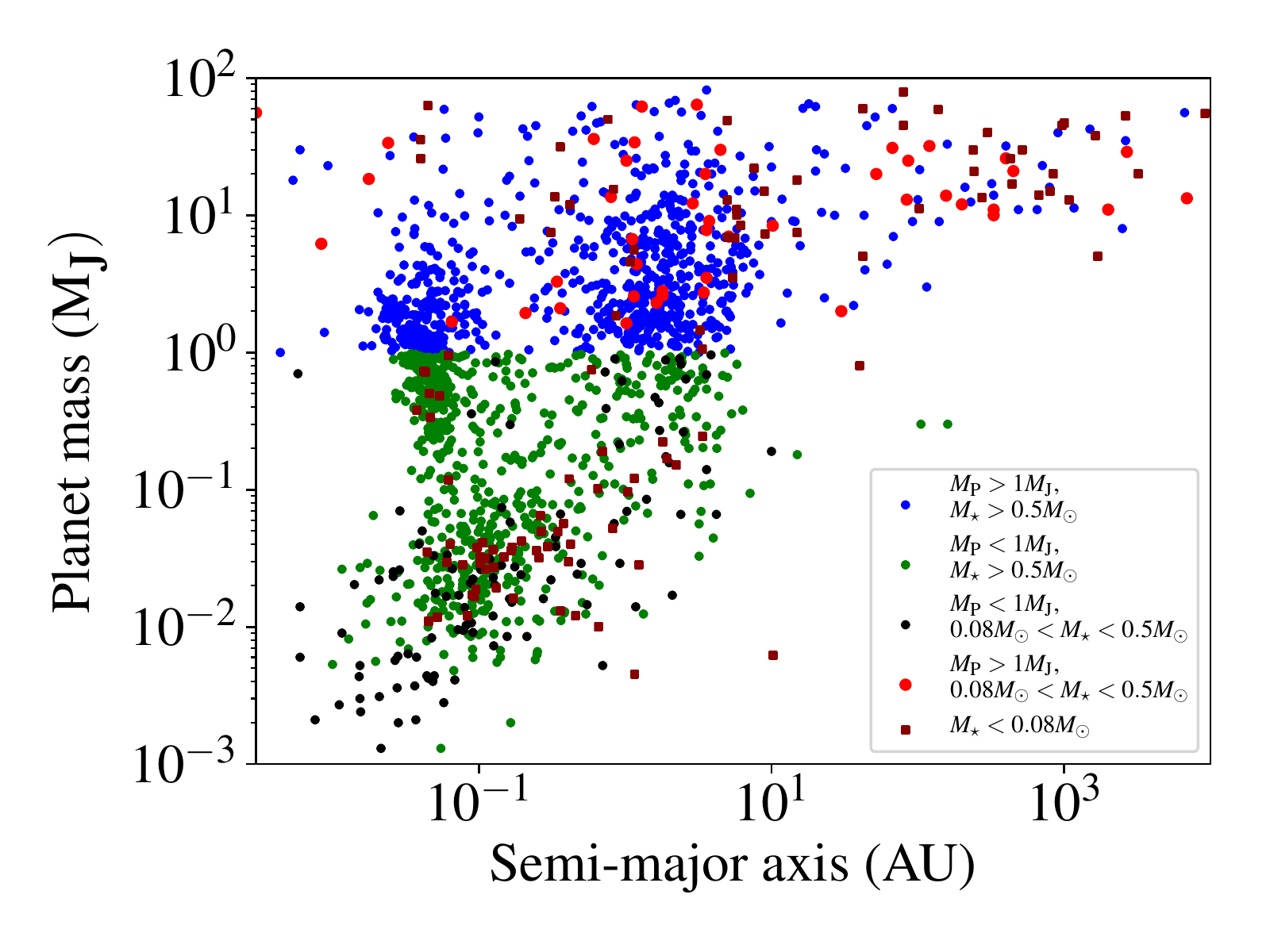}}}
  \caption
  {
    Properties of the observed exoplanets around M dwarfs compared with the properties of all observed exoplanets. Red:~companions (exoplanets and brown dwarfs up to $60\mjup$) with $M_{\rm P}>1\mjup$ around M dwarfs ($M_\star<0.5\msun$). Black: exoplanets with
    $M_{\rm P}<1\mjup$ around M dwarfs. Blue: companions with $M_{\rm P}>1\mjup$ around higher
    mass stars ($M_\star>0.5\msun$). Green: exoplanets with $M_{\rm P}<1\mjup$ around higher
    mass stars ($M_\star>0.5\msun$). Brown: companions around brown dwarfs ($M_\star<0.08\msun$). Data taken from  https://exoplanet.eu/\citep{Schneider:2011a}. This database uses the \cite{Hatzes:2015a} definition for planets (based on the mass-density relationship) and so it includes objects with masses  $13-60\mjup$, which would be classified as brown dwarfs according to their mass. The inclusion of these objects does not affect the discussion presented in this paper.
  }
  \label{fig:exoproperties1}
\end{center}
\begin{center}    
\subfloat{\resizebox{0.35\hsize}{!}
{\includegraphics{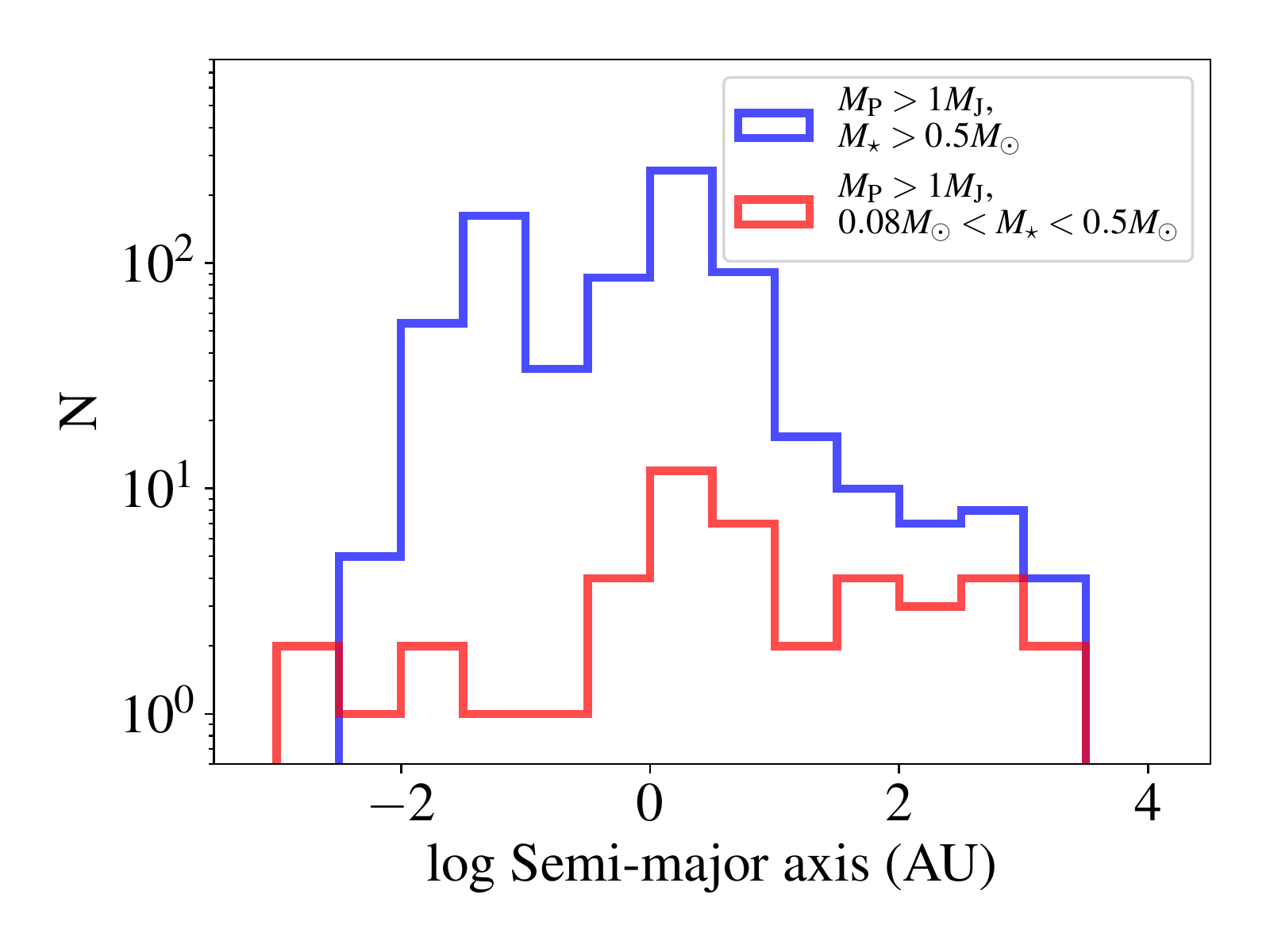}}}\hspace*{-1.em}
\subfloat{\resizebox{0.35\hsize}{!}
{\includegraphics{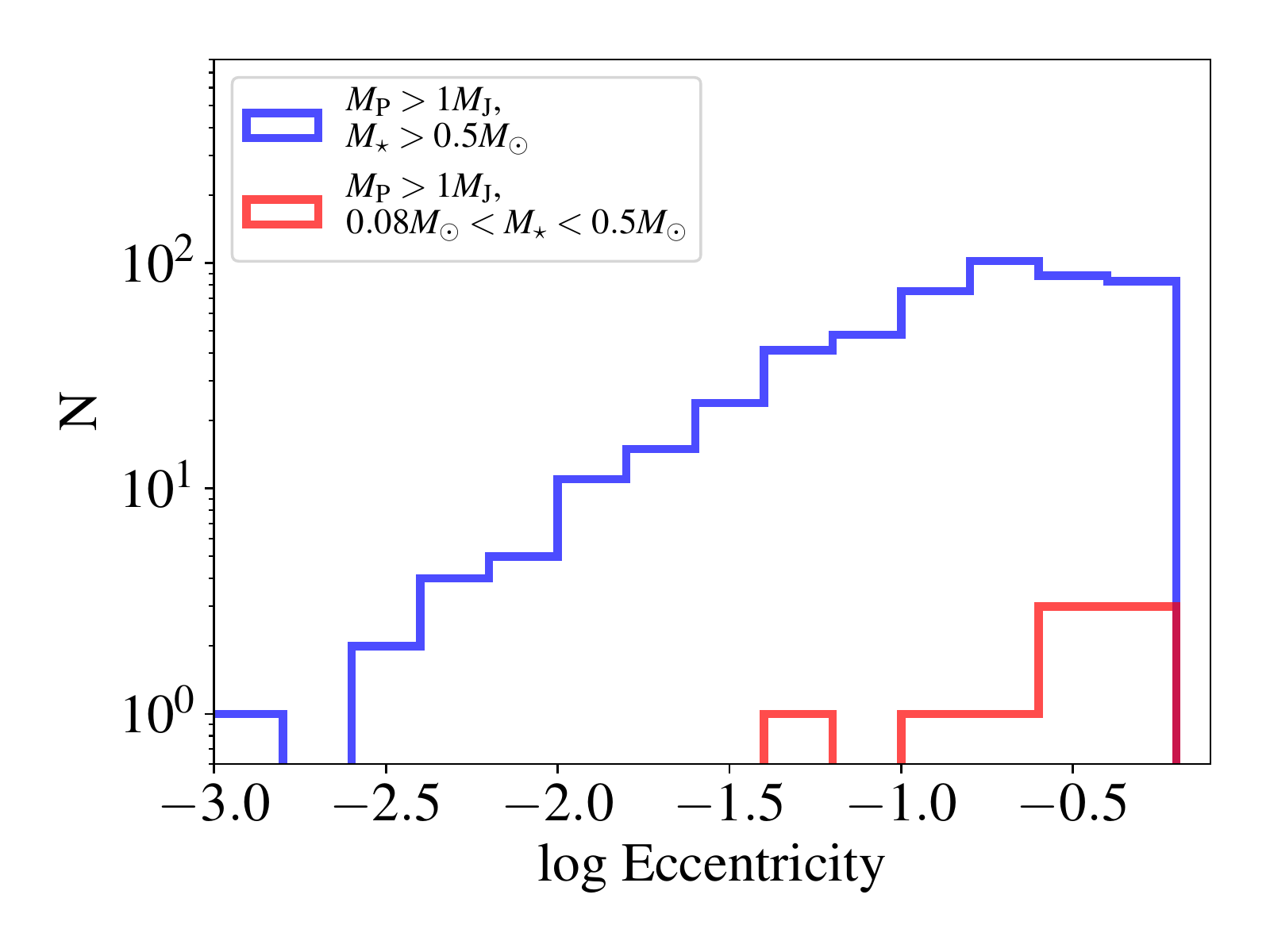}}}\hspace*{-1.em}
\subfloat{\resizebox{0.35\hsize}{!}
{\includegraphics{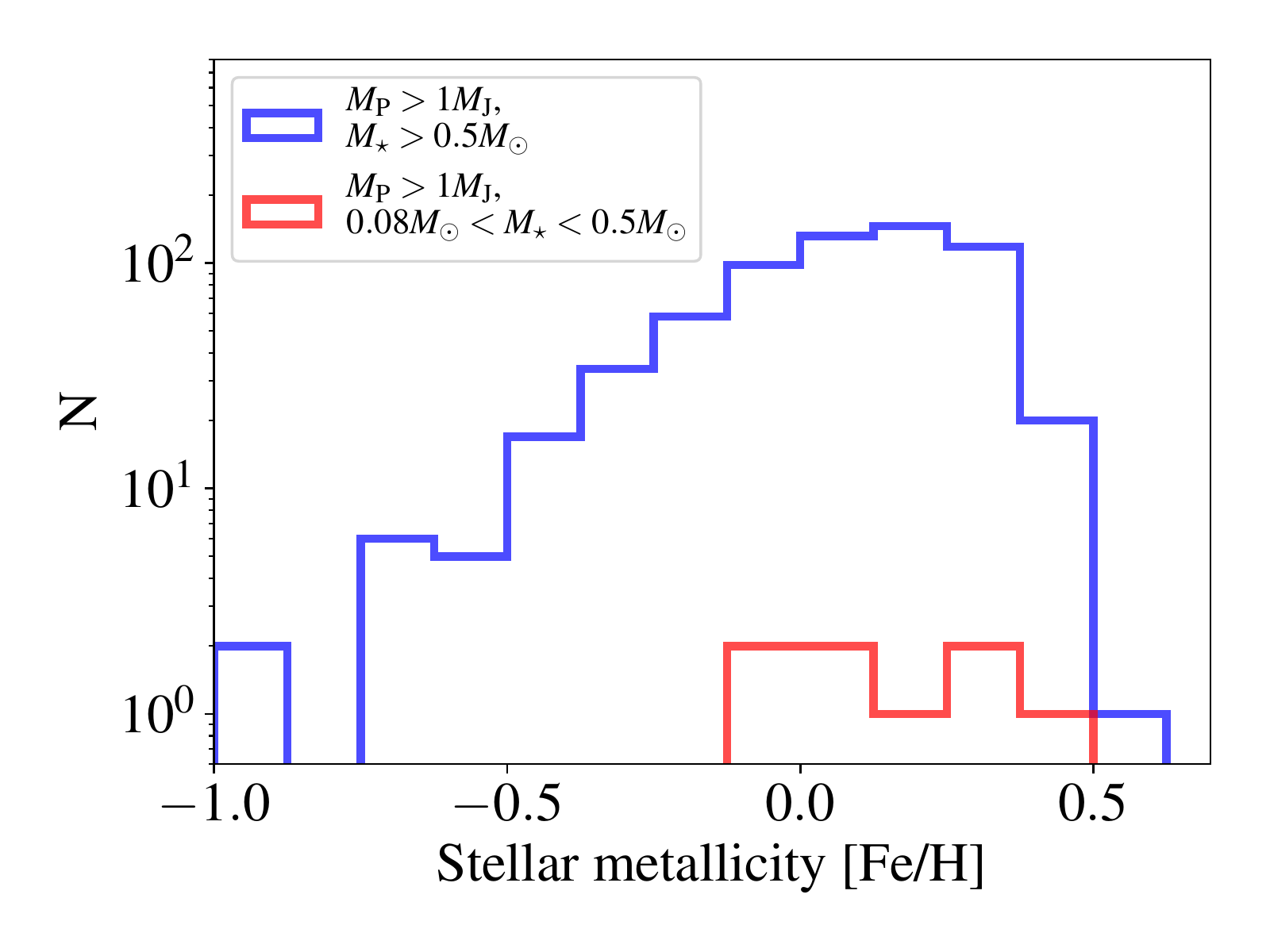}}}
\caption{Comparison of the properties (semi-major axis, eccentricity and stellar metallicity) of high-mass exoplanets around M dwarfs with the properties of high-mass exoplanets around higher-masss stars (colours as in Figure~\ref{fig:exoproperties1}). Exoplanets around M dwarfs (red histograms)
tend to have high eccentricities and high metallicities.}
\label{fig:exoproperties2}
\end{center}
  \begin{center}
  \subfloat{\resizebox{0.35\hsize}{!}
  {\includegraphics{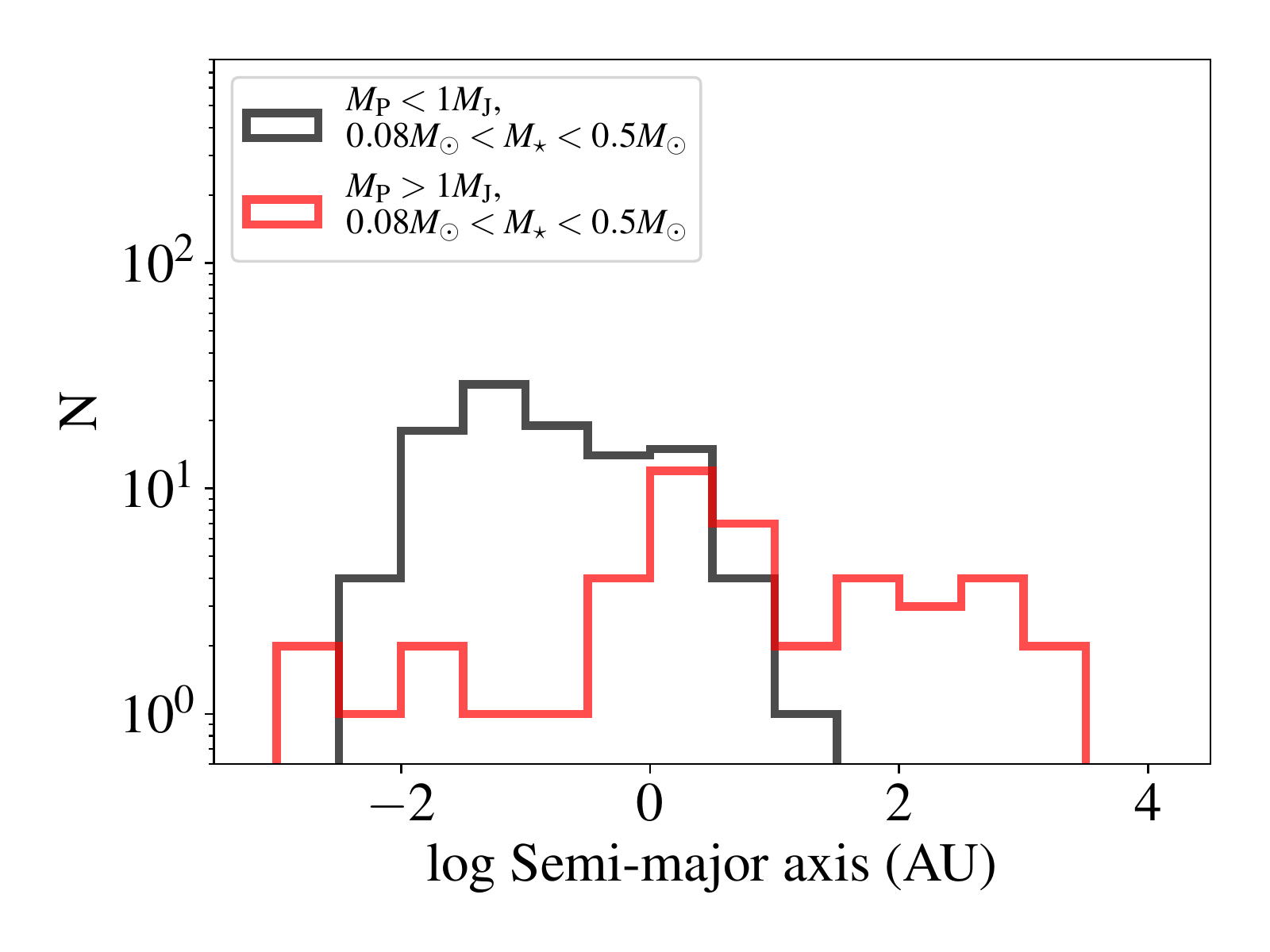}}}\hspace*{-1.em}
  \subfloat{\resizebox{0.35\hsize}{!}
  {\includegraphics{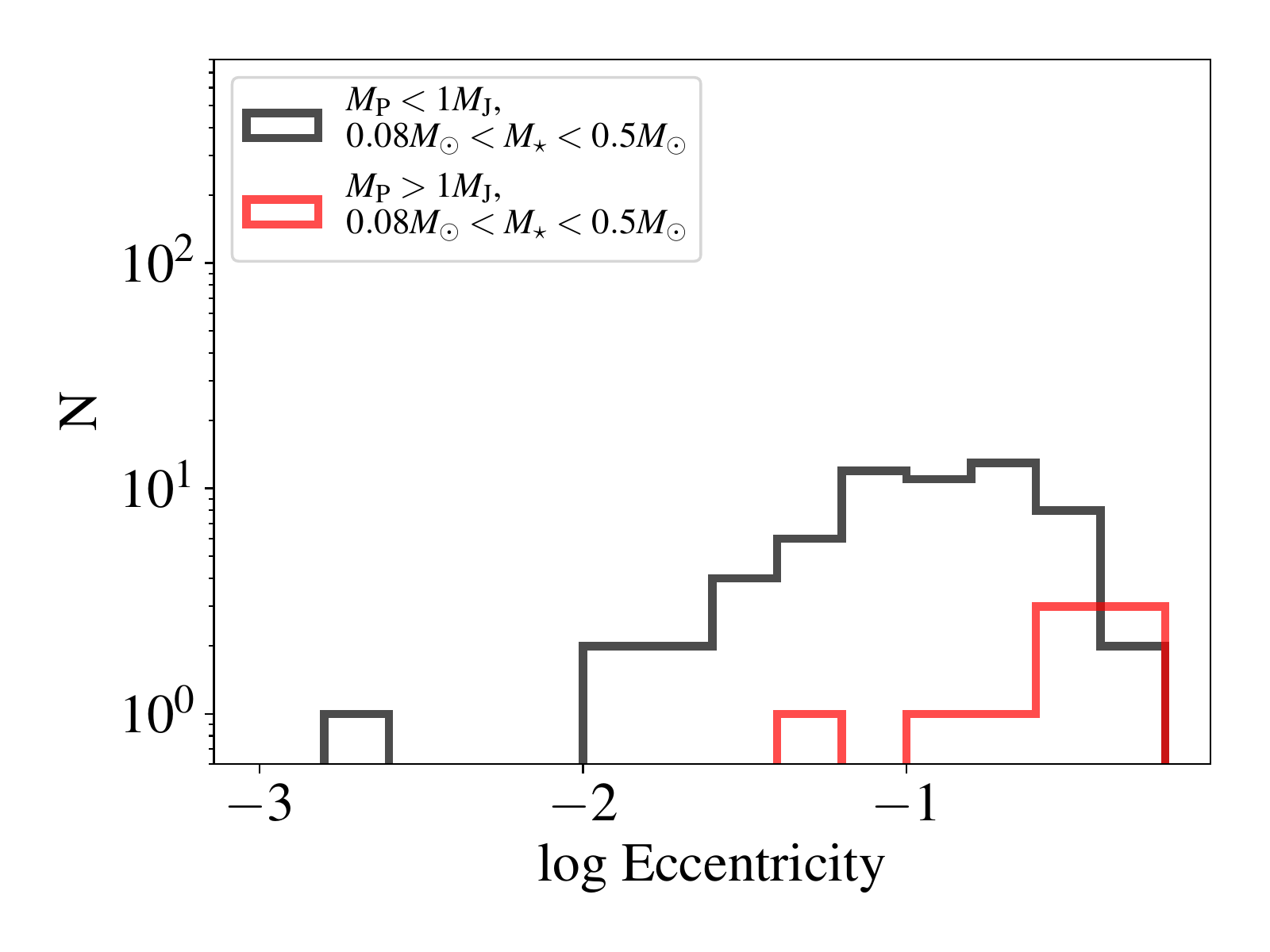}}}\hspace*{-1.em}
  \subfloat{\resizebox{0.35\hsize}{!}
  {\includegraphics{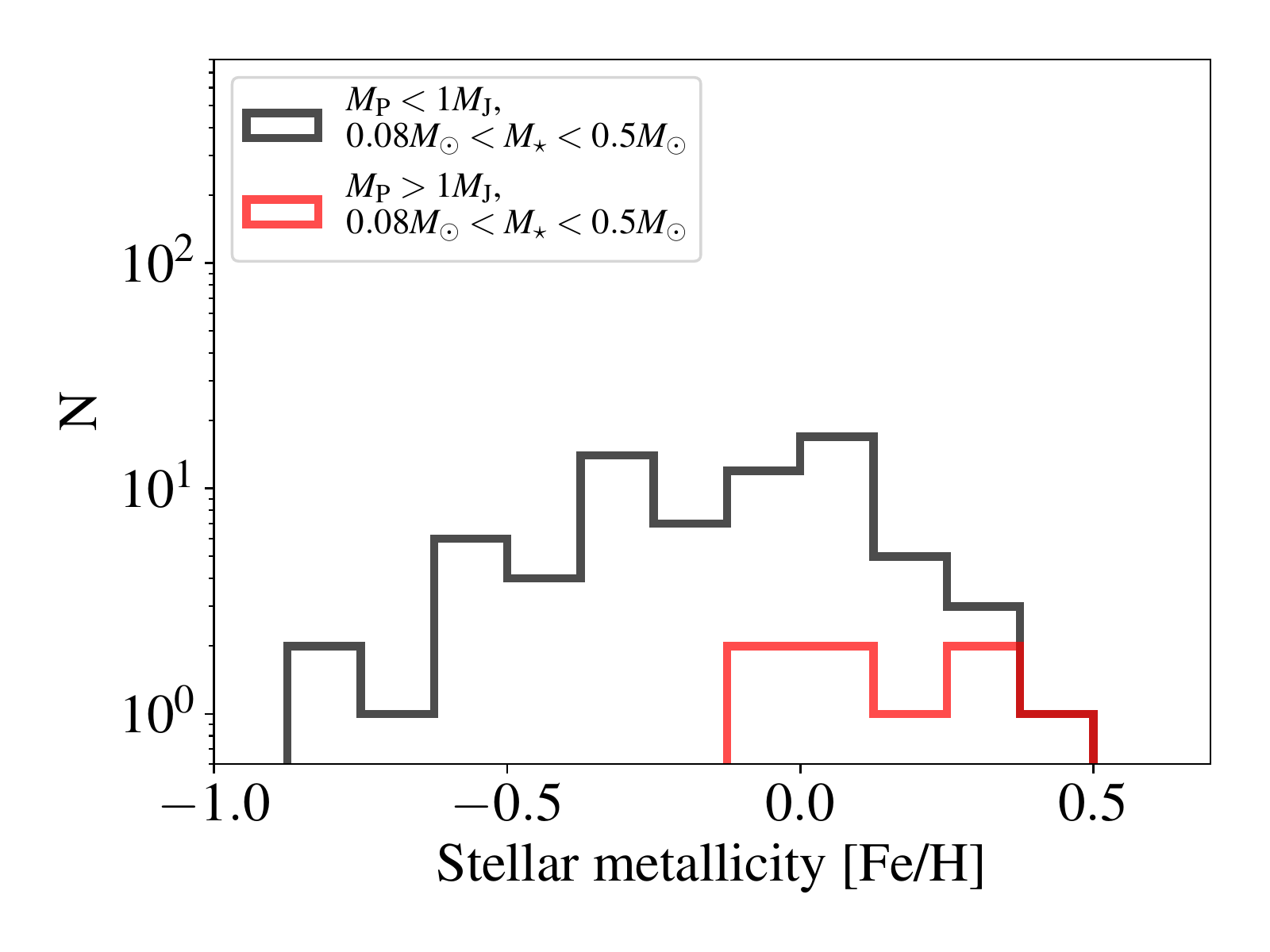}}}
  \caption{Comparison between the properties  (semi-major axis, eccentricity and stellar metallicity) of high-mass and low-mass exoplanets around M dwarfs (colours as in Figure~\ref{fig:exoproperties1}). High-mass exoplanets (red histograms) tend to have high eccentricities and high metallicities when compared to their lower-mass counterparts.}
  \label{fig:exoproperties3}  
\end{center}
  \end{figure*}

The planets around M dwarfs are diverse
(see Figures~\ref{fig:exoproperties1}-\ref{fig:exoproperties3}). They have  small to high masses (from Earth-mass planets to 13$\mjup$-mass planets) and narrow to wide separations from their host stars ($10^{-3}$ to $10^4$~AU) (see Figure~\ref{fig:exoproperties1}).  A fraction of those planets ($\sim 30$\%) are gas giants with a mass larger than  1$\mjup$. Such massive planets are observed both near and far from their host star (Figure~\ref{fig:exoproperties1}), whereas their eccentricities and metallicities seem to be rather high when they are compared with low-mass planets around M dwarfs and also when they are compared with high-mass planets
around more massive stars (see Figures~\ref{fig:exoproperties2}-\ref{fig:exoproperties3}).
It is therefore of interest to investigate how these giant planets around M dwarfs form.   

Planets are believed to form by the core accretion scenario in which dust
particles coagulate into progressively larger aggregates until a solid core
forms, which can then promote the accretion of a gaseous envelope
\citep{Safronov:1969a, Goldreich:1973a, Greenberg:1978a, Hayashi:1985a,
Lissauer:1993a}. In this scenario, the formation of giant planets needs a few Myr, a timescale that may exceed the lifetime of the disc \citep{Haisch:2001a,
Cieza:2007a}, although the process of pebble accretion may accelerate the
process \citep{Lambrechts:2012a}.

An alternative theory of planet formation is disc instability, that is, planet formation by the gravitational fragmentation of young protostellar discs \citep{Kuiper:1951a, Cameron:1978a, Boss:1997a}. Fragmentation happens  provided that the Toomre criterion \citep{Toomre:1964a} is satisfied, 
\begin{equation}
  Q \equiv \frac{c_{s} (R) \kappa (R)}{\pi G \Sigma (R)} \stackrel{<}{_\sim}1,\,
\label{eqn:toomre}
\end{equation}
where $c_{\rm s} (R)$ is the sound speed, $\kappa (R)$ is the epicyclic frequency, and $\Sigma (R)$ is the surface density of the disc at a given orbital radius $R$.
The gravitational instability leads
to the formation of spiral arms that transfer angular momentum radially
outwards. A spiral arm can evolve non-linearly
and collapse if the cooling rate is sufficiently short: typically 
$t_{\textup{cool}} < (0.5 - 2) t_{\textup{orb}}$, that is, of the order of a few orbital
periods \citep{Gammie:2001a, Johnson:2003a, Rice:2003a, Rice:2005a}. In this
scenario protoplanets form on a dynamical timescale (a few thousand years) and
have initial masses of a few $\MJUP$ (set by the opacity limit for
fragmentation). However, these planets can rapidly accrete gas, growing in mass to become brown-dwarfs or low-mass hydrogen-burning stars
\citep{Stamatellos:2009a, Kratter:2010a, Vorobyov:2013a, Kratter:2016a}. Those
objects that do end up as planets are typically the ones that are ejected from the disc through gravitational interactions \citep{Li:2015b, Li:2016a, Mercer:2017a}. 

Observations of young discs have revealed the presence of multiple gaps and
bright rings at mm wavelengths \citep[e.g. HL
Tau,][]{ALMA-partnership:2015a}. Such gaps may be due to young planets
\citep{Dipierro:2015b}, which opens up the possibility that planets may form on
a short timescale. This idea is also corroborated by observations of later phase
(T Tauri) discs, which show that at their present age (a few Myr) they do not
have enough mass to form the observed population of exoplanets
\citep{Greaves:2010a, Manara:2018a}. Rapid planet formation due to disc
instability has been boosted by ALMA observations of massive extended discs in
the Class 0 phase \citep{Tobin:2016a} and of discs with spiral arms indicative of gravitational instabilities \citep{Perez:2016a, Tobin:2016a}. 

The existence of massive planets on wide orbits around M~dwarfs poses challenges
to both planet formation theories. M~dwarf discs have lower masses than the discs around 
solar-type stars \citep[$M_{\rm d}\approx M_*^{-2.4}$, e.g.][]
{Andrews:2013a,Mohanty:2013a,Ansdell:2017a,Stamatellos:2015b}; disc masses are typically below a
few $\mjup$ \citep{Ansdell:2017a, Manara:2018a}, with evidence of quicker disc
dissipation \citep{Ansdell:2017a}. Such low mass discs are not susceptible to
disc fragmentation nor do they provide a good environment for pebble accretion \citep{Liu:2019a}.

It is possible though that these discs were more massive during
their early phases, maybe massive enough for disc instability to operate and form giant planets fast. There are observations of planetary systems with a massive exoplanet with massx M$\sim\mjup$ around  a $\sim100\mjup$ star (see Figure~\ref{fig:exoproperties1}). This means that the initial disc mass was at least 10\% of the  mass of the host star. However, this fraction could be much higher if one considers that  (i) there may be other planets in the system that have not been detected, (ii) the stellar mass increases with time as it accretes gas from the disc, and (iii) a significant fraction of the disc mass may be lost due to accretion onto the host star, photoevaporation or disc winds.
Therefore, disc instability may be a good candidate for explaining the formation of massive planets around M dwarfs, like for example, the planet around star GJ~3512 \citep{Morales:2019a}. 

Massive planets on wide orbits around M dwarfs are ideal candidates for
formation by disc instability as the conditions for the instability to happen
 are met in the outer disc regions
\citep[e.g.][]{Stamatellos:2007c, Stamatellos:2009a, Stamatellos:2011d}.
Observational surveys indicate that only a small fraction of M dwarfs (less than
$\sim10\%$) host wide orbit planets and this also holds for higher-mass stars
\citep{Brandt:2014a, Bowler:2015b, Lannier:2016a, Reggiani:2016a,
Galicher:2016a, Bowler:2016a, Vigan:2017a, Baron:2018a, Stone:2018a,
Wagner:2019a, Nielsen:2019a} \citep[see review by][]{Bowler:2018a}. These
surveys typically explore  a region out to a few hundred AU from the central star
\citep[or even a few thousand AU,][]{Durkan:2016a, Naud:2017a} and they are
sensitive down to Jupiter-mass planets. The occurrence rates of giant planets on
wide orbits are uncertain as they depend upon the sensitivity limits derived
from models of planet evolution and the assumptions made about the
planet mass-period distribution \citep[see][]{Bowler:2018a}. Nevertheless,  it seems
unlikely that giant planets on wide orbits are common. This implies a formation
process that operates only in a small subset of disc initial conditions.
Alternatively, the low-fraction of wide-orbit planets may be due to subsequent
dynamical evolution, either migration towards the star due to disc-planet
interactions \citep[e.g.][]{Stamatellos:2018a}, scattering farther away from the
central star and/or ejection due to 3-body interactions
\citep[e.g.][]{Mercer:2017a} or disruption and ejection due to interactions
within a cluster \citep{Hao:2013a, Cai:2017a}.

Disc instability in M dwarf discs has not been extensively studied.
\cite{Boss:2006a} suggests that the formation of Jupiter mass planets is
possible via fragmentation of discs around stars with masses $0.1$ and $0.5
\msun$. The discs that this author studies are small in extent ($4 < R < 20$~AU), so it is uncertain how the fast cooling needed for disc fragmentation is achieved.
\cite{Backus:2016a} perform simulations of discs around a $0.33 \msun$ star.
They find that only the discs which exhibit $Q_{\textup{crit}} \lesssim 0.9$,
fragment. The radii of the discs studied are between 0.3 and 30 AU, with masses
between 0.01 and 0.08$\msun$. This study focuses on locally isothermal discs,
which are more prone to fragmentation even at small radii given the fact that
fragments can cool to the background temperature  instantaneously,
therefore artificially satisfying the cooling criterion for fragmentation.

In this paper, we improve upon  previous studies by investigating the
fragmentation of discs around M dwarfs using radiative hydrodynamic simulations
with appropriate cooling. Our aim is two-fold: (i) to find the minimum disc mass
required for fragmentation to happen, and (ii) to determine the properties of the planets that form and provisionally compare them with the observed properties of
exoplanets around M dwarfs. 

The paper is laid out as follows. In Section
\ref{sec:numerical_method}, we describe the numerical methods employed within
the paper. Section \ref{sec:initial_conditions} outlines the initial conditions
of each simulation and Section~\ref{sec:mass_loading_test_and_convergence} presents the tests performed to check the validity of our method.
 In Section \ref{sec:fragmentation_of_m_dwarf_discs} we
discuss how different parameters affect the disc fragmentation mass. We investigate
the properties of the formed planets in Section
\ref{sec:the_properties_of_planets_formed_through_disc_fragmentation}, and in Section~\ref{sec:comparison_with_observations} we compare these properties with exoplanet observations. Finally,  the
work is summarised in Section \ref{sec:conclusions}.

\section{Numerical methods}
\label{sec:numerical_method}

We study the dynamics of fragmentation of protostellar discs around M dwarfs by
performing hydrodynamic simulations of initially gravitationally stable discs
that progressively increase in mass and fragment. In the following subsections
we describe in detail the methods that we used. 

\subsection{Hydrodynamics}
\label{sub:hydrodynamics}

We utilized the code \textsc{GANDALF} \citep{Hubber:2018a} to perform smoothed
particle hydrodynamical simulations. The code uses the conservative grad-h SPH scheme
\citep{Springel:2002a}. The \cite{Cullen:2010a} implementation of time-dependent viscosity was utilised 
in order to reduce artificial viscosity away from shocks. An M4 cubic spline
kernel \citep{Schoenberg:1946,Monaghan:1985a} was used as the smoothing function. 

The radiative transfer processes that regulate cooling and heating in the disc
were treated with the method of \cite{Lombardi:2015a}, which is based on the
method of \cite{Stamatellos:2007b} \citep[see also][]{Forgan:2009b}. { This method
uses the gas pressure scale-height of a particle $i$, $H_{\textup{P}, i}$ to
obtain the column density, through which heating and cooling happens. The pressure scale-height is calculated using
\begin{equation}
H_{\textup{P}, i} = \frac{P_{i}}{\rho_{i} \left|\vec a_{h, i}\right|},
\label{eqn:pressure_scale_height}
\end{equation}
where $P_{i}$ and $\rho_{i}$ are the pressure and density of the gas
respectively. $\vec a_{h, i}$ is the hydrodynamical acceleration of the gas
(i.e. the gravitational or viscous accelerations are not included). The column density of particle $i$ is then set to
\begin{equation}
\bar{\Sigma}_{i} = \zeta^\prime \rho_{i} H_{\textup{P}, i} ,
\label{eqn:sigma}
\end{equation}
where $\zeta^\prime = 1.014$ is a dimensionless coefficient with a weak dependence on
the polytropic index.} This formulation has been shown to yield a more accurate
estimate of the particle column density in the context of protostellar discs when compared to the method that uses the gravitational potential
\citep{Mercer:2018a}.

Once the column density is calculated the heating/cooling rate of a particle $i$
is set to
\begin{equation}
\frac{\dif u_{i}}{\dif t} = \frac{4 \sigma_{\textsc{sb}}
\left(T^{4}_{\textsc{bgr}} - T^{4}_{i}\right)}{\bar{\Sigma}^{2}_{i}
\kappa_{\textsc{r}}(\rho_{i}, T_{i}) +
\kappa_{\textsc{p}}^{-1}(\rho_{i}, T_{i})}.
\label{eqn:dudt}
\end{equation}
$\sigma_{\textsc{sb}}$ is the Stefan-Boltzmann constant and $T_{\textsc{bgr}}$
is a background temperature that particles cannot radiatively cool below.
$\kappa_{\textsc{r}}$ and $\kappa_{\textsc{p}}$ are the pseudo-mean Rosseland-
and Planck opacities \citep[see][ for details]{Lombardi:2015a}, respectively,
and are assumed to be the same. Equation~(\ref{eqn:dudt}) allows the calculation
of the cooling rate smoothly between the optically thin and thick regimes, whereas at the optically thick regime it reduces to the diffusion approximation
\citep{Mihalas:1970a}. 

 We used the \cite{Bell:1994a} opacities such that $\kappa(\rho, T) = \kappa_{0}
\rho^{a} T^{b}$, where $\kappa_{0}$, $a$ and $b$ are constants set depending on
the chemical species contributing to the opacity at a given density and temperature.
Ice melting, dust sublimation, bound-free, free-free and electron scattering
interactions are taken into account. We also used a detailed equation of state
for the gas that considers the rotational and vibrational degrees of freedom of
$\textup{H}_{2}$, the dissociation of $\textup{H}_{2}$, {\ and the ionisation of hydrogen and helium \citep[see][for details]{Stamatellos:2007b}.}

\subsection{Mass loading}
\label{sub:mass_loading}

In order to find the minimum fragmentation mass of an M dwarf disc, we started with a graviationally stable disc and slowly increased its mass at a constant
rate, employing a low mass accretion rate \citep[see][] {Zhu:2012a}. The method
can be conceptually thought as accretion onto the disc from an infalling
envelope, where material is distributed across the whole disc. We set the disc mass accretion rate to
\begin{equation}
\dot{M}_{\textup{disc}} = \frac{\chi M_{\textup{disc,0}}}{t_{\textup{orb}}},
\label{eqn:mass_accretion_rate}
\end{equation}
where $M_{\textup{disc, 0}}$ is the initial disc mass and $\chi$ is a factor
which regulates the magnitude of accretion. $t_{\textup{orb}}$ is the orbital
period of the disc at a radius $R = 100$~AU, where
\begin{equation}
t_{\textup{orb}} = 2 \pi \sqrt{\frac{R^{3}}{GM_{\star}}}.
\label{eqn:orbital_period}
\end{equation}
Therefore, $\chi$ represents the fraction of the increase of the disc mass
during approximately one rotation at its initial outer edge. The mass accretion
is simply performed by increasing the mass of every particle equally every
timestep. We refer to this method as mass loading. The discs were evolved until they fragmented, which we define as when a density of $\rho > 10^{-9} \textup{ g cm}^{-3}$ is attained. { This value is around the density where the first hydrostatic core forms during the collapse \citep{Larson:1969a, Stamatellos:2007b}. By chosing a relatively high density threshold like this, we ensured that bound objects formed (i.e. the disc {fragments}) rather than just transient over-densities. The density of the first core (when it forms) does not vary significantly with the core mass \citep{Stamatellos:2009d}, therefore the same value can be used for all fragments forming in the disc, irrespective of their mass. From a practical point of view, the highest density SPH particle wass identified at each timestep and its density was compared with the threshold density (the SPH particle density is calculated using $\sim 50$ neighbourghing particles). The centre of the clump was found by the position of the highest density particle within it (usually there were  $\sim 10^5$ particles within each clump, see Section~\ref{sec:the_properties_of_planets_formed_through_disc_fragmentation})}.

One caveat of the mass loading method is that higher density regions
of the disc (i.e.  where there are more particles) are preferentially
mass-loaded. For example, spiral arms may receive a higher proportion of the
accreted mass and the collapse of a dense region may be driven artificially, if
the accretion rate is set too high . We therefore used a relatively low
disc accretion rate (see tests below) so that accretion is not the key driver of the gravitational instability \citep[e.g.][]{Hennebelle:2016a}.

\begin{table*}
  \centering
    \caption
    {
      Initial conditions of the disc simulations: initial stellar mass
      ($M_{\star, 0}$), disc radius ($R_{\textup{init}}$), metallicity ($z$), disc
      mass ($M_{\textup{disc, 0}}$), and mass loading rate
      ($\dot{M}_{\textup{disc}}$). Disc masses were set such that the Toomre
      parameter at the outer disc radius is $Q = 10$ (i.e. the discs are
      initially gravitationally stable). The constant mass loading rate onto the
      disc was set from Equation \ref{eqn:mass_accretion_rate}, where $\chi = 0.1$.
      The disc metallicity was varied by modifying the opacities by a factor $z$.
    }
    \label{tab:ics}
    \centering
    \begin{tabular}{c c c c c c}
      \hline
      \hline
  
      Run & $M_{\star, 0}$ (\MSUN) & $R_{\textup{init}}$ (AU) & $z$ &
      $M_{\textup{disc, 0}}$ (\MSUN) & $\dot{M}_{\textup{disc}}~ (10^{-6}
      \msun \textup{ yr}^{-1}$)
      \\
      \hline
      01 & 0.2 & 60  & 1   & 0.040 & 1.80 \\
      02 & 0.2 & 60  & 0.1 & 0.040 & 1.80 \\
      03 & 0.2 & 60  & 10  & 0.040 & 1.80 \\
      \hline
      04 & 0.2 & 90  & 1   & 0.050 & 2.25 \\
      05 & 0.2 & 90  & 0.1 & 0.050 & 2.25 \\
      06 & 0.2 & 90  & 10  & 0.050 & 2.25 \\
      \hline
      07 & 0.2 & 120 & 1   & 0.059 & 2.63 \\
      08 & 0.2 & 120 & 0.1 & 0.059 & 2.63 \\
      09 & 0.2 & 120 & 10  & 0.059 & 2.63 \\
      \hline
      10 & 0.3 & 60  & 1   & 0.049 & 2.70 \\
      11 & 0.3 & 60  & 0.1 & 0.049 & 2.70 \\
      12 & 0.3 & 60  & 10  & 0.049 & 2.70 \\
      \hline
      13 & 0.3 & 90  & 1   & 0.062 & 3.38 \\
      14 & 0.3 & 90  & 0.1 & 0.062 & 3.38 \\
      15 & 0.3 & 90  & 10  & 0.062 & 3.38 \\
      \hline
      16 & 0.3 & 120 & 1   & 0.072 & 3.95 \\
      17 & 0.3 & 120 & 0.1 & 0.072 & 3.95 \\
      18 & 0.3 & 120 & 10  & 0.072 & 3.95 \\
      \hline
      19 & 0.4 & 60  & 1   & 0.057 & 3.60 \\
      20 & 0.4 & 60  & 0.1 & 0.057 & 3.60 \\
      21 & 0.4 & 60  & 10  & 0.057 & 3.60 \\
      \hline
      22 & 0.4 & 90  & 1   & 0.071 & 4.50 \\
      23 & 0.4 & 90  & 0.1 & 0.071 & 4.50 \\
      24 & 0.4 & 90  & 10  & 0.071 & 4.50 \\
      \hline
      25 & 0.4 & 120 & 1   & 0.083 & 5.26 \\
      26 & 0.4 & 120 & 0.1 & 0.083 & 5.26 \\
      27 & 0.4 & 120 & 10  & 0.083 & 5.26 \\
      \hline
    \end{tabular}
  \end{table*}

\section{Initial conditions}
\label{sec:initial_conditions}

\begin{figure}
  \begin{center}
    \resizebox{\hsize}{!}{\includegraphics{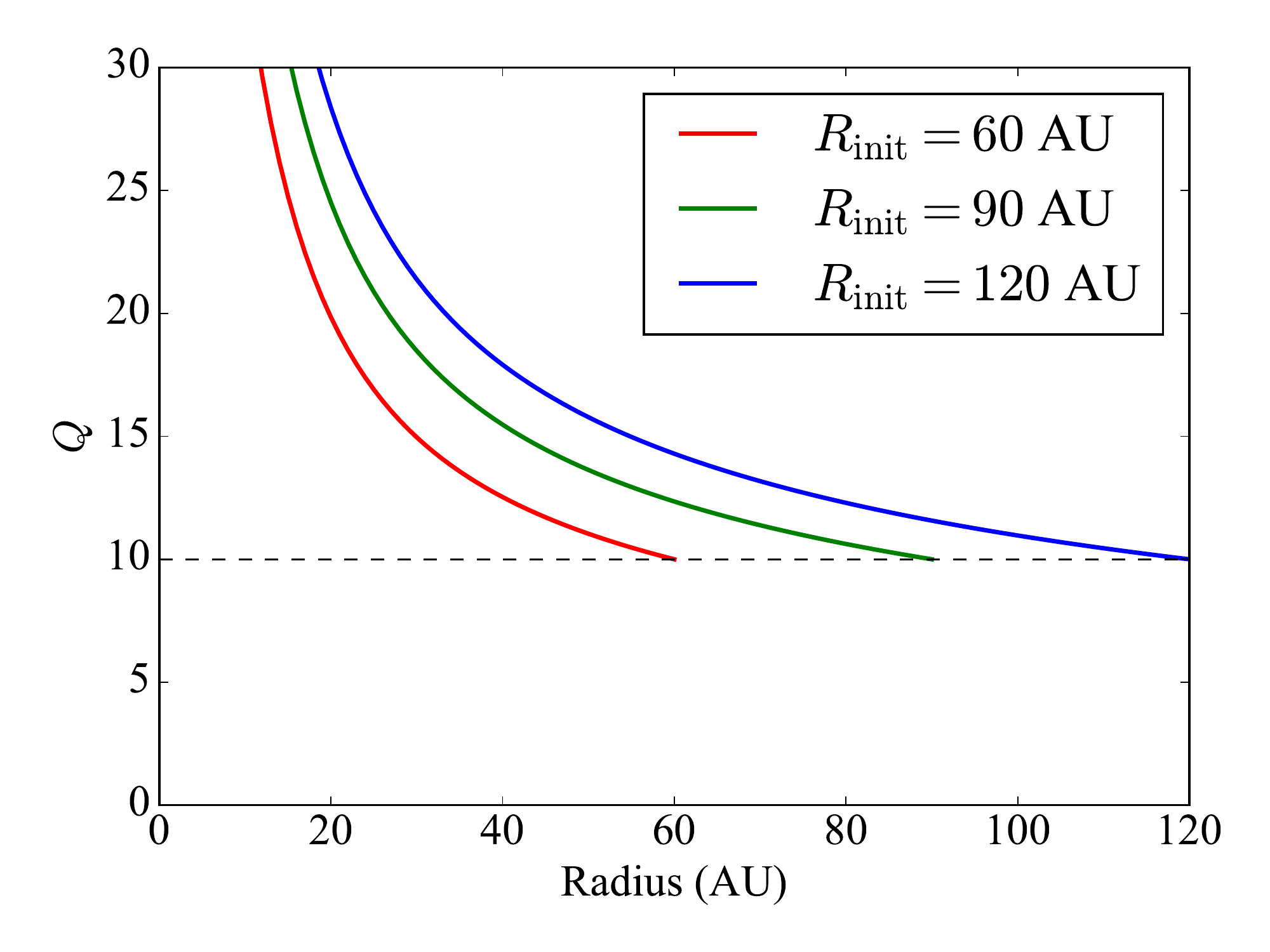}}
    \caption
    {
      Toomre parameter, $Q$ as a function of the radius for discs with outer 
      extents $R_{\textup{init}} = [60,~90,~120]$~AU. The dashed black line
      represents a value of $Q = 10$, the Toomre parameter value at the disc outer edge. Each disc is initially stable at all radii.
    }
    \label{fig:toomre}
  \end{center}
\end{figure}

We constructed protostellar systems consisting of M dwarfs attended by discs with
different stellar mass, disc radial extent and metallicity (see Table \ref{tab:ics}). The stellar masses
were set to $M_{\star} =[0.2,~0.3,~0.4]\msun$ exploring a range of masses for
M dwarfs. The initial disc radii were set to $R_{\textup{init}} = [60,~90,~120]$~AU,
whereas the discs' inner edge was set to $5$~AU. The metallicity was varied by
modifying the opacities by factors of $z = [0.1, 1, 10]$. The initial disc mass
was chosen such that the Toomre parameter has a fixed value at the outer radius
of the disc ($Q_{\textup{out}} = 10$). This ensures that the discs are initially gravitationally
stable (see Figure~\ref{fig:toomre}). Each disc was comprised of $N \approx 2 \times 10^{6}$ SPH particles, so that both the Jeans mass and the Toomre mass were well resolved \citep{Bate:1997a,
Nelson:2006a, Stamatellos:2009d}. { Similarly, the disc vertical structure was
adequately resolved; for the number of SPH particles that we used the disc scale height is generally at least $\sim 10$ smoothing lengths \citep[see][]{Stamatellos:2009d}.}

The surface density and temperature profiles of the disc were set to $\Sigma
\propto R^{-p}$ and $T \propto R^{-q}$, respectively. The surface density power
index $p$ is thought to lie between 1 and $3/2$ from semi-analytical studies of
cloud collapse and disc creation \citep{Lin:1990a, Tsukamoto:2015a,
Macfarlane:2017a}. The temperature power index $q$ ranges from $0.35$ to $0.8$
as derived from observations of pre-main sequence stars \citep{Andrews:2009b}.
Here, we adopted $p = 1$ and $q = 0.7$. The exact initial surface density profile that we used is
\begin{equation}
\Sigma(R) = \Sigma_{0} \left(\frac{R_{0}^{2}}{R^{2} + R_{0}^{2}} \right)^{1/2},
\label{eqn:sigma_profile}
\end{equation}
where $\Sigma_{0}$ is the surface density 1 AU away from the star and $R_{0} =
0.01$~AU is a smoothing radius to prevent unphysically high values close to the
star. The disc initial temperature profile was set to
\begin{equation}
T(R) = \left[T^{2}_{0} \left( \frac{R^{2} +
R^{2}_{0}}{\textup{AU}^{2}}\right)^{-0.7} + T_{\infty}^{2}\right]^{1/2}.
\label{eqn:stellar_heating}
\end{equation}
Here, $T_{0} = 100$~K is the temperature at 1 AU from the star and $T_{\infty} =
10$~K is the temperature far away from the star. This profile were also used
to provide a minimum temperature below which SPH particles cannot radiatively cool, that is, $T_{\textsc{bgr}}$ in Equation \ref{eqn:dudt}.

\begin{figure}
  \begin{center}
    \resizebox{\hsize}{!}{\includegraphics{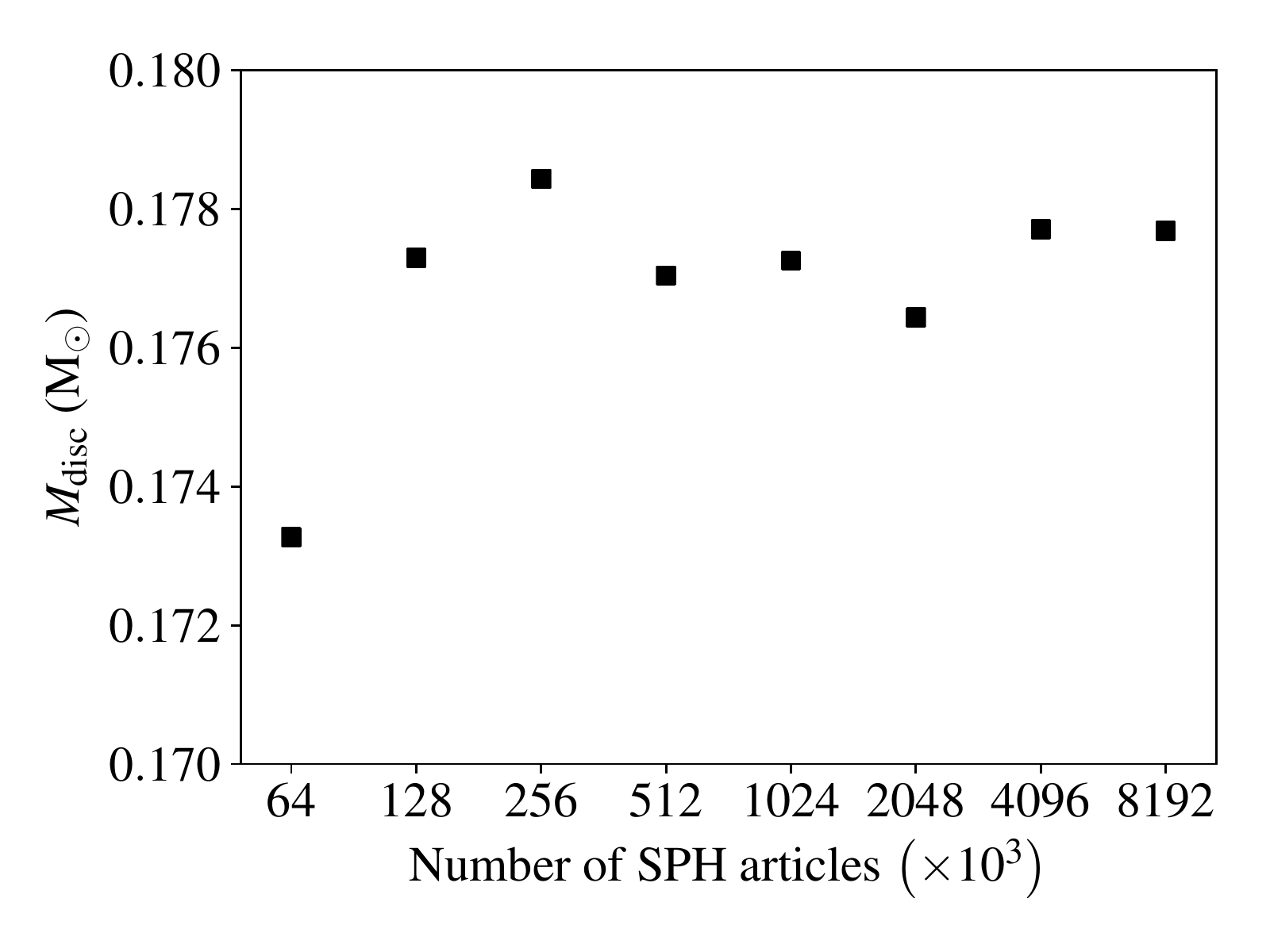}}
    \caption
    {
      Convergence test for the mass loading method described in Section
      \ref{sec:numerical_method}. We performed SPH simulations with an increasing
      number of particles and compared the disc mass at the point of
      fragmentation. We see that for $> 128 \times 10^{3}$ particles, there is
      little difference in the fragmentation mass of the disc. We therefore conclude that the method is well converged for $N
      \approx 2 \times 10^{6}$, the number of particles used for the simulations
      presented in this work. 
    }
    \label{fig:convergence}
  \end{center}
\end{figure} 
\section{Method tests}
\label{sec:mass_loading_test_and_convergence}

\subsection{Test 1: Disc fragmentation mass} 
\label{sub:test_1_disc_fragmentation_mass}

To check the validity of the mass loading method as a good way to estimate the
fragmentation minimum disc mass, we performed a simulation of a disc around an M
dwarf of mass $0.2\msun$, where the initial disc mass was set to $0.12 \msun$ and
the disc accretion rate to $3 \times 10^{-5} \msun \textup{ yr}^{-1}$ that is, $\chi=0.5$ (see Equation \ref{eqn:mass_accretion_rate}). We also performed a set of
simulations where the disc masses are fixed (i.e. without any mass loading), but in each simulation  the disc had mass from $0.15$ to $0.2 \msun$, in $0.01 \msun$ intervals. Each disc had surface
density and temperature profiles of $\Sigma \propto R^{-1}$ and $T \propto
R^{-0.7}$ respectively, extends from 5 to 90 AU, and were comprised of $N \approx
2 \times 10^{6}$ particles. We found that the disc with a fixed mass of $0.17
\msun$ does not undergo fragmentation whereas the disc with a fixed mass of
$0.18 \msun$ did. The disc in the simulation that included mass loading fragmented at a mass of
$0.176 \msun$, consistent with fixed-mass disc simulations.

\subsection{Test 2: Mass loading convergence}
\label{sub:test_2_mass_loading_convergence}

We performed a set of simulations with the same parameters as the simulation with mass loading described above, but with an increasing number of SPH particles in order to check for convergence. Figure
\ref{fig:convergence} shows the mass at which a disc fragmented under mass
loading with an increasing number of particles. Even for a relatively small
number of particles $\left(N = 128,000\right)$ we achieved convergence. Only
negligible differences were seen when the particle number is consequently
doubled, up to a maximum of $N \approx 8 \times 10^{6}$.

\subsection{Test 3: Mass loading and choice of accretion rate}
\label{sub:test_3_mass_loading_and_choice_of_accretion_rate}

\begin{figure}
  \begin{center}
    \resizebox{\hsize}{!}{\includegraphics{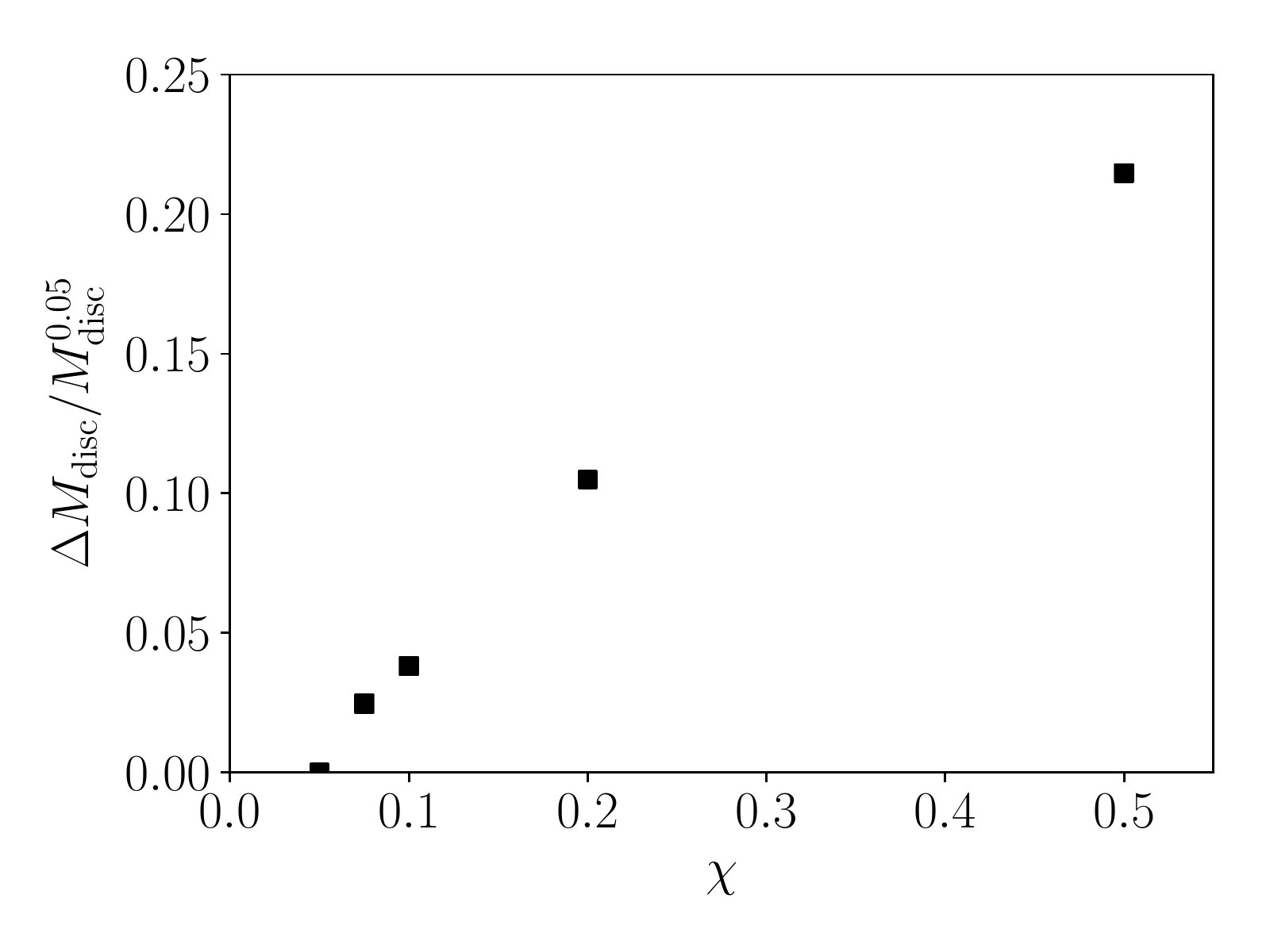}}
    \caption
    {
      Fractional difference in the disc fragmentation mass for different values of the parameter $\chi$, which regulates the disc accretion rate (see Equation  
      \ref{eqn:mass_accretion_rate}). The reference value is the disc fragmentation mass for $\chi=0.05$ (denoted as $M_{\mathrm{disc}}^{0.05}$).  The corresponding mass accretion rates for $\chi = [0.05, 0.075, 0.1, 0.2, 0.5]$ are $\dot{M}_{\textup{disc}} = [1.25, 1.88, 2.5, 5,
      12.5] \times 10^{-6} \msun \textup{ yr}^{-1}$, respectively. We show that
      for values of $\chi \leq 0.1$, there is only a small difference ($\leq 5\%$) in the disc fragmentation mass. As such, a value of $\chi = 0.1$ is adopted for the work presented here. The difference in the disc fragmentation mass between $\chi=0.05$ and $\chi=0.5$ is relatively small ($\sim20\%$ ). 
          } 
    \label{fig:chi_comparison}
  \end{center}
\end{figure}

We investigated the effect of the factor which regulates the amount of mass
loading, $\chi$ (see Equation \ref{eqn:mass_accretion_rate}). { We chose a disc with properties (mass, radius, metallicity) the same as the disc of Rxun 4 in Table \ref{tab:ics}, and we performed simulations with different accretion rates.} 
{ Figure \ref{fig:chi_comparison} shows the fractional difference in the disc fragmentation masses for 
$\chi = [0.05, 0.075, 0.1, 0.2, 0.5]$. 
The corresponding mass
accretion rates are $\dot{M}_{\textup{disc}} = [1.25, 1.88, 2.5, 5, 12.5] \times
10^{-6} \msun \textup{ yr}^{-1}$. We need to use a low accretion rate onto
the disc so that its evolution is not affected by the mass loading, whereas for
computational purposes we need to have a high accretion rate so that the
fragmentation mass is achieved quickly. There is little difference ($\leq 5\%$) in the
computed fragmentation mass for $\chi \leq 0.1$  and so this is the value we
adopted for the rest of the work presented in this paper.} It is important to note that the
difference in the disc fragmentation mass between $\chi=0.05$ and $\chi=0.5$ is relatively small ($\sim20\%$ ). 

\section{Fragmentation of M dwarf discs}
\label{sec:fragmentation_of_m_dwarf_discs}

We performed a set of 27 simulations, varying the initial disc mass, disc radius,
metallicity, and the mass of the host star. Each disc was initially
gravitationally stable, but its mass increased over time through
mass loading (see Section~\ref{sec:numerical_method}). As such, each disc
eventually became unstable and spiral arms developed. In the majority of cases,
continued mass loading caused the spiral arms to evolve non-linearly, and to
ultimately form gravitationally bound fragments.

The results of the disc
simulations are presented in Table \ref{tab:sim_results}.  When a disc fragments, its mass yields the minimum disc mass for fragmentation, which we denote as $M_{\textup{disc}}$. We  also calculated the time of fragmentation $t$, the disc-to-star mass ratio when fragmentation happens $q$, and the radius of the disc $R_{\textup{disc}}$, which encompasses $95 \%$ of the disc mass. The
distance between the central star and the formed fragment is denoted as
$a_{\textup{frag}}$. In the table we also state the stellar
mass and and the disc-to-star mass ratio when fragmentation happens.

\begin{table*}
  \centering
    \caption
    {
      Results for the disc simulations with the initial conditions listed in Table
      \ref{tab:ics} after 30~kyr of evolution. $M_{\star}$ and $M_{\textup{disc}}$
      are respectively the masses of the star and disc when the disc fragments. $t$ is the time
      at which fragmentation occurs, $q$ is the disc-to-star mass ratio at
      fragmentation and $R_{\textup{disc}}$ is the radius of the disc. $a_{\textup{frag}}$ is the distance of
      the fragment from the host star. Some discs did not fragment
      as noted by  dashes in the corresponding rows.
    }
    \label{tab:sim_results}
    \centering
    \begin{tabular}{c c c c c c c c}
      \hline
      \hline
  
      Run & $M_{\star}$ (\MSUN) & $M_{\textup{disc}}$ (\MSUN) & $\Delta
      M_{\textup{disc}}$ (\MJUP) & $t$ (kyr) & $q$ & $R_{\textup{disc}}$ (AU) &
      $a_{\textup{frag}}$ (AU)
      \\
      \hline
      01 & 0.205 & 0.075 & 36.6 & 22.1 & 0.37 & 75  & 49  \\
      02 & 0.205 & 0.077 & 38.9 & 23.3 & 0.38 & 72  & 35  \\
      03 & -     & -     & -    & -    & -    & -   & -   \\
      \hline
      04 & 0.205 & 0.104 & 56.2 & 25.9 & 0.51 & 92  & 54  \\
      05 & 0.206 & 0.105 & 57.2 & 27.6 & 0.51 & 137 & 55  \\
      06 & 0.207 & 0.106 & 58.1 & 28.0 & 0.51 & 169 & 32  \\
      \hline
      07 & 0.204 & 0.124 & 68.7 & 26.7 & 0.61 & 144 & 46  \\
      08 & 0.205 & 0.126 & 70.8 & 27.6 & 0.62 & 128 & 30  \\
      09 & 0.207 & 0.128 & 72.2 & 28.8 & 0.62 & 190 & 54  \\
      \hline
      10 & 0.305 & 0.094 & 46.9 & 18.5 & 0.31 & 96  & 40  \\
      11 & 0.305 & 0.097 & 50.2 & 19.6 & 0.32 & 68  & 32  \\
      12 & -     & -     & -    & -    & -    & -   & -   \\
      \hline
      13 & 0.305 & 0.122 & 63.0 & 19.2 & 0.40 & 89  & 59  \\
      14 & 0.305 & 0.125 & 66.3 & 20.2 & 0.41 & 89  & 28  \\
      15 & -     & -     & -    & -    & -    & -   & -   \\
      \hline
      16 & 0.304 & 0.146 & 77.8 & 20.0 & 0.48 & 131 & 55  \\
      17 & 0.305 & 0.150 & 81.7 & 21.0 & 0.49 & 126 & 60  \\
      18 & 0.307 & 0.155 & 86.7 & 22.7 & 0.50 & 159 & 84  \\
      \hline
      19 & -     & -     & -    & -    & -    & -   & -   \\
      20 & -     & -     & -    & -    & -    & -   & -   \\
      21 & -     & -     & -    & -    & -    & -   & -   \\
      \hline
      22 & 0.405 & 0.144 & 76.2 & 17.3 & 0.36 & 105 & 57  \\
      23 & 0.405 & 0.140 & 72.1 & 16.4 & 0.35 & 91  & 37  \\
      24 & -     & -     & -    & -    & -    & -   & -   \\
      \hline
      25 & 0.405 & 0.165 & 86.1 & 16.5 & 0.41 & 123 & 47  \\
      26 & 0.405 & 0.171 & 91.6 & 17.6 & 0.42 & 130 & 43  \\
      27 & 0.407 & 0.176 & 97.4 & 19.0 & 0.43 & 164 & 116 \\
      \hline
    \end{tabular}
  \end{table*}

\begin{figure}
  \begin{center}
    \resizebox{1.\hsize}{!}{\includegraphics{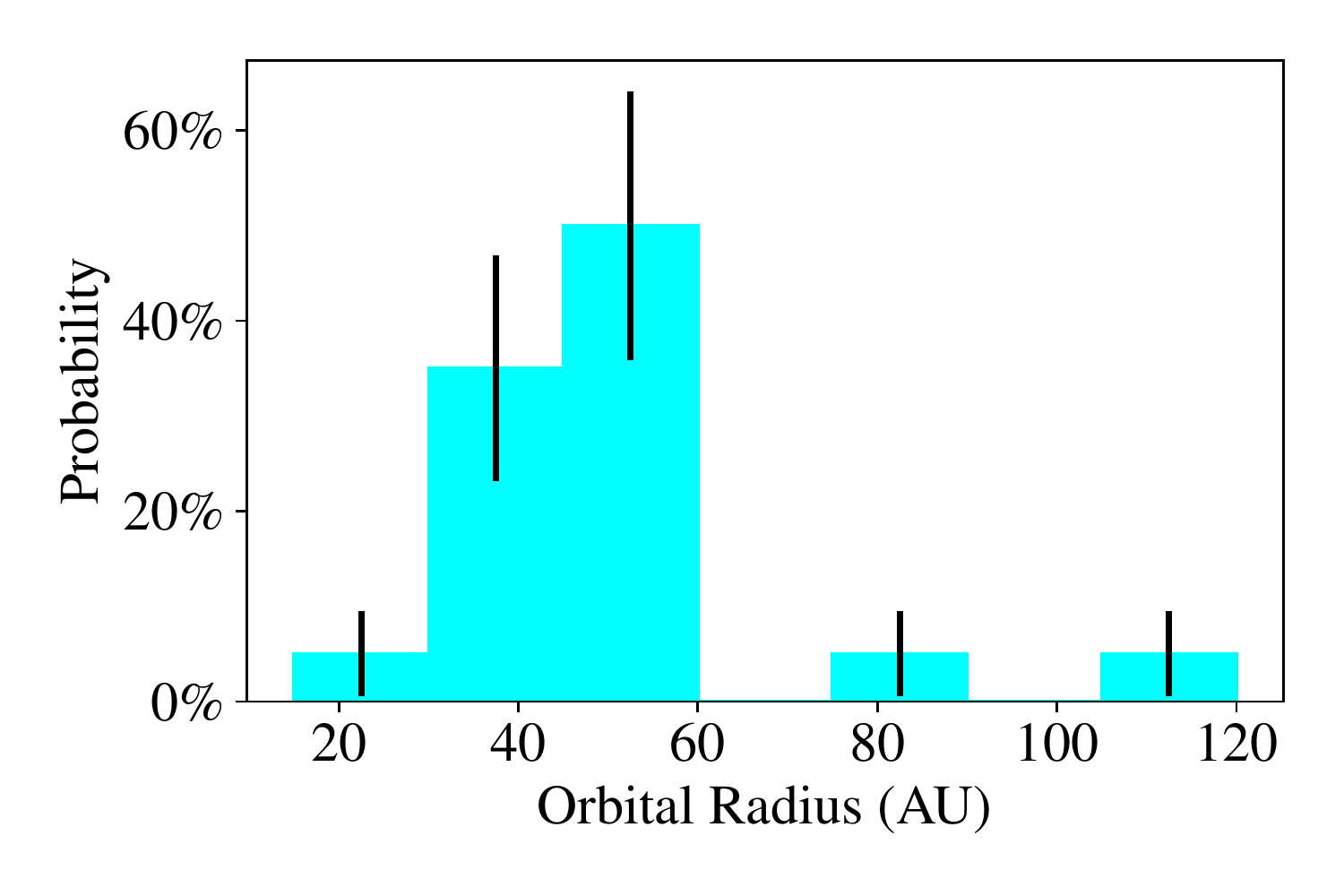}}
    \caption
    {Probability distribution of the orbital radius of the first fragment that forms in each simulation. M dwarf discs are most likely to fragment at distances $50-60$~AU from the host star. The error bars correspond to the Poisson (statistical) error. Only 20 fragments formed in the simulations, therefore the uncentainties are rather large.}  
    \label{fig:r_hist}
  \end{center}
\end{figure}

Fragmentation happens quite fast, within a few tens of kyr ($\sim 16-28$~kyr; see Table~\ref{tab:sim_results}). The discs generally fragment at distances $>30$~AU from the host star; the most likely distance for fragmentation to happen is $50-60$~AU (see Figure~\ref{fig:r_hist}). This is closer to the central star than for fragmentation of discs around more massive stars;  \cite{Stamatellos:2009a} find a most likely distance of $100-150$~AU, for massive discs around 0.7-\MSUN\ stars. This is consistent with the expectation that discs around less massive stars fragment closer to the central star, $a_{\rm fragm}\propto\left({M_\star}/{\MSUN}\right)^{1/3}$ \citep{Whitworth:2006a}. According to this relation one would expect the optimal region for fragmentation around M dwarfs to be around 75-100~AU, which is larger than what we find here. However, this is expected as in the simulations of \cite{Stamatellos:2009a}  a slightly different radiative transfer method is used (utilizing the gravitational potential of the particle to calculate the column density) which results in less efficient cooling, making fragmentation at a specific distance from the host star less likely than the method used here \citep[utilizing the pressure scale height; see][]{Mercer:2018a}.

\begin{figure*}
  \begin{center}
    \resizebox{0.98\hsize}{!}
    {\includegraphics{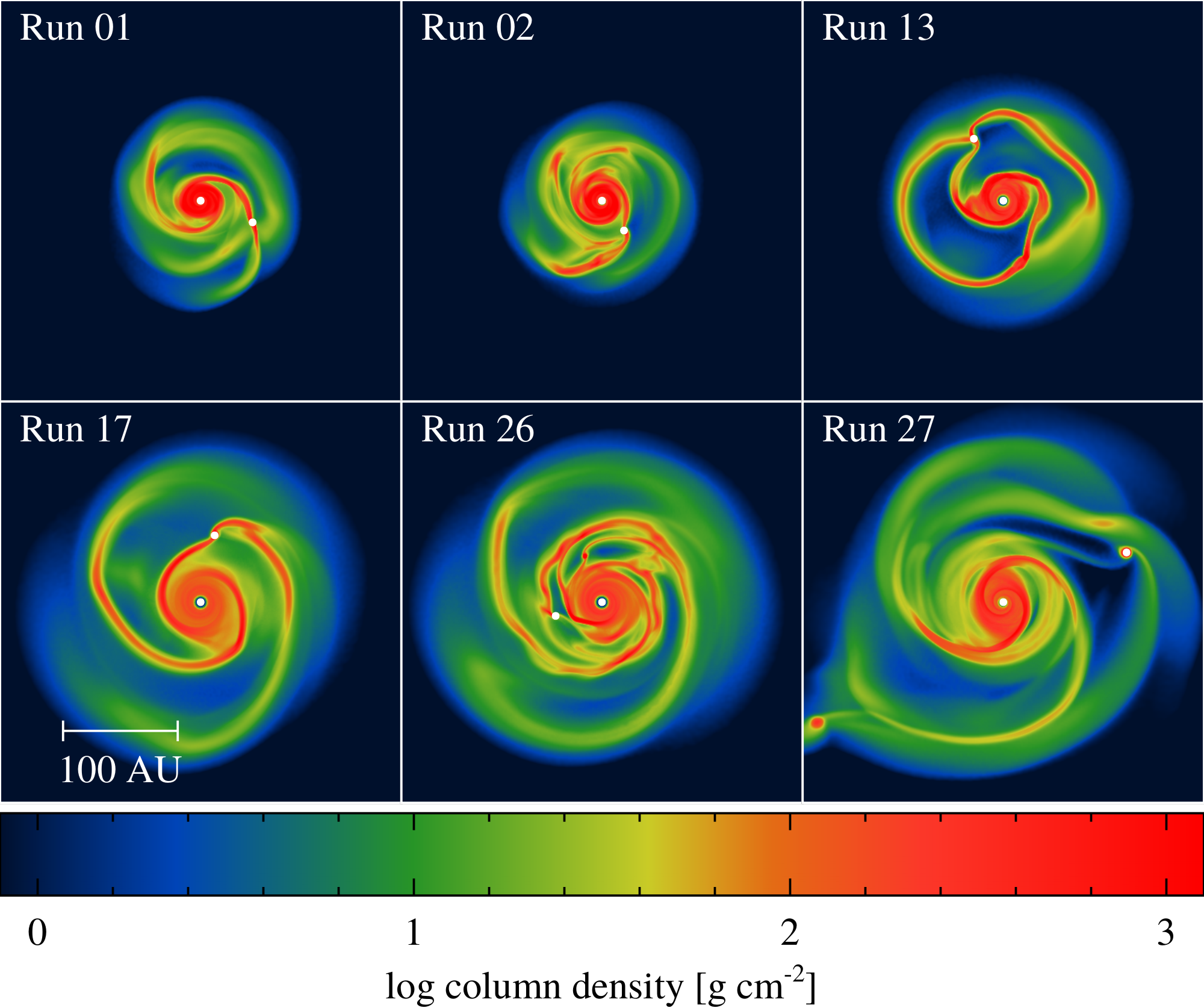}}
    \caption
    {
      Surface density plots for a selection of discs at the time of
      fragmentation (i.e. when a density of $10^{-9} \textup{ g cm}^{-3}$ is
      reached). The fragments are shown by the white points. The initial
      conditions for each run can be found in Table \ref{tab:ics} and the
      results in Table \ref{tab:sim_results}.
    }
    \label{fig:representative_snapshots}
  \end{center}
\end{figure*}

Figure \ref{fig:representative_snapshots} shows the surface density snapshots of six representative simulations at the time when the density at the center of the fragment is $10^{-9} \textup{ g cm}^{-3}$ (i.e. when, by definition,
fragmentation has happened). In Figures~\ref{fig:disc_stellar_mass}-\ref{fig:q_radius_z}, we present the relations between the different parameters that we investigated in this study (stellar mass, disc mass, disc radius, metallicity). In the following subsections we discuss these relations in detail.

\subsection{The effect of the stellar mass and the disc radius}
\label{sub:consquences_of_initial_stellar_mass_and_disc_radius}

\begin{figure}
  \centering
    \resizebox{\hsize}{!}{\includegraphics{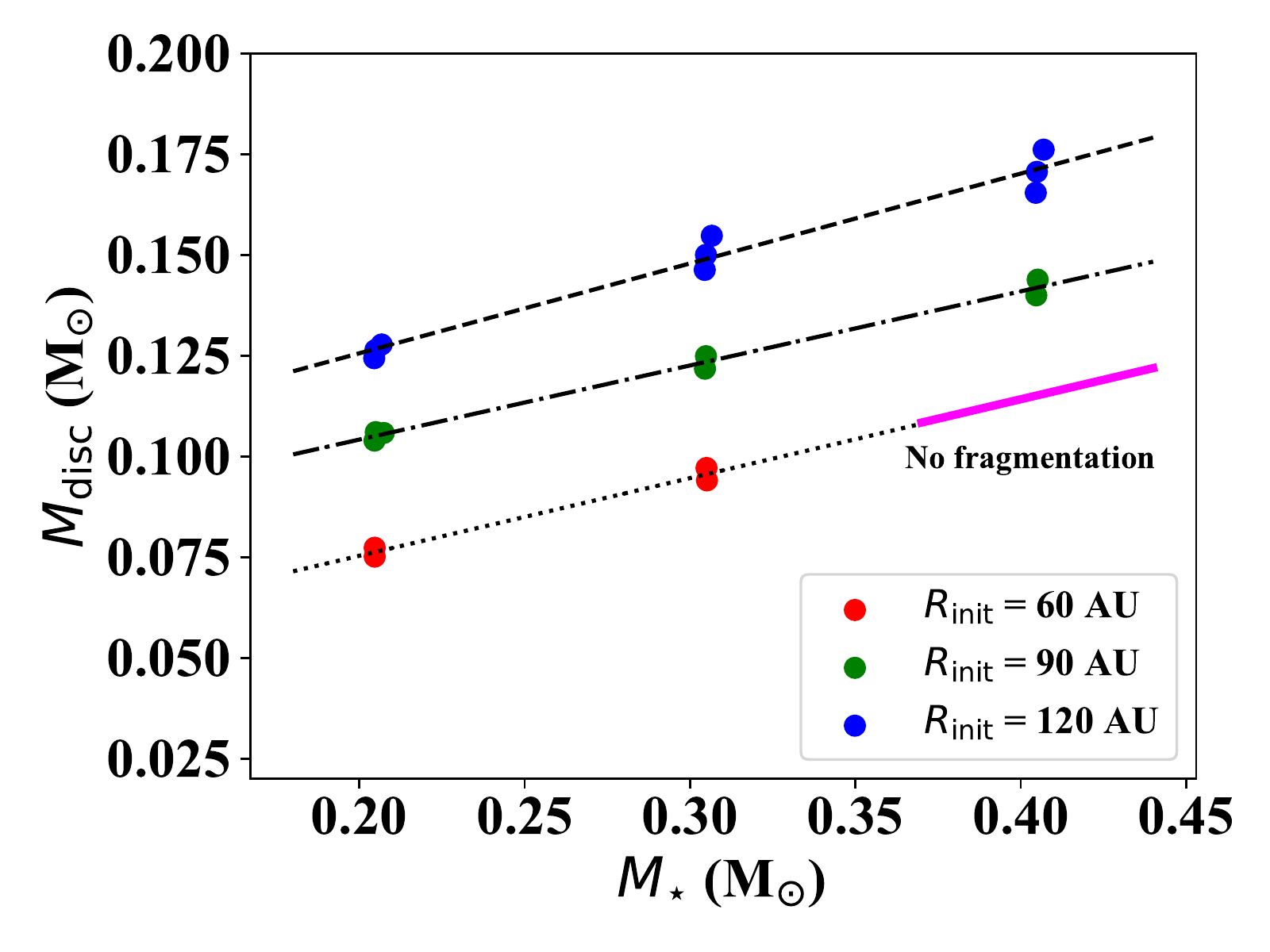}}
    \caption
    {
      Disc mass as a function of stellar mass  when the disc
      fragments. Different colours correspond to different initial disc radii
      (as marked on the graph). The relationship between the two quantities is
      linear for a given initial disc radius. Smaller discs fragment at a lower
      mass, as the average disc surface density is larger. The different lines correspond to the linear relations derived for simulations with the same initial disc radius (see text). The purple line corresponds to the area in the parameter space where no fragmentation occurs (small discs around more massive M dwarfs). 
    }
    \label{fig:disc_stellar_mass}
   \centering
  \subfloat{\resizebox{\hsize}{!}{\includegraphics{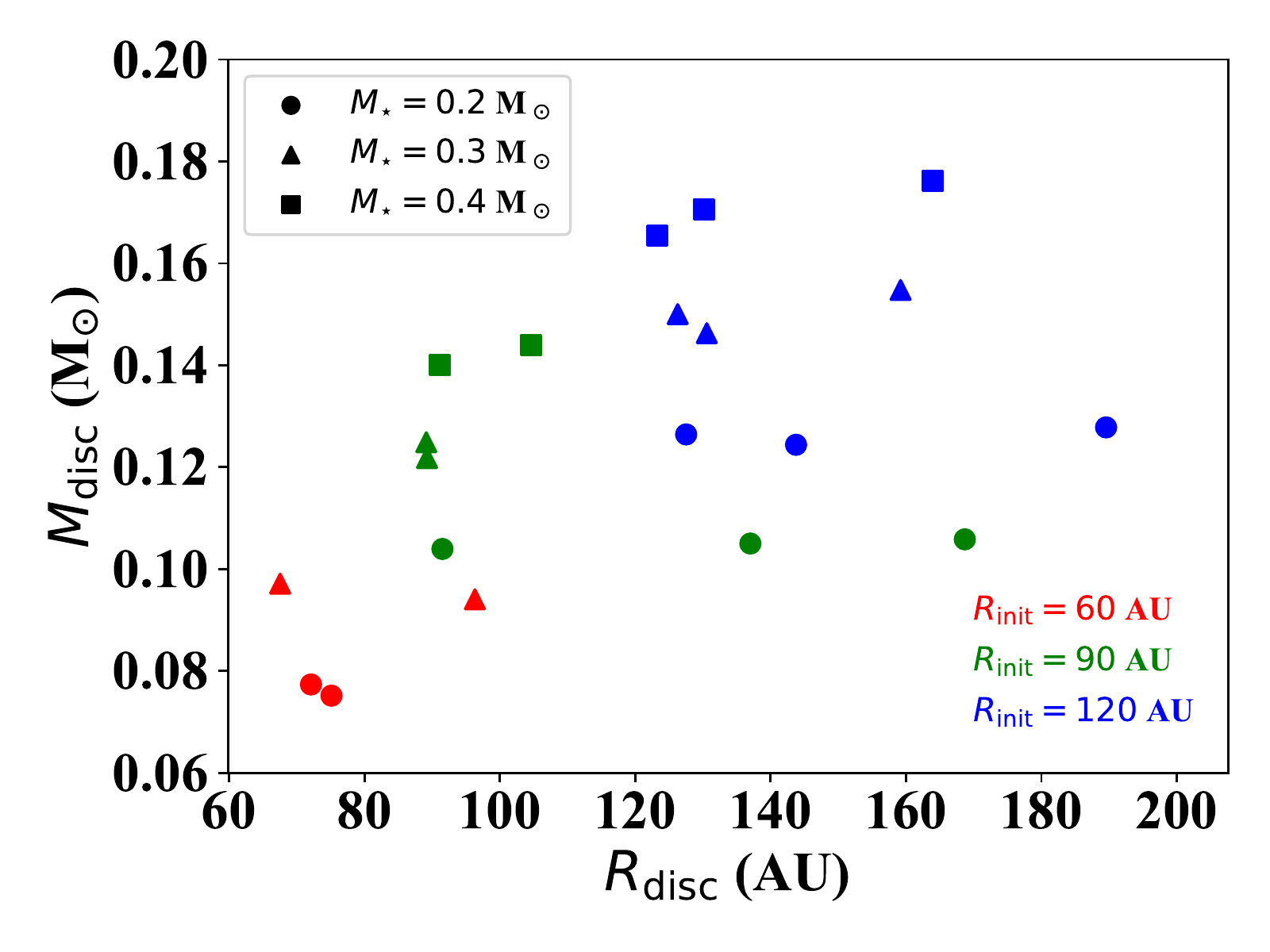}}}
  \caption
  {
    Disc mass and  as a function of disc
    radius at the time when the disc fragments. The disc radius is the radius which encompasses 95\% of the disc mass. Generally speaking,  a higher disc mass  is required for fragmentation of more extended discs. Initial disc radii of $R_{\textup{init}} = [60, 90, 120]$~AU are shown
    by the red, green and blue points, respectively. The initial stellar
    masses of $M_{\star} = [0.2, 0.3, 0.4] \msun$ are denoted by the circles,
    triangles and squares, respectively.  We note the difference between { initial} $R_{\textup{init}}$ and { final} disc radius $R_{\textup{disc}}$ (i.e. the disc radius when it fragments).
  }
  \label{fig:mass_radius}
\end{figure}

The disc fragmentation mass is shown as a function of the stellar mass in Figure
\ref{fig:disc_stellar_mass}, which demonstrates that for a given initial disc
radius (60, 90, 120~AU), the disc fragmentation mass increases linearly with the stellar mass:
 $M^{\rm 60 AU}_{\rm disc} = 0.04 + 0.19 M_\star$,
 $M^{\rm 90 AU}_{\rm disc} = 0.07 + 0.18 M_\star$, 
 $M^{\rm 120 AU}_{\rm disc} = 0.08 + 0.22 M_\star$.
A more massive central star results in a more stable disc as $Q \propto \Omega$
and $\Omega \propto M_{\star}^{1/2}$. The disc fragmentation mass also increases
with the initial disc radius as the average surface density of smaller discs is larger for the same disc mass. Hence, smaller discs (but still discs with size $>70$~AU) fragment at a lower mass, as
$\Sigma \propto R^{-1}$ and $Q \propto \Sigma^{-1}$ (see also Figure~\ref{fig:mass_radius}).

\begin{figure}
  \centering
    \subfloat{\resizebox{\hsize}{!}{\includegraphics{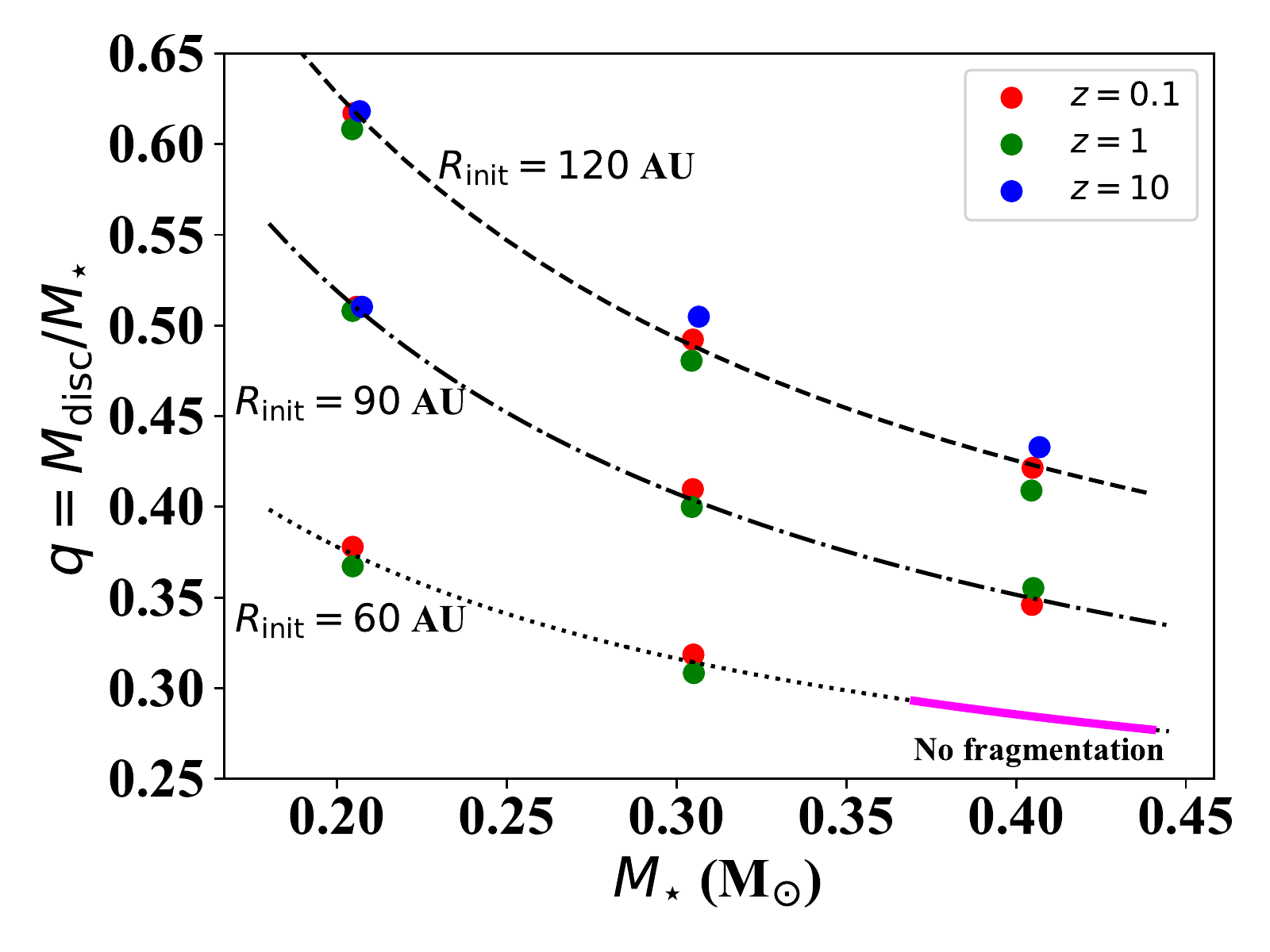}}}
  \caption
  {
    Disc-to-star mass ratio, $q$, at the time of disc fragmentation as a function of stellar mass for metallicities $z=[0.1, 1, 10]$
    marked by the red, green and blue points, respectively. Each group of points (3 or 2 points) correspond to simulations of different metallicity discs (that have the same initial radius) around the stars with the same mass. We note that discs (with the same initial extent) around more massive stars fragment at a lower disc-to-star mass ratio.  The disc-to-star mass ratio required for fragmentation varies from $\sim 0.3$ (for small discs) to $\sim 0.6$ (for more extended discs). The different lines correspond to the  hyperbolic relations derived for simulations with the same initial disc radius. The purple line corresponds to the area in the parameter space where no fragmentation occurs (small discs around more massive M dwarfs). 
  }
  \label{fig:q_stellar_mass}
  \centering
  \subfloat{\resizebox{\hsize}{!}{\includegraphics{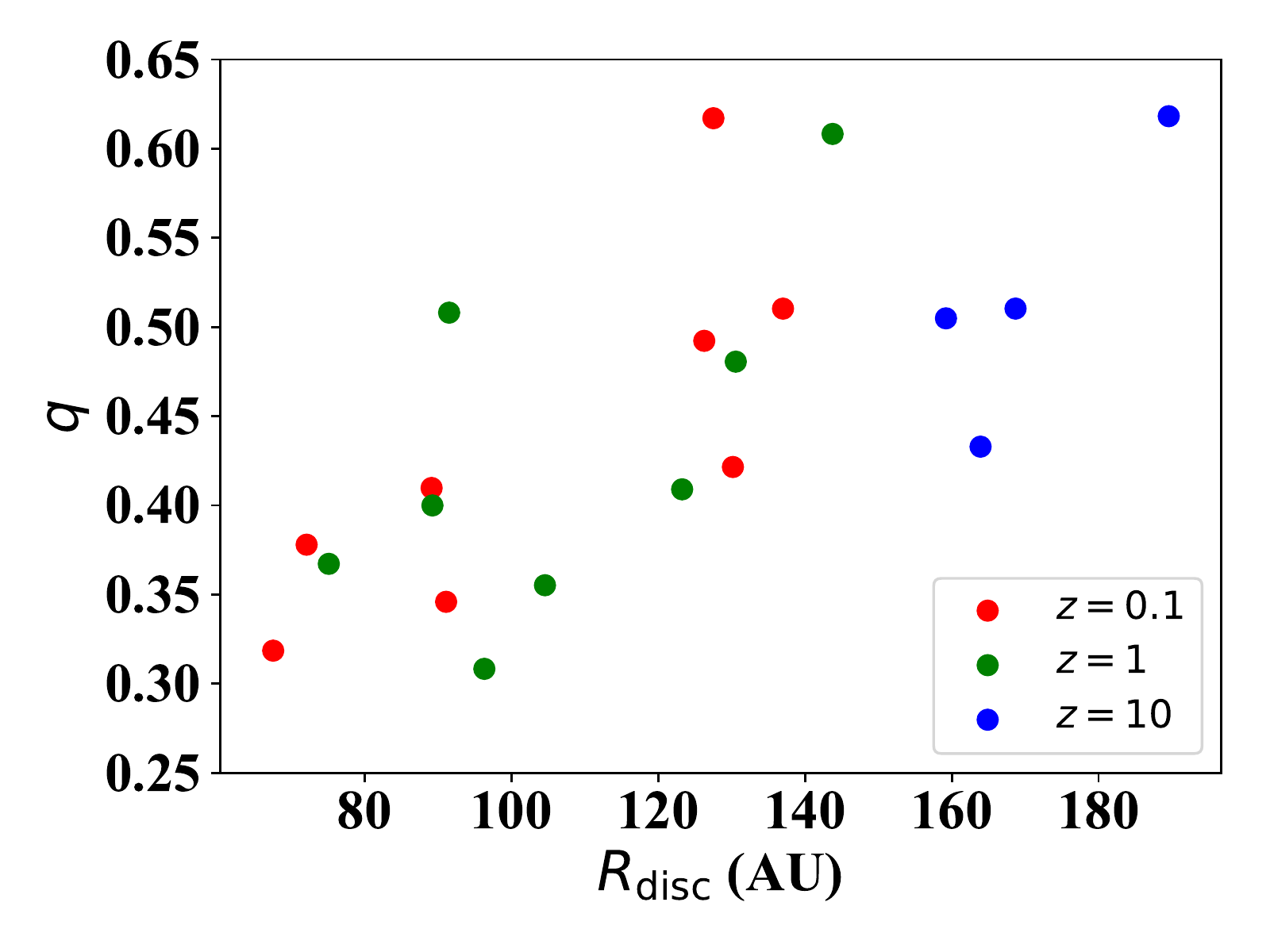}}}
  \caption
  {
   Disc-to-star mass ratio, $q$, at the time of disc fragmentation as a function of the disc radius, for metallicities $z=[0.1, 1, 10]$
    marked by the red, green and blue points, respectively. 
    Discs with higher metallicity are larger when they fragment, suggesting a period of expansion.  The required disc-to-mass ratio for fragmentation increases with disc size.
  }
  \label{fig:q_radius_z}
\end{figure}

The disc-to-star mass ratio needed for fragmentation  varies from $\sim 0.3$ (for small discs) to $\sim 0.6$ (for more extended discs) (see Figures~\ref{fig:q_stellar_mass}-\ref{fig:q_radius_z}). Therefore relatively
large disc masses are needed for fragmentation to happen around M dwarfs. Such high disc
masses have not been observed \citep{Andrews:2013a, Mohanty:2013a,
Ansdell:2017a}, but it may be possible that M dwarf discs are more massive at their younger phase, as for example, the discs around solar-mass Class 0 objects
\citep{Dunham:2014b, Tobin:2016a}. We also find that discs (with the same initial size) around more massive stars fragment at a lower disc-to-star mass ratio as the disc fragmentation mass increases slower than the stellar mass (see Figure~\ref{fig:q_stellar_mass}).

The discs in runs 19 - 21, which correspond to small discs ($R=60$~AU) around more
massive M dwarfs ($M_{\star} = 0.4 \msun$) do not fragment. This also true for
runs 3, 12, 15, and 24, which correspond to discs with high metallicity. We
attribute this behaviour to a period of rapid disc expansion, a result of strong
spiral arm formation and efficient outward transport of angular momentum which
reduces the surface density and stabilises the discs. To demonstrate the effect
of disc expansion, we compare runs 1 - 3; the disc in runs 1 and 2 undergo
fragmentation, whereas the disc in run 3 exhibits rapid expansion. Figure
\ref{fig:expansion} shows azimuthally-averaged Toomre parameter (a) and the
cooling time in units of the local orbital period (b). Although in each case the
cooling time is short enough to allow for a fragment to condense out, the disc
in run 3 does not fragment due to rapid expansion (the spiral arms
efficiently distribute angular momentum outwards).

\subsection{The effect of metallicity}
\label{sub:effect_of_metallicity}

We examined three different values of the metallicity by modifying the opacities used in Equation \ref{eqn:dudt} by factors of $z = [0.1, 1, 10]$. We find that changing the metallicity has little effect on the disc fragmentation mass for the disc with the same extent (see Figure~\ref{fig:q_stellar_mass}) (although in some cases the disc does not fragment, see discussion below).

\begin{figure*}
  \centering
      \subfloat{\resizebox{0.51\hsize}{!}{
        \includegraphics{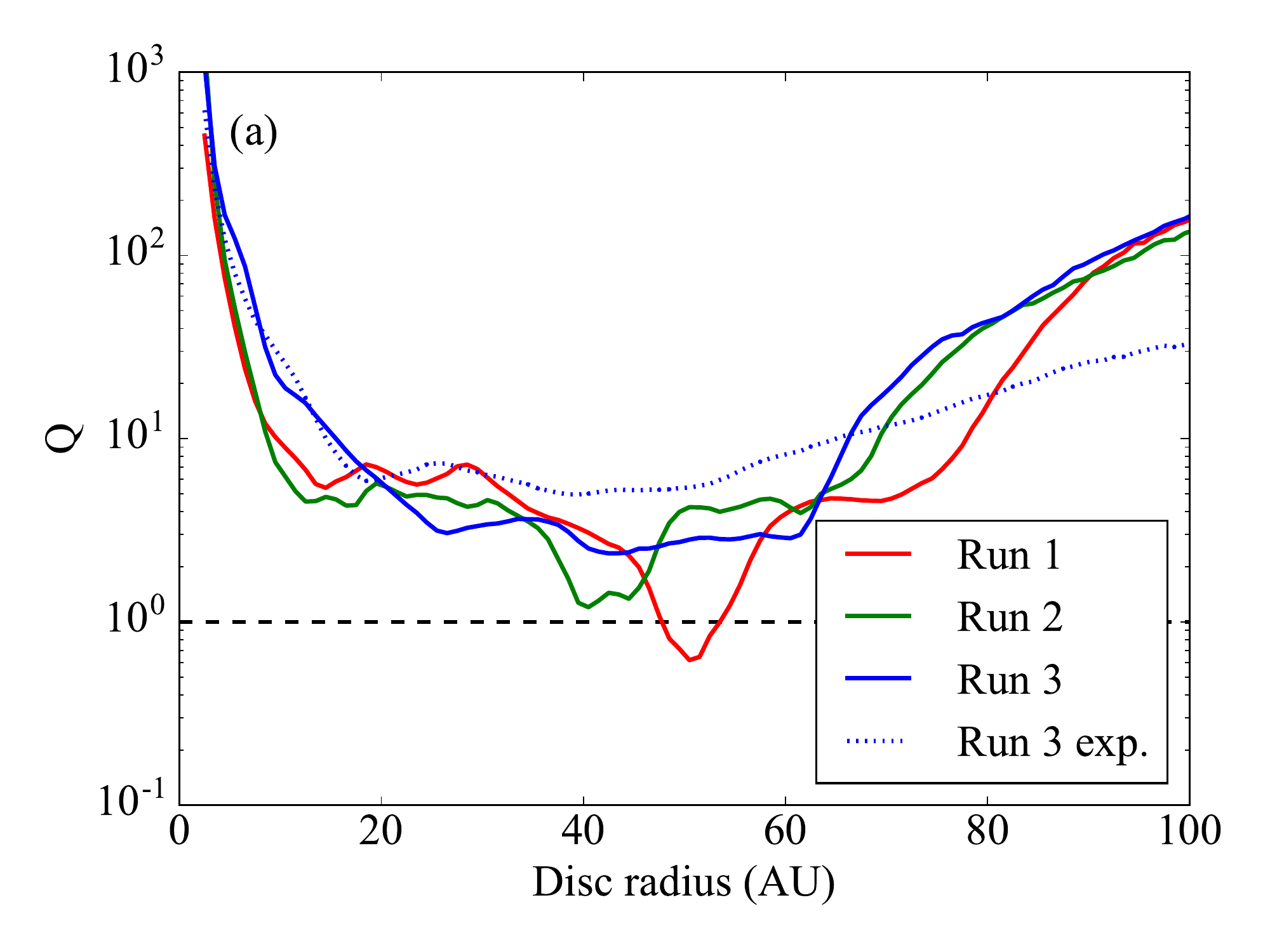}}}
    \subfloat{\resizebox{0.51\hsize}{!}{
        \includegraphics{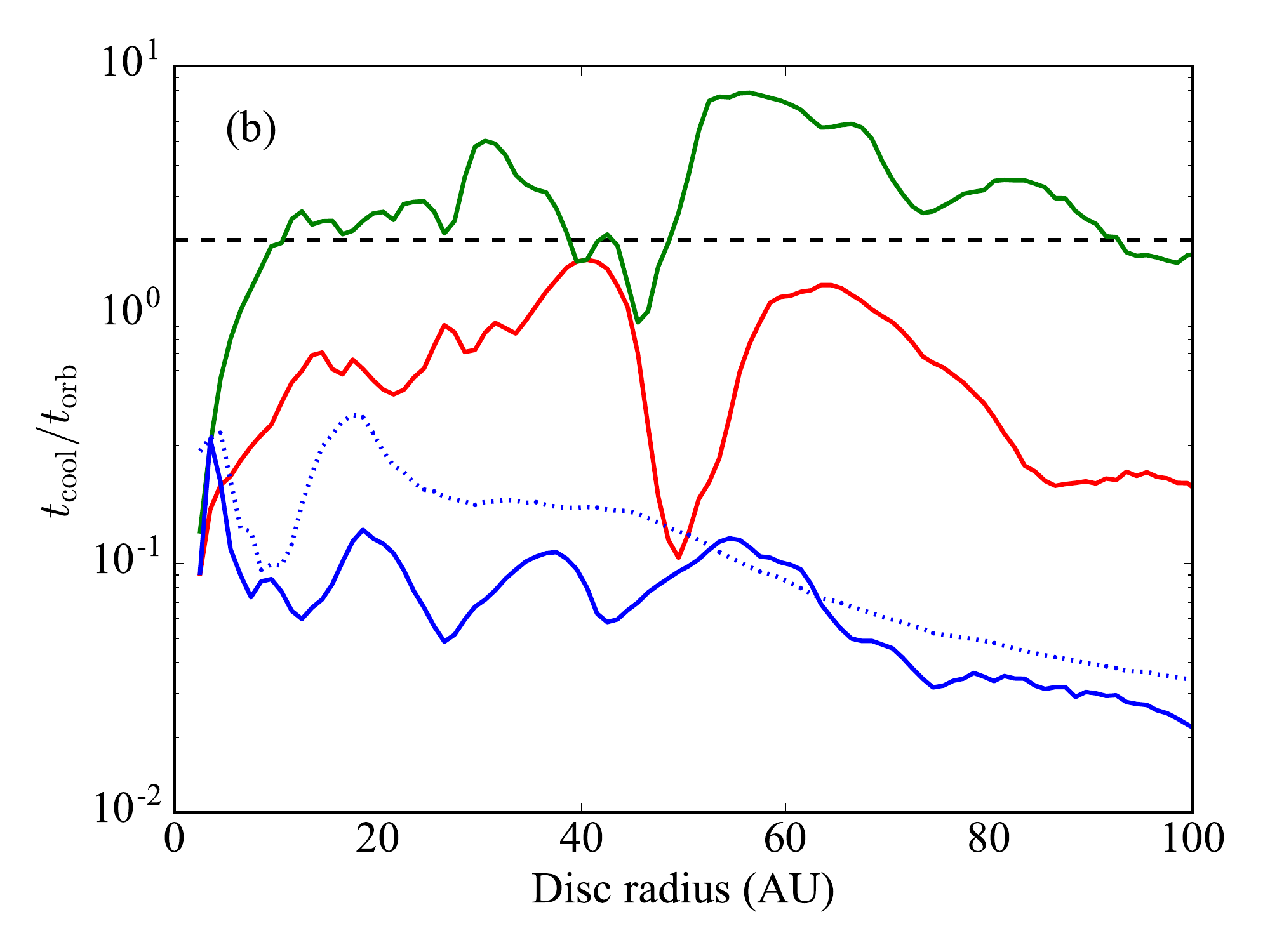}}}
    \caption
    {
      Azimuthally-averaged Toomre parameter (a) and the cooling time in units of
      the local orbital period (b) for runs 1 - 3 (red, green and blue,
      respectively). The time at which these quantities are shown are just prior
      to fragmentation (runs 1 and 2), and just prior to a period of disc
      expansion (run 3). The dashed blue line shows run 3 after the expansion.
      Each disc is gravitationally unstable such that spiral arms form, but only
      in runs 1 and 2 does the Toomre parameter fall below unity so that bound
      fragments form. In all cases, the cooling time is sufficiently short for a
      condensed fragment to collapse. The expansion of the disc in run 3 (and
      characteristic of most runs with an increased metallicity) acts to
      stabilise it.
    }
    \label{fig:expansion}
    \vspace{0.3cm}
  \centering
      \resizebox{1\hsize}{!}{\includegraphics{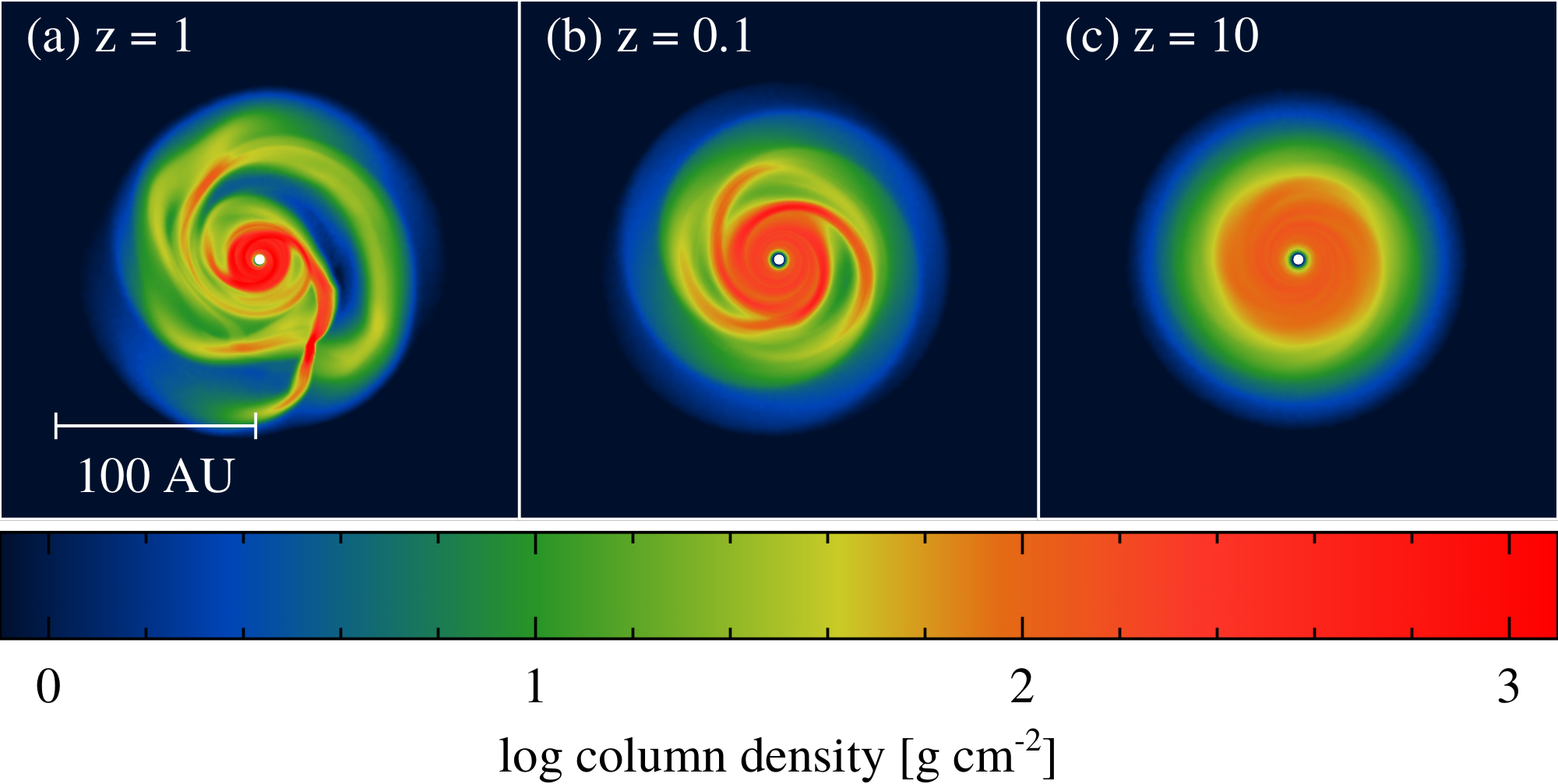}}
    \caption
    {
      Surface density of the discs in runs 1, 2, and 3, at 22 kyr (see Table
      \ref{tab:sim_results}). The disc in panel (a) has $z=1$ (solar
      metallicity), and is shown just prior to the formation of a bound
      fragment. The disc in panel (b) has a metallicity reduced by an order of
      magnitude ($z=0.1$). The disc is unstable but the spiral arms are not as
      strong as in disc in run 1. The disc in panel (c) has a metallicity
      increased by an order of magnitude ($z=10$). No strong spirals have
      formed yet. Spirals do eventually form, but the disc does not fragment due to a period of rapid expansion.
     }
    \label{fig:snapshot_comparison}
\end{figure*}

On the other hand, the disc evolution is affected by metallicity; from the onset of the
gravitational instability, to the collapse of dense fragments (see Figures~\ref{fig:expansion}-\ref{fig:snapshot_comparison}). In Figure
\ref{fig:snapshot_comparison}, we present surface density snapshots of runs 1 - 3 (panels a, b and c respectively) at a time of 22 kyr. Figure
\ref{fig:snapshot_comparison}a shows the disc shortly before fragmentation (for
run 1, i.e. disc with solar metallicity, $z=1$). When the disc metallicity is
lower ($z=0.1$; Figure~\ref{fig:snapshot_comparison}b), the disc exhibits
weaker, but well defined spiral features. Given that the optical depth $\tau =
\Sigma \kappa$, and the metallicity has been reduced, more gas is required for
the spiral arms to attain $\tau \sim 1$, where cooling is most efficient. As
such, the spirals in this case take longer to fragment (see
Table~\ref{tab:sim_results}). However, once a sufficient surface density is
reached, fragmentation occurs as cooling is efficient. When the disc
metallicity is higher ($z=10$; Figure~\ref{fig:snapshot_comparison}c), the disc
does not fragment as it cannot cool fast enough; instead it expands and becomes gravitationally stable ($Q>1$), although it can then cool fast enough, as the surface density has decreased (see Figure~\ref{fig:expansion}).

In Figure~\ref{fig:q_stellar_mass}, we present the relationship between the
stellar mass and the disc-to-star mass ratio, and in Figure~\ref{fig:q_radius_z} the relationship between the disc fragmentation mass and the disc-to-star mass ratio, for different disc metallicities. In general, we find that metal rich discs fragment at a slightly higher disc-to-star mass ratio (Figure~\ref{fig:q_stellar_mass}) and their corresponding discs are more extended when they fragment (Figure~\ref{fig:q_radius_z}). We also find
that the smaller  ($R = [60, 90]$~AU) discs with metallicity $z=10$ do not fragment (apart from run 6). This is due to period of fast disc expansion,  when the spiral features become strong, combined with inefficient cooling (during the expansion phase). Runs 3, 12 and 15 are examples of this. This is the reason why the discs in the runs 19 - 21 as well as 24 do not fragment (see discussion in Section \ref{sub:consquences_of_initial_stellar_mass_and_disc_radius}).

\subsection{Accretion onto the central star}
\label{sub:accretion_relation}

\begin{figure}
  \begin{center}
    \resizebox{\hsize}{!}{\includegraphics{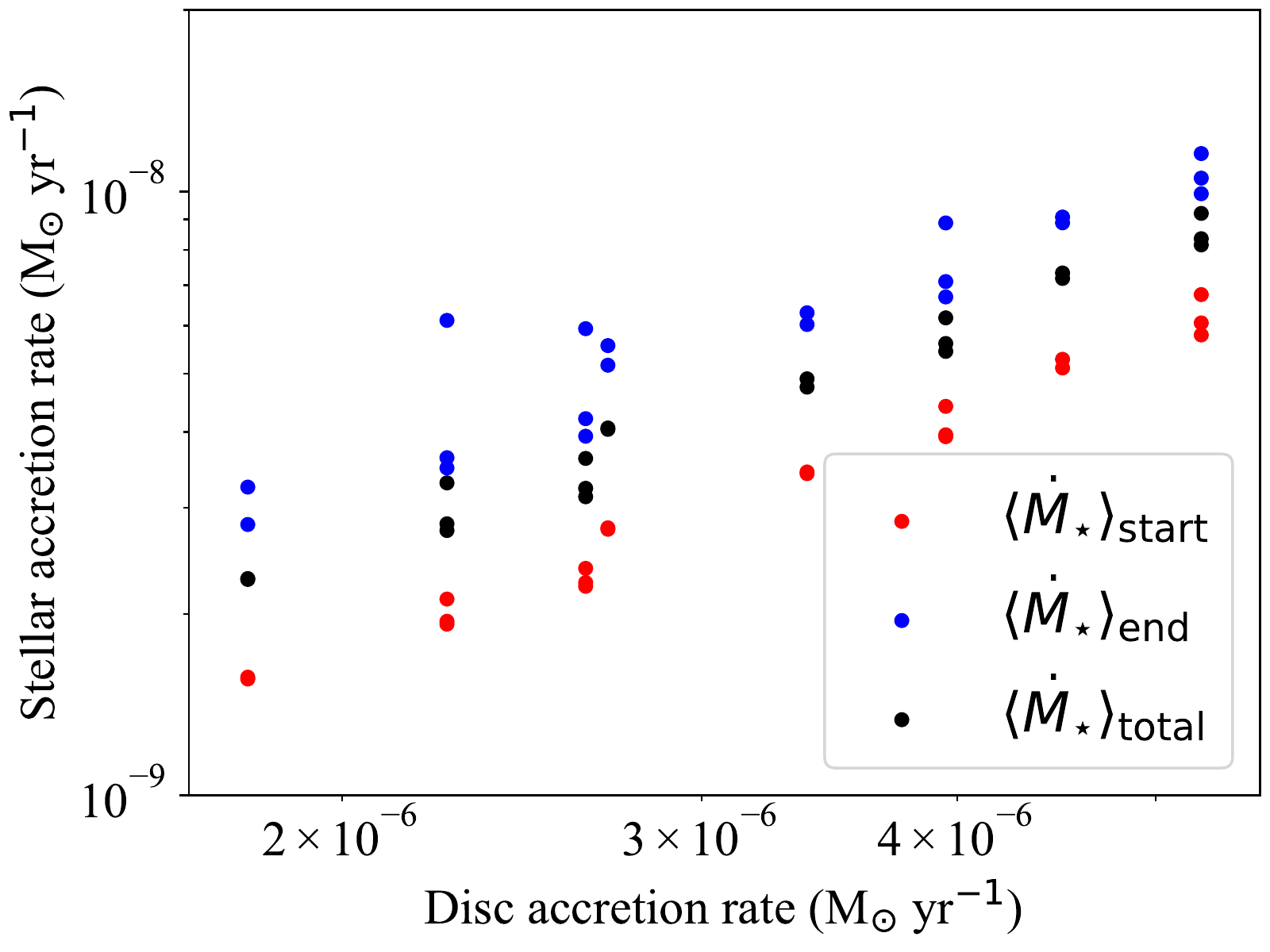}}
    \caption
    {
      Relationship between the accretion rate onto the disc (Equation \ref{eqn:mass_accretion_rate}) and the accretion rate onto the
      central star. The black points show the average
      accretion for the entire simulation,  whereas the red and blue points the average stellar rate during the first and last $10\%$ of the 
      simulated time, respectively. 
    }
    \label{fig:accretion_relation}
  \end{center}
\end{figure}

Typically, the mass accretion rate of the central star scales with the disc
accretion rate, albeit $\sim 3$ orders of magnitude smaller. Figure
\ref{fig:accretion_relation} shows this relation. The disc accretion rate for
each disc is set by Equation \ref{eqn:mass_accretion_rate} and listed in Table
\ref{tab:ics}. We show the average stellar accretion rate throughout the whole
simulation, as well as the beginning and end of each simulation. We find that
the stellar accretion rate is smaller at the start of each simulation and larger
at the end, as compared to the total average accretion rate. Towards the end of the simulations, the discs are gravitationally unstable and accretion is enabled by outward angular momentum transfer due to gravitational torques. Prior to the onset of the instability, angular momentum transport outwards is inefficient, and material only moves inwards slowly.

\section{The properties of  protoplanets formed around M dwarfs by disc instability}
\label{sec:the_properties_of_planets_formed_through_disc_fragmentation}

\begin{table*}
  \centering
    \caption
    {
      Properties of Type I and Type II protoplanets (see discussion in text). We list the Run ID,  the type of protoplanet,  the density at the centre of the protoplanet $\rho_c$,  the time $\Delta t_c$ it took the protoplanet to collapse from density $10^{-9}{\rm g\ cm}^{-3}$ to its final central density,  the distance of the protoplanet from the star, $a$, the first and second hydrostatic core radii ($R_{\textup{fc}}$ and $R_{\textup{sc}}$, respectively), the first and second hydrostatic core masses ($M_{\textup{fc}}$ and $M_{\textup{sc}}$, respectively), the number of SPH particles of the first core, which is indicative of how well the first core and its collapse are resolved, 
      the ratios of rotational-to-gravitational $\beta_{\rm rot}$ and thermal-to-gravitational energies $\alpha_{\rm therm}$, for the first and second
      cores, and finally the specific angular momenta of the first and second cores.
    }
    \label{tab:protoplanet_properties}
    \centering
    \begin{tabular}{c c c c c c c c c c c c c c c c}
      \hline
      \hline
  
  
      Run & Type &$\rho_c$  & $\Delta t_c$ & $a$ & $R_{\textup{fc}}$ & $R_{\textup{sc}}$
      & $M_{\textup{fc}}$ & $N_{\textup{fc}}$ & $M_{\textup{sc}}$ &
      $\beta_{\rm rot}^{\textup{fc}}$ & $\alpha_{\rm therm}^{\textup{fc}}$ & $\beta_{\rm rot}^{\textup{sc}}$ &
      $\alpha_{\rm therm}^{\textup{sc}}$ & $J_{\textup{fc}}$ & $J_{\textup{sc}}$ \\
  
      & & (${\rm g\ cm}^{-3}$) & (kyr)& (AU) & (AU) & ({\RSUN}) & (\MJUP)& ($10^5$) &
      (\MJUP) & & & & & (cm$^{2}$ s$^{-1}$ & $\times 10^{18}$) \\
      \hline
      01 & IIb & $ 8.5\times10^{-9}$ & 0.4 & 49  & 6.6 & -   & 9.5 & 2.3 & -   & 0.22 & 0.46 & -    & -    & 59  & -   \\
      02 & IIb& $ 1.8\times10^{-7}$ & 0.3 & 24  & 2.3 & -   & 6.5 & 1.6 & -   & 0.21 & 0.48 & -    & -    & 18  & -   \\
      \hline
      04 & IIb&$ 6.0\times10^{-8}$ & 0.8 & 104 & 14  & -   & 9.0 & 1.6 & -   & 0.12 & 0.60 & -    & -    & 32  & -   \\
      05 & Ia &$10^{-3}$           & 1.0 & 27  & 3.2 & 7.3 & 9.2 & 1.6 & 2.6 & 0.12 & 0.80 & 0.07 & 0.95 & 21  & 0.3 \\
      06 & Ia& $10^{-3}$           & 0.3 & 32  & 4.5 & 28  & 21  & 3.6 & 11  & 0.06 & 0.92 & 0.05 & 0.96 & 25  & 2.6 \\
      \hline
      07 & Ia& $10^{-3}$           & 0.7 & 27  & 3.7 & 5.8 & 10  & 1.5 & 2.8 & 0.08 & 0.86 & 0.05 & 1.02 & 18  & 0.2 \\
      08 & Ia& $10^{-3}$           & 0.1 & 14  & 3.2 & 5.5 & 6.0 & 0.9 & 2.9 & 0.05 & 0.90 & 0.04 & 1.06 & 4.1 & 0.2 \\
      09 & Ia& $10^{-3}$           & 1.1 & 105 & 7.1 & 6.1 & 13  & 1.8 & 5.0 & 0.08 & 0.88 & 0.06 & 0.99 & 24  & 0.5 \\
      \hline
      10 & IIb&$ 7.1\times10^{-7}$ & 0.6 & 38  & 4.2 & -   & 9.5 & 1.8 & -   & 0.10 & 0.67 & -    & -    & 25  & -   \\
      11 & Ia&$10^{-3}$           & 0.3 & 24  & 2.3 & 5.9 & 6.9 & 1.3 & 2.4 & 0.10 & 0.84 & 0.06 & 0.99 & 10  & 0.2 \\
     \hline
      13 & IIa &$ 3.1\times10^{-6}$ & 0.3 & 69  & 2.7 & 29  & 14  & 2.2 & 5.3 & 0.08 & 0.79 & 0.04 & 1.17 & 20  & 1.4 \\
      14 & Ib &$10^{-3}$           & 0.2 & 18  & 2.8 & -   & 6.1 & 0.9 & -   & 0.10 & 0.86 & -    & -    & 4.3 & -   \\
     \hline 
      16 & Ib &$10^{-3}$           & 0.4 & 63  & 6.5 & -   & 21  & 2.8 & -   & 0.11 & 0.82 & -    & -    & 60  & -   \\
      17 & Ia &$10^{-3}$           & 0.5 & 51  & 4.5 & 8.8 & 11  & 1.4 & 2.6 & 0.13 & 0.78 & 0.08 & 0.93 & 26  & 0.2 \\
      18 & Ib&$10^{-3}$           & 1.0 & 72  & 1.0 & -   & 15  & 1.8 & -   & 0.09 & 0.90 & -    & -    & 16  & -   \\
      \hline
      22 & Ia &$10^{-3}$           & 0.6 & 40  & 3.1 & 6.4 & 9.7 & 1.3 & 3.5 & 0.07 & 0.87 & 0.04 & 1.05 & 10  & 0.2 \\
      23 & Ib&$10^{-3}$           & 0.3 & 33  & 3.2 & -   & 8.9 & 1.2 & -   & 0.11 & 0.82 & -    & -    & 12  & -   \\
     \hline
      25 & Ib&$10^{-3}$           & 0.6 & 49  & 6.2 & -   & 16  & 1.8 & -   & 0.11 & 0.83 & -    & -    & 46  & -   \\
      26 & Ia&$10^{-3}$           & 0.3 & 46  & 5.1 & 5.5 & 5.6 & 0.6 & 1.9 & 0.07 & 0.86 & 0.04 & 1.00 & 7.7 & 0.1 \\
      27 &Ia &$10^{-3}$           & 1.7 & 79  & 8.9 & 7.6 & 17  & 1.7 & 4.9 & 0.10 & 0.86 & 0.09 & 0.94 & 49  & 0.5 \\
      \hline
    \end{tabular}
  \end{table*}

The evolution of the discs that fragment was followed until the first  fragment formed in each simulation collapsed further to densities higher than $10^{-9} \textup{ g cm}^{-3}$ (i.e. the limit we set for fragmentation in the previous section). These fragments  are collectively referred to as {protoplanets}. It is expected that some of them evolve to become planets, whereas others get tidally disrupted and disperse or accumulate too much gas and become brown dwarfs \citep{Stamatellos:2009a,Kratter:2010a,Zhu:2012a}.

\begin{figure*}
  \begin{center}
  \subfloat{\resizebox{0.45\hsize}{!} 
  {\includegraphics{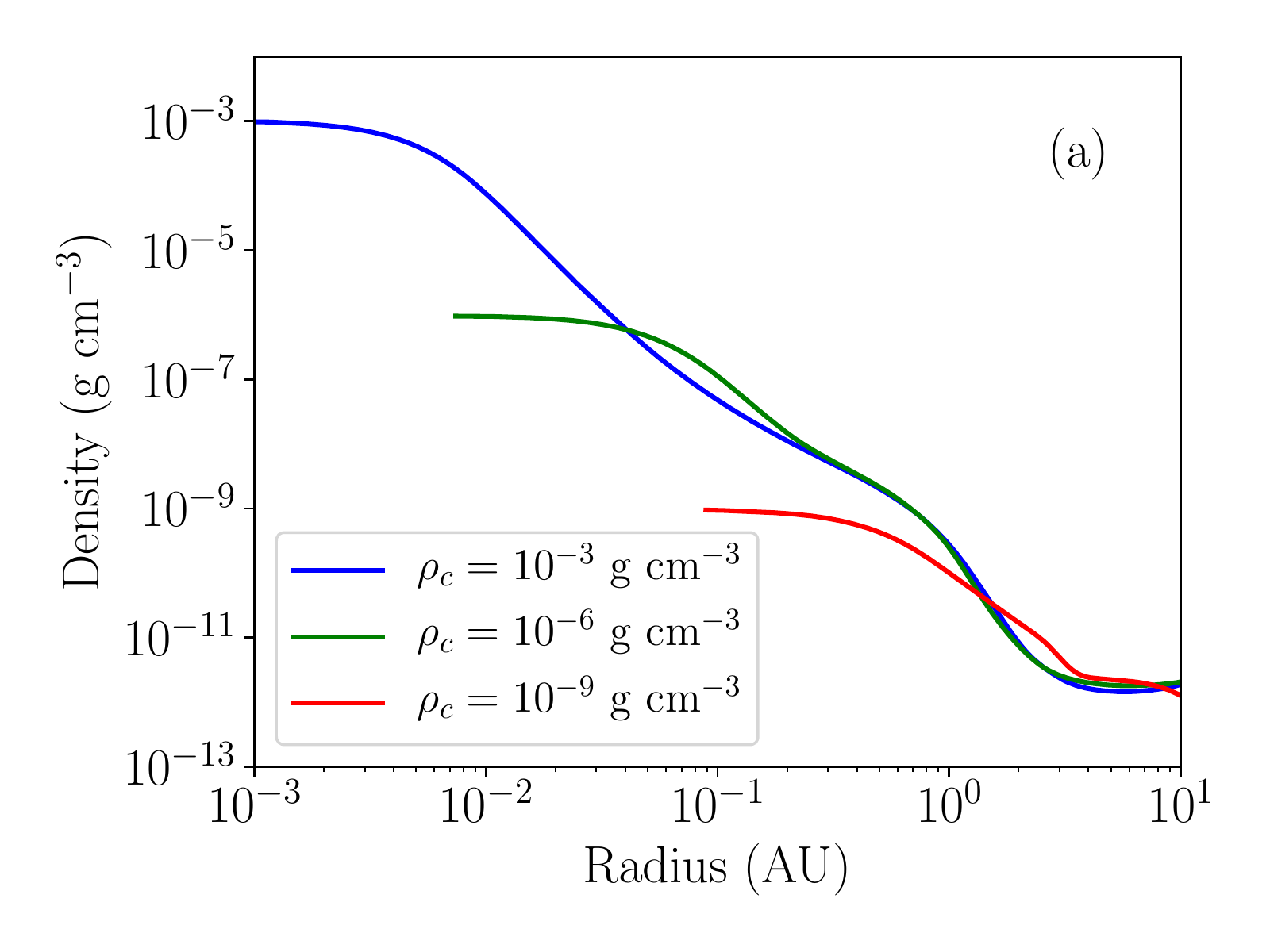}}}
  \subfloat{\resizebox{0.45\hsize}{!}
  {\includegraphics{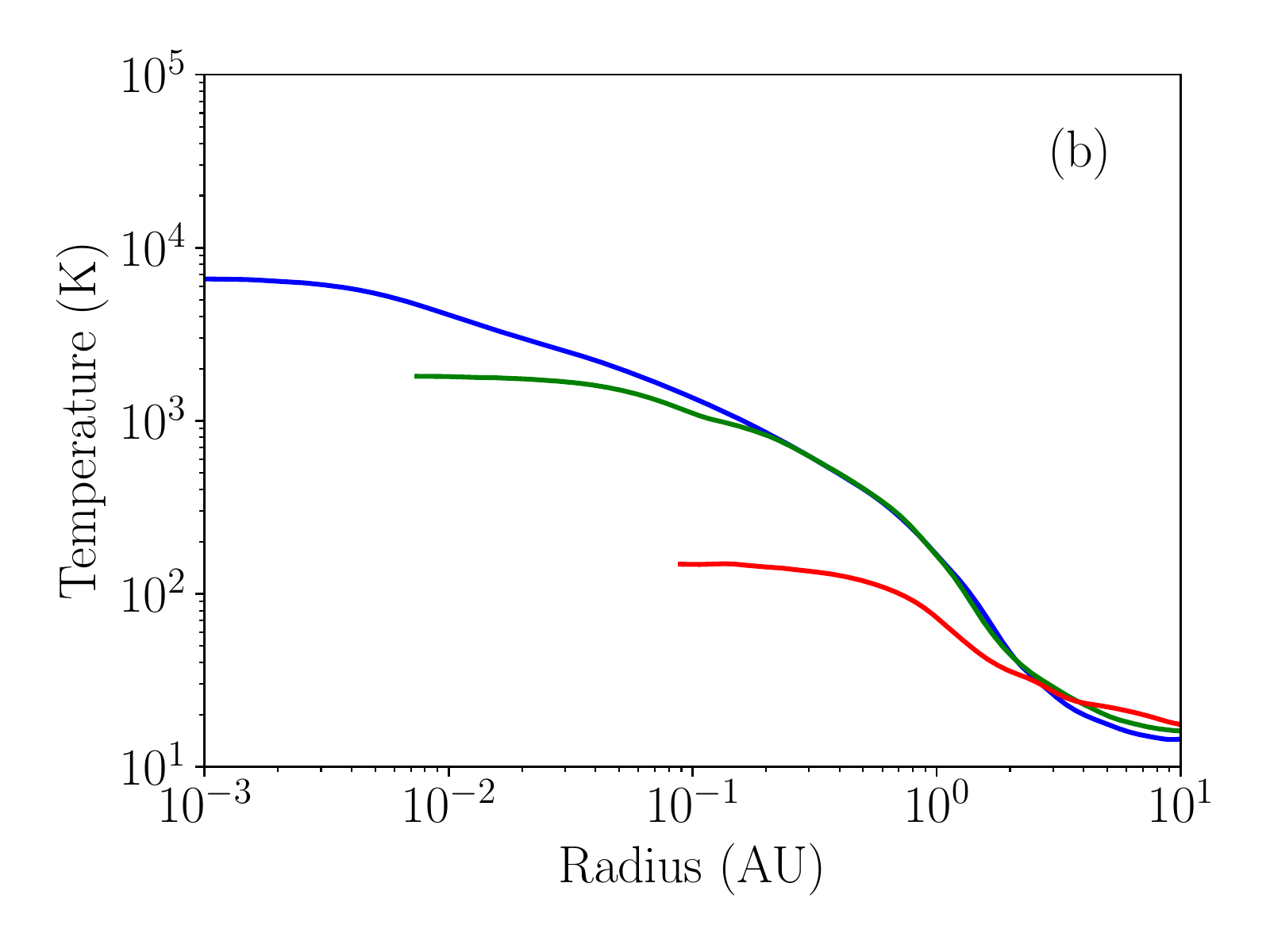}}} \\
  \subfloat{\resizebox{0.45\hsize}{!}
  {\includegraphics{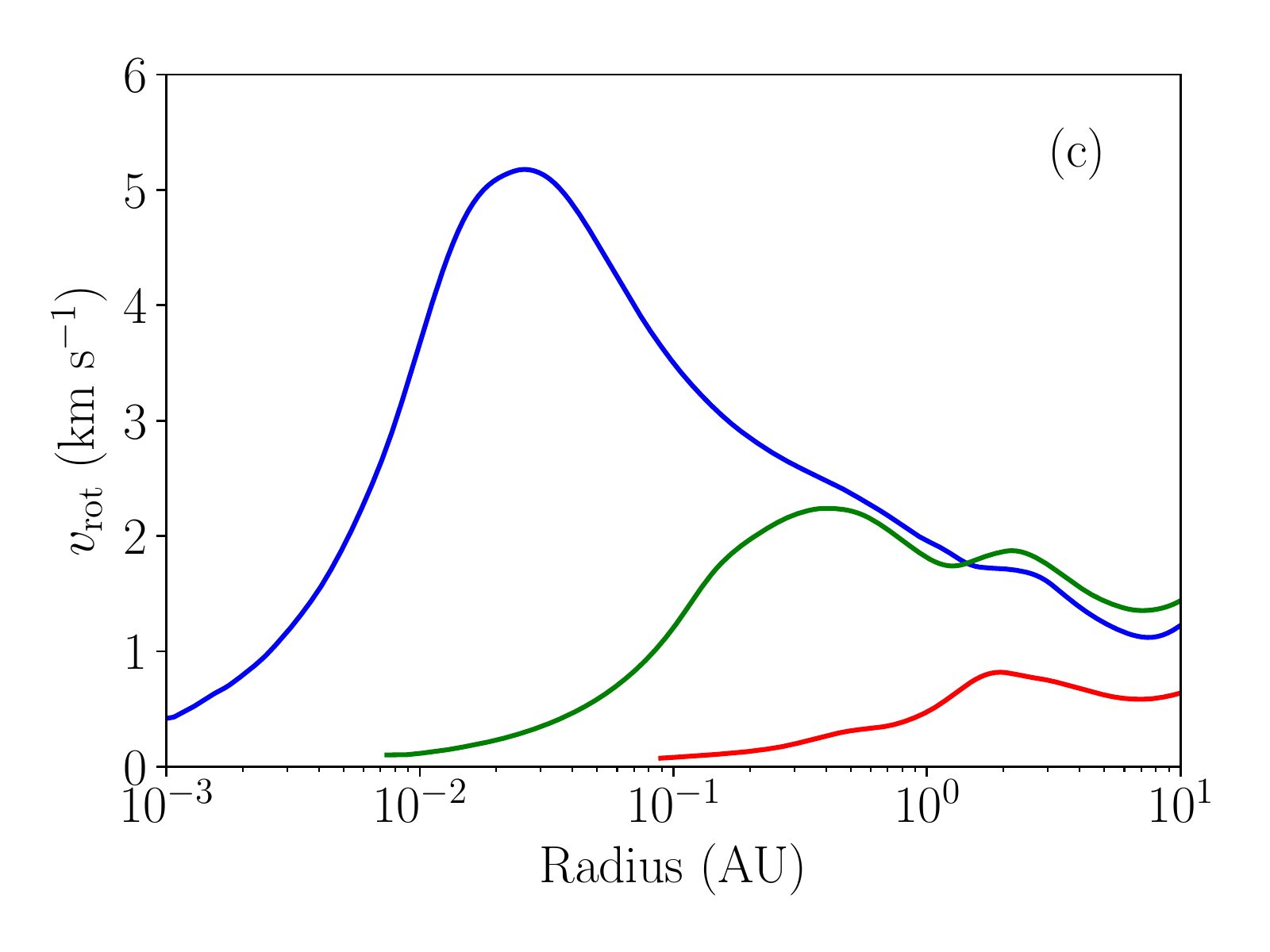}}}
  \subfloat{\resizebox{0.45\hsize}{!}
  {\includegraphics{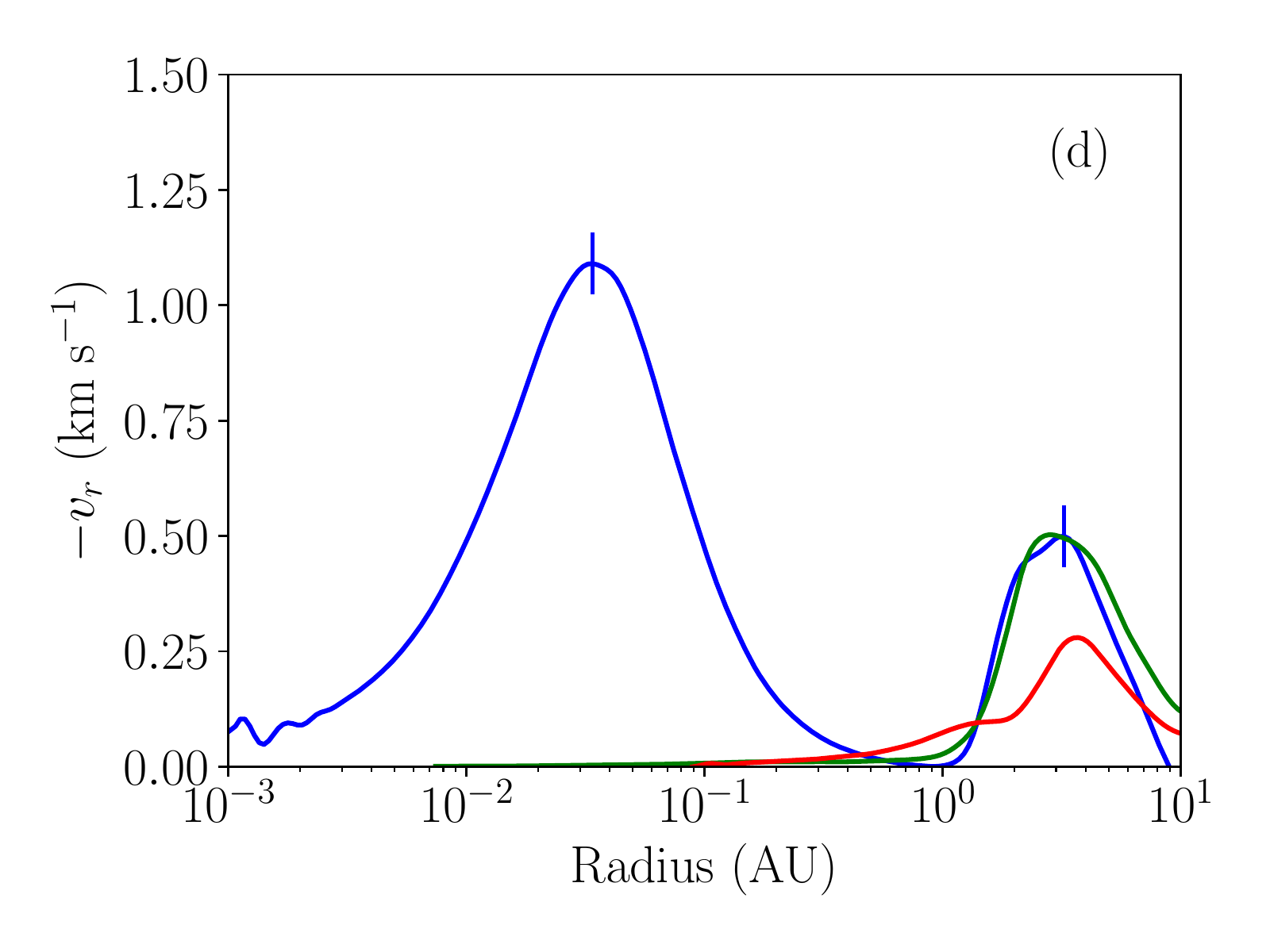}}} \\
  \subfloat{\resizebox{0.45\hsize}{!} 
  {\includegraphics{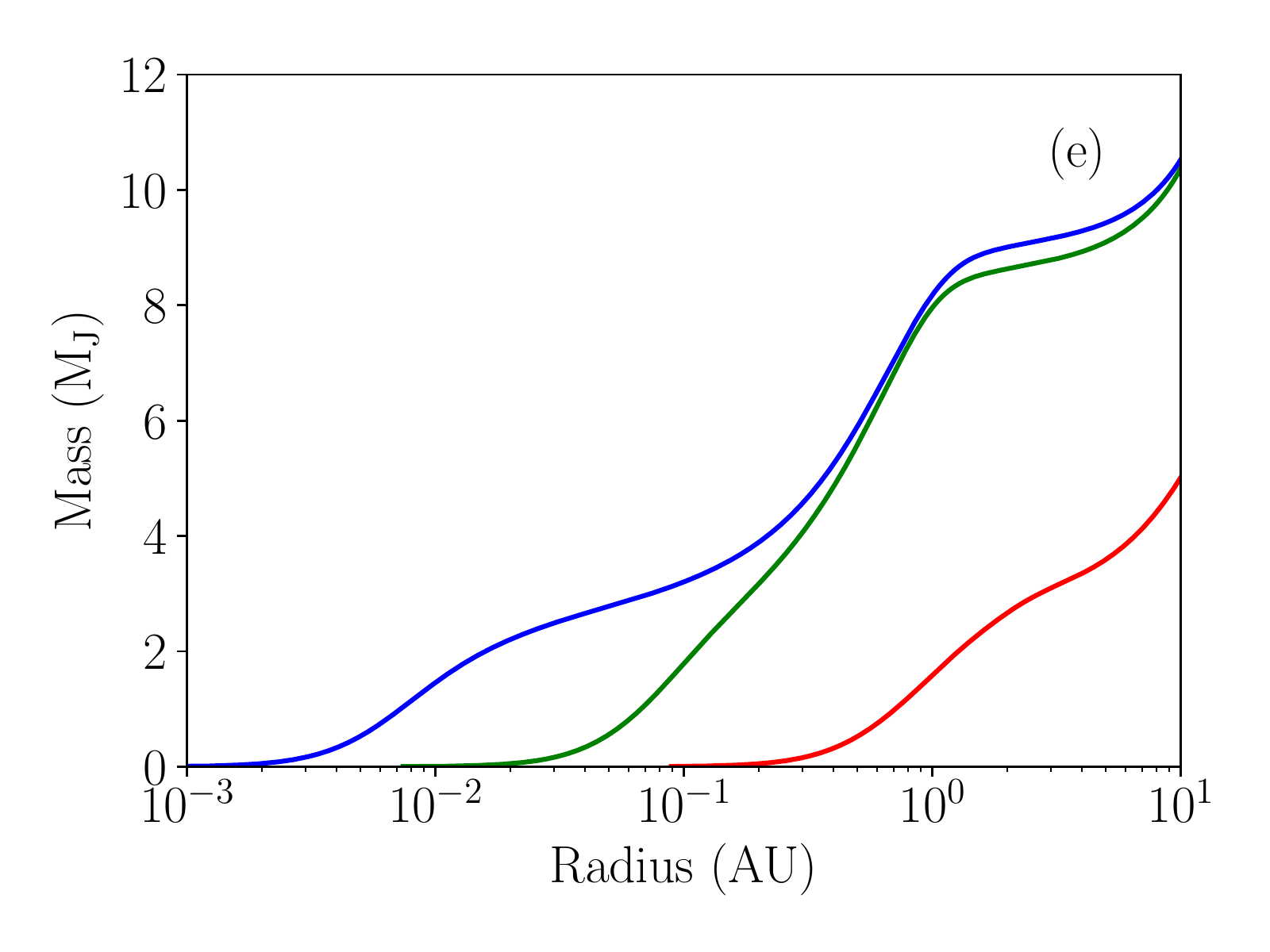}}}
  \subfloat{\resizebox{0.45\hsize}{!}  
  {\includegraphics{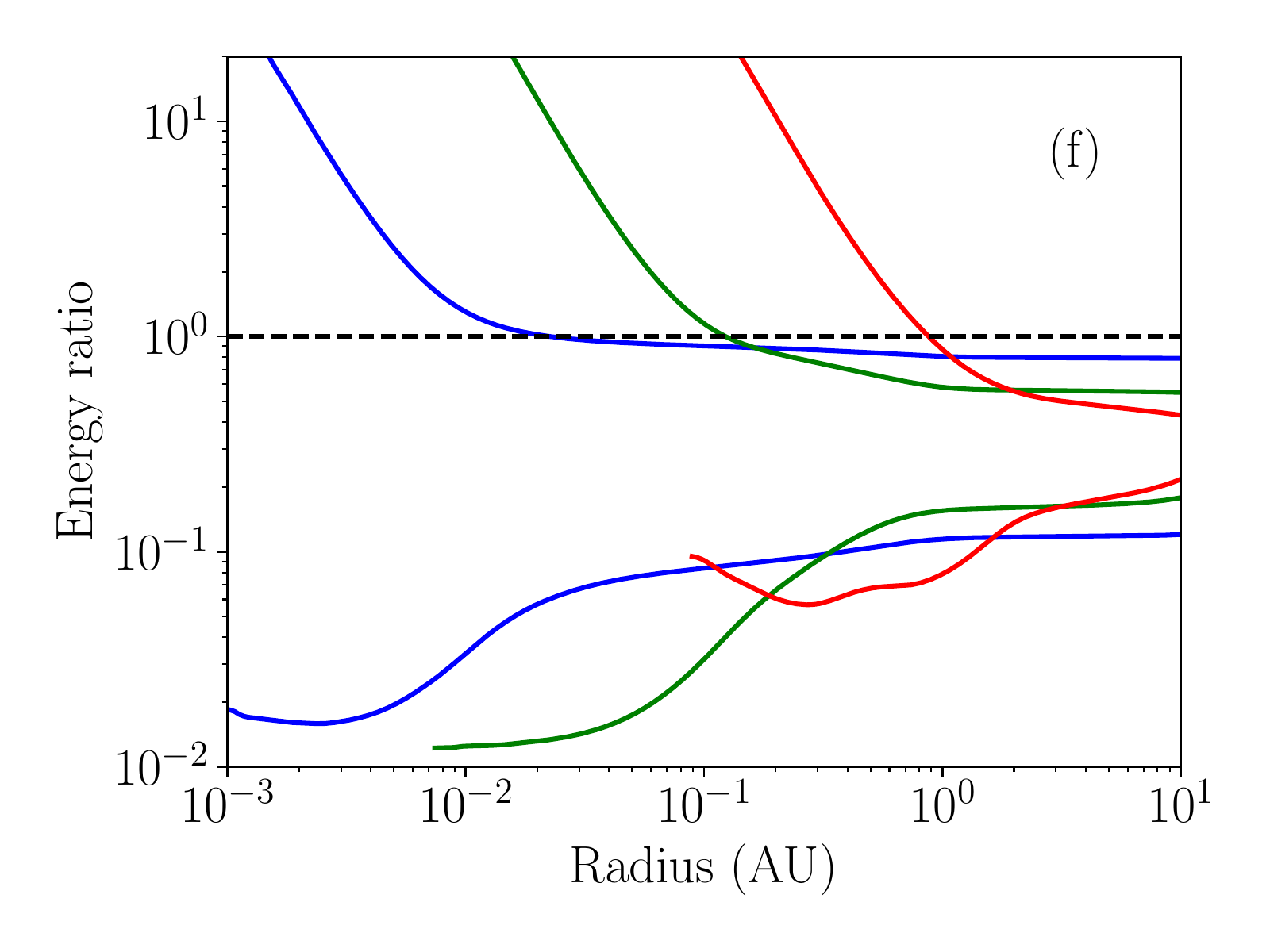}}}
  \caption{Evolution of a representative protoplanet (Run 5).
    Panels (a) and (b) show the spherically-averaged density and temperature, respectively. Panels (c)
    and (d) show rotational (azimuthally-averaged) and radial infall velocity
    (spherically-averaged). { The first and second hydrostatic cores boundaries are
    identified by the peaks in the infall velocity profiles (the positions of the boundaries are marked with the short vertical lines)}. Panel (e) shows the
    mass within a given radius within each fragment. Panel (f) shows the ratio
    of total energies interior to a given radius: $\alpha_{\rm therm}=E_{\textup{ther}} /
    E_{\textup{grav}}$ (top set of lines) and $\beta_{\rm rot}=E_{\textup{rot}} /
    E_{\textup{grav}}$ (bottom set of lines). The rotational energy is
    significant only in the outer parts of the fragment. }
  \label{fig:properties_run5}
\end{center}
\end{figure*}

The evolution of the density, temperature, rotational velocity, infall velocity,
mass, and ratios of the thermal-to-gravitational and
rotational-to-gravitational energy as a typical fragment collapses (in Run 5) are shown in Figure~\ref{fig:properties_run5}. The fragment generally goes through the phase of first collapse,
first core formation, second collapse, and second core formation
\citep{Stamatellos:2009d}, just like a solar-mass collapsing core
\citep{Larson:1969a, Masunaga:1998a, Masunaga:2000a, Stamatellos:2007b,
Vaytet:2017a, Bhandare:2018a}. Initially, during the first collapse, the
temperature increases slowly as the fragment is optically thin, but when it
becomes optically thick the first hydrostatic core forms (as evidenced by the
fragment infall velocity profile showing the accretion shock on the boundary of
the first core) and the collapse slows down, proceeding quasi-statically and
almost adiabatically. The temperature at the centre of the fragment eventually
gets high enough ($\sim 2,000$~K) for molecular hydrogen to start dissociating, a
process that acts as an energy sink. Then, the second collapse is initiated and
the second core forms (as evidenced again by the accretion shock in the infall velocity profile).

The simulations terminated once the density at the centre of the fragments reached  $10^{-3} \textup{ g cm}^{-3}$ (although for a few of the simulations this density was not reached).  We note however, that due to the rotation of the fragments and interactions with the disc and other fragments there were deviations from this general behaviour. We have therefore grouped the protoplanets that were formed in these simulations into 2 types (each with 2 sub-types).

\begin{figure*}
  \centering
  \subfloat{\resizebox{0.38\hsize}{!}
  {\includegraphics{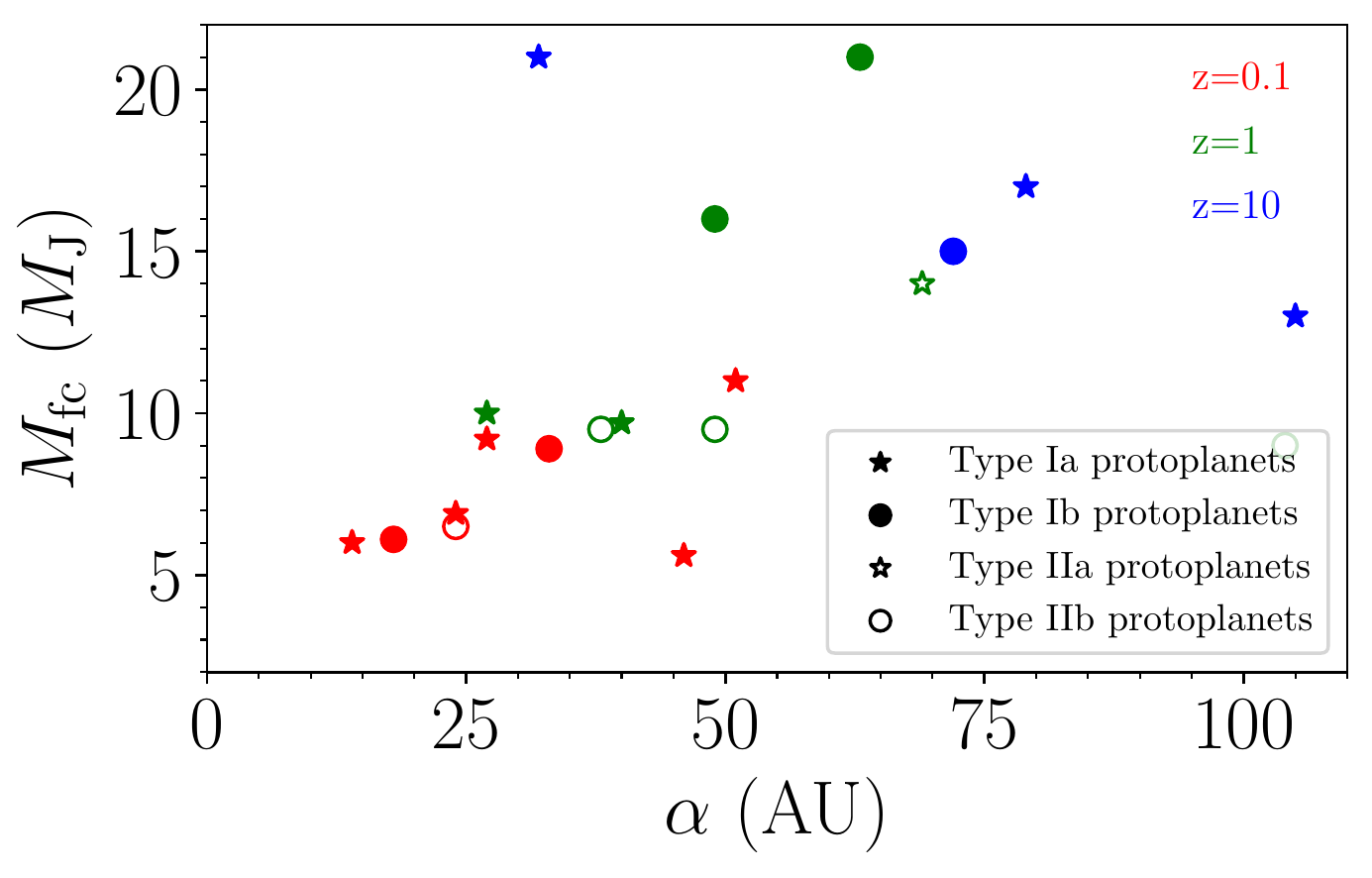}}}
  \subfloat{\resizebox{0.39\hsize}{!}
  {\includegraphics{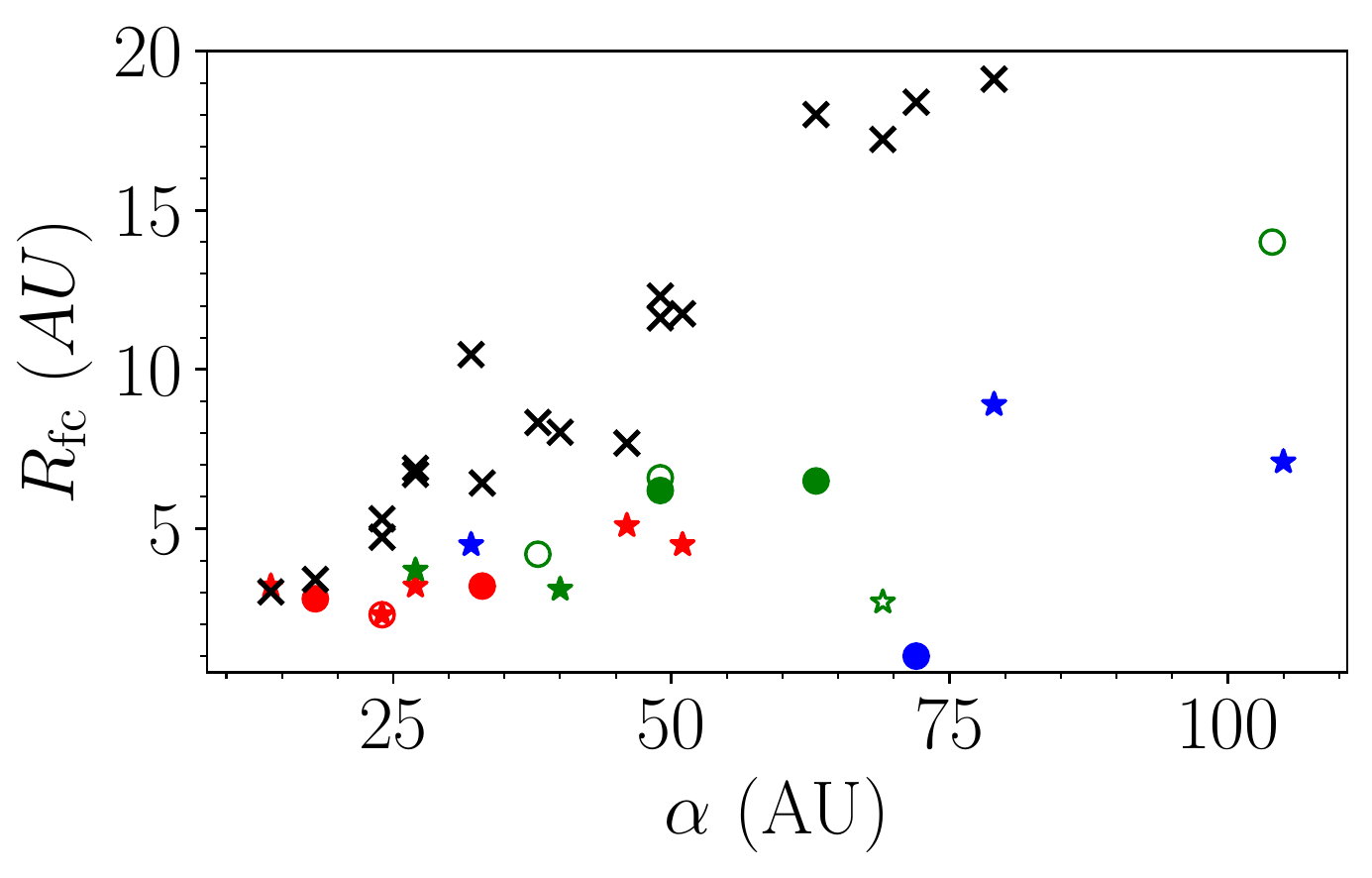}}}
  \caption
  { Mass (left) and radius (right) of the first cores formed in the simulations in Table~\ref{tab:protoplanet_properties}. Different symbols correspond to different type of protoplanets (see discussion in text). Colours correspond to different opacities (red: $z=0.1$, green: $z=1$, blue: $z=10$). Black crosses correspond to the Hill radius of each fragment.
  }
  \label{fig:protoplanets-fc}
 \centering
   \subfloat{\resizebox{0.39\hsize}{!}
  {\includegraphics{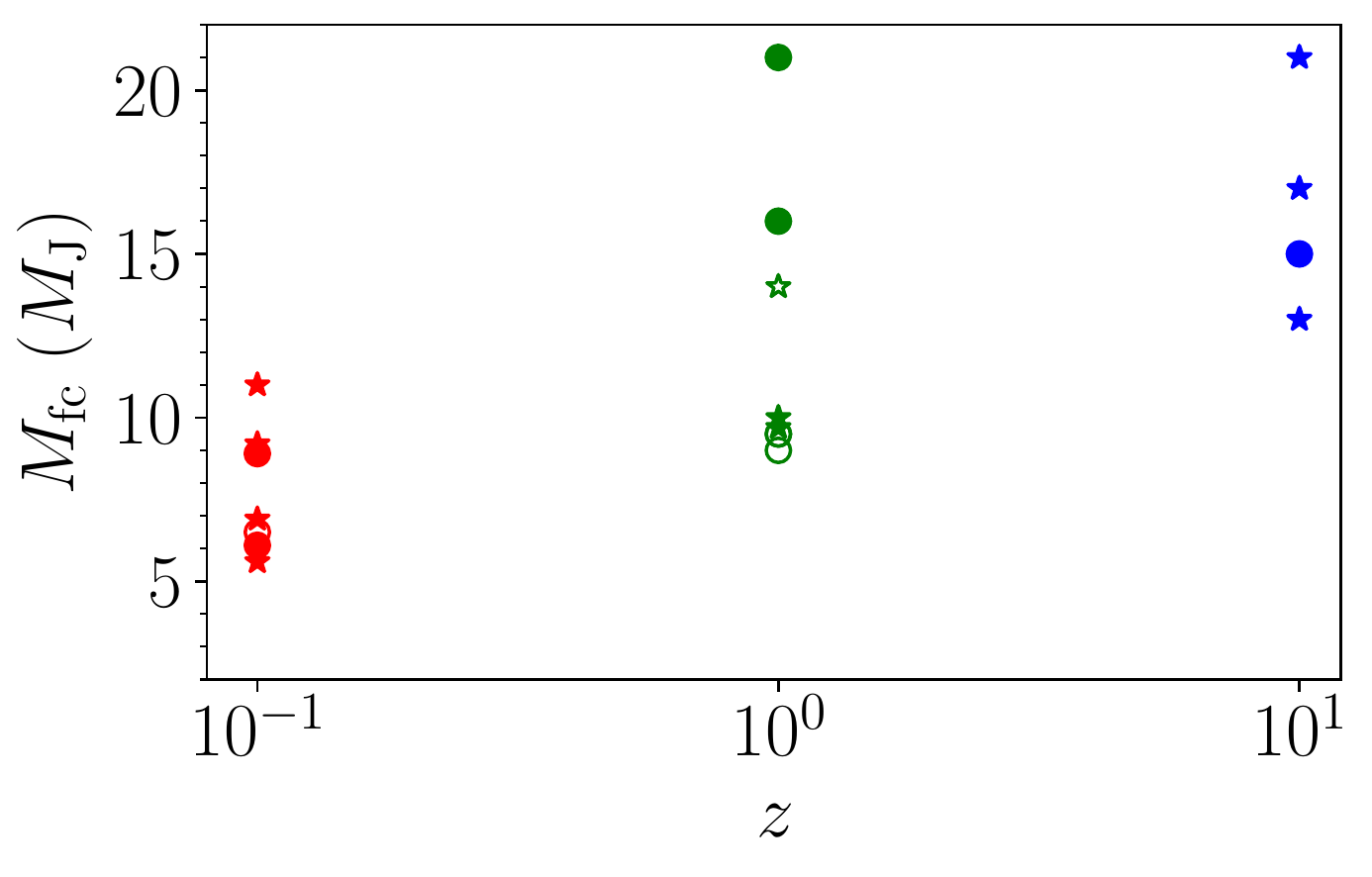}}}
  \subfloat{\resizebox{0.38\hsize}{!}
  {\includegraphics{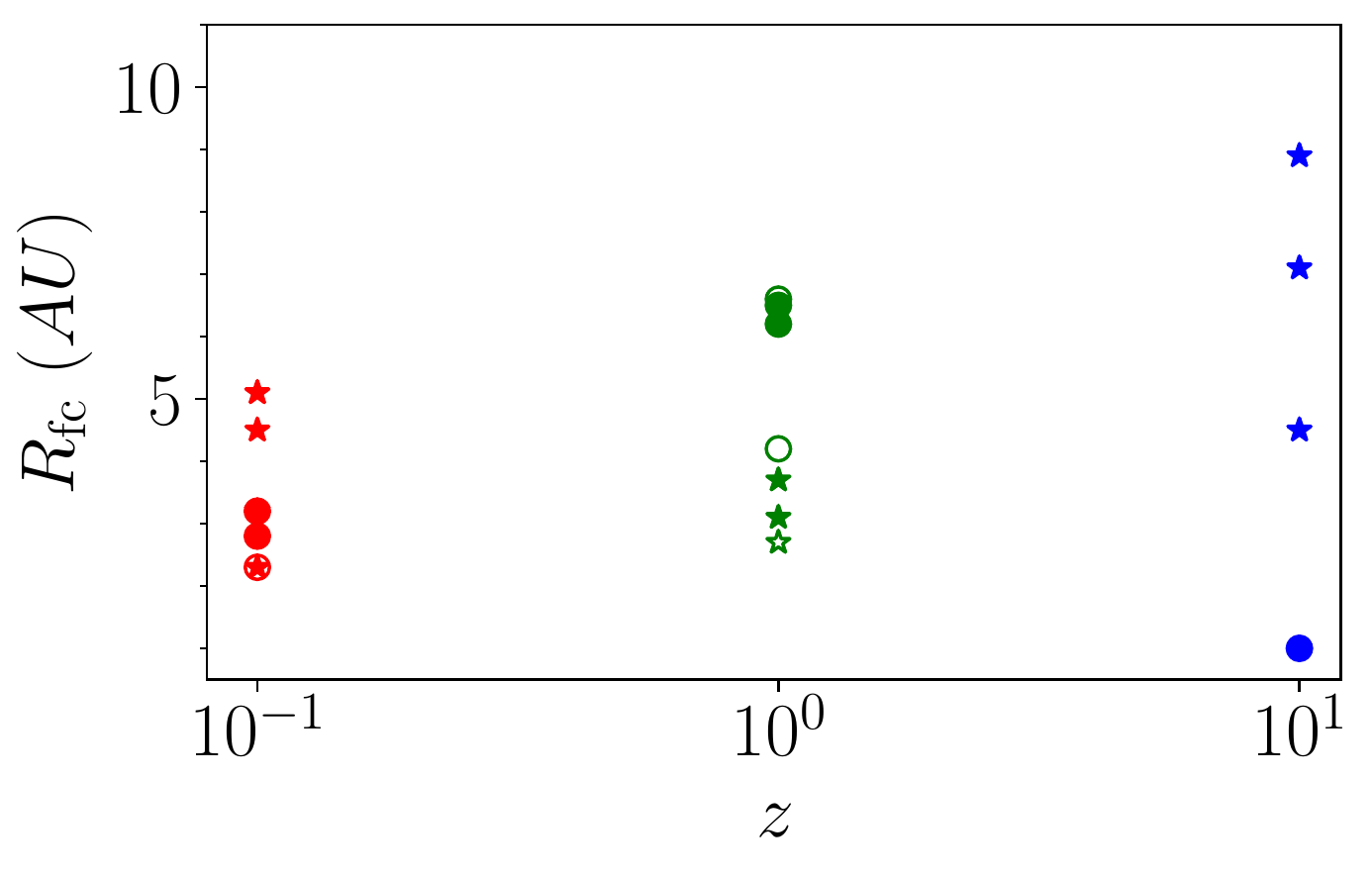}}}
  \caption
  {Mass (left) and radius (right) of the first cores  versus metallicity $z$. Symbols are the same as in Figure~\ref{fig:protoplanets-fc}.
  }
  \label{fig:protoplanets-fc-z}
  \centering
  \subfloat{\resizebox{0.37\hsize}{!}
  {\includegraphics{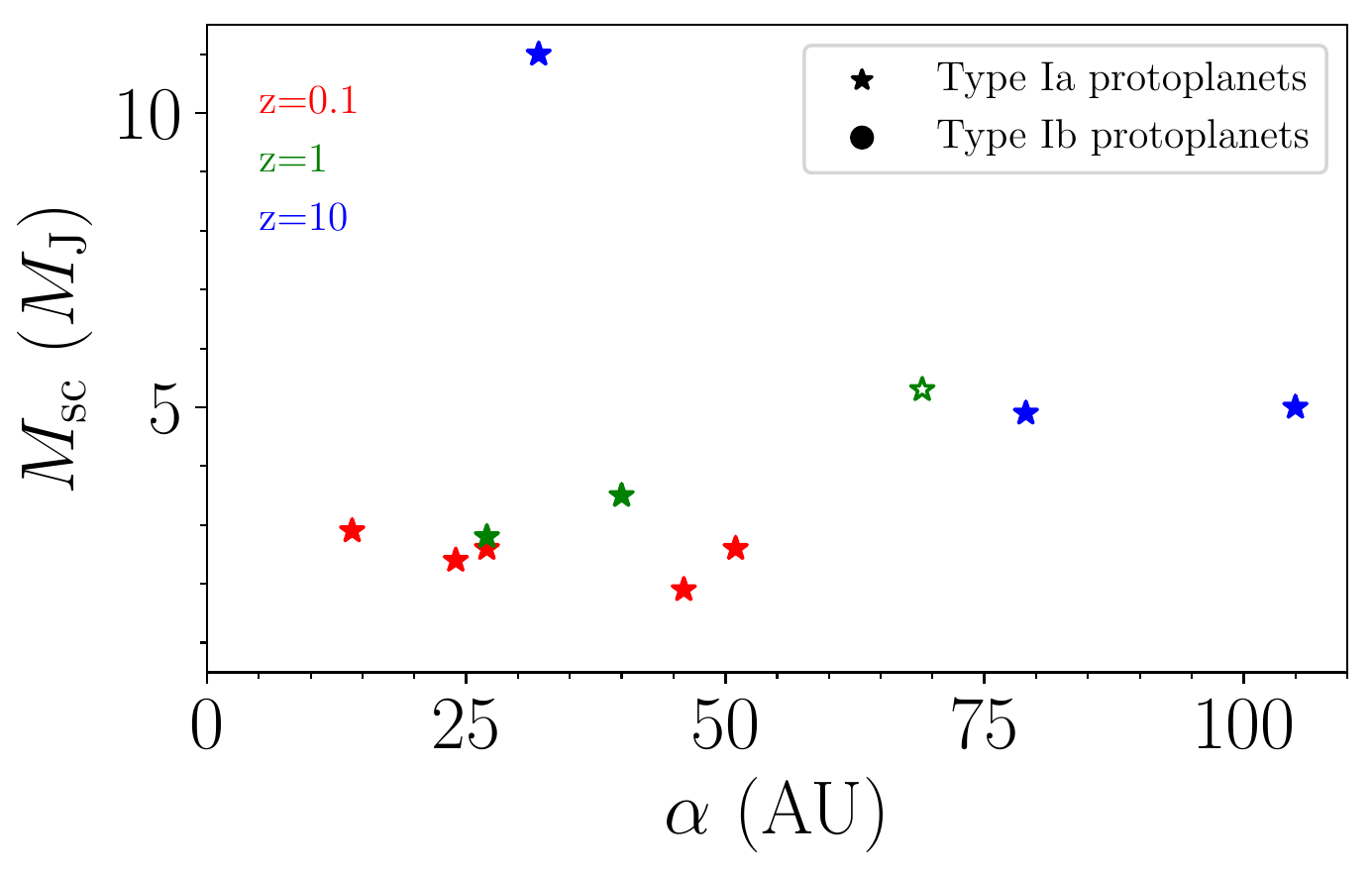}}}
  \subfloat{\resizebox{0.39\hsize}{!}
  {\includegraphics{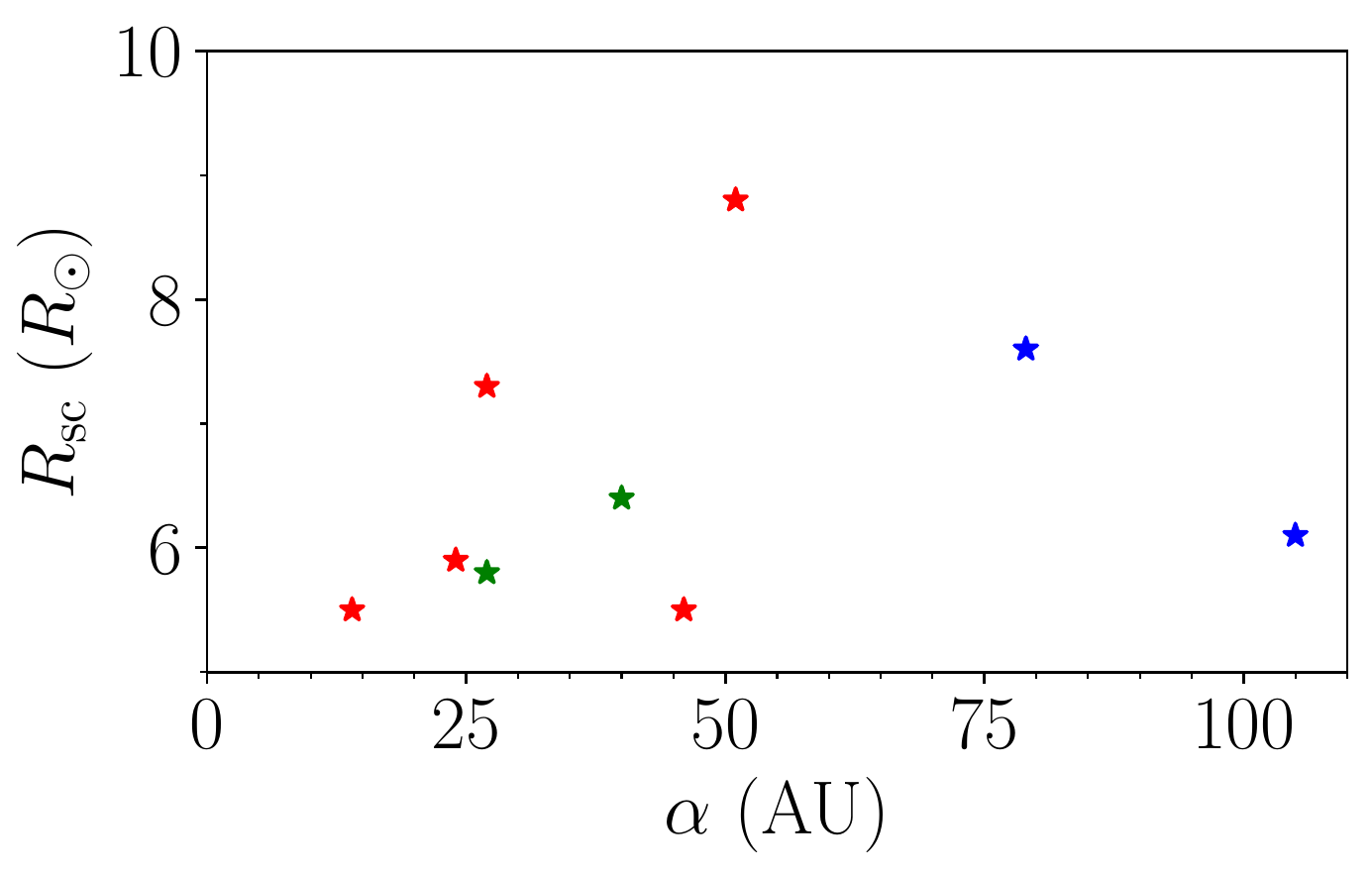}}}
  \caption
  {
    Mass (left) and radius (right) of the second cores. Symbols and colours are the same as in Figure~\ref{fig:protoplanets-fc}.
  }
  \label{fig:protoplanets-sc}
 \centering
   \subfloat{\resizebox{0.37\hsize}{!}
  {\includegraphics{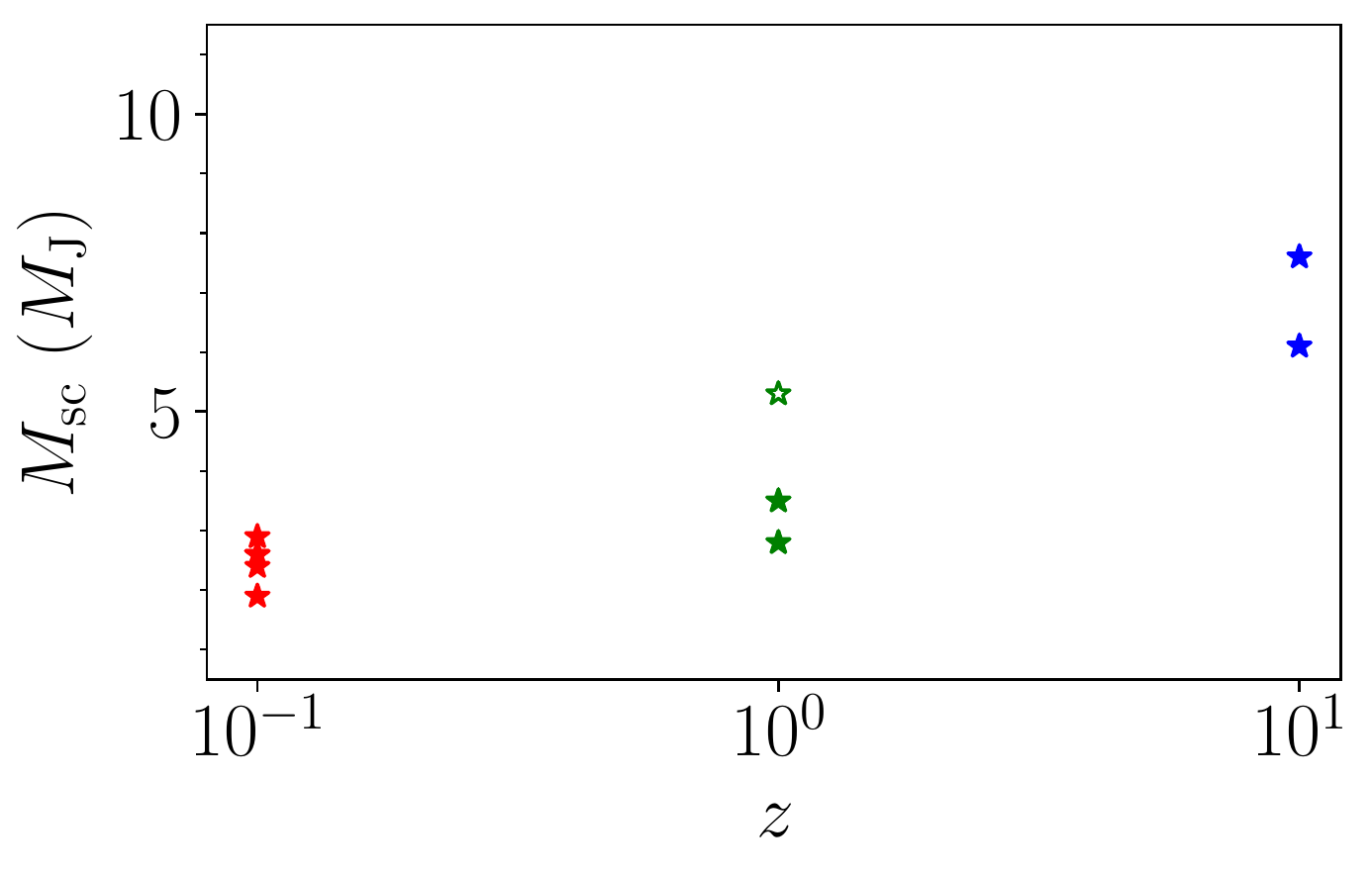}}}
  \subfloat{\resizebox{0.39\hsize}{!}
  {\includegraphics{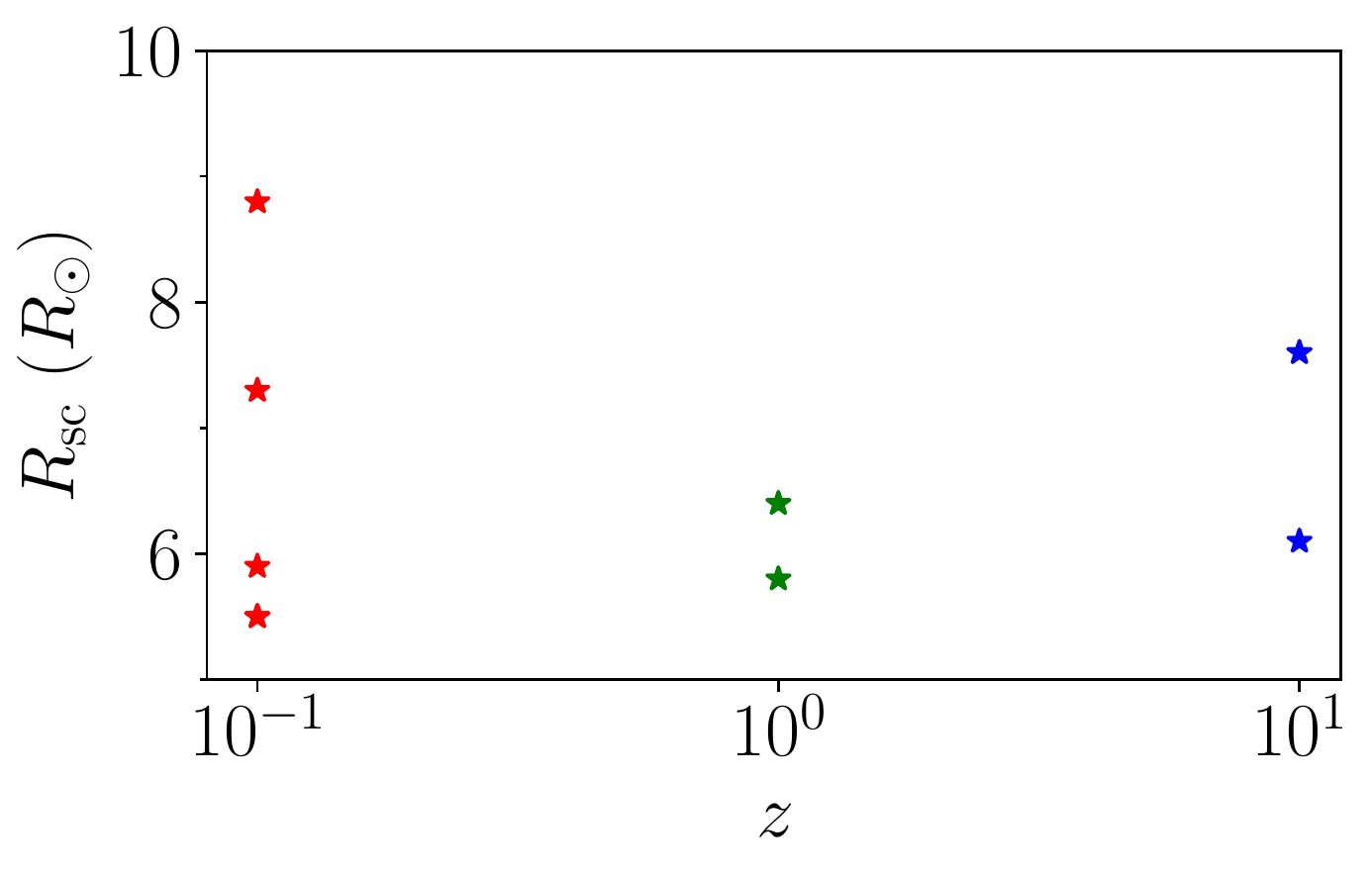}}}
  \caption
  {
    Mass (left) and radius (right) of the second cores versus metallicity $z$. Symbols and colours are the same as in Figure~\ref{fig:protoplanets-fc}.
  }
  \label{fig:protoplanets-sc-z}
\end{figure*}

\begin{figure*}
 \centering
   \subfloat{\resizebox{0.38\hsize}{!} 
  {\includegraphics{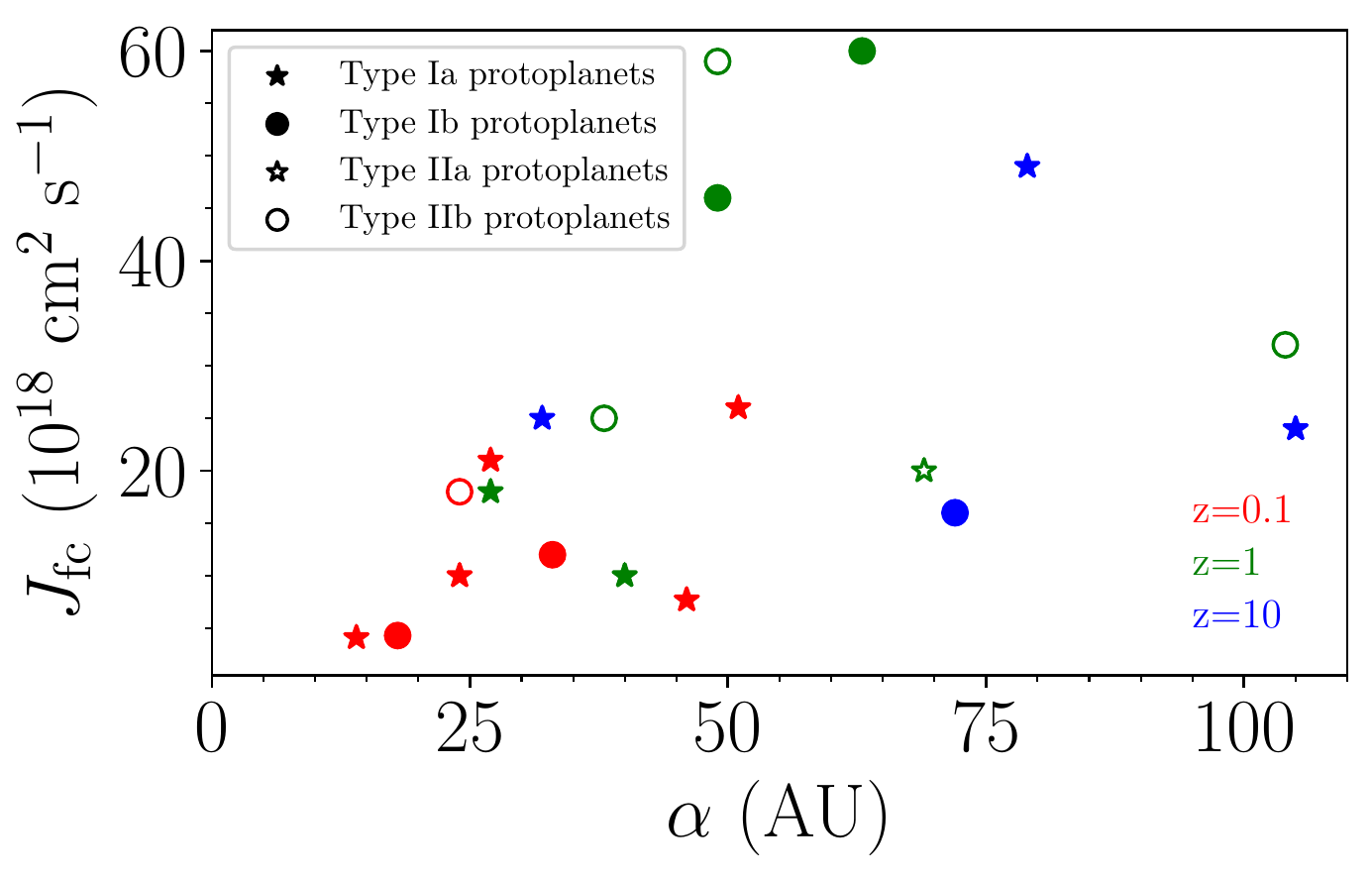}}}
  \subfloat{\resizebox{0.4\hsize}{!}
  {\includegraphics{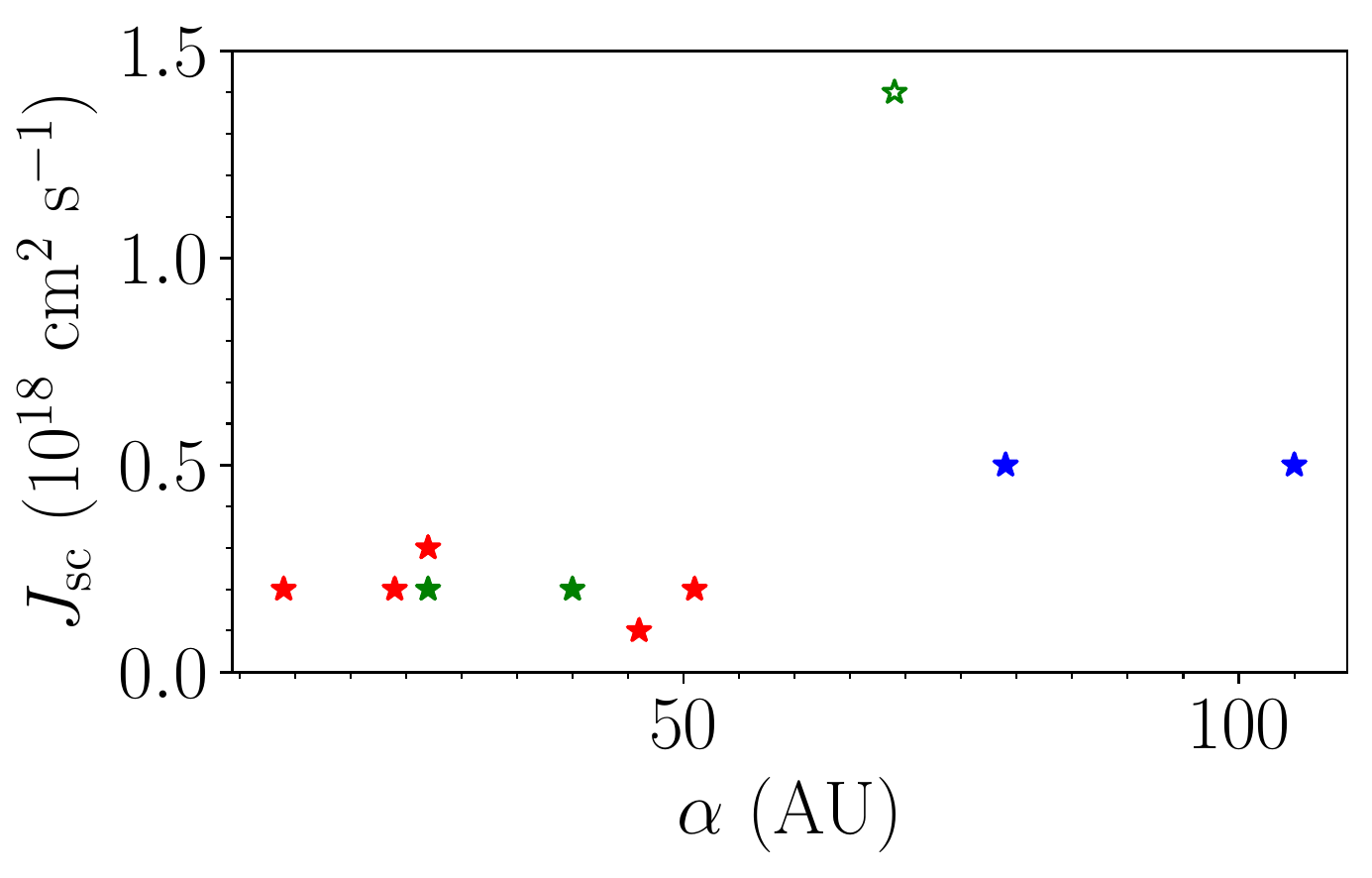}}}
  \caption
  {
   Specific angular momenta of  the first (left) and second (right) cores of protoplanets formed by disc instability. Fragments that do not undergo a second collapse (open circles) tend to have higher specific angular momentum. Symbols are the same as in Figure~\ref{fig:protoplanets-fc}.
  } 
  \label{fig:protoplanets-sam}
 \centering
   \subfloat{\resizebox{0.39\hsize}{!} 
  {\includegraphics{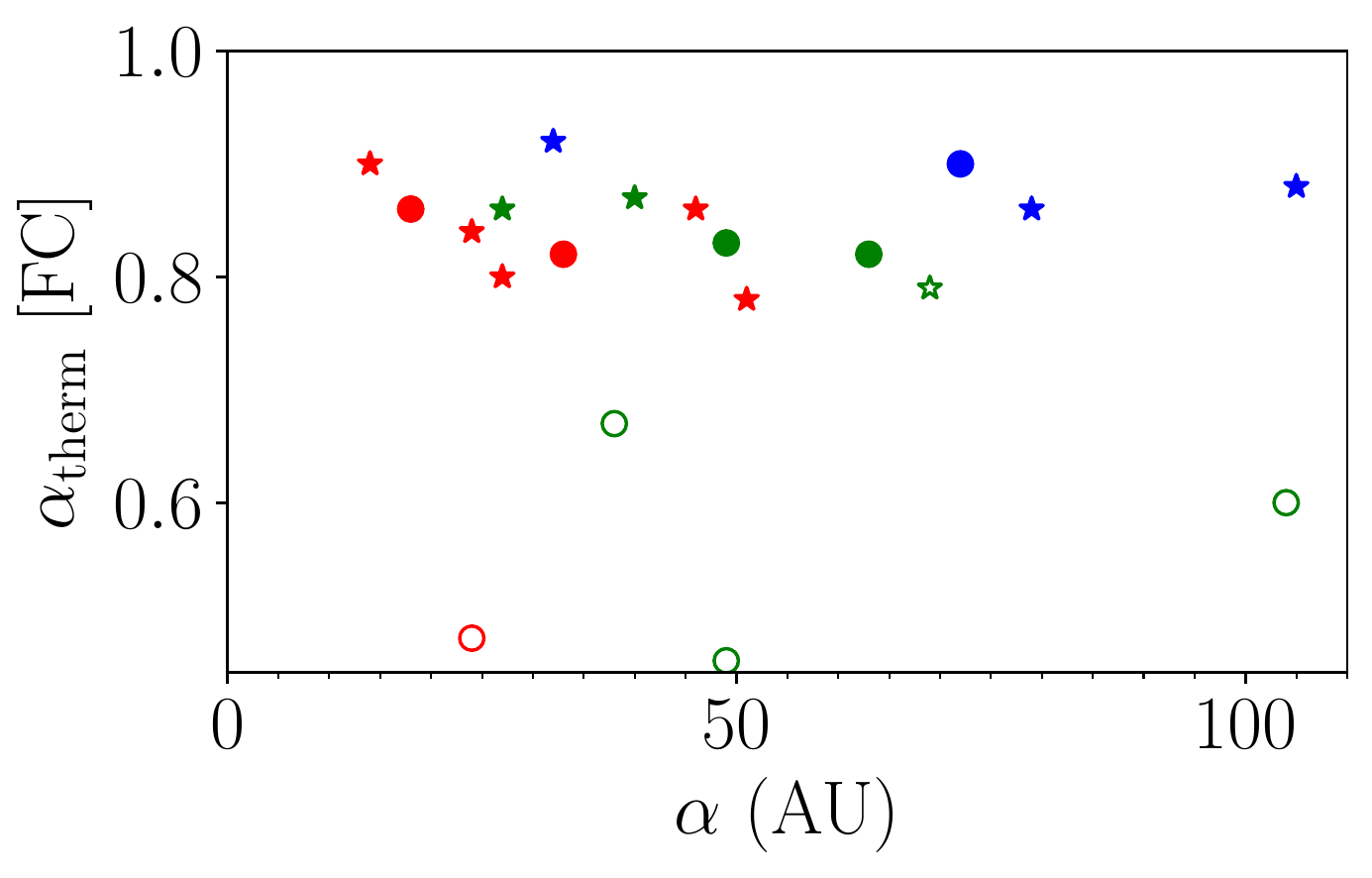}}}
  \subfloat{\resizebox{0.39\hsize}{!} 
  {\includegraphics{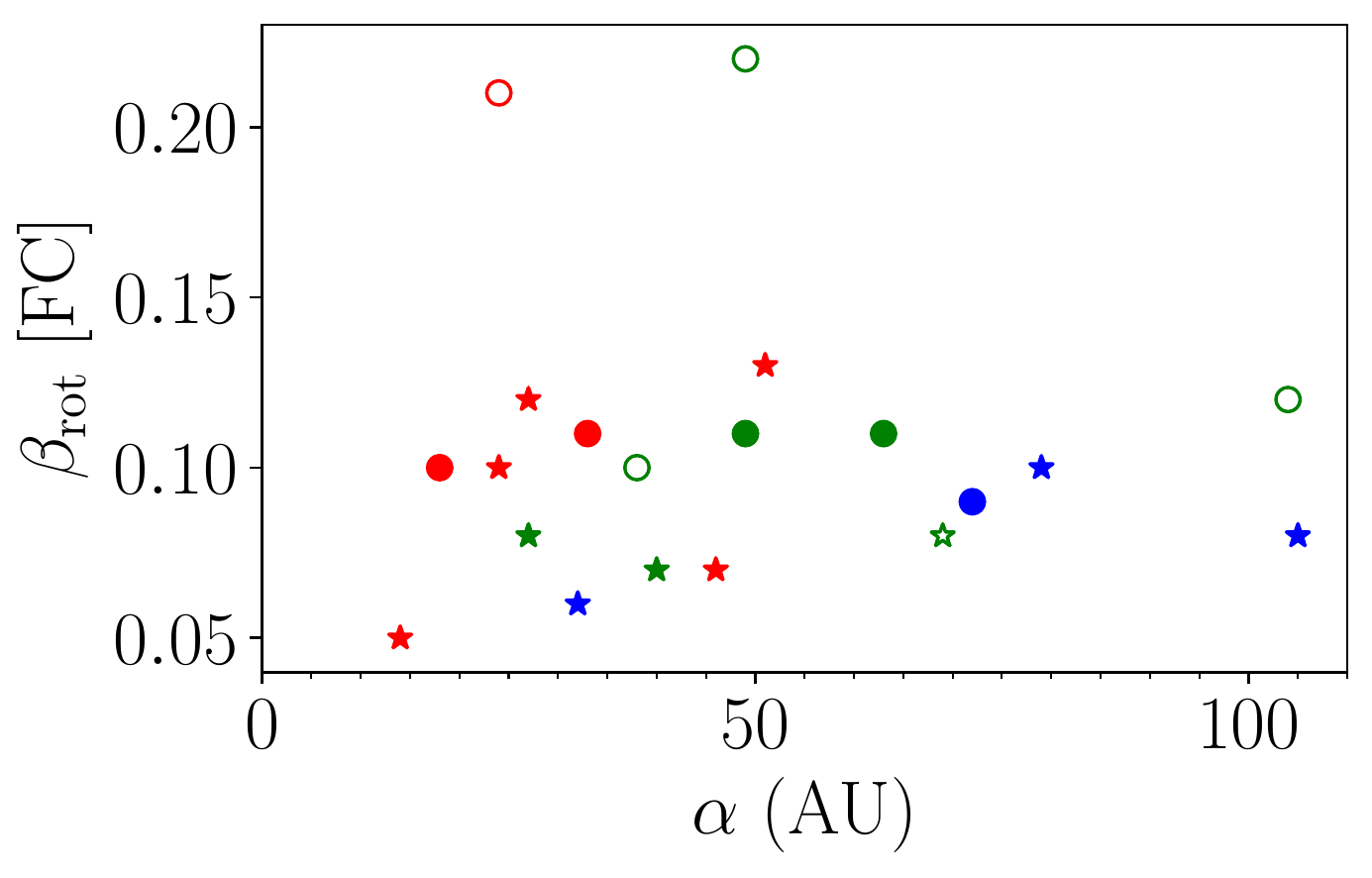}}} \\
  \subfloat{\resizebox{0.39\hsize}{!} 
  {\includegraphics{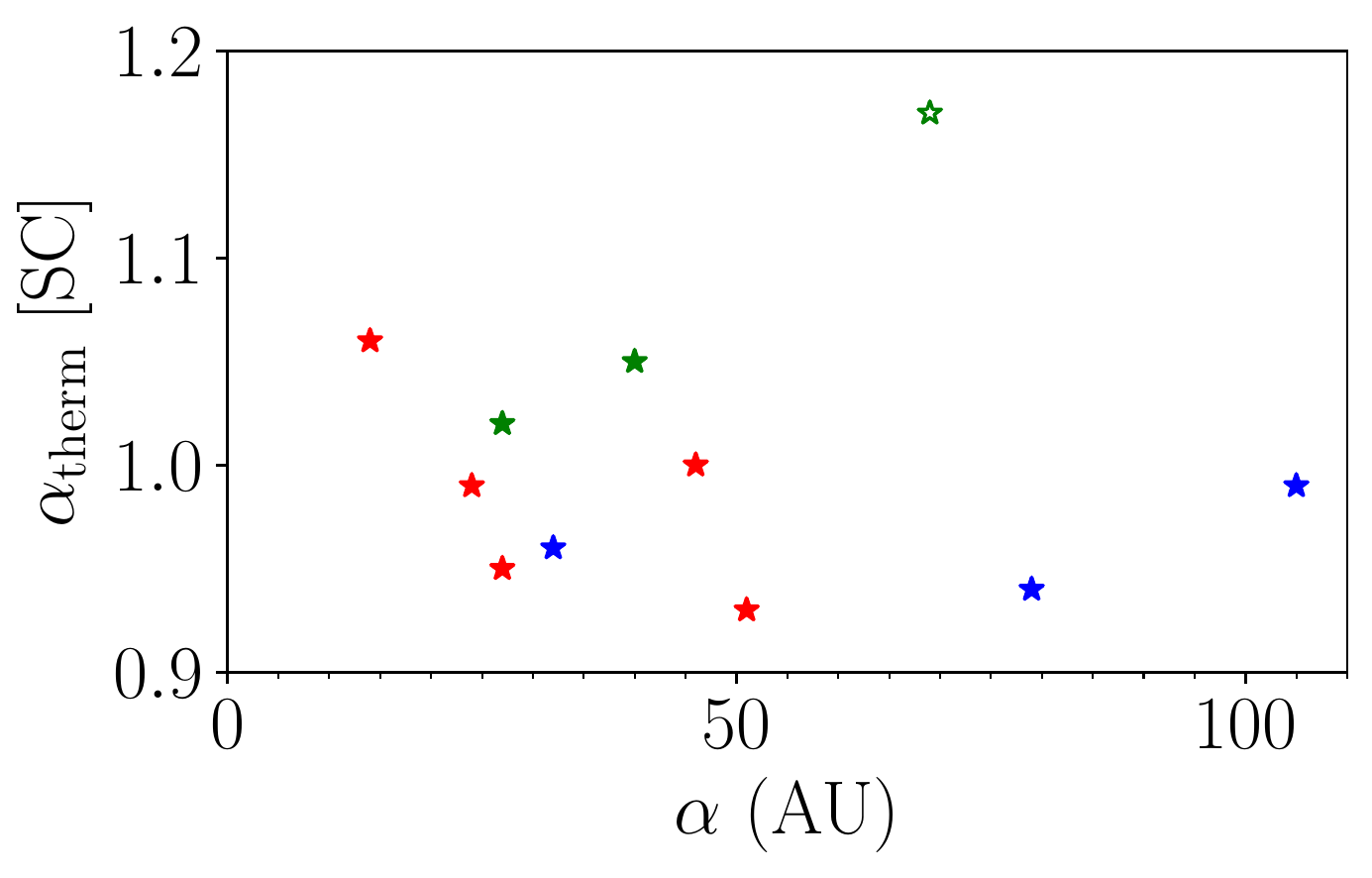}}}
  \subfloat{\resizebox{0.39\hsize}{!} 
  {\includegraphics{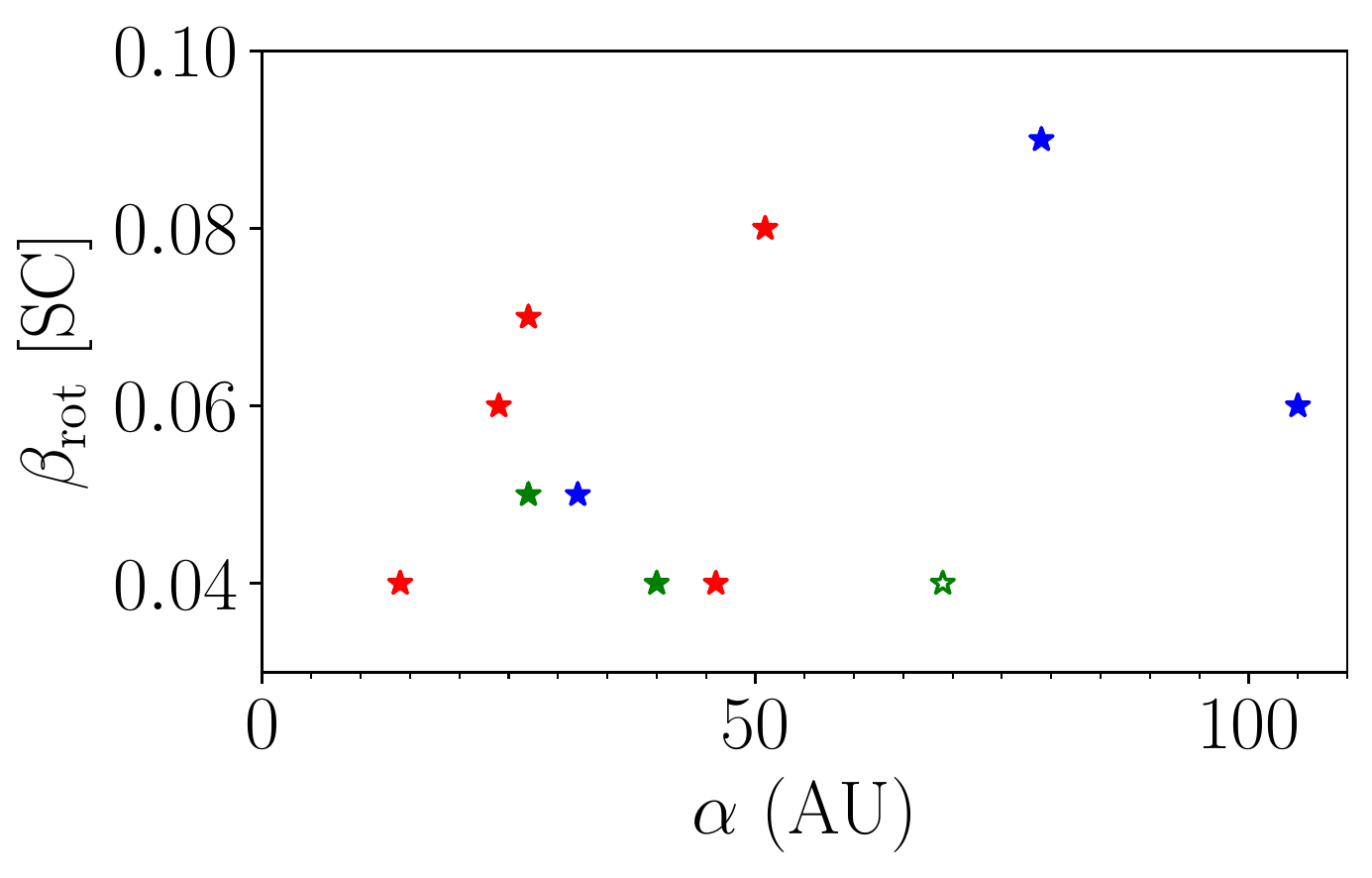}}}    
  \caption
  { Ratios of thermal-to-gravitational $\alpha_{\rm therm}={E_{\textup{ther}}}/{E_{\textup{grav}}}$ (left) and rotational-to-gravitational 
    $\beta_{\rm rot}={E_{\textup{rot}}}/{E_{\textup{grav}}}$ (right), for the first (top) and second (bottom) cores. Fragments that do not undergo a second collapse (open circles) tend to have high fractions of rotational energy. Symbols are the same as in Figure~\ref{fig:protoplanets-fc}.
} 
  \label{fig:protoplanets-ratios}
 \centering
     \resizebox{0.45\hsize}{!}{\includegraphics{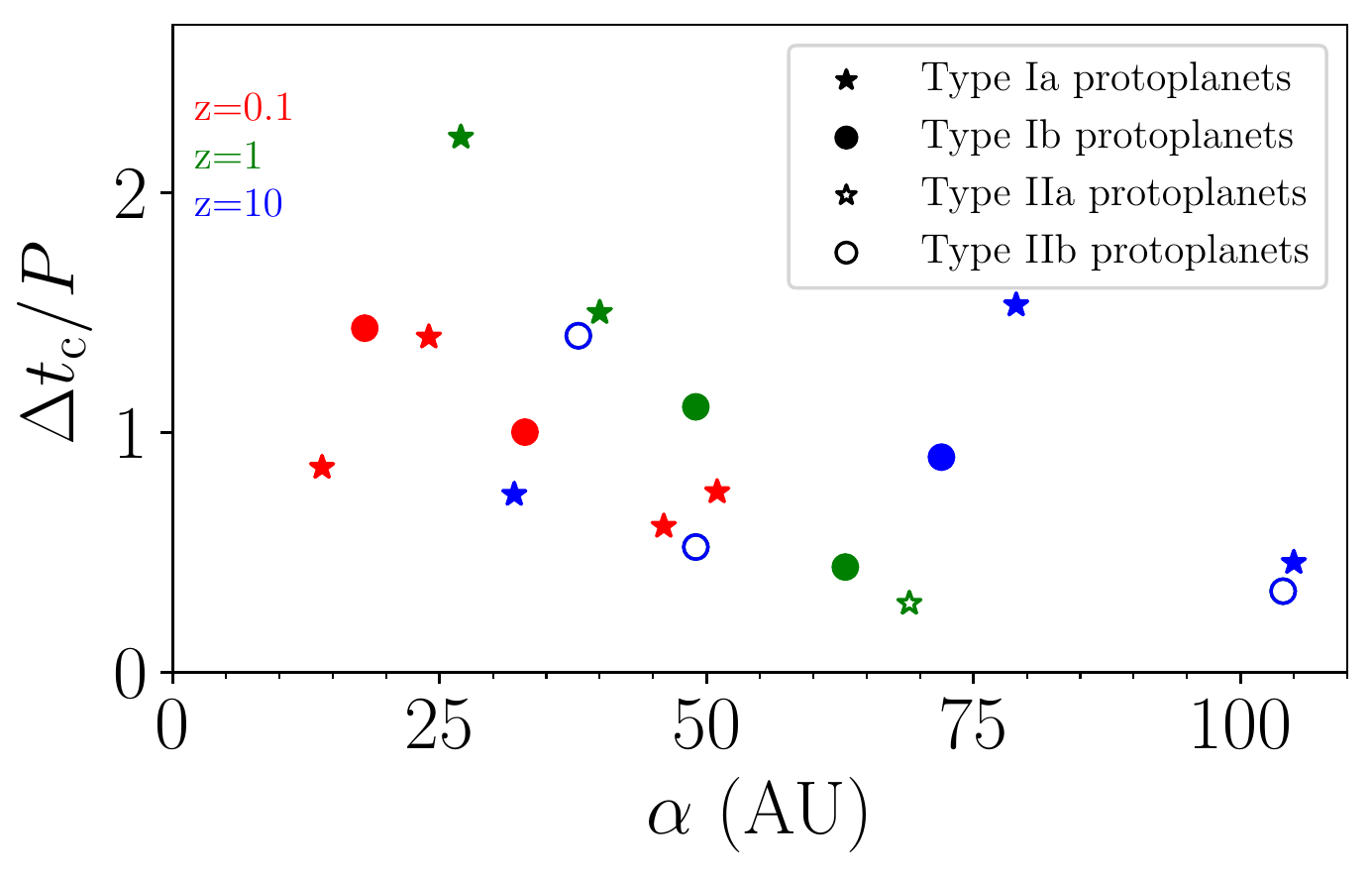}} 
    \caption
    {Time (in units of the local orbital period, P) it takes for a protoplanet to collapse from $10^{-9}{\rm g\ cm}^{-3}$ to its final central density ($10^{-3}{\rm g\ cm}^{-3}$ for Type I protoplanets). Symbols are the same as in Figure~\ref{fig:protoplanets-fc}.  
    } 
    \label{fig:alpha-time-p}
\end{figure*}

"Type I protoplanets" are defined as those protoplanets that undergo a second collapse (the temperature at their centre rises above $2,000$~K) reaching densities  $10^{-3} \textup{ g cm}^{-3}$ at their centres. Most of these protoplanets ({Type Ia protoplanets}) have a second core as it evidenced by an accretion shock (seen in the infall velocity profile, Figure~\ref{fig:properties_run5}d). These protoplanets are depicted by filled stars in Figures~\ref{fig:protoplanets-fc}-\ref{fig:alpha-time-p}. The radial profiles of their properties are shown in Figure~\ref{fig:typeIa}. A few of these protoplanets (Type Ib protoplanets) show no signature of a second core in the infall velocity profiles. These are depicted by filled circles in Figures~\ref{fig:protoplanets-fc}-\ref{fig:alpha-time-p} and the radial profiles of their properties are shown in Figure~\ref{fig:typeIb}. 

"Type II protoplanets" are defined as those protoplanets that do not reach density  $10^{-3} \textup{ g cm}^{-3}$ at their centres (at least for the time we follow their evolution). One of these protoplanets ({Type IIa protoplanets}; Run 13) undergoes a second collapse and shows evidence of a second core in the radial infall profile. This is depicted by an open star in Figures~\ref{fig:protoplanets-fc}-\ref{fig:alpha-time-p}. However,  the rest of these protoplanets do not undergo second collapse ({Type IIb protoplanets}). These are depicted by open circles in Figures~\ref{fig:protoplanets-fc}-\ref{fig:alpha-time-p}. The radial profiles of the properties of Type II  protoplanets are shown in Figure~\ref{fig:typeII}. It is expected that Type II protoplanets eventually evolve to Type I as more gas is accreted from the disc initiating the second collapse.

The properties of the protoplanets formed in the simulations are presented in
Table~\ref{tab:protoplanet_properties}. We define the boundaries of the first and second cores as the maxima in the infall velocity profiles. It is important to note that a few of the protoplanets (generally the ones that rotate faster) do not show clear velocity signatures of a second core (Type Ib; see discussion in Appendix~\ref{sec:appendix}, Figure~\ref{fig:v_ratio}). In the table we also list the number of SPH particles within the first core, which is indicative of how well the first core is resolved. Generally, each first core is represented by more than $\sim 10^5$ SPH particles, which ensures that its collapse is well-resolved up to densities of $10^{-3} \textup{ g cm}^{-3}$ \citep[see][]{Stamatellos:2007b}. Due to computational constraints (the timestep becomes too short) we are unable to follow the evolution of the protoplanets after the second collapse. Previous studies \citep[e.g.][]{Stamatellos:2009a} typically introduce sink particles to represent the protoplanets once they form. However, a more detailed treatment is required to accurately capture the internal evolution of these protoplanets as they interact with their parent disc in order to determine their final properties. Therefore, here we discuss only the initial properties of the protoplanets (i.e. when they form in the disc) and leave their subsequent evolution for a follow-up study.

Most of the protoplanets that form in the simulations presented here are Type I, that is,  they have undergone second collapse and have reached central densities of   $10^{-3} \textup{ g cm}^{-3}$. These protoplanets have reached high temperatures ($\sim6,000-12,000~$K; also seen in the lower-resolution simulations of \cite{Stamatellos:2009d}), and therefore correspond to the hot-start model of planet formation \citep[e.g.][]{Marley:2007a, Mordasini:2012a, Mordasini:2013a, Baruteau:2016a}. The estimated temperatures are similar to the temperatures of the accretion shock around planets formed by core accretion \citep{Marleau:2017a,Szulagyi:2017d, Szulagyi:2018a, Szulagyi:2017b,Marleau:2019a} and therefore their circumplanetary discs are also expected to be relatively hot. These high temperatures contradict the results of the disc instability model presented in \cite{Szulagyi:2017a}, as in the simulations presented here we were able to follow the collapse of a fragment at much higher densities  and capture the formation of the first and second core.

The first core masses are super-Jovian ($\stackrel{>}{_\sim}5\mjup$; see Figure~\ref{fig:protoplanets-fc}), and in some cases, are higher (up to $20 \mjup$)  than the deuterium burning mass limit of $\sim 13 \mjup$, that is, they are in the brown dwarf mass regime. They have radii between 1 and 10 AU, and in all cases their sizes are smaller than their corresponding Hill radii as expected (see Figure~\ref{fig:protoplanets-fc}, black crosses on the left graph). The first cores form at distances from 15 to 100 AU, that is on relatively wide orbits. The masses and radii of the first cores tend to increase with metallicity (see Figure~\ref{fig:protoplanets-fc-z}), although there is a rather considerable spread for each metallicity. This dependence is expected as at the high optical depth regime the cooling rate of the protoplanet decreases with increasing opacity and therefore the first core mass and radius increase \citep[e.g.][]{Masunaga:1998a,Masunaga:1999a,Masunaga:2000a}.

{ The second cores have masses  of the order of a few
Jupiter masses ($\sim 2-6\mjup$;  Figure~\ref{fig:protoplanets-sc}, left) and radii  of the order of a few solar radii ($\sim 5-9\;\rsun$; Figure~\ref{fig:protoplanets-sc}, right). These masses and radii are similar to the ones of second cores formed in  solar-mass collapsing cores 
\citep{Larson:1969a, Masunaga:2000a,Stamatellos:2007b,Tomida:2013a, Vaytet:2013a, Bate:2014a, Tsukamoto:2015a,Bhandare:2018a}. As with the first cores, the second core mass tends to be higher for higher metallicity, but there is no apparent relation between metallicity and the size of the second core (Figure~\ref{fig:protoplanets-sc-z}). }

In Figure~\ref{fig:protoplanets-sam} we plot the specific angular momenta of the first and second cores, and in Figure~\ref{fig:protoplanets-ratios} we plot the ratios of thermal-to-gravitational $\alpha_{\rm therm}={E_{\textup{ther}}}/{E_{\textup{grav}}}$ (left) and rotational-to-gravitational $\beta_{\rm rot}={E_{\textup{rot}}}/{E_{\textup{grav}}}$ (right), for the first (top) and second (bottom) cores. Fragments that do not undergo a second collapse (Type IIb protoplanets, open circles) or undergo a second collapse but without forming an accretion shock around the second core (Type Ia protoplanets, filled circles) tend to have high specific angular momentum and high rotational energy. This is similar to the behaviour of first and second cores forming in higher-mass (i.e. solar-mass) rotating cores \citep{Saigo:2006a,Saigo:2008a}. However, we note that these graphs depict these properties at  the final stage of the collapse. To determine the relation between fragment rotation and the presence or not of a second core, the pre-collapse properties of each fragment need to be examined and put in context with the movement of the fragment within the disc.  We will investigate this issue in a subsequent paper.

The time it takes a protoplanet to collapse from a central density of $10^{-9}{\rm g\ cm}^{-3}$ to its final central density ($10^{-3}{\rm g\ cm}^{-3}$ for Type I protoplanets) is shown in Figure~\ref{fig:alpha-time-p}. It varies from 0.3 to 1.5 times the local orbital period which allows for possible interactions (and maybe disruption) before a bound second core forms.

\clearpage

\section{Comparison with the observed properties of exoplanets around M dwarfs}
\label{sec:comparison_with_observations}

\begin{figure}
  \begin{center}
    \resizebox{\hsize}{!}{\includegraphics{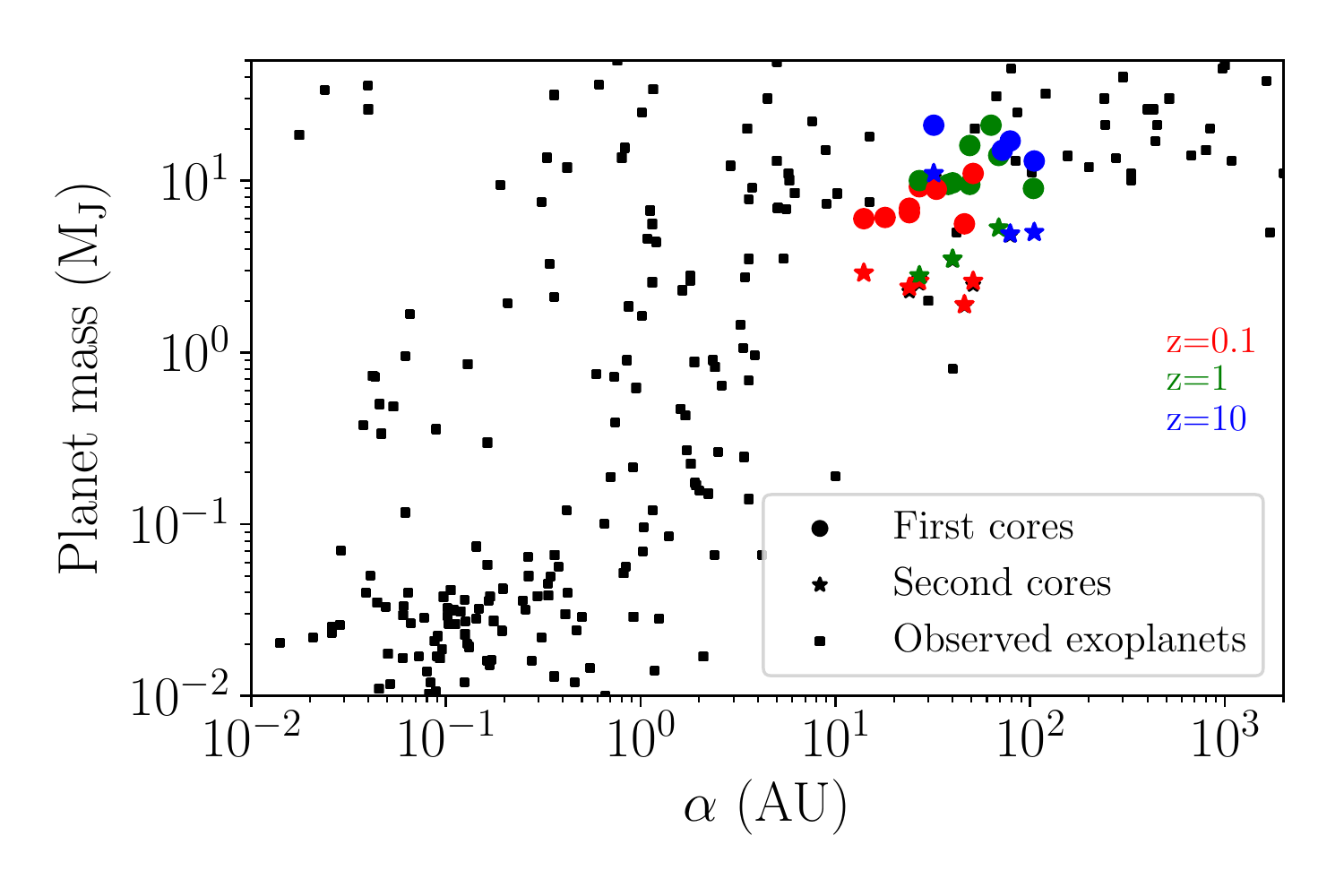}}
    \caption
    {
      Masses ($M_p \sin(i)$, where $i$ is the planet orbit orientation) of planets around M dwarfs ($M_{\star} < 0.5 \msun$) as a function of their semi-major axis. Black points correspond to the observed exoplanets. The coloured symbols correspond to the protoplanets formed in the simulations presented here. Circles correspond to first cores, whereas stars to the second cores. Colours correspond to different opacities (red: $z=0.1$, green: $z=1$, blue: $z=10$). As these protoplanets are still embedded within their protostellar discs, they may  migrate inwards or outwards, changing their final semi-major axis. Similarly, they may undergo gas accretion or tidal stripping, changing their final mass.
    }
    \label{fig:planet_comparison}
  \end{center}
\end{figure}

The initial masses of the protoplanets formed in our simulations by disc instability and their distances from their host star are shown in Figure \ref{fig:planet_comparison}, where they are compared against the properties of the observed exoplanets around M dwarfs. We plot the properties of both first and second cores. The disc instability exoplanets occupy the high-mass, wide-orbit region of the graph. Protoplanets formed through disc
instability have super-Jovian masses ($2 - 5 \mjup$) and orbit at distances $10 - 100$~AU.

These protoplanet properties are expected to  change due to interactions with the disc and with other protoplanets \citep{Forgan:2013a, Nayakshin:2017b, Nayakshin:2017a,Hall:2017a, Forgan:2018b, Stamatellos:2018a,Fletcher:2019a}. Protoplanets  may migrate inwards rapidly  until they open up a gap \citep{Stamatellos:2015a, Stamatellos:2018a}. Thereafter they may continue to migrate inwards slowly or start migrating outward s, if the edges of the gap within which the planet resides are gravitationally unstable. Additionally, stochastic migration of young protoplanets may happen due to gravitational interactions with other protoplanets in the disc \citep[e.g.][]{Veras:2012b}. The protoplanet mass may also increase significantly as they can accrete gas from the relatively massive disc \citep{Stamatellos:2018a}. This could increase the protoplanet's mass so that it may become a brown dwarf ($M> 13 \mjup$) or a hydrogen-burning star ($M> 80 \mjup$)
\citep{Stamatellos:2009a, Kratter:2010b, Zhu:2012a}. The gas accretion rate onto the protoplanet can be reduced if the planet is hot enough to  heat the neighbouring disc \citep[e.g.][]{Stamatellos:2018a, Mercer:2017a}. Alternatively, a protoplanet
may undergo tidal downsizing, that is, tidal stripping via disruption from another protoplanet  or  during migration, reducing its mass even potentially in the terrestrial mass regime
\citep{Nayakshin:2010a, Nayakshin:2011b,Humphries:2019a, Humphries:2019b}. More studies are needed to determine the final properties of protoplanets formed by disc instability \citep{Muller:2018a}, taking into account computational issues \citep{Fletcher:2019a}.

\section{Conclusions}
\label{sec:conclusions}

We have performed a set of hydrodynamic simulations of protostellar discs around
M dwarf  stars. We varied the initial stellar mass such that $M_{\star}
= [0.2, 0.3, 0.4] \msun$, as well as the initial disc radius,
$R_{\textup{out}} = [60, 90, 120]$~AU. Additionally, we investigated the effect
of metallicity, $z = [0.1, 1, 10]$. The discs that we studied were initially
stable, but their masses were steadily increased through the method of mass loading, which can be notionally thought as accretion from an envelope during the early stage of star and disc formation. Most of the discs eventually became gravitationally unstable, spiral arms developed, and in the majority of cases, a protoplanet formed via fragmentation. The formation of protoplanets happens fast on a dynamical timescale  (within 30 kyr). The density requirement for fragment formation was chosen to be $\rho > 10^{-9} \textup{ g cm}^{-3}$ that is, a threshold typically reached during gravitational collapse after the formation of the first hydrostatic core \citep{Larson:1969a}.  From
the simulations of discs that do fragment, we determined the minimum disc mass necessary for fragmentation to occur and  the properties of the resulting protoplanets.

The fragmentation of protostellar discs around M dwarfs requires a
  disc-to-star mass ratio of at least $q \sim 0.3$ for  smaller discs, increasing to $q \sim 0.6$ for larger discs. These mass ratios are relatively high. However, there are observed systems with planet-to-star mass ratio of $\sim0.2$ (see the exoplanet.eu database at https://exoplanet.eu), which confirms that the discs in which they have formed must have had at least 20\% the mass of their host stars. In fact this fraction could have been much higher, considering that there may be other planets in the system not yet detected and that a significant fraction of the disc mass is lost due to accretion onto the central star and due to disc winds. 

The mass at which a disc fragments increases with  the size of the
  disc and the mass of the central star. However, no fragmentation occurs for
  small discs (initial radius $R_{\textup{init}} = 60$~AU)  around  more massive M dwarfs (mass $0.4\msun$). This is likely due to rapid disc expansion because of  the formation of strong spiral features, combined with stronger rotational support to the smaller disc and  inefficient cooling closer to the central star.  This is in agreement with previous analytical  \citep{Whitworth:2006a} and numerical \citep[e.g.][]{Stamatellos:2009a, Mercer:2018a}  studies  that show that fragmentation can happen only in the outer regions of extended discs. We find that the optimal region for fragmentation around M dwarfs is around 50~AU, that is, closer to the host star than what expected for higher mass (e.g. solar-type) stars \citep[e.g.][]{Stamatellos:2009a}. We find that the small discs  (but still with size $>75$~AU) around lower mass M dwarfs are most susceptible to gravitational  fragmentation.

 The disc metallicity does not significantly affect the mass at which a disc fragments, but in some cases fragmentation may be suppressed. In the cases where the metallicity is an order of magnitude smaller, spiral arms take more time to fragment. When the metallicity is increased by an order of
  magnitude, spiral arms take longer to develop, and the disc may not undergo gravitational fragmentation at all  due to a period of rapid expansion combined with inefficient cooling.

\begin{figure}
  \begin{center}
    \resizebox{\hsize}{!}{\includegraphics{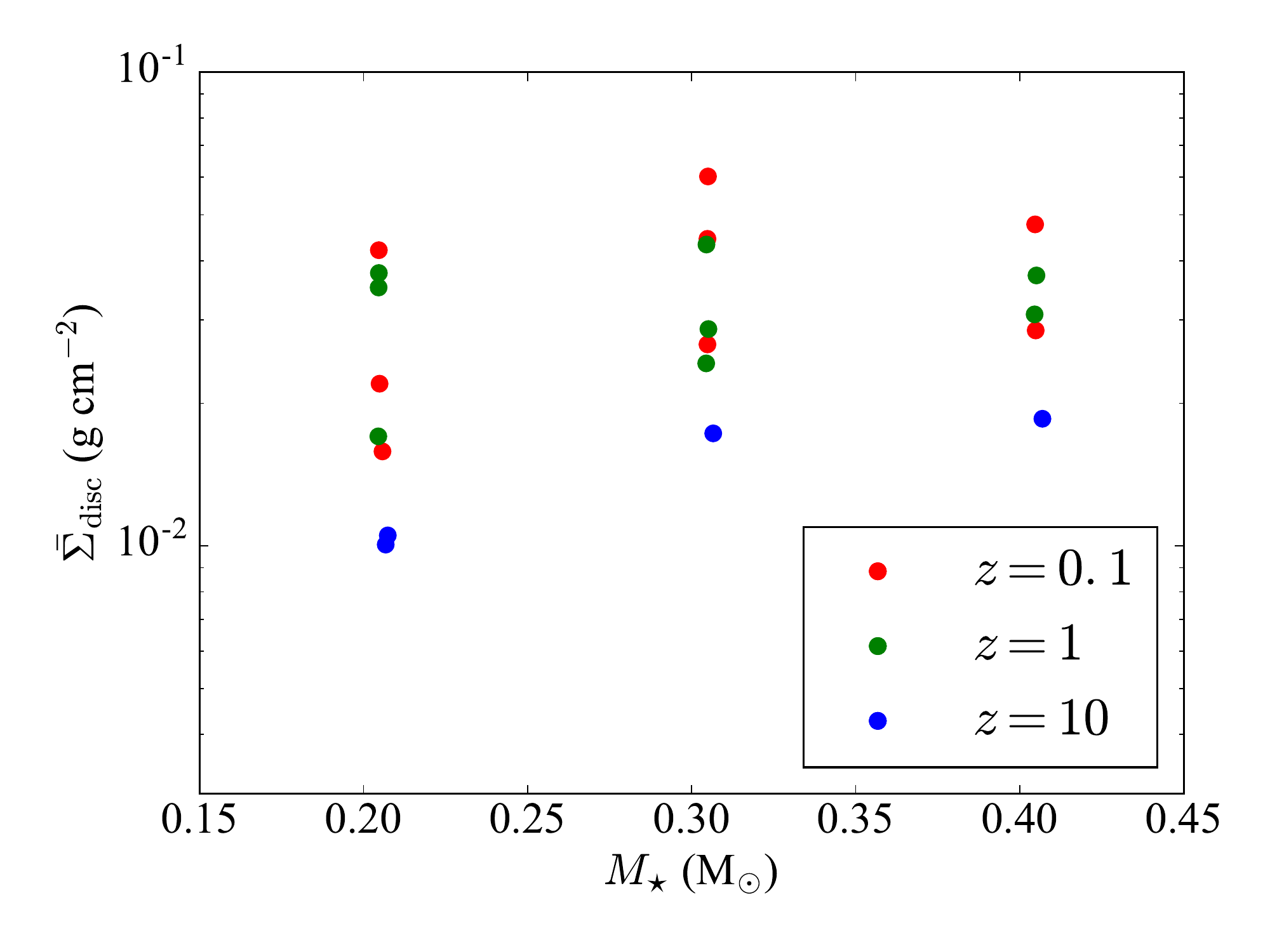}}
    \caption
    {
      Average surface density of the discs at the time they fragment where,
      $\bar{\Sigma}_{\textup{disc}} = M_{\textup{disc}} / \pi
      R_{\textup{disc}}^{2}$. We find that a lower average surface density is
      required for fragmentation when the disc metallicity is higher. 
    }
    \label{fig:sigma_stellar_mass}
  \end{center}
\end{figure}

  To facilitate comparisons with disc observations, we have calculated the average surface density $\bar{\Sigma}_{\textup{disc}}$ for the discs that fragment for a variety of stellar masses, shown in Figure \ref{fig:sigma_stellar_mass}. We find that disc fragmentation requires an average surface density $\bar{\Sigma}_{\textup{disc}}>0.01 \textup{ g cm}^{-2}$ and that higher metallicity discs can fragment at a lower average density (although small-sized, high-metallicity discs may not fragment due to a period of rapid expansion). We note however  that the minimum surface density needed for fragmentation does not increase monotonically with the metallicity (at least for the parameter space investigated in this paper).

  Protoplanets due to disc instability around M~dwarfs form very fast, on a dynamical timescale (within a few thousand years). Initially  they are massive ($2 - 6 \mjup$) and on wide orbits ($15  - 105$~AU). Those that form in high metallicity discs are more massive and form on initially wider orbits. However, both masses and orbital radii are expected to  evolve as the protoplanets interact with their discs; therefore, their long term evolution must be studied in order to compare these with the corresponding properties of the observed exoplanets around M dwarfs. All protoplanets formed in the simulations presented in this paper have similar density and temperature profiles, and possess significant rotational energy, which in some cases may delay or even suppress the second collapse of the protoplanet. Nevertheless, most of the exoplanets undergo the second collapse phase and therefore attain high central temperatures (6,000-12,000~K). These temperatures are similar to the temperatures at the accretion shocks around planets formed by core accretion \citep{Szulagyi:2017a, Marleau:2019a}, therefore the temperature alone cannot provide a way to distinguish between these two formation scenarios.    
  
  We conclude that disc instability may be a viable way to quickly form gas giant planets on wide orbits around M~dwarfs that are difficult to form by core accretion, provided that discs around M~dwarfs have significant mass when compared to the mass of their host star. Future observations of massive young discs embedded in their parental clouds or of planets that have formed fast around very young proto- M dwarfs could provide evidence that disc instability occurs.  Wide orbit planets formed in this way may migrate inwards or outwards, contributing  to the observed population of planets around M dwarfs at various orbital radii.

\begin{acknowledgements}
\label{sec:acknowledgements}
We would like to thank A. Bhandare, C. Hall and S. Nayaskshin  for useful discussions, and the anonymous referee for his/her suggestions. Surface density plots were produced using the \textsc{splash} software package
\citep{Price:2007b}. AM is supported by STFC grant ST/N504014/1. DS is partly
supported by STFC grant ST/M000877/1. This work used the DiRAC Complexity
system, operated by the University of Leicester IT Services, which forms part of
the STFC DiRAC HPC Facility ({\url{http://www.dirac.ac.uk}}). This equipment is
funded by BIS National E-Infrastructure capital grant ST/K000373/1 and STFC
DiRAC Operations grant ST/K0003259/1. DiRAC is part of the UK National
E-Infrastructure.

\end{acknowledgements}

\bibpunct{(}{)}{;}{a}{}{,}
\bibliographystyle{aa}
\bibliography{mddf}

\begin{thebibliography}{122}
\expandafter\ifx\csname natexlab\endcsname\relax\def\natexlab#1{#1}\fi

\bibitem[{{ALMA Partnership} {et~al.}(2015){ALMA Partnership}, {Brogan},
  {P{\'e}rez}, {Hunter}, {Dent}, {Hales}, {Hills}, {Corder}, {Fomalont},
  {Vlahakis}, {Asaki}, {Barkats}, {Hirota}, {Hodge}, {Impellizzeri}, {Kneissl},
  {Liuzzo}, {Lucas}, {Marcelino}, {Matsushita}, {Nakanishi}, {Phillips},
  {Richards}, {Toledo}, {Aladro}, {Broguiere}, {Cortes}, {Cortes}, {Espada},
  {Galarza}, {Garcia-Appadoo}, {Guzman-Ramirez}, {Humphreys}, {Jung}, {Kameno},
  {Laing}, {Leon}, {Marconi}, {Mignano}, {Nikolic}, {Nyman}, {Radiszcz},
  {Remijan}, {Rod{\'o}n}, {Sawada}, {Takahashi}, {Tilanus}, {Vila Vilaro},
  {Watson}, {Wiklind}, {Akiyama}, {Chapillon}, {de Gregorio-Monsalvo}, {Di
  Francesco}, {Gueth}, {Kawamura}, {Lee}, {Nguyen Luong}, {Mangum}, {Pietu},
  {Sanhueza}, {Saigo}, {Takakuwa}, {Ubach}, {van Kempen}, {Wootten},
  {Castro-Carrizo}, {Francke}, {Gallardo}, {Garcia}, {Gonzalez}, {Hill},
  {Kaminski}, {Kurono}, {Liu}, {Lopez}, {Morales}, {Plarre}, {Schieven},
  {Testi}, {Videla}, {Villard}, {Andreani}, {Hibbard}, \&
  {Tatematsu}}]{ALMA-partnership:2015a}
{ALMA Partnership}, {Brogan}, C.~L., {P{\'e}rez}, L.~M., {et~al.} 2015, \apjl,
  808, L3

\bibitem[{{Andrews} {et~al.}(2013){Andrews}, {Rosenfeld}, {Kraus}, \&
  {Wilner}}]{Andrews:2013a}
{Andrews}, S.~M., {Rosenfeld}, K.~A., {Kraus}, A.~L., \& {Wilner}, D.~J. 2013,
  \apj, 771, 129

\bibitem[{Andrews {et~al.}(2009)Andrews, Wilner, Hughes, Qi, \&
  Dullemond}]{Andrews:2009b}
Andrews, S.~M., Wilner, D.~J., Hughes, A.~M., Qi, C., \& Dullemond, C.~P. 2009,
  \apj, 700, 1502

\bibitem[{{Ansdell} {et~al.}(2017){Ansdell}, {Williams}, {Manara}, {Miotello},
  {Facchini}, {van der Marel}, {Testi}, \& {van Dishoeck}}]{Ansdell:2017a}
{Ansdell}, M., {Williams}, J.~P., {Manara}, C.~F., {et~al.} 2017, \aj, 153, 240

\bibitem[{{Backus} \& {Quinn}(2016)}]{Backus:2016a}
{Backus}, I. \& {Quinn}, T. 2016, \mnras, 463, 2480

\bibitem[{Baron {et~al.}(2018)Baron, Artigau, Rameau, Lafreni{\`{e}}re,
  Gagn{\'{e}}, Malo, Albert, Naud, Doyon, Janson, Delorme, \&
  Beichman}]{Baron:2018a}
Baron, F., Artigau, {\'{E}}., Rameau, J., {et~al.} 2018, The Astronomical
  Journal, 156, 137

\bibitem[{{Baruteau} {et~al.}(2016){Baruteau}, {Bai}, {Mordasini}, \&
  {Molli{\`e}re}}]{Baruteau:2016a}
{Baruteau}, C., {Bai}, X., {Mordasini}, C., \& {Molli{\`e}re}, P. 2016, \ssr,
  205, 77

\bibitem[{{Bate}(2014)}]{Bate:2014a}
{Bate}, M.~R. 2014, \mnras, 442, 285

\bibitem[{Bate \& Burkert(1997)}]{Bate:1997a}
Bate, M.~R. \& Burkert, A. 1997, \mnras, 288, 1060, (c) 1997 The Royal
  Astronomical Society

\bibitem[{Bell \& Lin(1994)}]{Bell:1994a}
Bell, K.~R. \& Lin, D. N.~C. 1994, \apj, 427, 987

\bibitem[{{Bhandare} {et~al.}(2018){Bhandare}, {Kuiper}, {Henning}, {Fendt},
  {Marleau}, \& {K{\"o}lligan}}]{Bhandare:2018a}
{Bhandare}, A., {Kuiper}, R., {Henning}, T., {et~al.} 2018, \aap, 618, A95

\bibitem[{{Bonfils} {et~al.}(2013){Bonfils}, {Lo Curto}, {Correia}, {Laskar},
  {Udry}, {Delfosse}, {Forveille}, {Astudillo-Defru}, {Benz}, {Bouchy},
  {Gillon}, {H{\'e}brard}, {Lovis}, {Mayor}, {Moutou}, {Naef}, {Neves}, {Pepe},
  {Perrier}, {Queloz}, {Santos}, \& {S{\'e}gransan}}]{Bonfils:2013a}
{Bonfils}, X., {Lo Curto}, G., {Correia}, A.~C.~M., {et~al.} 2013, \aap, 556,
  A110

\bibitem[{Boss(1997)}]{Boss:1997a}
Boss, A.~P. 1997, Science, 276, 1836

\bibitem[{Boss(2006)}]{Boss:2006a}
Boss, A.~P. 2006, \apj, 644, L79

\bibitem[{{Bowler}(2016)}]{Bowler:2016a}
{Bowler}, B.~P. 2016, \pasp, 128, 102001

\bibitem[{{Bowler} {et~al.}(2015){Bowler}, {Liu}, {Shkolnik}, \&
  {Tamura}}]{Bowler:2015b}
{Bowler}, B.~P., {Liu}, M.~C., {Shkolnik}, E.~L., \& {Tamura}, M. 2015, \apjs,
  216, 7

\bibitem[{{Bowler} \& {Nielsen}(2018)}]{Bowler:2018a}
{Bowler}, B.~P. \& {Nielsen}, E.~L. 2018, Occurrence Rates from Direct Imaging
  Surveys, 155

\bibitem[{{Brandt} {et~al.}(2014){Brandt}, {McElwain}, {Turner}, {Mede},
  {Spiegel}, {Kuzuhara}, {Schlieder}, {Wisniewski}, {Abe}, {Biller},
  {Brandner}, {Carson}, {Currie}, {Egner}, {Feldt}, {Golota}, {Goto}, {Grady},
  {Guyon}, {Hashimoto}, {Hayano}, {Hayashi}, {Hayashi}, {Henning}, {Hodapp},
  {Inutsuka}, {Ishii}, {Iye}, {Janson}, {Kandori}, {Knapp}, {Kudo}, {Kusakabe},
  {Kwon}, {Matsuo}, {Miyama}, {Morino}, {Moro-Mart{\'{\i}}n}, {Nishimura},
  {Pyo}, {Serabyn}, {Suto}, {Suzuki}, {Takami}, {Takato}, {Terada}, {Thalmann},
  {Tomono}, {Watanabe}, {Yamada}, {Takami}, {Usuda}, \&
  {Tamura}}]{Brandt:2014a}
{Brandt}, T.~D., {McElwain}, M.~W., {Turner}, E.~L., {et~al.} 2014, \apj, 794,
  159

\bibitem[{{Cai} {et~al.}(2017){Cai}, {Kouwenhoven}, {Portegies Zwart}, \&
  {Spurzem}}]{Cai:2017a}
{Cai}, M.~X., {Kouwenhoven}, M.~B.~N., {Portegies Zwart}, S.~F., \& {Spurzem},
  R. 2017, \mnras, 470, 4337

\bibitem[{{Cameron}(1978)}]{Cameron:1978a}
{Cameron}, A.~G.~W. 1978, Moon and Planets, 18, 5

\bibitem[{Chabrier(2003)}]{Chabrier:2003a}
Chabrier, G. 2003, The Publications of the Astronomical Society of the Pacific,
  115, 763

\bibitem[{{Cieza} {et~al.}(2007){Cieza}, {Padgett}, {Stapelfeldt}, {Augereau},
  {Harvey}, {Evans}, {Mer{\'{\i}}n}, {Koerner}, {Sargent}, {van Dishoeck},
  {Allen}, {Blake}, {Brooke}, {Chapman}, {Huard}, {Lai}, {Mundy}, {Myers},
  {Spiesman}, \& {Wahhaj}}]{Cieza:2007a}
{Cieza}, L., {Padgett}, D.~L., {Stapelfeldt}, K.~R., {et~al.} 2007, \apj, 667,
  308

\bibitem[{Cullen \& Dehnen(2010)}]{Cullen:2010a}
Cullen, L. \& Dehnen, W. 2010, \mnras, 408, 669

\bibitem[{{Dipierro} {et~al.}(2015){Dipierro}, {Price}, {Laibe}, {Hirsh},
  {Cerioli}, \& {Lodato}}]{Dipierro:2015b}
{Dipierro}, G., {Price}, D., {Laibe}, G., {et~al.} 2015, \mnras, 453, L73

\bibitem[{{Dunham} {et~al.}(2014){Dunham}, {Vorobyov}, \&
  {Arce}}]{Dunham:2014b}
{Dunham}, M.~M., {Vorobyov}, E.~I., \& {Arce}, H.~G. 2014, \mnras, 444, 887

\bibitem[{{Durkan} {et~al.}(2016){Durkan}, {Janson}, \&
  {Carson}}]{Durkan:2016a}
{Durkan}, S., {Janson}, M., \& {Carson}, J.~C. 2016, \apj, 824, 58

\bibitem[{{Fletcher} {et~al.}(2019){Fletcher}, {Nayakshin}, {Stamatellos},
  {Dehnen}, {Meru}, {Mayer}, {Deng}, \& {Rice}}]{Fletcher:2019a}
{Fletcher}, M., {Nayakshin}, S., {Stamatellos}, D., {et~al.} 2019, \mnras, 486,
  4398

\bibitem[{{Forgan} \& {Rice}(2013)}]{Forgan:2013a}
{Forgan}, D. \& {Rice}, K. 2013, \mnras, 432, 3168

\bibitem[{Forgan {et~al.}(2009)Forgan, Rice, Stamatellos, \&
  Whitworth}]{Forgan:2009b}
Forgan, D., Rice, K., Stamatellos, D., \& Whitworth, A.~P. 2009, \mnras, 394,
  882

\bibitem[{{Forgan} {et~al.}(2018){Forgan}, {Hall}, {Meru}, \&
  {Rice}}]{Forgan:2018b}
{Forgan}, D.~H., {Hall}, C., {Meru}, F., \& {Rice}, W.~K.~M. 2018, \mnras, 474,
  5036

\bibitem[{{Galicher} {et~al.}(2016){Galicher}, {Marois}, {Macintosh},
  {Zuckerman}, {Barman}, {Konopacky}, {Song}, {Patience}, {Lafreni{\`e}re},
  {Doyon}, \& {Nielsen}}]{Galicher:2016a}
{Galicher}, R., {Marois}, C., {Macintosh}, B., {et~al.} 2016, \aap, 594, A63

\bibitem[{Gammie(2001)}]{Gammie:2001a}
Gammie, C.~F. 2001, \apj, 553, 174

\bibitem[{Goldreich \& Ward(1973)}]{Goldreich:1973a}
Goldreich, P. \& Ward, W.~R. 1973, \apj, 183, 1051

\bibitem[{Greaves \& Rice(2010)}]{Greaves:2010a}
Greaves, J.~S. \& Rice, W. K.~M. 2010, \mnras, 407, 1981

\bibitem[{{Greenberg} {et~al.}(1978){Greenberg}, {Wacker}, {Hartmann}, \&
  {Chapman}}]{Greenberg:1978a}
{Greenberg}, R., {Wacker}, J.~F., {Hartmann}, W.~K., \& {Chapman}, C.~R. 1978,
  \icarus, 35, 1

\bibitem[{{Haisch} {et~al.}(2001){Haisch}, {Lada}, \& {Lada}}]{Haisch:2001a}
{Haisch}, Jr., K.~E., {Lada}, E.~A., \& {Lada}, C.~J. 2001, \apjl, 553, L153

\bibitem[{{Hall} {et~al.}(2017){Hall}, {Forgan}, \& {Rice}}]{Hall:2017a}
{Hall}, C., {Forgan}, D., \& {Rice}, K. 2017, \mnras, 470, 2517

\bibitem[{{Hao} {et~al.}(2013){Hao}, {Kouwenhoven}, \& {Spurzem}}]{Hao:2013a}
{Hao}, W., {Kouwenhoven}, M.~B.~N., \& {Spurzem}, R. 2013, \mnras, 433, 867

\bibitem[{{Hatzes} \& {Rauer}(2015)}]{Hatzes:2015a}
{Hatzes}, A.~P. \& {Rauer}, H. 2015, \apjl, 810, L25

\bibitem[{{Hayashi} {et~al.}(1985){Hayashi}, {Nakazawa}, \&
  {Nakagawa}}]{Hayashi:1985a}
{Hayashi}, C., {Nakazawa}, K., \& {Nakagawa}, Y. 1985, in Protostars and
  Planets II, ed. D.~C. {Black} \& M.~S. {Matthews}, 1100--1153

\bibitem[{{Hennebelle} {et~al.}(2016){Hennebelle}, {Lesur}, \&
  {Fromang}}]{Hennebelle:2016a}
{Hennebelle}, P., {Lesur}, G., \& {Fromang}, S. 2016, \aap, 590, A22

\bibitem[{{Hubber} {et~al.}(2018){Hubber}, {Rosotti}, \&
  {Booth}}]{Hubber:2018a}
{Hubber}, D.~A., {Rosotti}, G.~P., \& {Booth}, R.~A. 2018, \mnras, 473, 1603

\bibitem[{{Humphries} \& {Nayakshin}(2019)}]{Humphries:2019b}
{Humphries}, J. \& {Nayakshin}, S. 2019, \mnras, 489, 5187

\bibitem[{{Humphries} {et~al.}(2019){Humphries}, {Vazan}, {Bonavita}, {Helled},
  \& {Nayakshin}}]{Humphries:2019a}
{Humphries}, J., {Vazan}, A., {Bonavita}, M., {Helled}, R., \& {Nayakshin}, S.
  2019, \mnras, 488, 4873

\bibitem[{Johnson \& Gammie(2003)}]{Johnson:2003a}
Johnson, B.~M. \& Gammie, C.~F. 2003, \apj, 597, 131

\bibitem[{{Kratter} \& {Lodato}(2016)}]{Kratter:2016a}
{Kratter}, K. \& {Lodato}, G. 2016, \araa, 54, 271

\bibitem[{{Kratter} {et~al.}(2010){Kratter}, {Matzner}, {Krumholz}, \&
  {Klein}}]{Kratter:2010a}
{Kratter}, K.~M., {Matzner}, C.~D., {Krumholz}, M.~R., \& {Klein}, R.~I. 2010,
  \apj, 708, 1585

\bibitem[{Kratter {et~al.}(2010)Kratter, Murray-Clay, \&
  Youdin}]{Kratter:2010b}
Kratter, K.~M., Murray-Clay, R.~A., \& Youdin, A.~N. 2010, \apj, 710, 1375

\bibitem[{Kroupa(2001)}]{Kroupa:2001a}
Kroupa, P. 2001, \mnras, 322, 231

\bibitem[{Kuiper(1951)}]{Kuiper:1951a}
Kuiper, G.~P. 1951, Proceedings of the National Academy of Sciences of the
  United States of America, 37, 1

\bibitem[{{Lambrechts} \& {Johansen}(2012)}]{Lambrechts:2012a}
{Lambrechts}, M. \& {Johansen}, A. 2012, \aap, 544, A32

\bibitem[{{Lannier} {et~al.}(2016){Lannier}, {Delorme}, {Lagrange}, {Borgniet},
  {Rameau}, {Schlieder}, {Gagn{\'e}}, {Bonavita}, {Malo}, {Chauvin},
  {Bonnefoy}, \& {Girard}}]{Lannier:2016a}
{Lannier}, J., {Delorme}, P., {Lagrange}, A.~M., {et~al.} 2016, \aap, 596, A83

\bibitem[{{Larson}(1969)}]{Larson:1969a}
{Larson}, R.~B. 1969, \mnras, 145, 271

\bibitem[{{Li} {et~al.}(2015){Li}, {Kouwenhoven}, {Stamatellos}, \&
  {Goodwin}}]{Li:2015b}
{Li}, Y., {Kouwenhoven}, M.~B.~N., {Stamatellos}, D., \& {Goodwin}, S.~P. 2015,
  \apj, 805, 116

\bibitem[{{Li} {et~al.}(2016){Li}, {Kouwenhoven}, {Stamatellos}, \&
  {Goodwin}}]{Li:2016a}
{Li}, Y., {Kouwenhoven}, M.~B.~N., {Stamatellos}, D., \& {Goodwin}, S.~P. 2016,
  \apj, 831, 166

\bibitem[{{Lin} \& {Pringle}(1990)}]{Lin:1990a}
{Lin}, D.~N.~C. \& {Pringle}, J.~E. 1990, \apj, 358, 515

\bibitem[{{Lissauer}(1993)}]{Lissauer:1993a}
{Lissauer}, J.~J. 1993, \araa, 31, 129

\bibitem[{{Liu} {et~al.}(2019){Liu}, {Lambrechts}, {Johansen}, \&
  {Liu}}]{Liu:2019a}
{Liu}, B., {Lambrechts}, M., {Johansen}, A., \& {Liu}, F. 2019, arXiv e-prints
  [\eprint[arXiv]{1909.00759}]

\bibitem[{{Lombardi} {et~al.}(2015){Lombardi}, {McInally}, \&
  {Faber}}]{Lombardi:2015a}
{Lombardi}, J.~C., {McInally}, W.~G., \& {Faber}, J.~A. 2015, \mnras, 447, 25

\bibitem[{{MacFarlane} \& {Stamatellos}(2017)}]{Macfarlane:2017a}
{MacFarlane}, B.~A. \& {Stamatellos}, D. 2017, \mnras, 472, 3775

\bibitem[{{Manara} {et~al.}(2018){Manara}, {Morbidelli}, \&
  {Guillot}}]{Manara:2018a}
{Manara}, C.~F., {Morbidelli}, A., \& {Guillot}, T. 2018, \aap, 618, L3

\bibitem[{{Marleau} {et~al.}(2017){Marleau}, {Klahr}, {Kuiper}, \&
  {Mordasini}}]{Marleau:2017a}
{Marleau}, G.-D., {Klahr}, H., {Kuiper}, R., \& {Mordasini}, C. 2017, \apj,
  836, 221

\bibitem[{{Marleau} {et~al.}(2019){Marleau}, {Mordasini}, \&
  {Kuiper}}]{Marleau:2019a}
{Marleau}, G.-D., {Mordasini}, C., \& {Kuiper}, R. 2019, \apj, 881, 144

\bibitem[{{Marley} {et~al.}(2007){Marley}, {Fortney}, {Hubickyj},
  {Bodenheimer}, \& {Lissauer}}]{Marley:2007a}
{Marley}, M.~S., {Fortney}, J.~J., {Hubickyj}, O., {Bodenheimer}, P., \&
  {Lissauer}, J.~J. 2007, \apj, 655, 541

\bibitem[{Marois {et~al.}(2008)Marois, Macintosh, Barman, Zuckerman, Song,
  Patience, Lafreni{\`e}re, \& Doyon}]{Marois:2008a}
Marois, C., Macintosh, B., Barman, T., {et~al.} 2008, Science, 322, 1348

\bibitem[{Masunaga \& Inutsuka(1999)}]{Masunaga:1999a}
Masunaga, H. \& Inutsuka, S. 1999, \apj, 510, 822

\bibitem[{Masunaga \& Inutsuka(2000)}]{Masunaga:2000a}
Masunaga, H. \& Inutsuka, S. 2000, \apj, 531, 350

\bibitem[{Masunaga {et~al.}(1998)Masunaga, Miyama, \&
  Inutsuka}]{Masunaga:1998a}
Masunaga, H., Miyama, S.~M., \& Inutsuka, S. 1998, \apj, 495, 346

\bibitem[{{Mayor} \& {Queloz}(1995)}]{Mayor:1995a}
{Mayor}, M. \& {Queloz}, D. 1995, \nat, 378, 355

\bibitem[{{Mercer} \& {Stamatellos}(2017)}]{Mercer:2017a}
{Mercer}, A. \& {Stamatellos}, D. 2017, \mnras, 465, 2

\bibitem[{{Mercer} {et~al.}(2018){Mercer}, {Stamatellos}, \&
  {Dunhill}}]{Mercer:2018a}
{Mercer}, A., {Stamatellos}, D., \& {Dunhill}, A. 2018, \mnras, 478, 3478

\bibitem[{{Mihalas}(1970)}]{Mihalas:1970a}
{Mihalas}, D. 1970, {Stellar atmospheres} (Series of Books in Astronomy and
  Astrophysics, San Francisco: Freeman)

\bibitem[{{Mohanty} {et~al.}(2013){Mohanty}, {Greaves}, {Mortlock}, {Pascucci},
  {Scholz}, {Thompson}, {Apai}, {Lodato}, \& {Looper}}]{Mohanty:2013a}
{Mohanty}, S., {Greaves}, J., {Mortlock}, D., {et~al.} 2013, \apj, 773, 168

\bibitem[{{Monaghan} \& {Lattanzio}(1985)}]{Monaghan:1985a}
{Monaghan}, J.~J. \& {Lattanzio}, J.~C. 1985, \aap, 149, 135

\bibitem[{{Morales} {et~al.}(2019){Morales}, {Mustill}, {Ribas}, {Davies},
  {Reiners}, {Bauer}, {Kossakowski}, {Herrero}, {Rodr{\'\i}guez},
  {L{\'o}pez-Gonz{\'a}lez}, {Rodr{\'\i}guez-L{\'o}pez}, {B{\'e}jar},
  {Gonz{\'a}lez-Cuesta}, {Luque}, {Pall{\'e}}, {Perger}, {Baroch}, {Johansen},
  {Klahr}, {Mordasini}, {Anglada-Escud{\'e}}, {Caballero},
  {Cort{\'e}s-Contreras}, {Dreizler}, {Lafarga}, {Nagel}, {Passegger},
  {Reffert}, {Rosich}, {Schweitzer}, {Tal-Or}, {Trifonov}, {Zechmeister},
  {Quirrenbach}, {Amado}, {Guenther}, {Hagen}, {Henning}, {Jeffers},
  {Kaminski}, {K{\"u}rster}, {Montes}, {Seifert}, {Abell{\'a}n}, {Abril},
  {Aceituno}, {Aceituno}, {Alonso-Floriano}, {Ammler-von Eiff}, {Antona},
  {Arroyo-Torres}, {Azzaro}, {Barrado}, {Becerril-Jarque}, {Ben{\'\i}tez},
  {Berdi{\~n}as}, {Bergond}, {Brinkm{\"o}ller}, {del Burgo}, {Burn},
  {Calvo-Ortega}, {Cano}, {C{\'a}rdenas}, {Guill{\'e}n}, {Carro}, {Casal},
  {Casanova}, {Casasayas-Barris}, {Chaturvedi}, {Cifuentes}, {Claret},
  {Colom{\'e}}, {Czesla}, {D{\'\i}ez-Alonso}, {Dorda}, {Emsenhuber},
  {Fern{\'a}ndez}, {Fern{\'a}ndez-Mart{\'\i}n}, {Ferro}, {Fuhrmeister},
  {Galad{\'\i}-Enr{\'\i}quez}, {Cava}, {Vargas}, {Garcia-Piquer}, {Gesa},
  {Gonz{\'a}lez-{\'A}lvarez}, {Hern{\'a}ndez}, {Gonz{\'a}lez-Peinado},
  {Gu{\`a}rdia}, {Guijarro}, {de Guindos}, {Hatzes}, {Hauschildt}, {Hedrosa},
  {Hermelo}, {Arabi}, {Otero}, {Hintz}, {Holgado}, {Huber}, {Huke}, {Johnson},
  {de Juan}, {Kehr}, {Kemmer}, {Kim}, {Kl{\"u}ter}, {Klutsch}, {Labarga},
  {Labiche}, {Lalitha}, {Lamp{\'o}n}, {Lara}, {Launhardt}, {L{\'a}zaro},
  {Lizon}, {Llamas}, {Lodieu}, {L{\'o}pez del Fresno}, {Salas},
  {L{\'o}pez-Santiago}, {Madinabeitia}, {Mall}, {Mancini}, {Mand el}, {Marfil},
  {Molina}, {Mart{\'\i}n}, {Mart{\'\i}n-Fern{\'a}ndez}, {Mart{\'\i}n-Ruiz},
  {Mart{\'\i}nez-Rodr{\'\i}guez}, {Marvin}, {Mirabet}, {Moya}, {Naranjo},
  {Nelson}, {Nortmann}, {Nowak}, {Ofir}, {Pascual}, {Pavlov}, {Pedraz},
  {Medialdea}, {P{\'e}rez-Calpena}, {Perryman}, {Rabaza}, {Ballesta}, {Rebolo},
  {Redondo}, {Rix}, {Rodler}, {Trinidad}, {Sabotta}, {Sadegi}, {Salz},
  {S{\'a}nchez-Blanco}, {Carrasco}, {S{\'a}nchez-L{\'o}pez}, {Sanz-Forcada},
  {Sarkis}, {Sarmiento}, {Sch{\"a}fer}, {Schlecker}, {Schmitt}, {Sch{\"o}fer},
  {Solano}, {Sota}, {Stahl}, {Stock}, {Stuber}, {St{\"u}rmer}, {Su{\'a}rez},
  {Tabernero}, {Tulloch}, {Veredas}, {Vico-Linares}, {Vilardell}, {Wagner},
  {Winkler}, {Wolthoff}, {Yan}, \& {Osorio}}]{Morales:2019a}
{Morales}, J.~C., {Mustill}, A.~J., {Ribas}, I., {et~al.} 2019, Science, 365,
  1441

\bibitem[{{Mordasini}(2013)}]{Mordasini:2013a}
{Mordasini}, C. 2013, \aap, 558, A113

\bibitem[{{Mordasini} {et~al.}(2012){Mordasini}, {Alibert}, {Klahr}, \&
  {Henning}}]{Mordasini:2012a}
{Mordasini}, C., {Alibert}, Y., {Klahr}, H., \& {Henning}, T. 2012, \aap, 547,
  A111

\bibitem[{{M{\"u}ller} {et~al.}(2018){M{\"u}ller}, {Helled}, \&
  {Mayer}}]{Muller:2018a}
{M{\"u}ller}, S., {Helled}, R., \& {Mayer}, L. 2018, \apj, 854, 112

\bibitem[{{Naud} {et~al.}(2017){Naud}, {Artigau}, {Doyon}, {Malo}, {Gagn{\'e}},
  {Lafreni{\`e}re}, {Wolf}, \& {Magnier}}]{Naud:2017a}
{Naud}, M.-E., {Artigau}, {\'E}., {Doyon}, R., {et~al.} 2017, \aj, 154, 129

\bibitem[{Nayakshin(2010)}]{Nayakshin:2010a}
Nayakshin, S. 2010, \mnras: Letters, 408, L36

\bibitem[{Nayakshin(2011)}]{Nayakshin:2011b}
Nayakshin, S. 2011, \mnras, 416, 2974

\bibitem[{{Nayakshin}(2017{\natexlab{a}})}]{Nayakshin:2017b}
{Nayakshin}, S. 2017{\natexlab{a}}, \mnras, 470, 2387

\bibitem[{{Nayakshin}(2017{\natexlab{b}})}]{Nayakshin:2017a}
{Nayakshin}, S. 2017{\natexlab{b}}, \pasa, 34, e002

\bibitem[{Nelson(2006)}]{Nelson:2006a}
Nelson, A.~F. 2006, \mnras, 373, 1039

\bibitem[{{Nielsen} {et~al.}(2019){Nielsen}, {De Rosa}, {Macintosh}, {Wang},
  {Ruffio}, {Chiang}, {Marley}, {Saumon}, {Savransky}, {Ammons}, {Bailey},
  {Barman}, {Blain}, {Bulger}, {Chilcote}, {Cotten}, {Czekala}, {Doyon},
  {Duchene}, {Esposito}, {Fabrycky}, {Fitzgerald}, {Follette}, {Fortney},
  {Gerard}, {Goodsell}, {Graham}, {Greenbaum}, {Hibon}, {Hinkley}, {Hirsch},
  {Hom}, {Hung}, {Dawson}, {Ingraham}, {Kalas}, {Konopacky}, {Larkin}, {Lee},
  {Lin}, {Maire}, {Marchis}, {Marois}, {Metchev}, {Millar-Blanchaer},
  {Morzinski}, {Oppenheimer}, {Palmer}, {Patience}, {Perrin}, {Poyneer},
  {Pueyo}, {Rafikov}, {Rajan}, {Rameau}, {Rantakyr o}, {Ren}, {Schneider},
  {Sivaramakrishnan}, {Song}, {Soummer}, {Tallis}, {Thomas}, {Ward-Duong}, \&
  {Wolff}}]{Nielsen:2019a}
{Nielsen}, E.~L., {De Rosa}, R.~J., {Macintosh}, B., {et~al.} 2019, arXiv
  e-prints [\eprint[arXiv]{1904.05358}]

\bibitem[{{P{\'e}rez} {et~al.}(2016){P{\'e}rez}, {Carpenter}, {Andrews},
  {Ricci}, {Isella}, {Linz}, {Sargent}, {Wilner}, {Henning}, {Deller},
  {Chandler}, {Dullemond}, {Lazio}, {Menten}, {Corder}, {Storm}, {Testi},
  {Tazzari}, {Kwon}, {Calvet}, {Greaves}, {Harris}, \& {Mundy}}]{Perez:2016a}
{P{\'e}rez}, L.~M., {Carpenter}, J.~M., {Andrews}, S.~M., {et~al.} 2016,
  Science, 353, 1519

\bibitem[{Price(2007)}]{Price:2007b}
Price, D.~J. 2007, Publications of the Astronomical Society of Australia, 24,
  159

\bibitem[{{Reggiani} {et~al.}(2016){Reggiani}, {Meyer}, {Chauvin}, {Vigan},
  {Quanz}, {Biller}, {Bonavita}, {Desidera}, {Delorme}, {Hagelberg}, {Maire},
  {Boccaletti}, {Beuzit}, {Buenzli}, {Carson}, {Covino}, {Feldt}, {Girard},
  {Gratton}, {Henning}, {Kasper}, {Lagrange}, {Mesa}, {Messina}, {Montagnier},
  {Mordasini}, {Mouillet}, {Schlieder}, {Segransan}, {Thalmann}, \&
  {Zurlo}}]{Reggiani:2016a}
{Reggiani}, M., {Meyer}, M.~R., {Chauvin}, G., {et~al.} 2016, \aap, 586, A147

\bibitem[{{Reiners} {et~al.}(2018){Reiners}, {Zechmeister}, {Caballero},
  {Ribas}, {Morales}, {Jeffers}, {Sch{\"o}fer}, {Tal-Or}, {Quirrenbach},
  {Amado}, {Kaminski}, {Seifert}, {Abril}, {Aceituno}, {Alonso-Floriano},
  {Ammler-von Eiff}, {Antona}, {Anglada-Escud{\'e}}, {Anwand-Heerwart},
  {Arroyo-Torres}, {Azzaro}, {Baroch}, {Barrado}, {Bauer}, {Becerril},
  {B{\'e}jar}, {Ben{\'{\i}}tez}, {Berdi{\~n}as}, {Bergond}, {Bl{\"u}mcke},
  {Brinkm{\"o}ller}, {del Burgo}, {Cano}, {C{\'a}rdenas V{\'a}zquez}, {Casal},
  {Cifuentes}, {Claret}, {Colom{\'e}}, {Cort{\'e}s-Contreras}, {Czesla},
  {D{\'{\i}}ez-Alonso}, {Dreizler}, {Feiz}, {Fern{\'a}ndez}, {Ferro},
  {Fuhrmeister}, {Galad{\'{\i}}-Enr{\'{\i}}quez}, {Garcia-Piquer},
  {Garc{\'{\i}}a Vargas}, {Gesa}, {G{\'o}mez Galera}, {Gonz{\'a}lez
  Hern{\'a}ndez}, {Gonz{\'a}lez-Peinado}, {Gr{\"o}zinger}, {Grohnert},
  {Gu{\`a}rdia}, {Guenther}, {Guijarro}, {de Guindos}, {Guti{\'e}rrez-Soto},
  {Hagen}, {Hatzes}, {Hauschildt}, {Hedrosa}, {Helmling}, {Henning}, {Hermelo},
  {Hern{\'a}ndez Arab{\'{\i}}}, {Hern{\'a}ndez Casta{\~n}o}, {Hern{\'a}ndez
  Hernando}, {Herrero}, {Huber}, {Huke}, {Johnson}, {de Juan}, {Kim}, {Klein},
  {Kl{\"u}ter}, {Klutsch}, {K{\"u}rster}, {Lafarga}, {Lamert}, {Lamp{\'o}n},
  {Lara}, {Laun}, {Lemke}, {Lenzen}, {Launhardt}, {L{\'o}pez del Fresno},
  {L{\'o}pez-Gonz{\'a}lez}, {L{\'o}pez-Puertas}, {L{\'o}pez Salas},
  {L{\'o}pez-Santiago}, {Luque}, {Mag{\'a}n Madinabeitia}, {Mall}, {Mancini},
  {Mandel}, {Marfil}, {Mar{\'{\i}}n Molina}, {Maroto Fern{\'a}ndez},
  {Mart{\'{\i}}n}, {Mart{\'{\i}}n-Ruiz}, {Marvin}, {Mathar}, {Mirabet},
  {Montes}, {Moreno-Raya}, {Moya}, {Mundt}, {Nagel}, {Naranjo}, {Nortmann},
  {Nowak}, {Ofir}, {Oreiro}, {Pall{\'e}}, {Panduro}, {Pascual}, {Passegger},
  {Pavlov}, {Pedraz}, {P{\'e}rez-Calpena}, {P{\'e}rez Medialdea}, {Perger},
  {Perryman}, {Pluto}, {Rabaza}, {Ram{\'o}n}, {Rebolo}, {Redondo}, {Reffert},
  {Reinhart}, {Rhode}, {Rix}, {Rodler}, {Rodr{\'{\i}}guez},
  {Rodr{\'{\i}}guez-L{\'o}pez}, {Rodr{\'{\i}}guez Trinidad}, {Rohloff},
  {Rosich}, {Sadegi}, {S{\'a}nchez-Blanco}, {S{\'a}nchez Carrasco},
  {S{\'a}nchez-L{\'o}pez}, {Sanz-Forcada}, {Sarkis}, {Sarmiento},
  {Sch{\"a}fer}, {Schmitt}, {Schiller}, {Schweitzer}, {Solano}, {Stahl},
  {Strachan}, {St{\"u}rmer}, {Su{\'a}rez}, {Tabernero}, {Tala}, {Trifonov},
  {Tulloch}, {Ulbrich}, {Veredas}, {Vico Linares}, {Vilardell}, {Wagner},
  {Winkler}, {Wolthoff}, {Xu}, {Yan}, \& {Zapatero Osorio}}]{Reiners:2018b}
{Reiners}, A., {Zechmeister}, M., {Caballero}, J.~A., {et~al.} 2018, \aap, 612,
  A49

\bibitem[{Rice {et~al.}(2003)Rice, Armitage, Bonnell, Bate, Jeffers, \&
  Vine}]{Rice:2003a}
Rice, W. K.~M., Armitage, P.~J., Bonnell, I.~A., {et~al.} 2003, \mnras, 346,
  L36

\bibitem[{Rice {et~al.}(2005)Rice, Lodato, \& Armitage}]{Rice:2005a}
Rice, W. K.~M., Lodato, G., \& Armitage, P.~J. 2005, \mnras: Letters, 364, L56

\bibitem[{Safronov \& Zvjagina(1969)}]{Safronov:1969a}
Safronov, V.~S. \& Zvjagina, E.~V. 1969, Icarus, 10, 109

\bibitem[{{Saigo} \& {Tomisaka}(2006)}]{Saigo:2006a}
{Saigo}, K. \& {Tomisaka}, K. 2006, \apj, 645, 381

\bibitem[{{Saigo} {et~al.}(2008){Saigo}, {Tomisaka}, \&
  {Matsumoto}}]{Saigo:2008a}
{Saigo}, K., {Tomisaka}, K., \& {Matsumoto}, T. 2008, \apj, 674, 997

\bibitem[{{Schneider} {et~al.}(2011){Schneider}, {Dedieu}, {Le Sidaner},
  {Savalle}, \& {Zolotukhin}}]{Schneider:2011a}
{Schneider}, J., {Dedieu}, C., {Le Sidaner}, P., {Savalle}, R., \&
  {Zolotukhin}, I. 2011, \aap, 532, A79

\bibitem[{Schoenberg(1946)}]{Schoenberg:1946}
Schoenberg, I.~J. 1946, Quarterly of Applied Mathematics, 4, 45

\bibitem[{{Springel} \& {Hernquist}(2002)}]{Springel:2002a}
{Springel}, V. \& {Hernquist}, L. 2002, \mnras, 333, 649

\bibitem[{{Stamatellos}(2015)}]{Stamatellos:2015a}
{Stamatellos}, D. 2015, \apjl, 810, L11

\bibitem[{{Stamatellos} \& {Herczeg}(2015)}]{Stamatellos:2015b}
{Stamatellos}, D. \& {Herczeg}, G.~J. 2015, \mnras, 449, 3432

\bibitem[{Stamatellos {et~al.}(2007{\natexlab{a}})Stamatellos, Hubber, \&
  Whitworth}]{Stamatellos:2007c}
Stamatellos, D., Hubber, D.~A., \& Whitworth, A.~P. 2007{\natexlab{a}}, \mnras:
  Letters, 382, L30

\bibitem[{{Stamatellos} \& {Inutsuka}(2018)}]{Stamatellos:2018a}
{Stamatellos}, D. \& {Inutsuka}, S.-i. 2018, \mnras, 477, 3110

\bibitem[{Stamatellos {et~al.}(2011)Stamatellos, Maury, Whitworth, \&
  Andr{\'e}}]{Stamatellos:2011d}
Stamatellos, D., Maury, A., Whitworth, A., \& Andr{\'e}, P. 2011, \mnras, 413,
  1787

\bibitem[{Stamatellos \& Whitworth(2009{\natexlab{a}})}]{Stamatellos:2009d}
Stamatellos, D. \& Whitworth, A. 2009{\natexlab{a}}, \mnras, 400, 1563

\bibitem[{Stamatellos \& Whitworth(2009{\natexlab{b}})}]{Stamatellos:2009a}
Stamatellos, D. \& Whitworth, A. 2009{\natexlab{b}}, \mnras, 392, 413

\bibitem[{Stamatellos {et~al.}(2007{\natexlab{b}})Stamatellos, Whitworth,
  Bisbas, \& Goodwin}]{Stamatellos:2007b}
Stamatellos, D., Whitworth, A.~P., Bisbas, T., \& Goodwin, S.
  2007{\natexlab{b}}, \aap, 475, 37

\bibitem[{{Stone} {et~al.}(2018){Stone}, {Skemer}, {Hinz}, {Bonavita},
  {Kratter}, {Maire}, {Defrere}, {Bailey}, {Spalding}, {Leisenring},
  {Desidera}, {Bonnefoy}, {Biller}, {Woodward}, {Henning}, {Skrutskie},
  {Eisner}, {Crepp}, {Patience}, {Weigelt}, {De Rosa}, {Schlieder}, {Brandner},
  {Apai}, {Su}, {Ertel}, {Ward-Duong}, {Morzinski}, {Schertl}, {Hofmann},
  {Close}, {Brems}, {Fortney}, {Oza}, {Buenzli}, \& {Bass}}]{Stone:2018a}
{Stone}, J.~M., {Skemer}, A.~J., {Hinz}, P.~M., {et~al.} 2018, \aj, 156, 286

\bibitem[{{Szul{\'a}gyi}(2017)}]{Szulagyi:2017d}
{Szul{\'a}gyi}, J. 2017, \apj, 842, 103

\bibitem[{{Szul{\'a}gyi} {et~al.}(2017){Szul{\'a}gyi}, {Mayer}, \&
  {Quinn}}]{Szulagyi:2017a}
{Szul{\'a}gyi}, J., {Mayer}, L., \& {Quinn}, T. 2017, \mnras, 464, 3158

\bibitem[{{Szul{\'a}gyi} \& {Mordasini}(2017)}]{Szulagyi:2017b}
{Szul{\'a}gyi}, J. \& {Mordasini}, C. 2017, \mnras, 465, L64

\bibitem[{{Szul{\'a}gyi} {et~al.}(2018){Szul{\'a}gyi}, {Plas}, {Meyer}, {Pohl},
  {Quanz}, {Mayer}, {Daemgen}, \& {Tamburello}}]{Szulagyi:2018a}
{Szul{\'a}gyi}, J., {Plas}, G. v.~d., {Meyer}, M.~R., {et~al.} 2018, \mnras,
  473, 3573

\bibitem[{{Tobin} {et~al.}(2016){Tobin}, {Looney}, {Li}, {Chandler}, {Dunham},
  {Segura-Cox}, {Sadavoy}, {Melis}, {Harris}, {Kratter}, \&
  {Perez}}]{Tobin:2016a}
{Tobin}, J.~J., {Looney}, L.~W., {Li}, Z.-Y., {et~al.} 2016, \apj, 818, 73

\bibitem[{{Tomida} {et~al.}(2013){Tomida}, {Tomisaka}, {Matsumoto}, {Hori},
  {Okuzumi}, {Machida}, \& {Saigo}}]{Tomida:2013a}
{Tomida}, K., {Tomisaka}, K., {Matsumoto}, T., {et~al.} 2013, \apj, 763, 6

\bibitem[{Toomre(1964)}]{Toomre:1964a}
Toomre, A. 1964, \apj, 139, 1217

\bibitem[{{Tsukamoto} {et~al.}(2015){Tsukamoto}, {Takahashi}, {Machida}, \&
  {Inutsuka}}]{Tsukamoto:2015a}
{Tsukamoto}, Y., {Takahashi}, S.~Z., {Machida}, M.~N., \& {Inutsuka}, S. 2015,
  \mnras, 446, 1175

\bibitem[{{Vaytet} {et~al.}(2013){Vaytet}, {Chabrier}, {Audit}, {Commer{\c
  c}on}, {Masson}, {Ferguson}, \& {Delahaye}}]{Vaytet:2013a}
{Vaytet}, N., {Chabrier}, G., {Audit}, E., {et~al.} 2013, \aap, 557, A90

\bibitem[{{Vaytet} \& {Haugb{\o}lle}(2017)}]{Vaytet:2017a}
{Vaytet}, N. \& {Haugb{\o}lle}, T. 2017, \aap, 598, A116

\bibitem[{Veras \& Raymond(2012)}]{Veras:2012b}
Veras, D. \& Raymond, S.~N. 2012, \mnras: Letters, 421, L117

\bibitem[{{Vigan} {et~al.}(2017){Vigan}, {Bonavita}, {Biller}, {Forgan},
  {Rice}, {Chauvin}, {Desidera}, {Meunier}, {Delorme}, {Schlieder}, {Bonnefoy},
  {Carson}, {Covino}, {Hagelberg}, {Henning}, {Janson}, {Lagrange}, {Quanz},
  {Zurlo}, {Beuzit}, {Boccaletti}, {Buenzli}, {Feldt}, {Girard}, {Gratton},
  {Kasper}, {Le Coroller}, {Mesa}, {Messina}, {Meyer}, {Montagnier},
  {Mordasini}, {Mouillet}, {Moutou}, {Reggiani}, {Segransan}, \&
  {Thalmann}}]{Vigan:2017a}
{Vigan}, A., {Bonavita}, M., {Biller}, B., {et~al.} 2017, \aap, 603, A3

\bibitem[{{Vorobyov}(2013)}]{Vorobyov:2013a}
{Vorobyov}, E.~I. 2013, \aap, 552, A129

\bibitem[{{Wagner} {et~al.}(2019){Wagner}, {Apai}, \& {Kratter}}]{Wagner:2019a}
{Wagner}, K., {Apai}, D., \& {Kratter}, K.~M. 2019, \apj, 877, 46

\bibitem[{Whitworth \& Stamatellos(2006)}]{Whitworth:2006a}
Whitworth, A.~P. \& Stamatellos, D. 2006, \aap, 458, 817

\bibitem[{Zhu {et~al.}(2012)Zhu, Hartmann, Nelson, \& Gammie}]{Zhu:2012a}
Zhu, Z., Hartmann, L., Nelson, R.~P., \& Gammie, C.~F. 2012, \apj, 746, 110

\end{thebibliography}

\appendix
\section{Fragment and protoplanet properties}
\label{sec:appendix}

We present plots of the properties of all fragments and protoplanets formed in the simulations we have performed, as these correspond to the initial stages of disc instability planets and therefore need to be further investigated.

\begin{figure*}
  \begin{center}
  \subfloat{\resizebox{0.45\hsize}{!}
  {\includegraphics{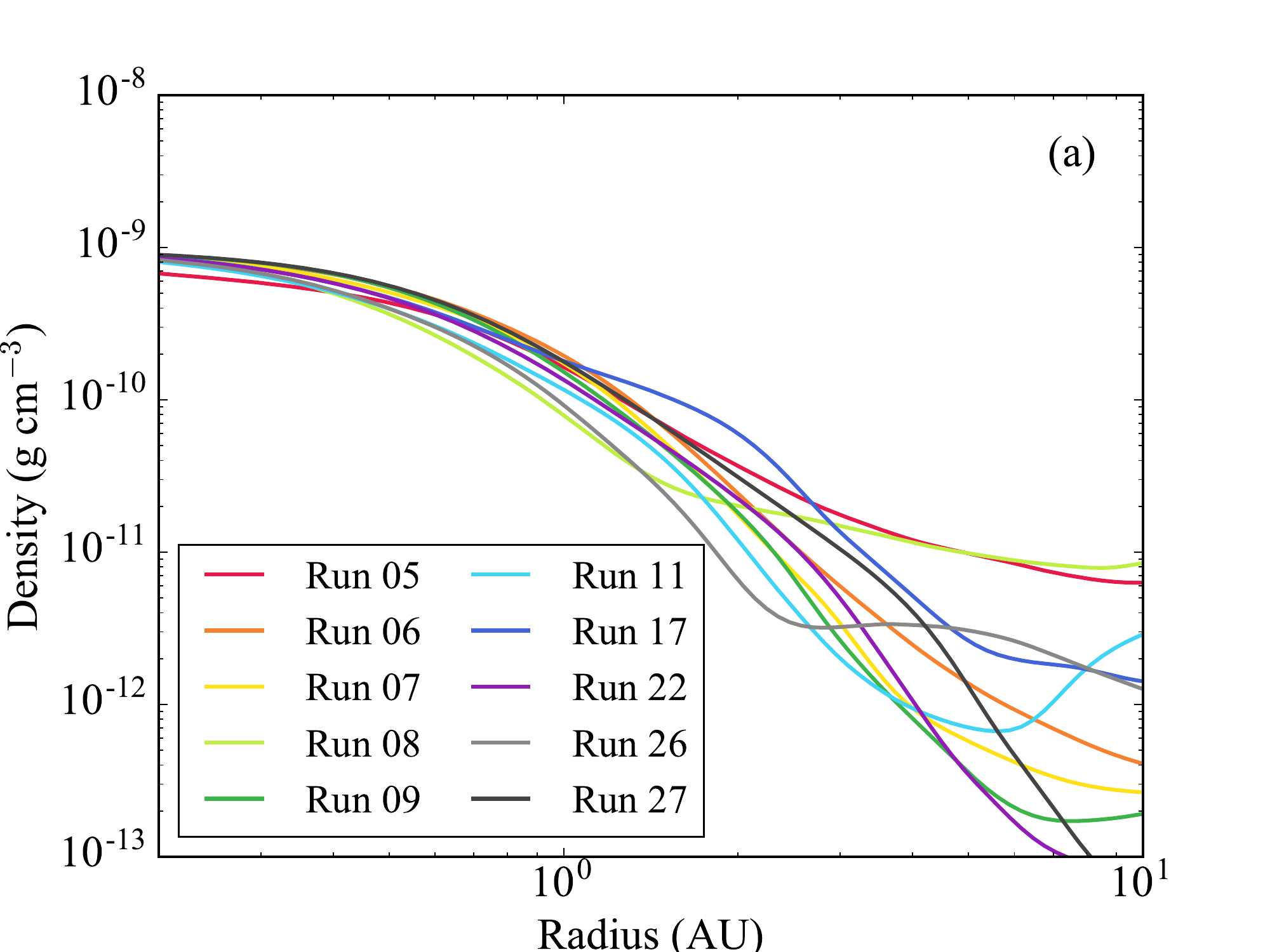}}}
  \subfloat{\resizebox{0.45\hsize}{!}
  {\includegraphics{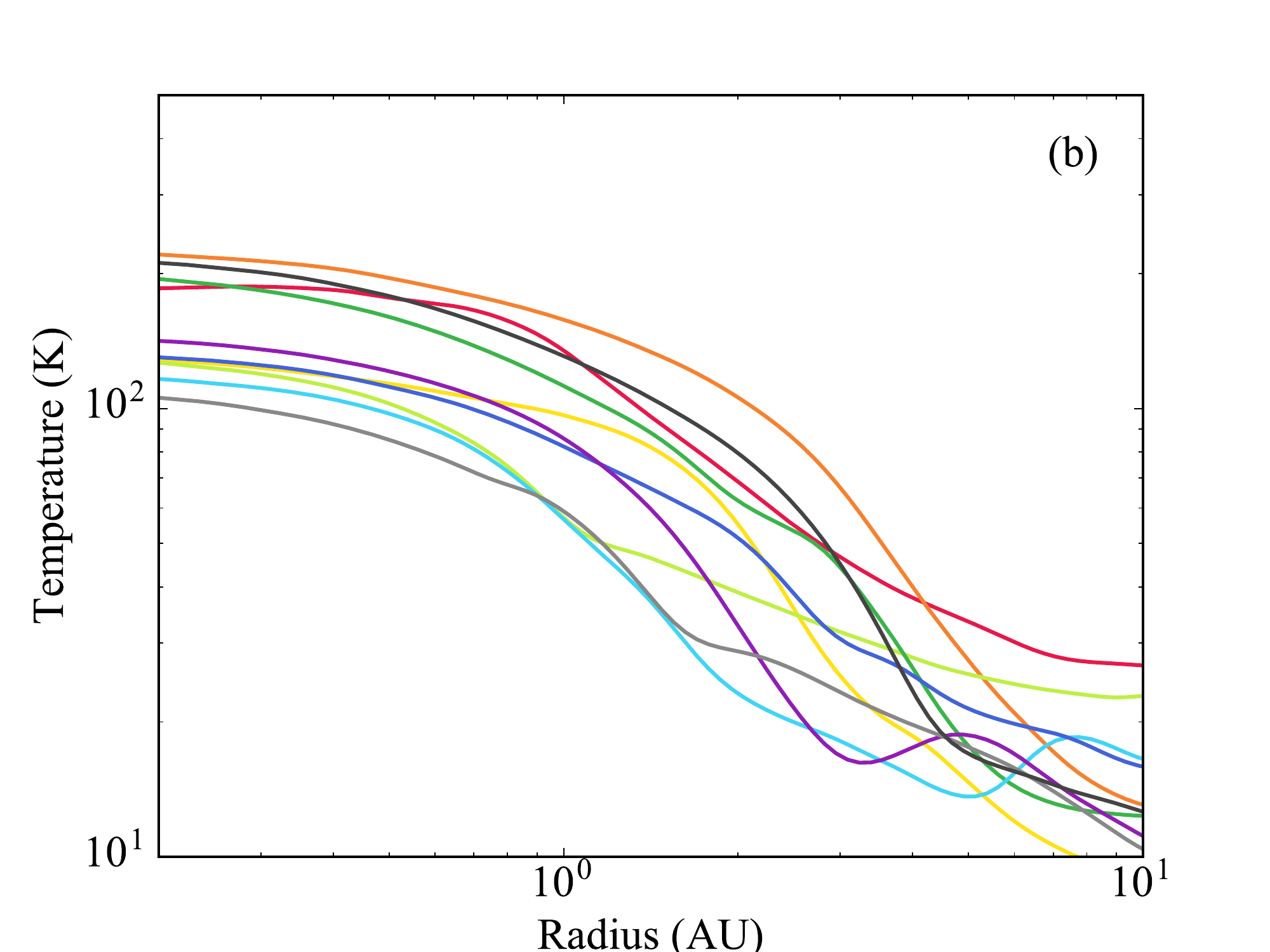}}} \\
  \subfloat{\resizebox{0.45\hsize}{!}
  {\includegraphics{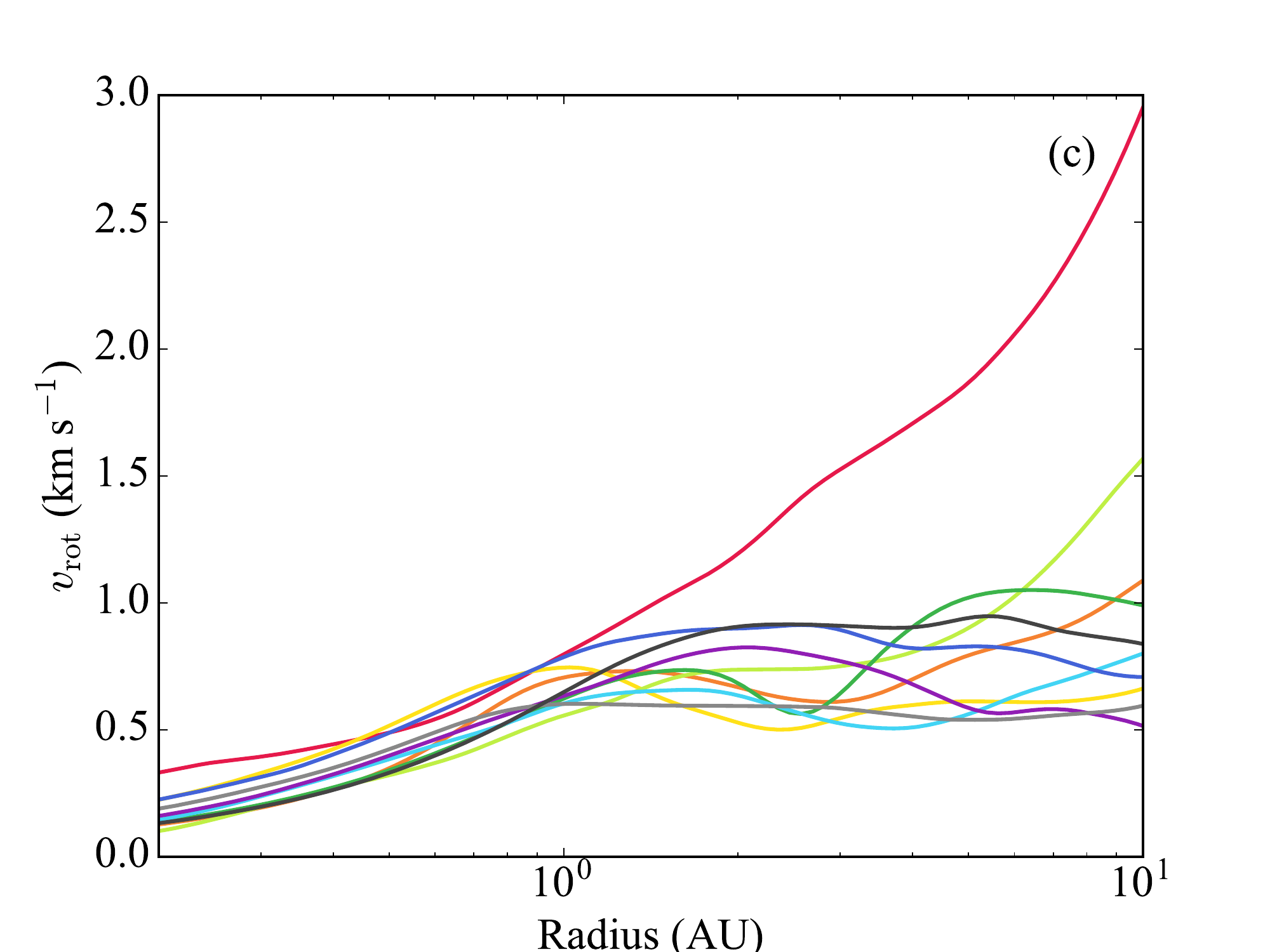}}}
  \subfloat{\resizebox{0.45\hsize}{!}
  {\includegraphics{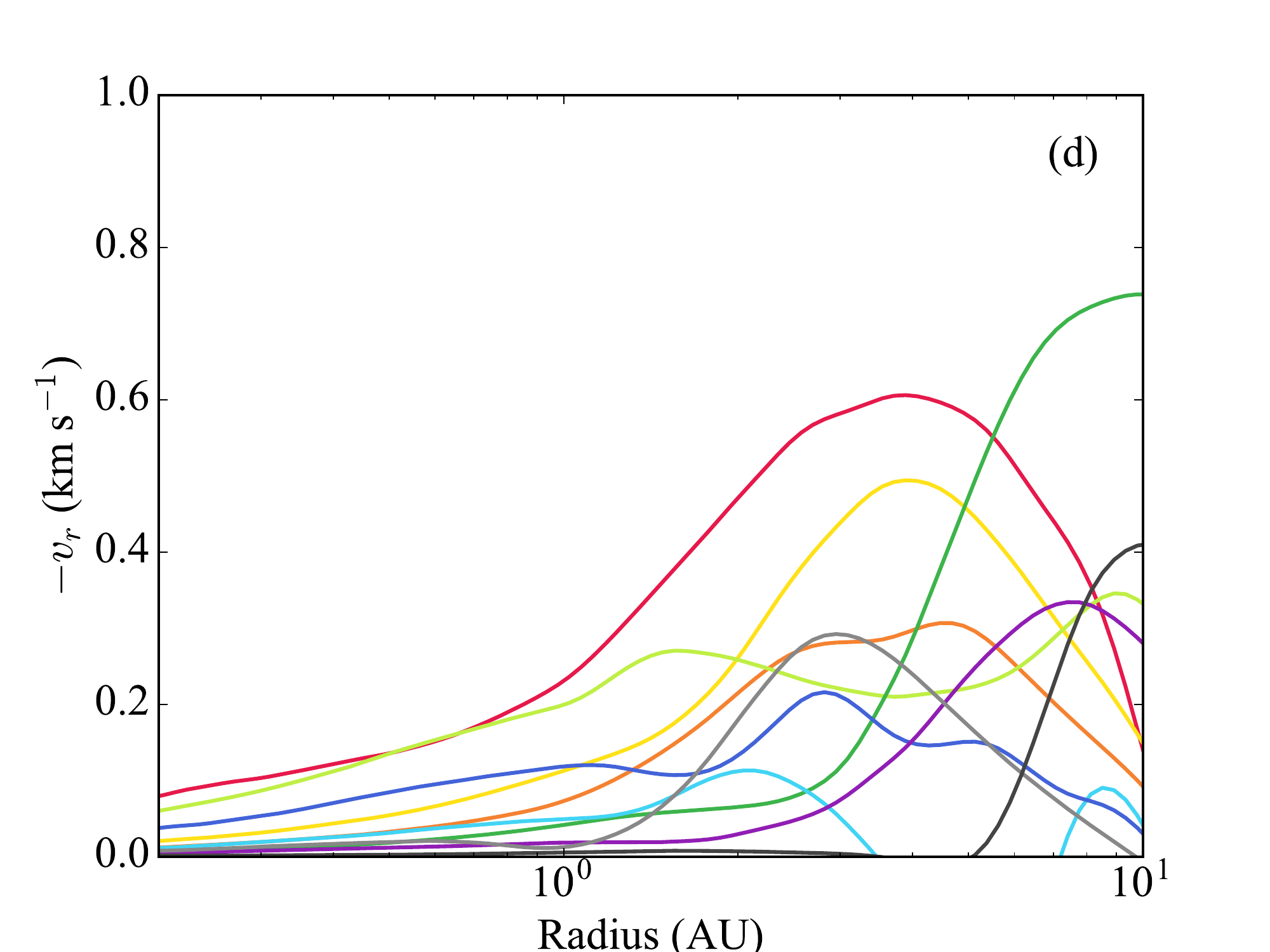}}} \\
  \subfloat{\resizebox{0.45\hsize}{!}
  {\includegraphics{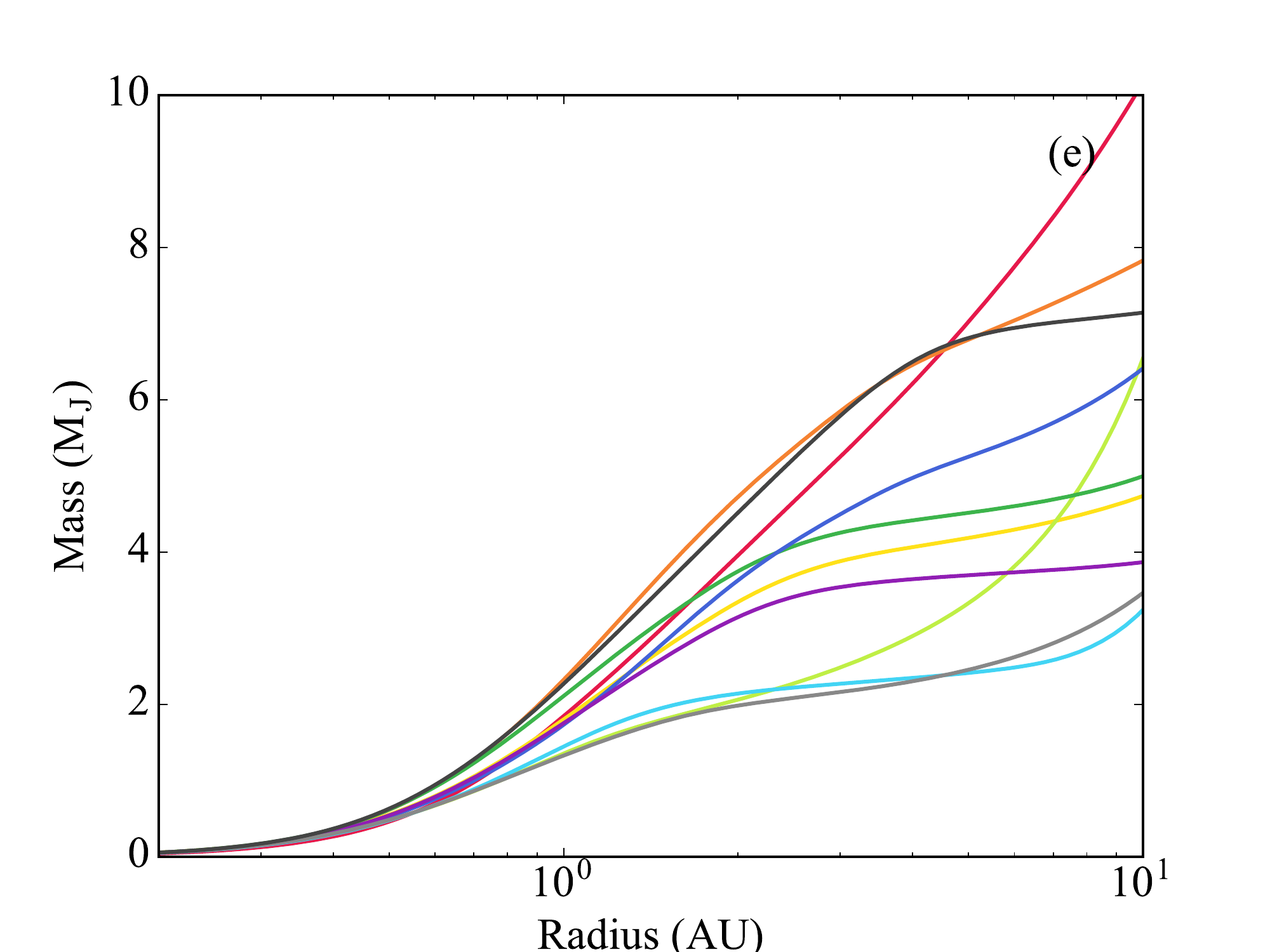}}}
  \subfloat{\resizebox{0.45\hsize}{!}
  {\includegraphics{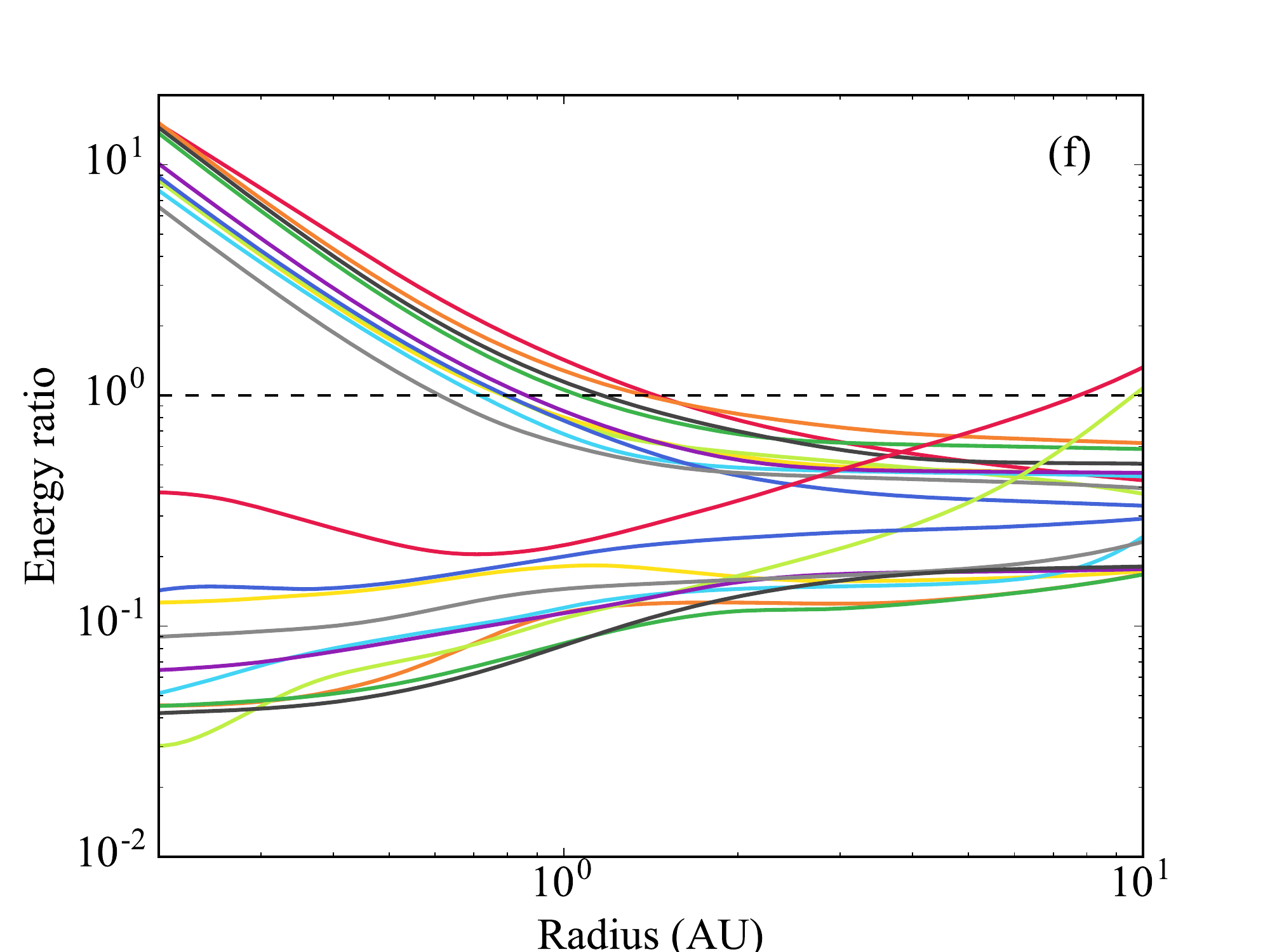}}}
  \caption{Properties of a set of fragments when they have attained a central density of $10^{-9} \textup{ g cm} ^{-3}$. Panels are the same as in Figure~\ref{fig:properties_run5}. The thermal-to-gravitational
    energy ratios are comparable for different fragments. Rotational energy is
    significant only in the outer parts of each fragment.}
  \label{fig:fragment_properties}
\end{center}
\end{figure*}

\begin{figure*}
  \begin{center}
  \subfloat{\resizebox{0.45\hsize}{!}
  {\includegraphics{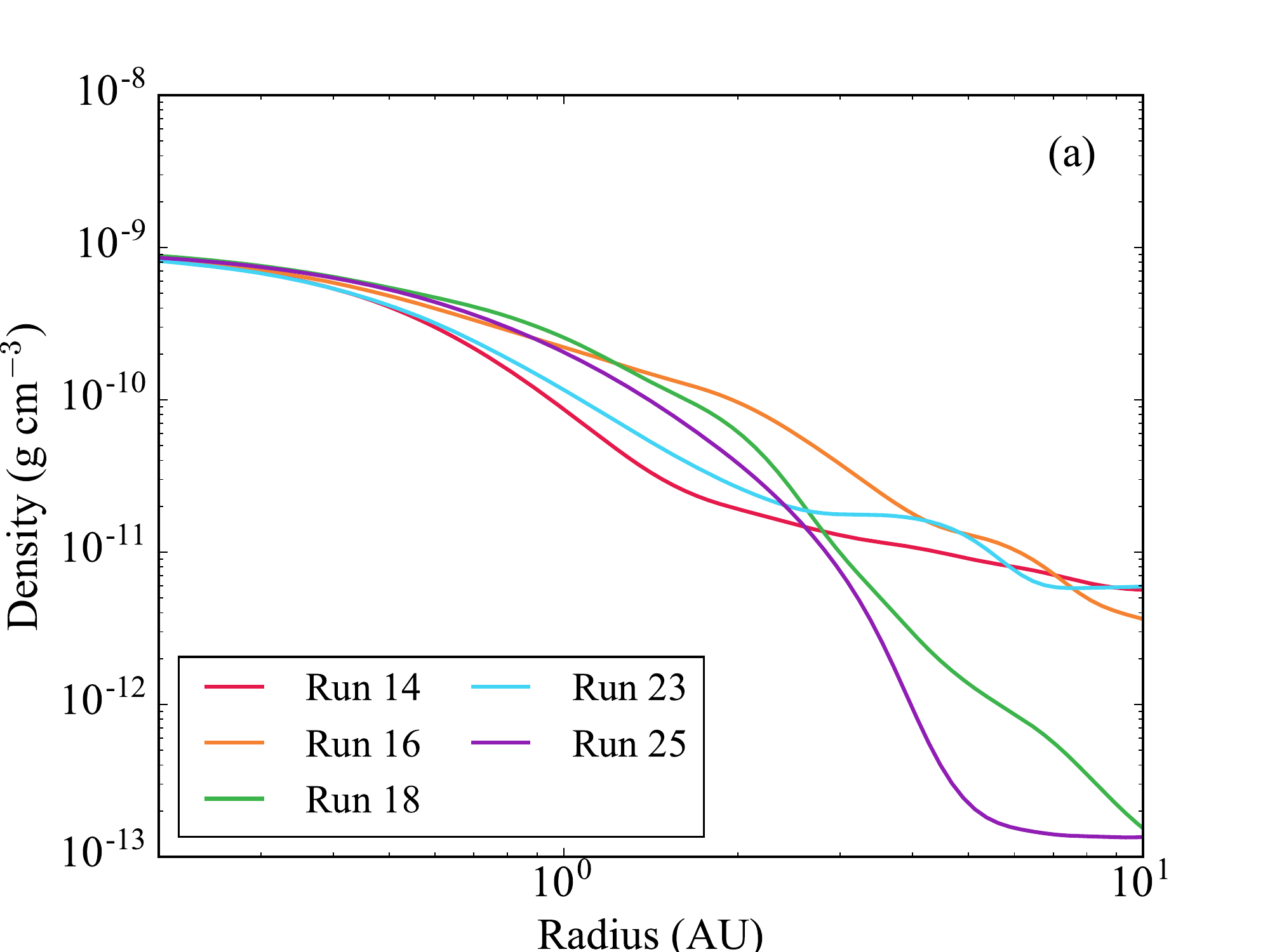}}}
  \subfloat{\resizebox{0.45\hsize}{!}
  {\includegraphics{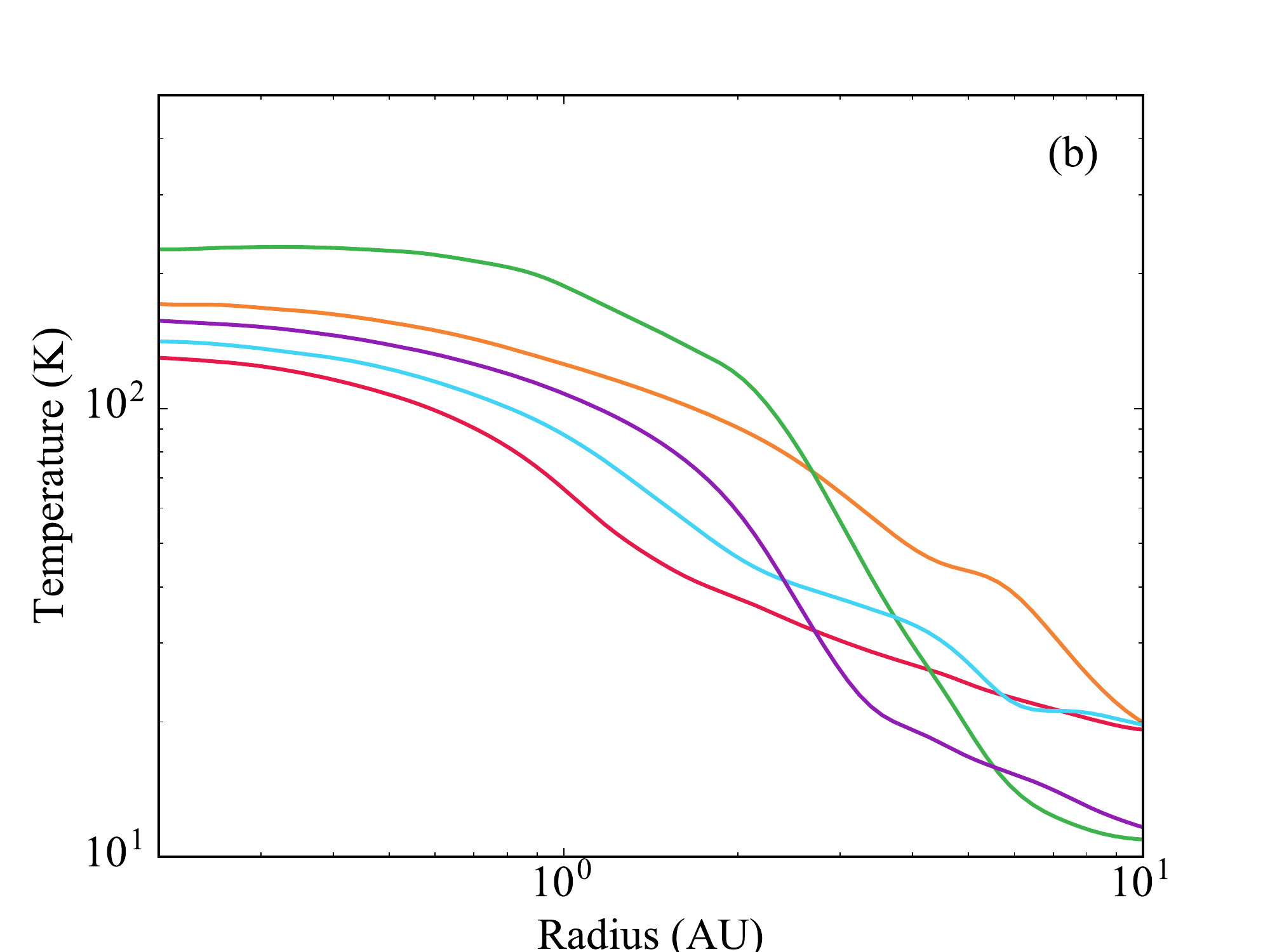}}} \\
  \subfloat{\resizebox{0.45\hsize}{!}
  {\includegraphics{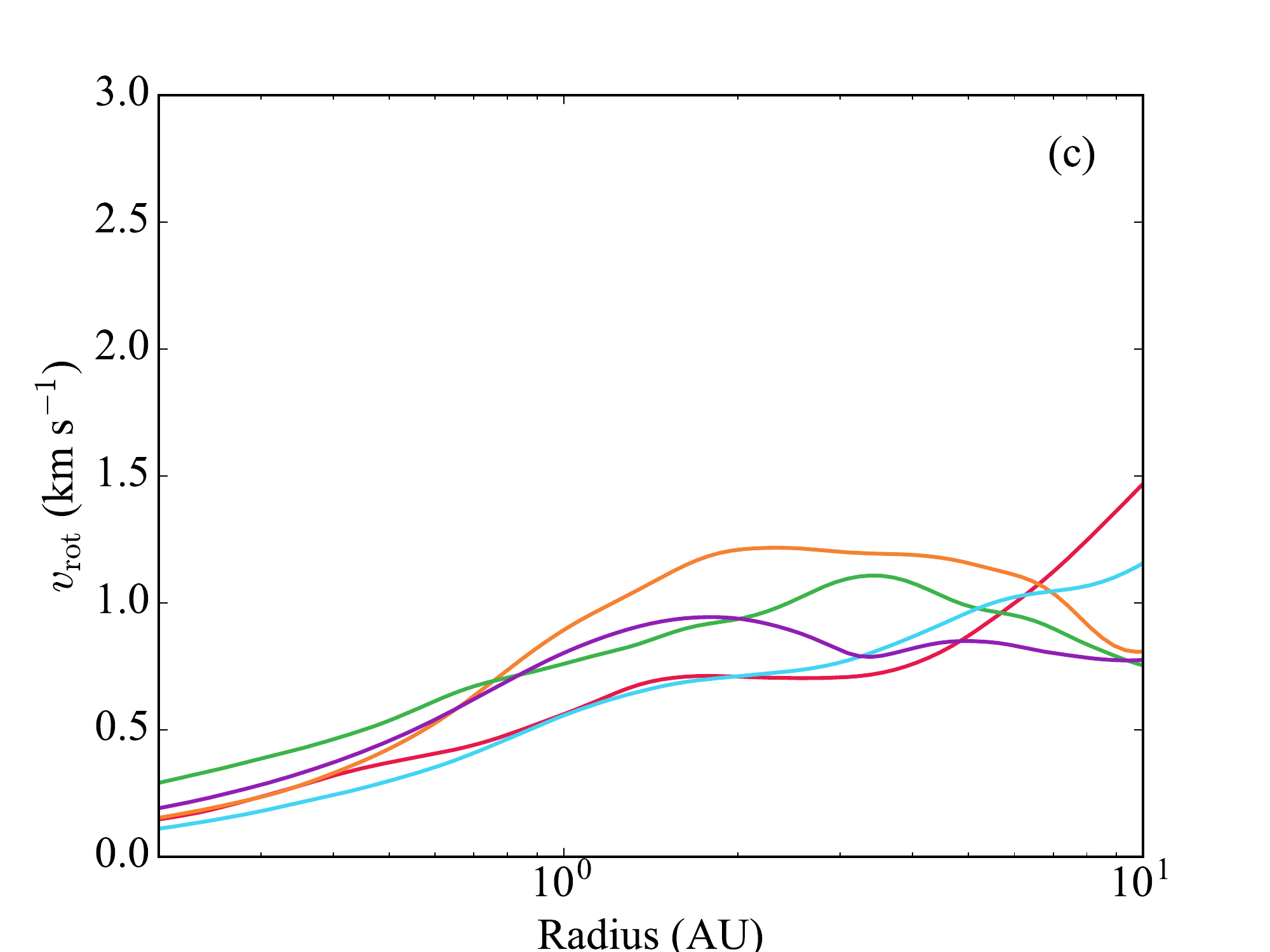}}}
  \subfloat{\resizebox{0.45\hsize}{!}
  {\includegraphics{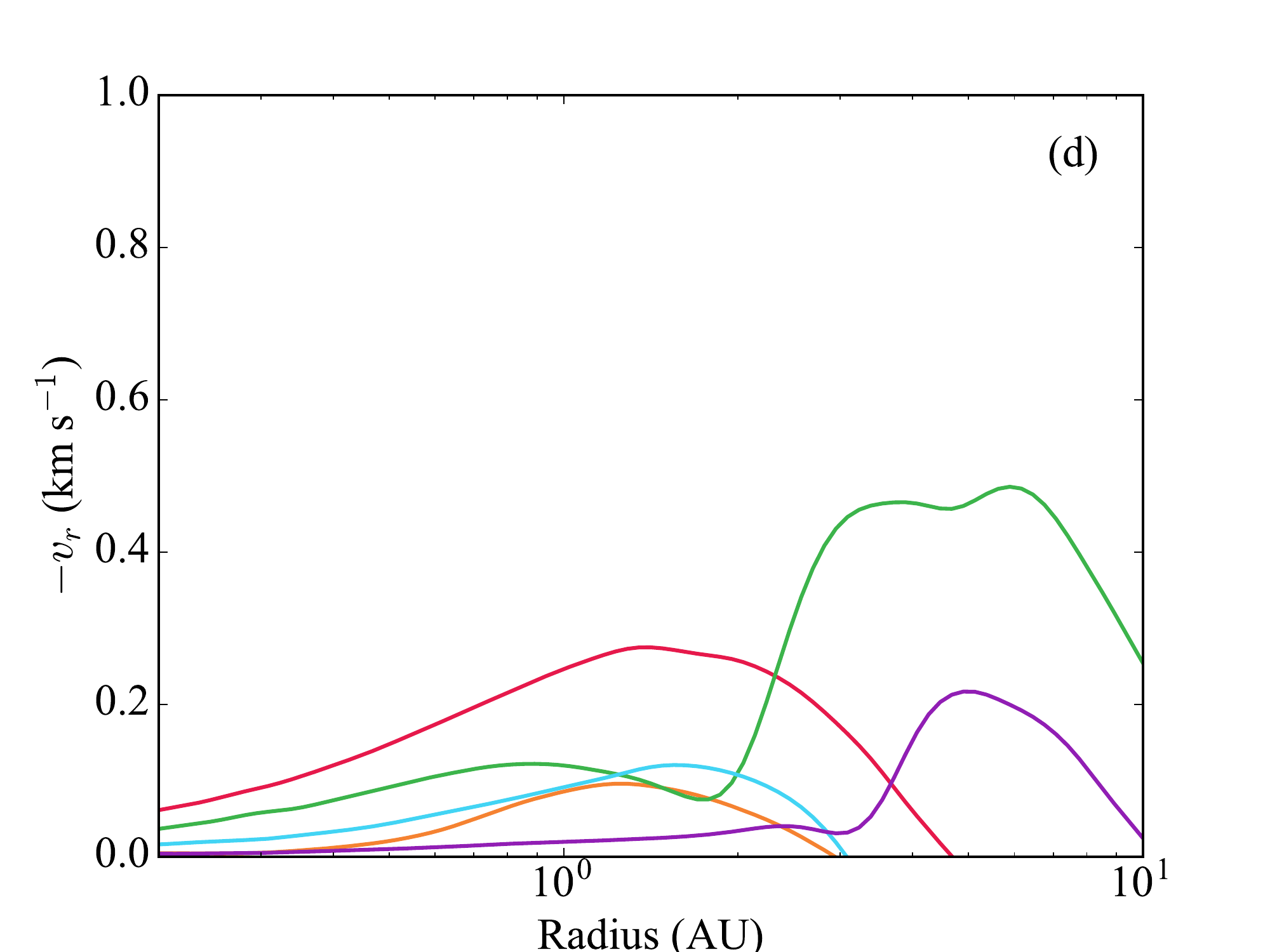}}} \\
  \subfloat{\resizebox{0.45\hsize}{!}
  {\includegraphics{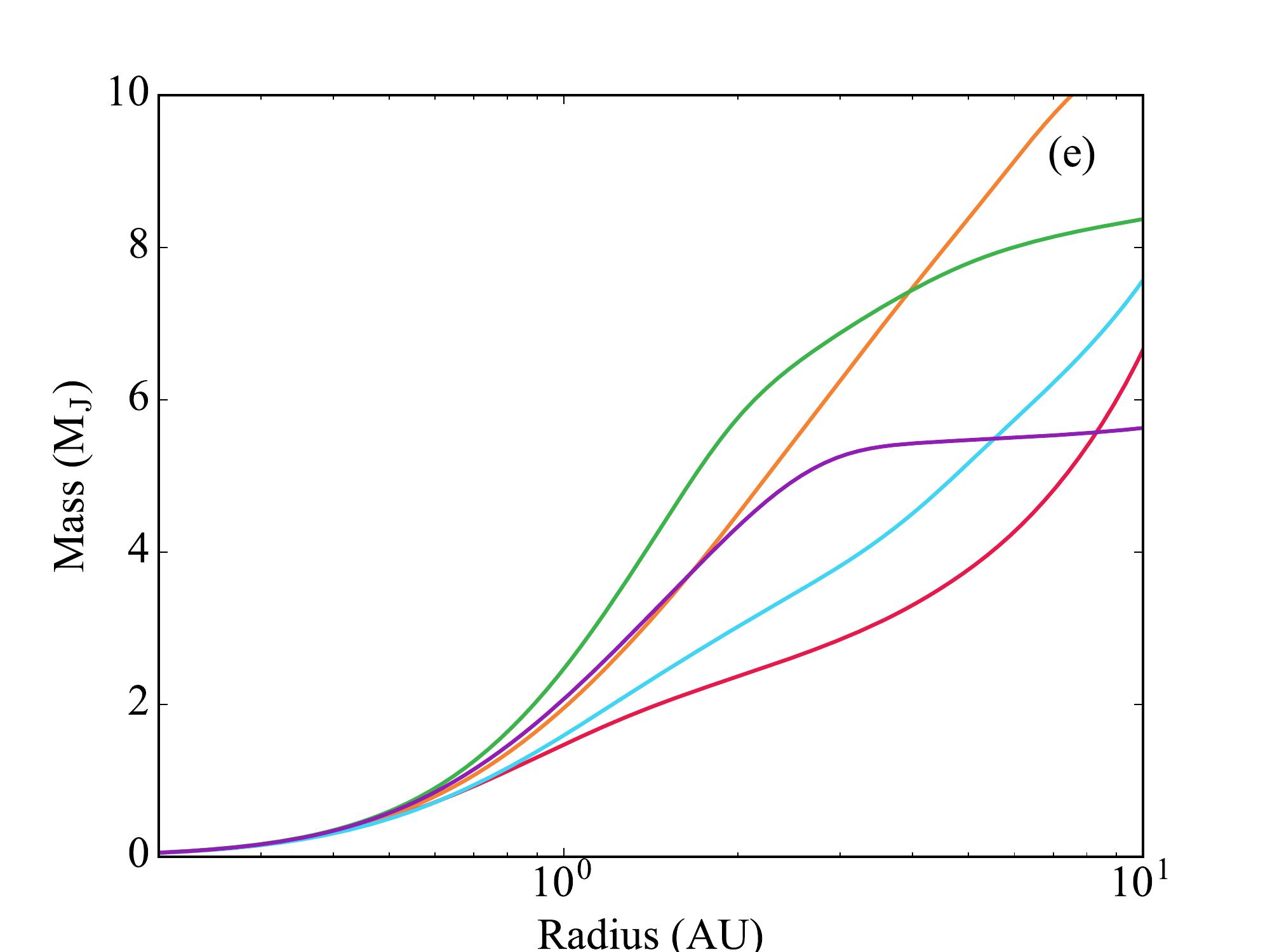}}}
  \subfloat{\resizebox{0.45\hsize}{!}
  {\includegraphics{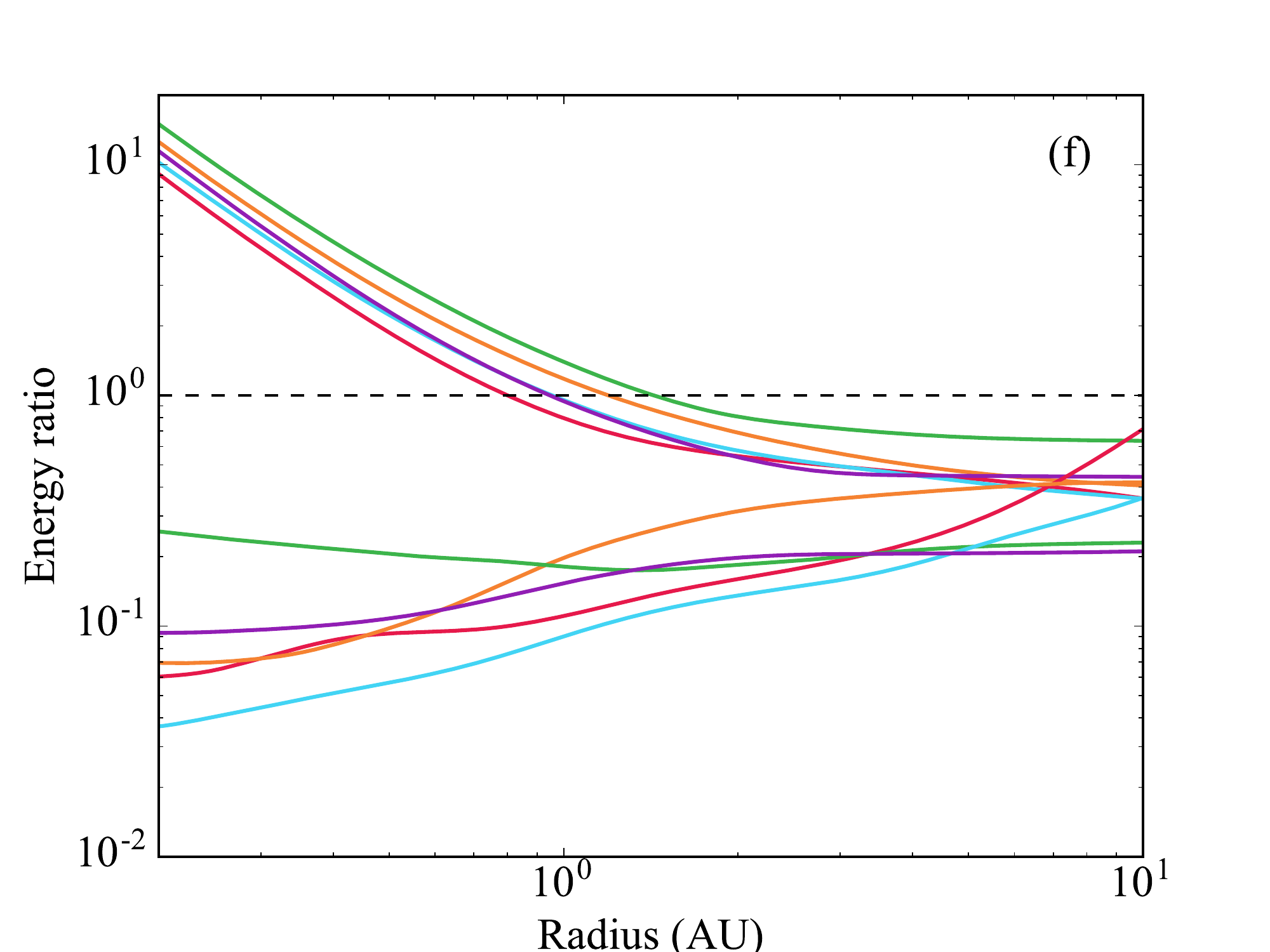}}}
  \caption
  { Properties of a set of fragments formed in the simulations (same as in
    Figure \ref{fig:fragment_properties}, but for a different set of fragments).
  }
  \label{fig:fragment_coreless_properties}
\end{center}
\end{figure*}

\begin{figure*}
  \begin{center}
  \subfloat{\resizebox{0.45\hsize}{!}
  {\includegraphics{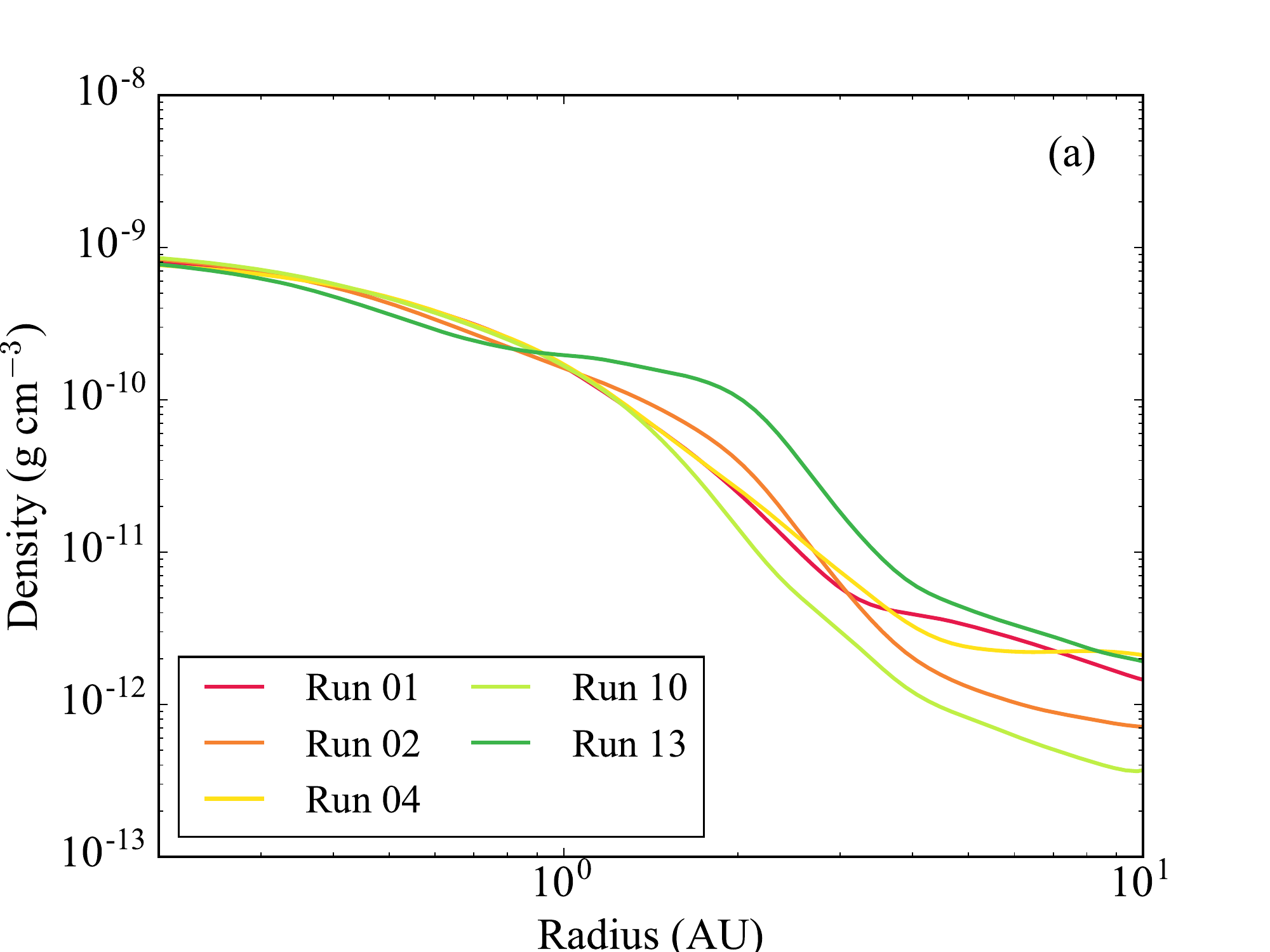}}}
  \subfloat{\resizebox{0.45\hsize}{!}
  {\includegraphics{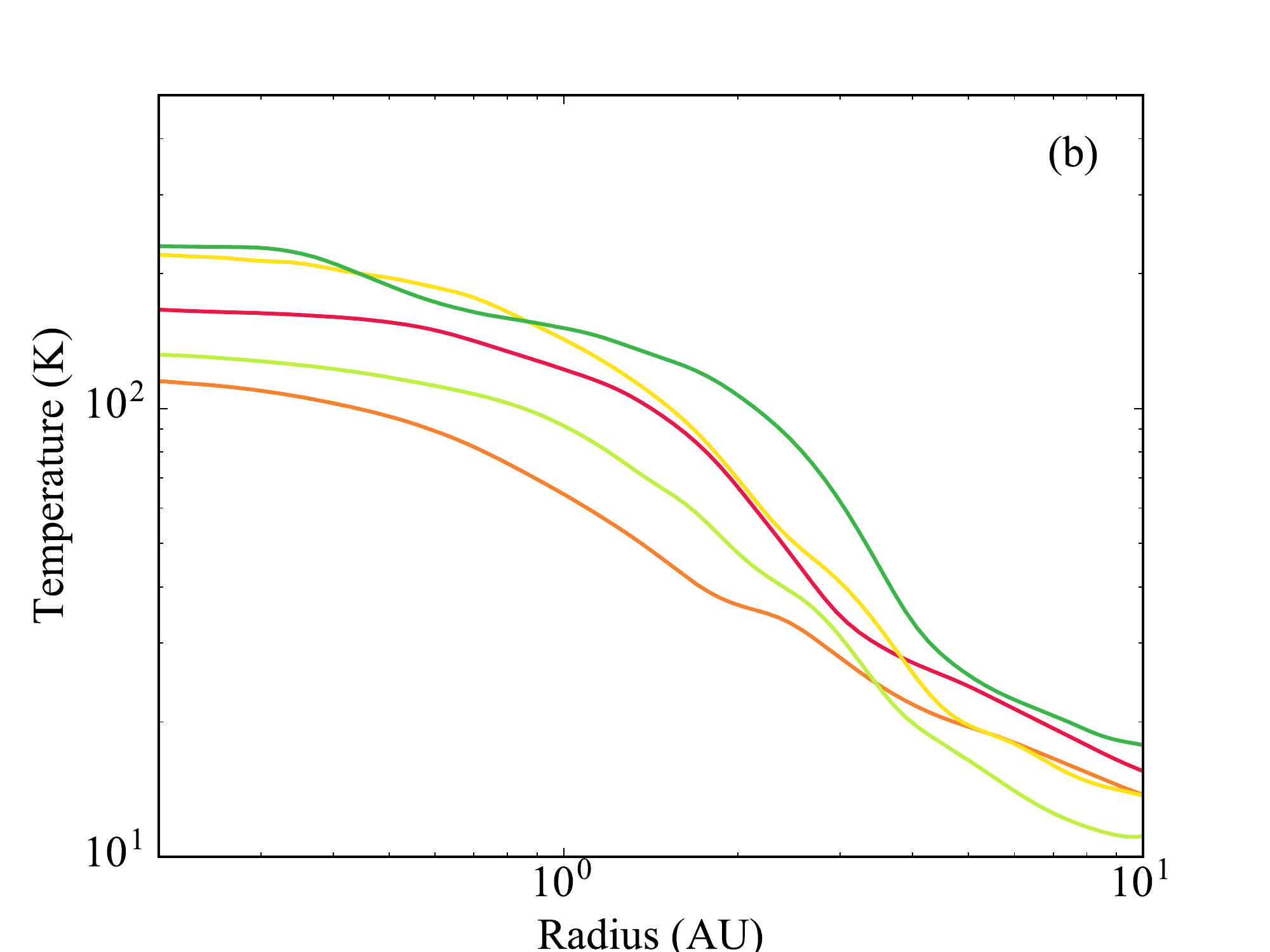}}} \\
  \subfloat{\resizebox{0.45\hsize}{!}
  {\includegraphics{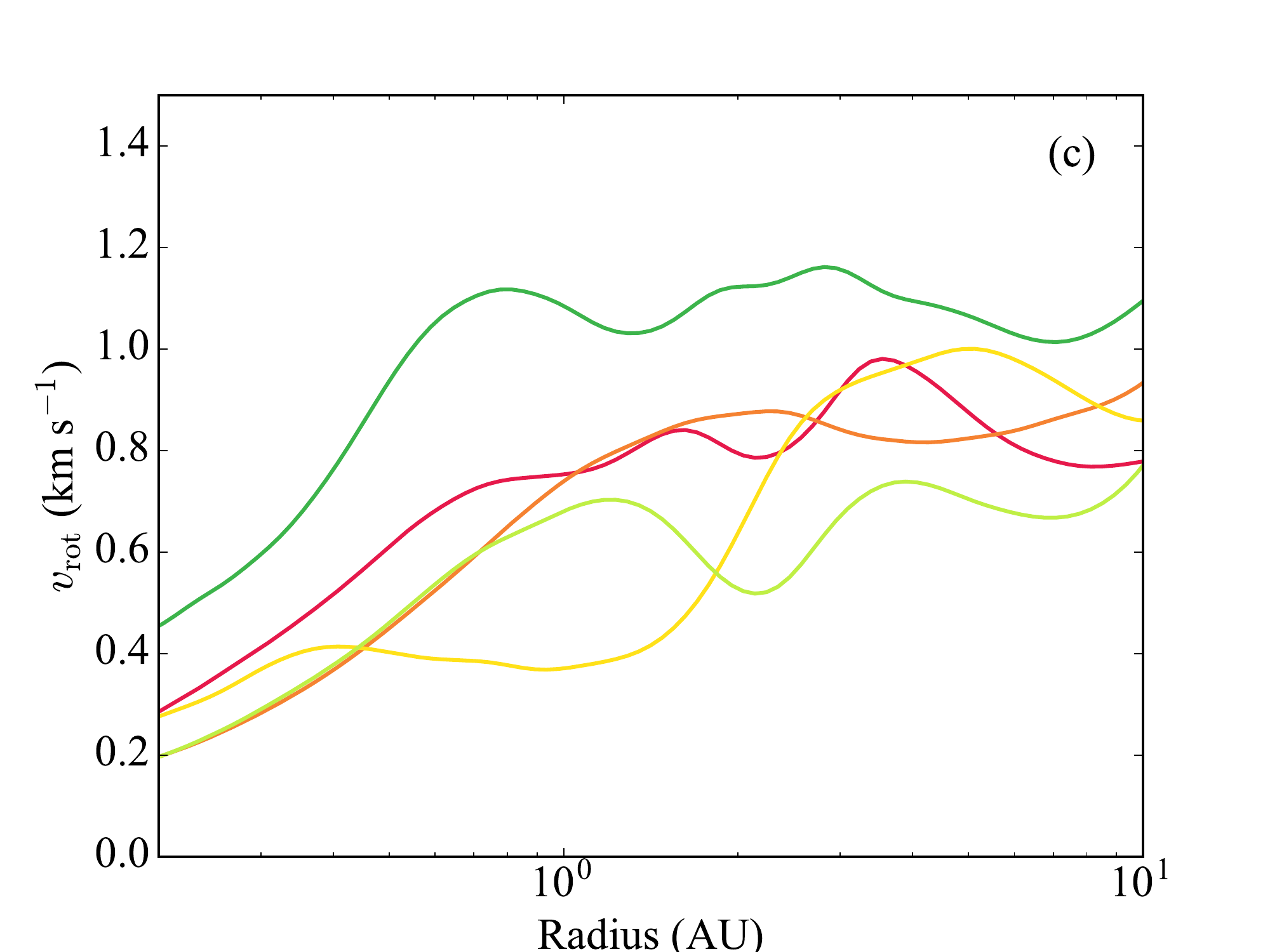}}}
  \subfloat{\resizebox{0.45\hsize}{!}
  {\includegraphics{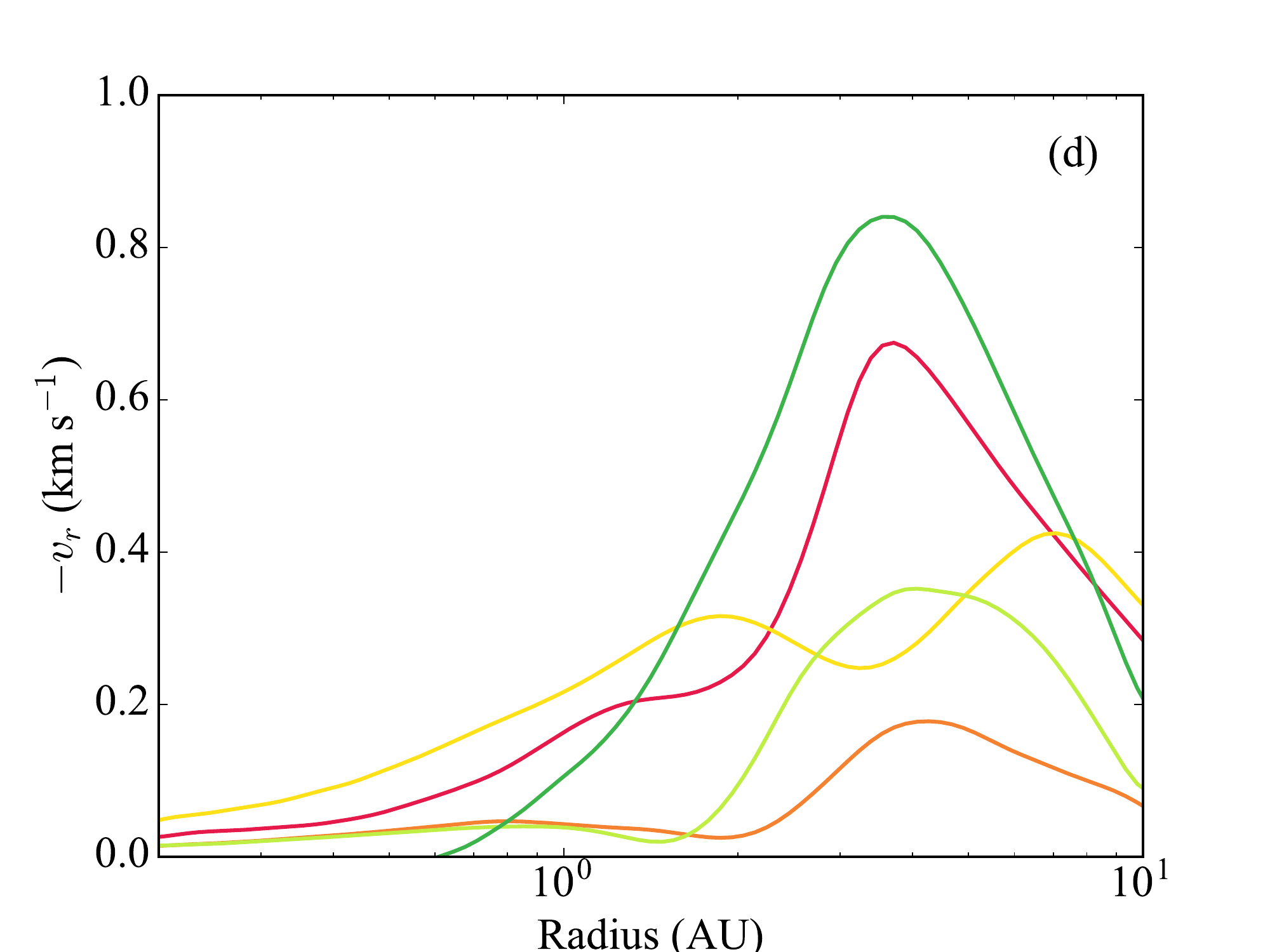}}} \\
  \subfloat{\resizebox{0.45\hsize}{!}
  {\includegraphics{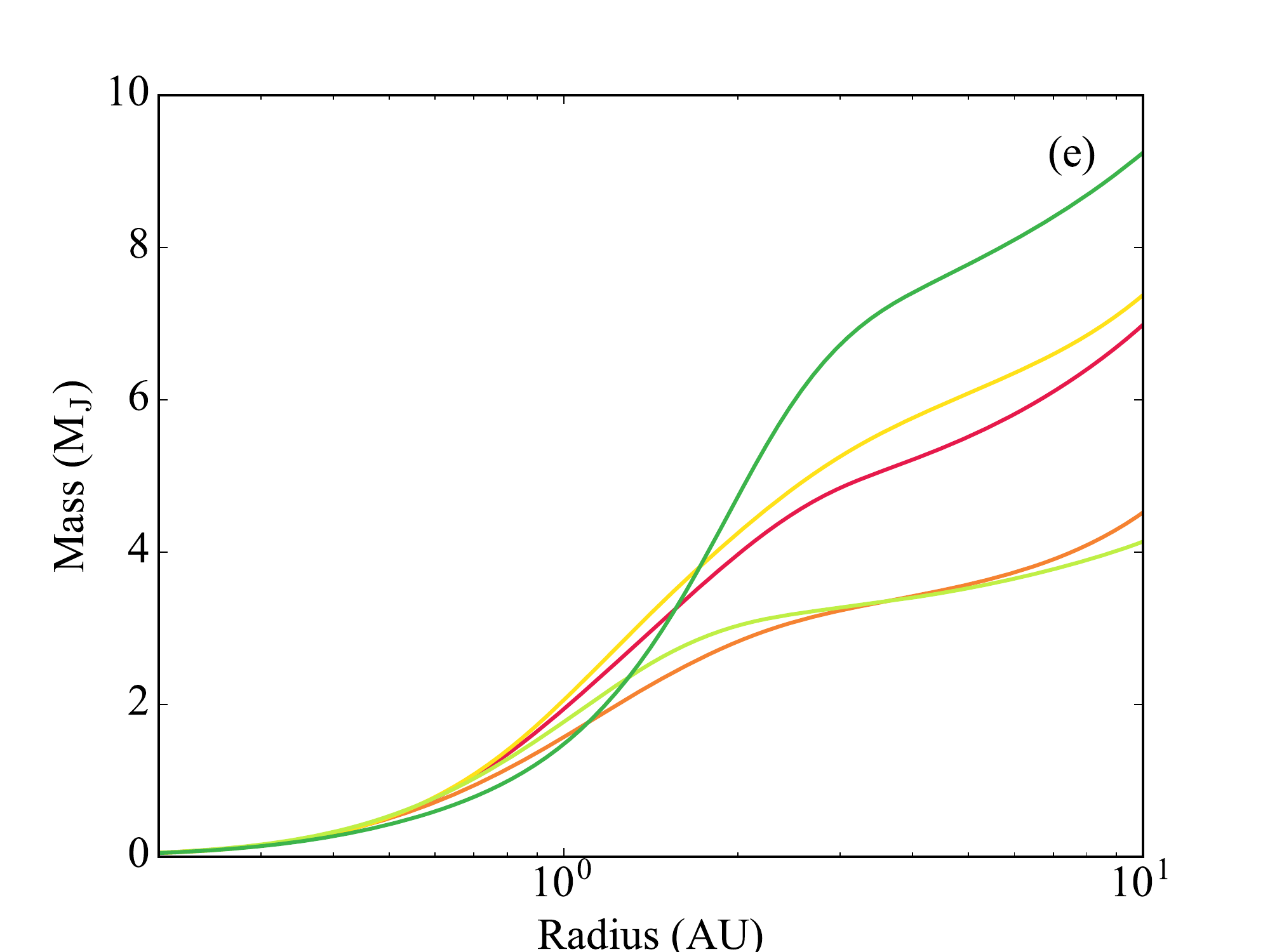}}}
  \subfloat{\resizebox{0.45\hsize}{!}
  {\includegraphics{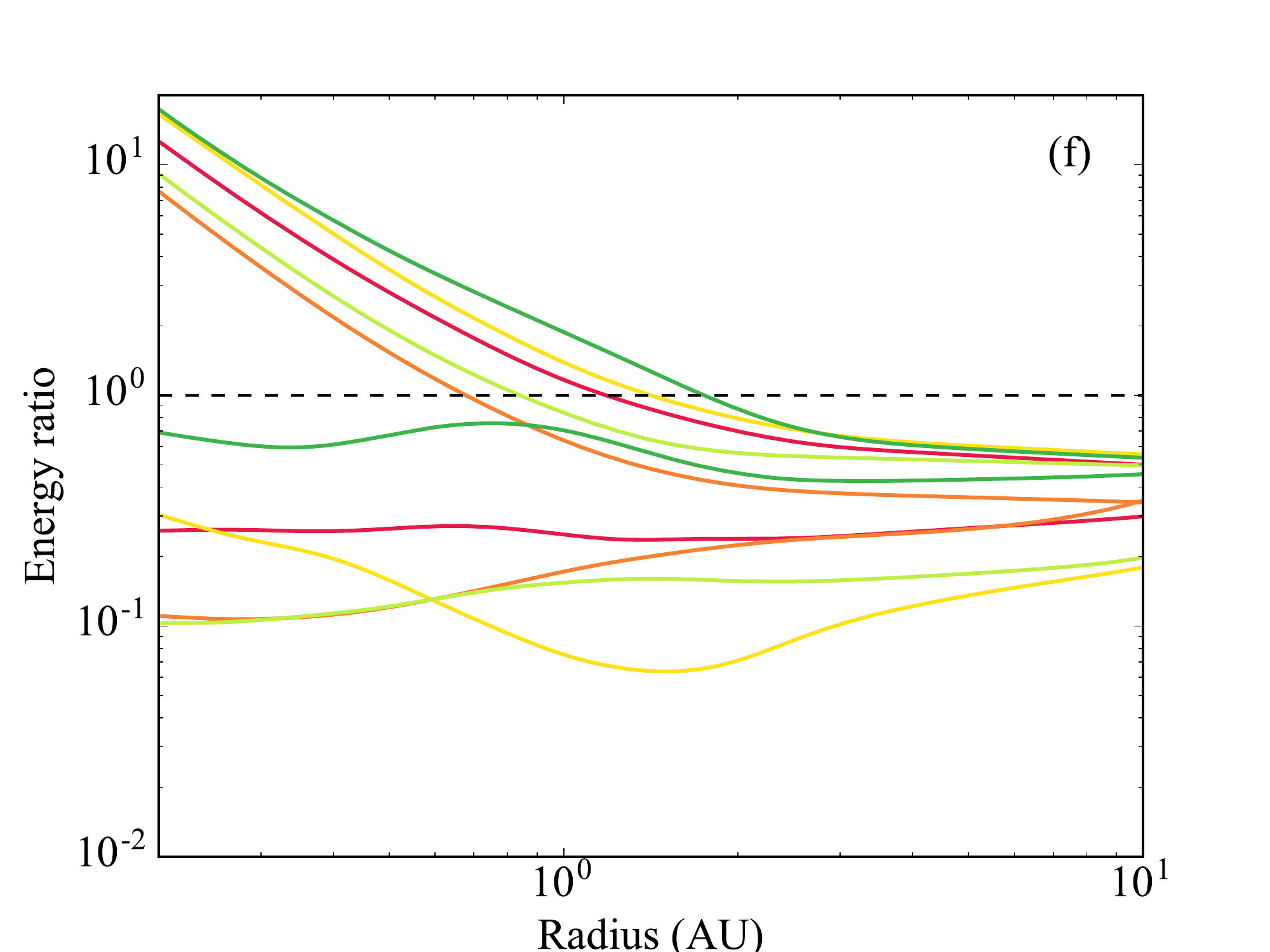}}}
  \caption
  { Properties of a set of fragments formed in the simulations (same as in
    Figure \ref{fig:fragment_properties}, but for a different set of fragments).
  }
  \label{fig:noncollapsed_fragment_properties}
\end{center}
\end{figure*}

\begin{figure*}
  \begin{center}
  \subfloat{\resizebox{0.45\hsize}{!}
  {\includegraphics{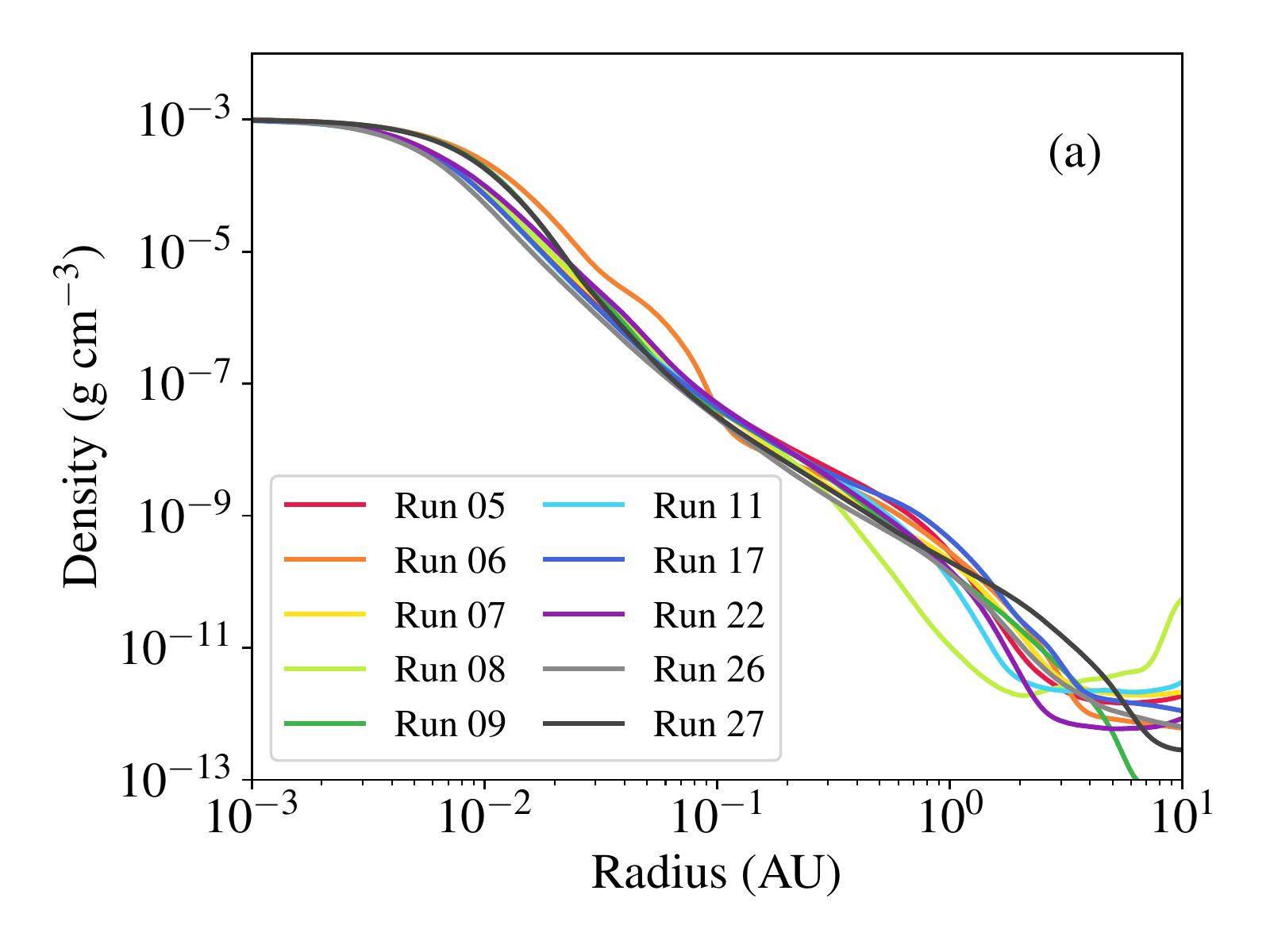}}}
  \subfloat{\resizebox{0.45\hsize}{!}
  {\includegraphics{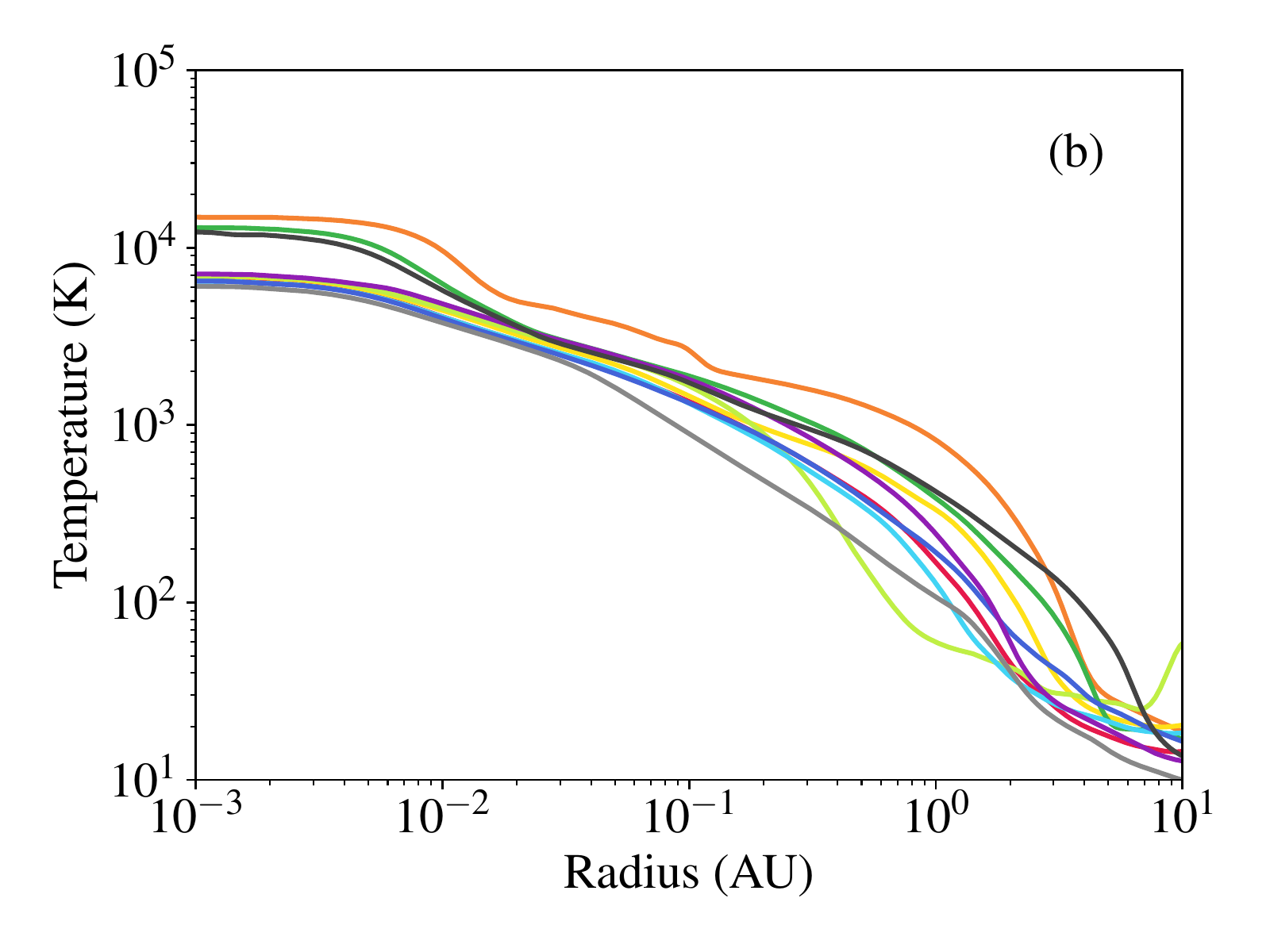}}} \\
  \subfloat{\resizebox{0.45\hsize}{!}
  {\includegraphics{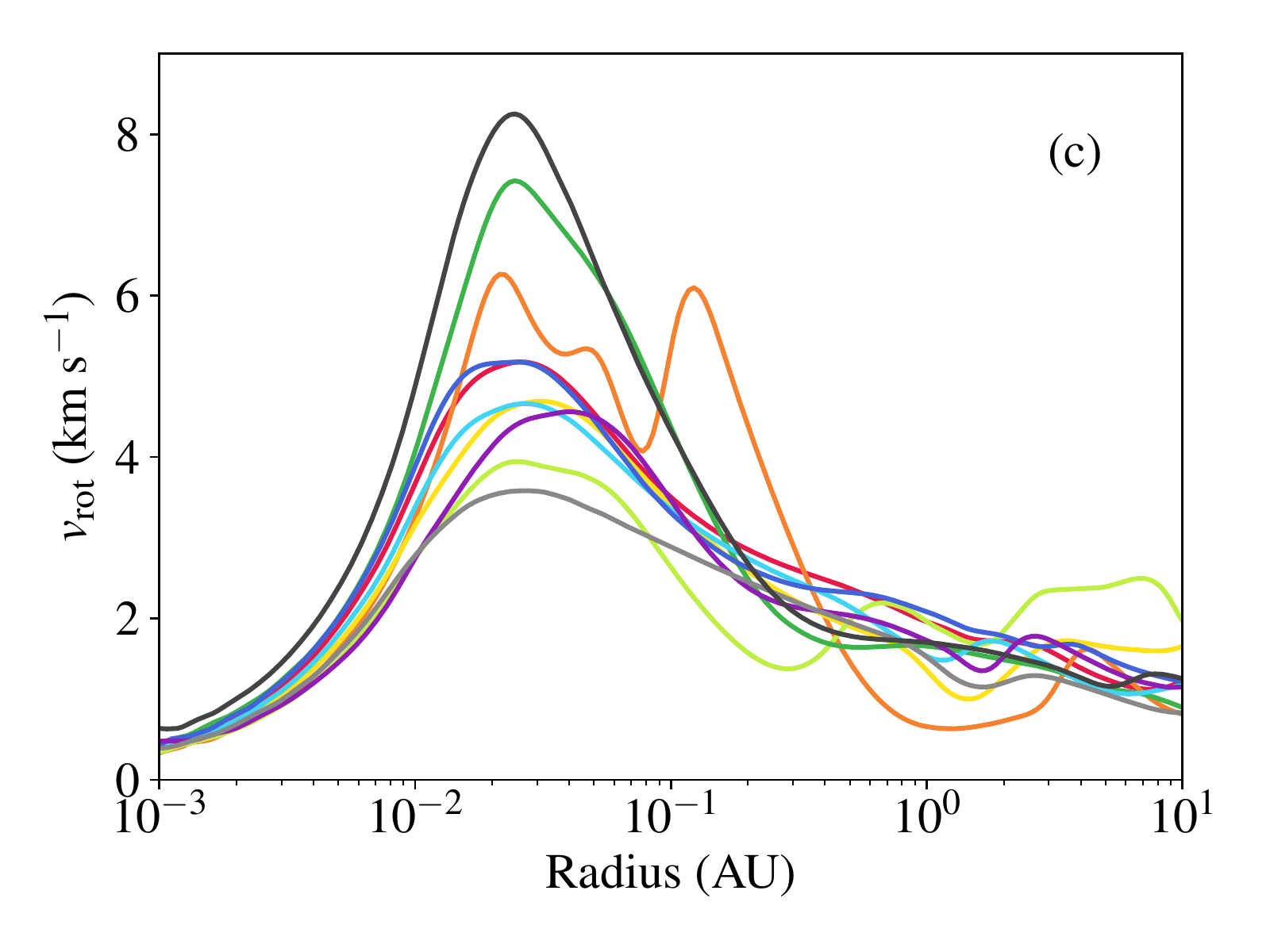}}}
  \subfloat{\resizebox{0.45\hsize}{!}
  {\includegraphics{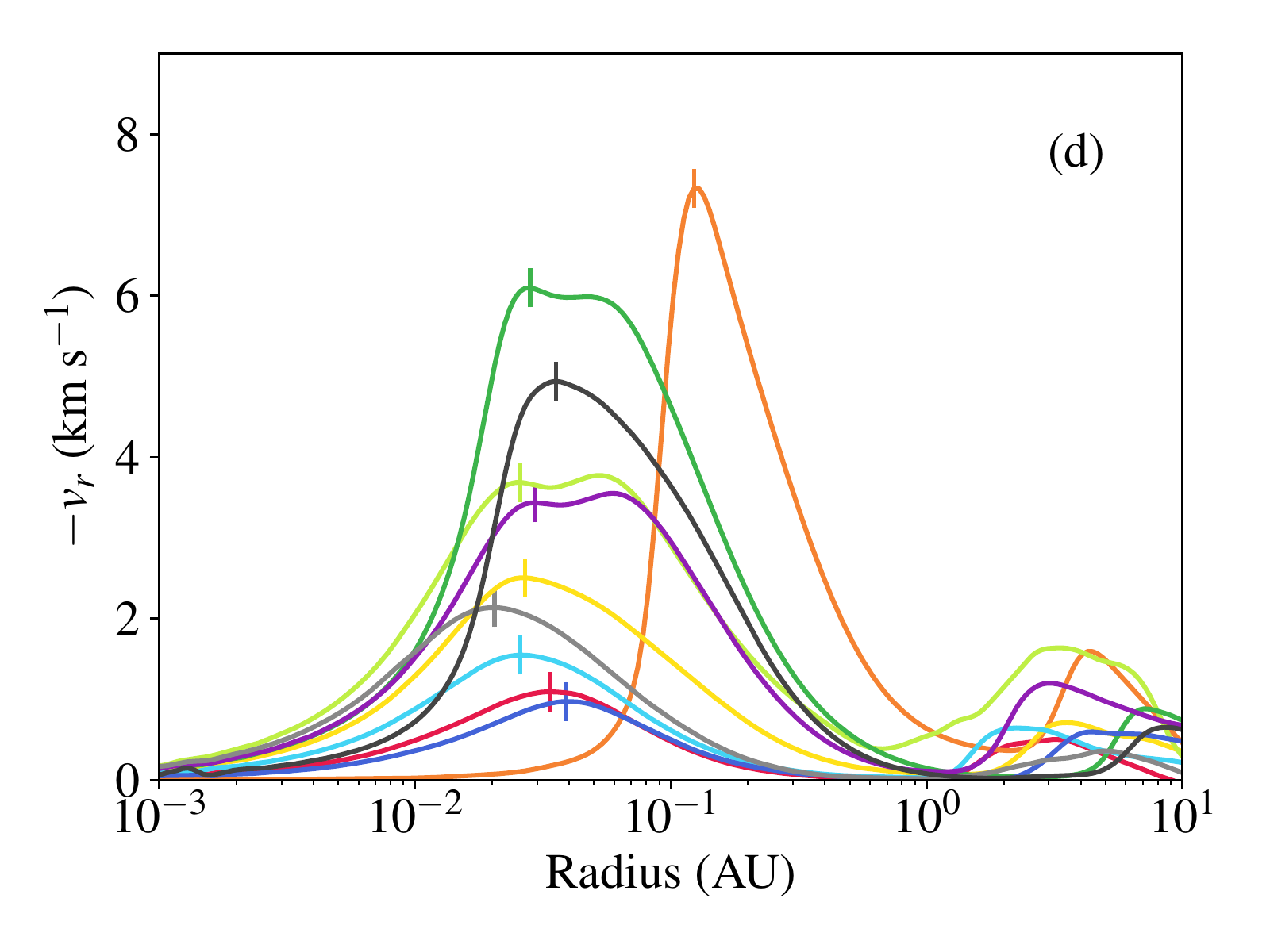}}} \\
  \subfloat{\resizebox{0.45\hsize}{!}
  {\includegraphics{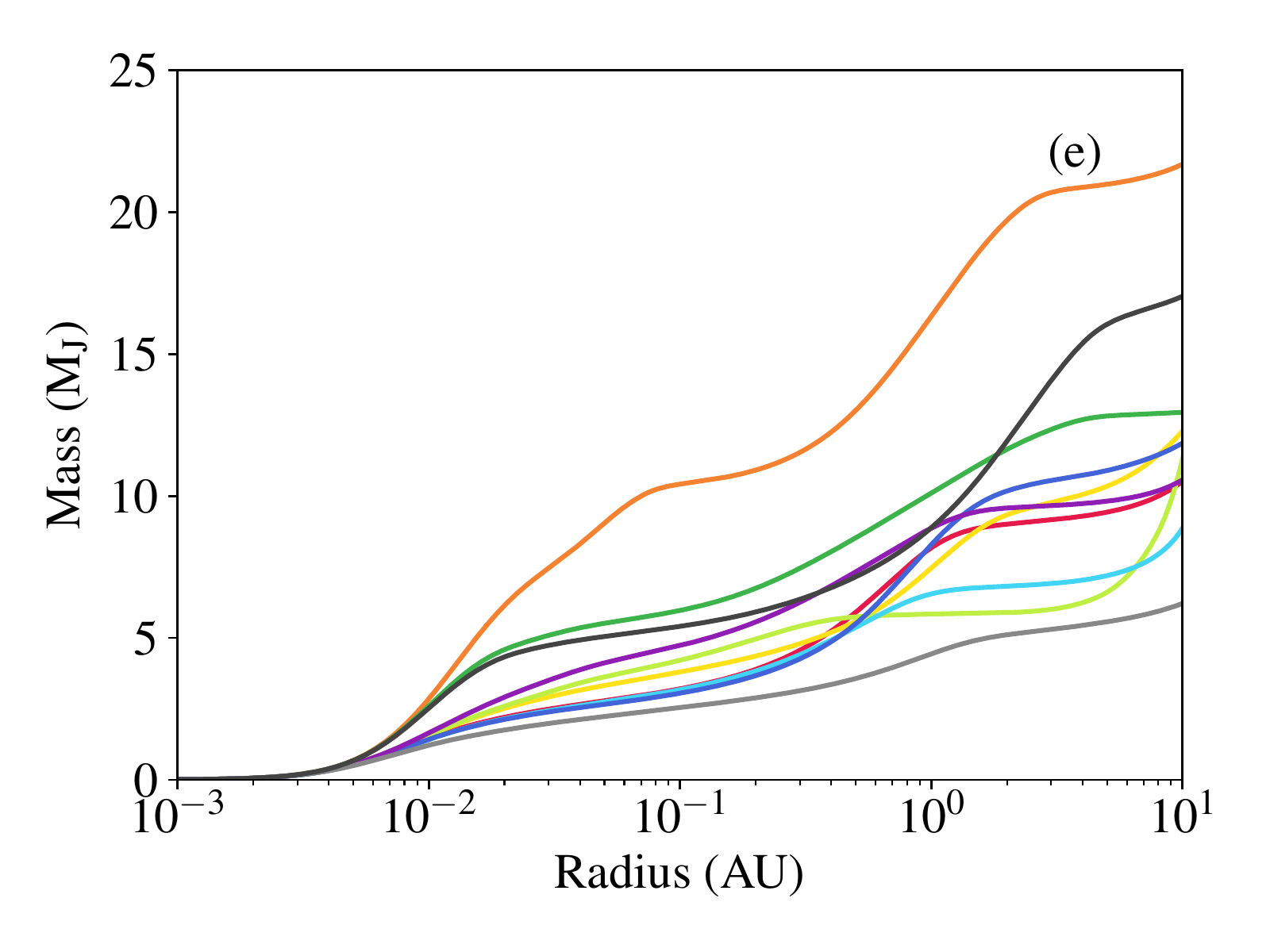}}}
  \subfloat{\resizebox{0.45\hsize}{!}
  {\includegraphics{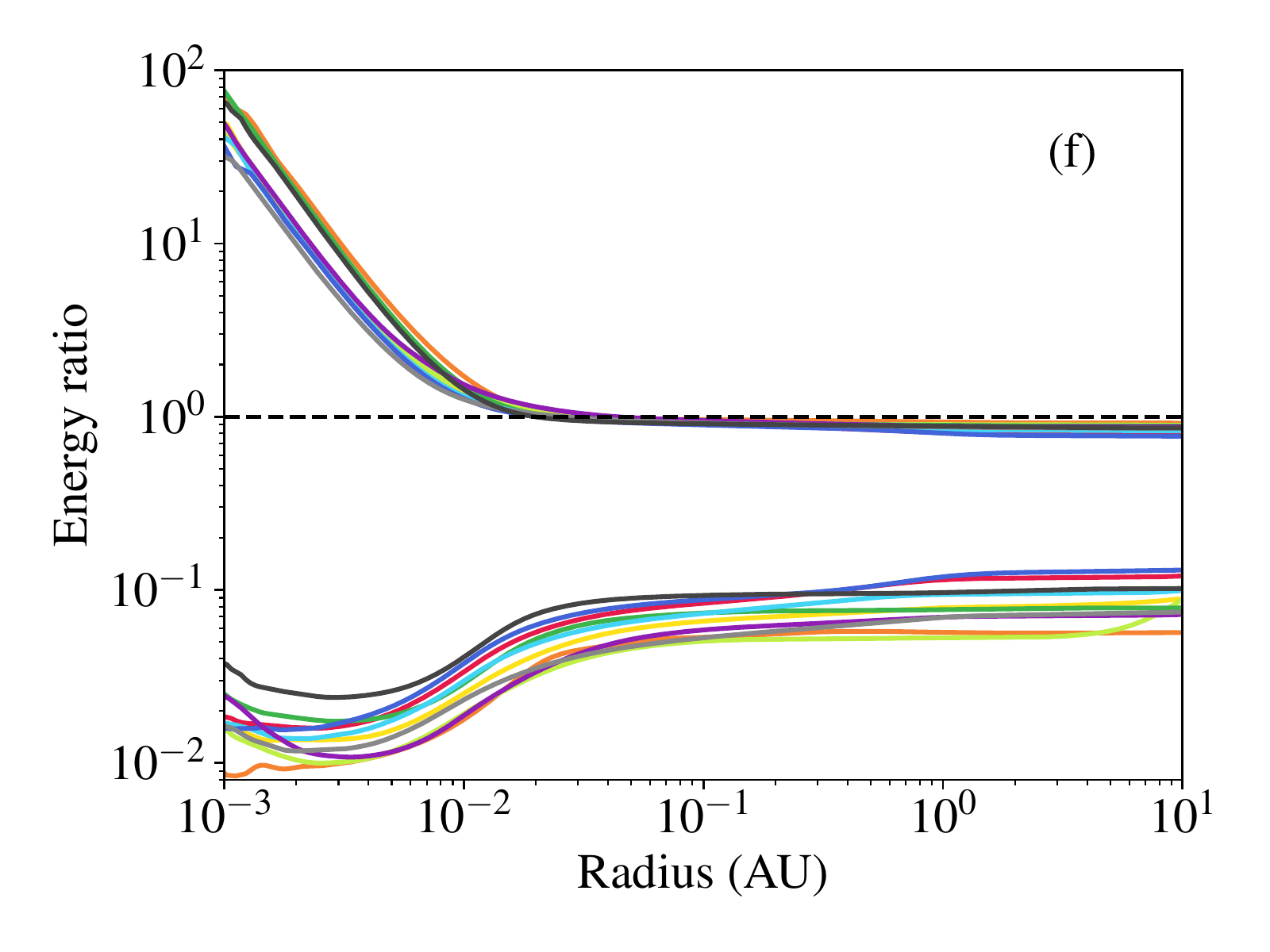}}}
  \caption
  {
    Properties of Type Ia protoplanets, that is, fragments which have undergone
    second collapse and attained a central density of $10^{-3} \textup{ g cm}
    ^{-3}$. Panels (a) and (b) show the spherically-averaged density and
    temperature, respectively. They do not vary significantly from protoplanet
    to protoplanet, though the protoplanets in runs 6, 9 and 27 posses denser
    and hotter central regions due to their higher mass. Panels (c) and (d) show
    rotational (azimuthally-averaged) and infall velocity (spherically-averaged),
    the former of which is significant as the protoplanets reside in a rotating
    disc. The peaks in infall velocity are indicative boundaries where gas
    begins to decelerate. The second core boundaries are at $R = 10^{-2} -
    10^{-1}$~AU and the first core boundaries at $R = 1 - 10$~AU. Panel (e)
    shows the mass of the protoplanet within a given radius, demonstrating that
    even in low mass discs, the mass of formed objects is of the order of a few
    $\MJUP$ or higher. Panel (f) shows the ratio of energies interior to a given
    radius: $E_{\textup{ther}} / E_{\textup{grav}}$ (top set of lines) and
    $E_{\textup{rot}} / E_{\textup{grav}}$ (bottom set of lines). Rotational
    energy is generally much lower than the gravitational energy. The short vertical lines in (d) indicate the positions of the second cores.
  }
  \label{fig:typeIa}
\end{center}
\end{figure*}

\begin{figure*}
  \begin{center}
  \subfloat{\resizebox{0.45\hsize}{!}
  {\includegraphics{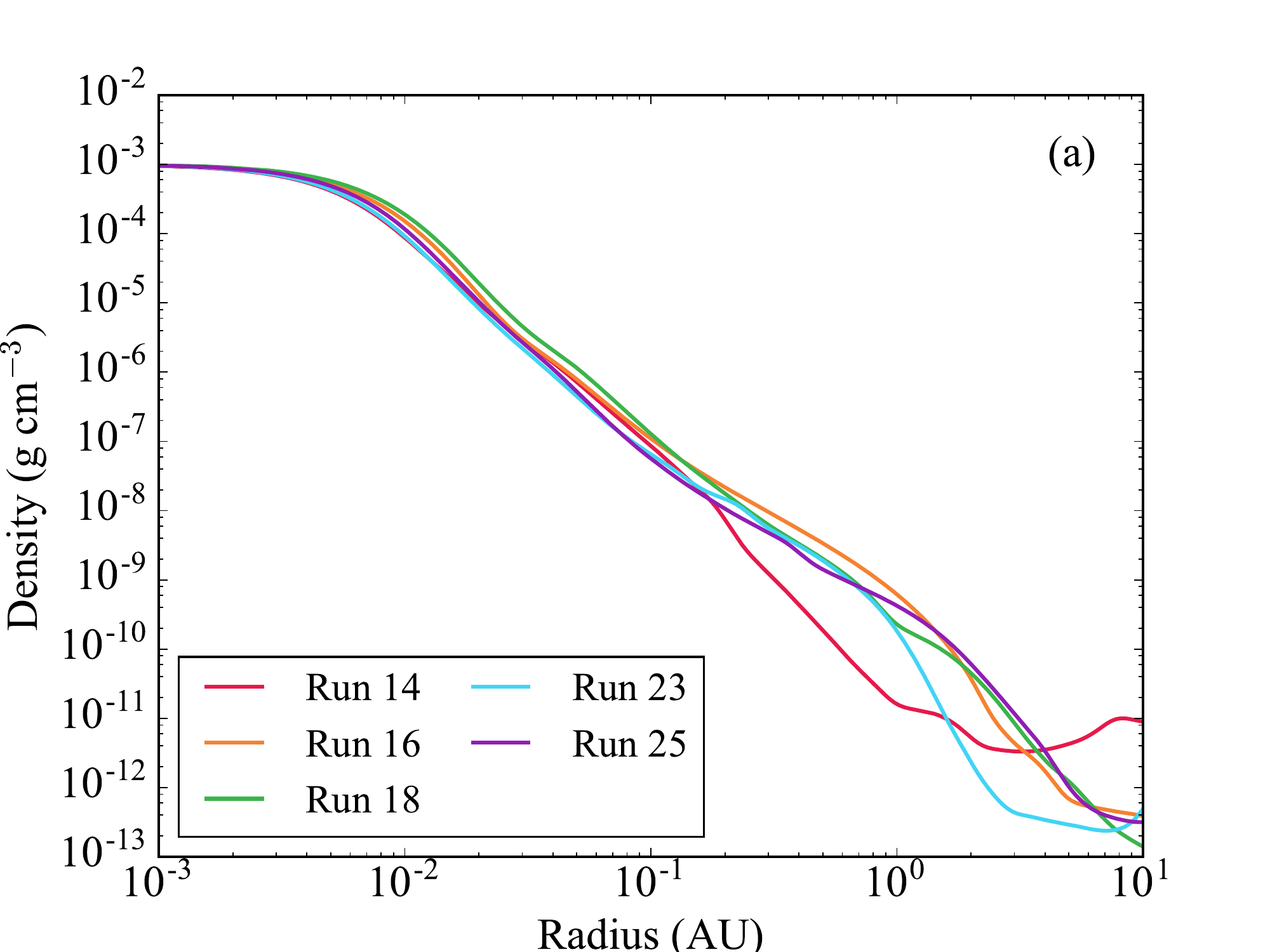}}}
  \subfloat{\resizebox{0.45\hsize}{!}
  {\includegraphics{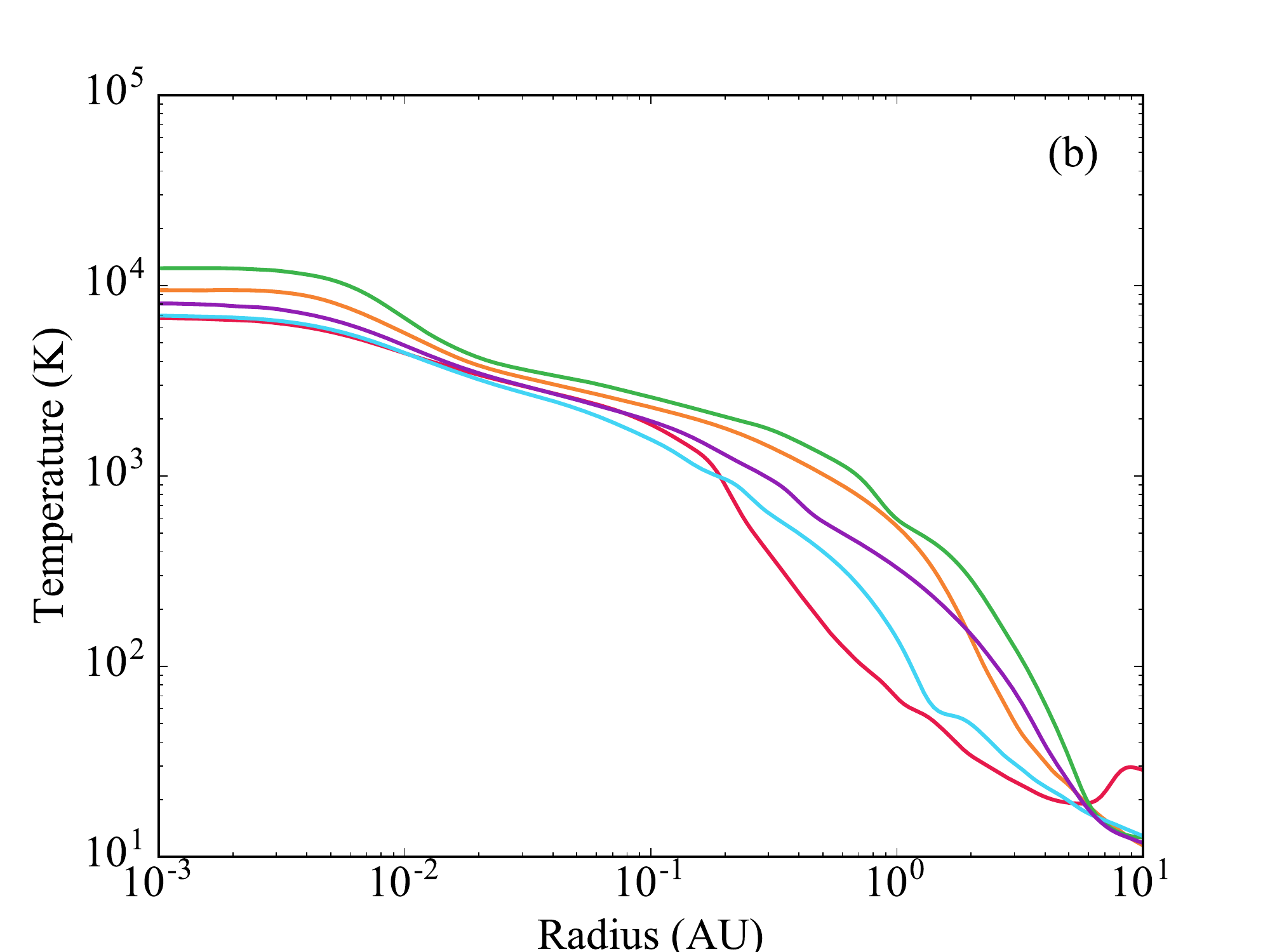}}} \\
  \subfloat{\resizebox{0.45\hsize}{!}
  {\includegraphics{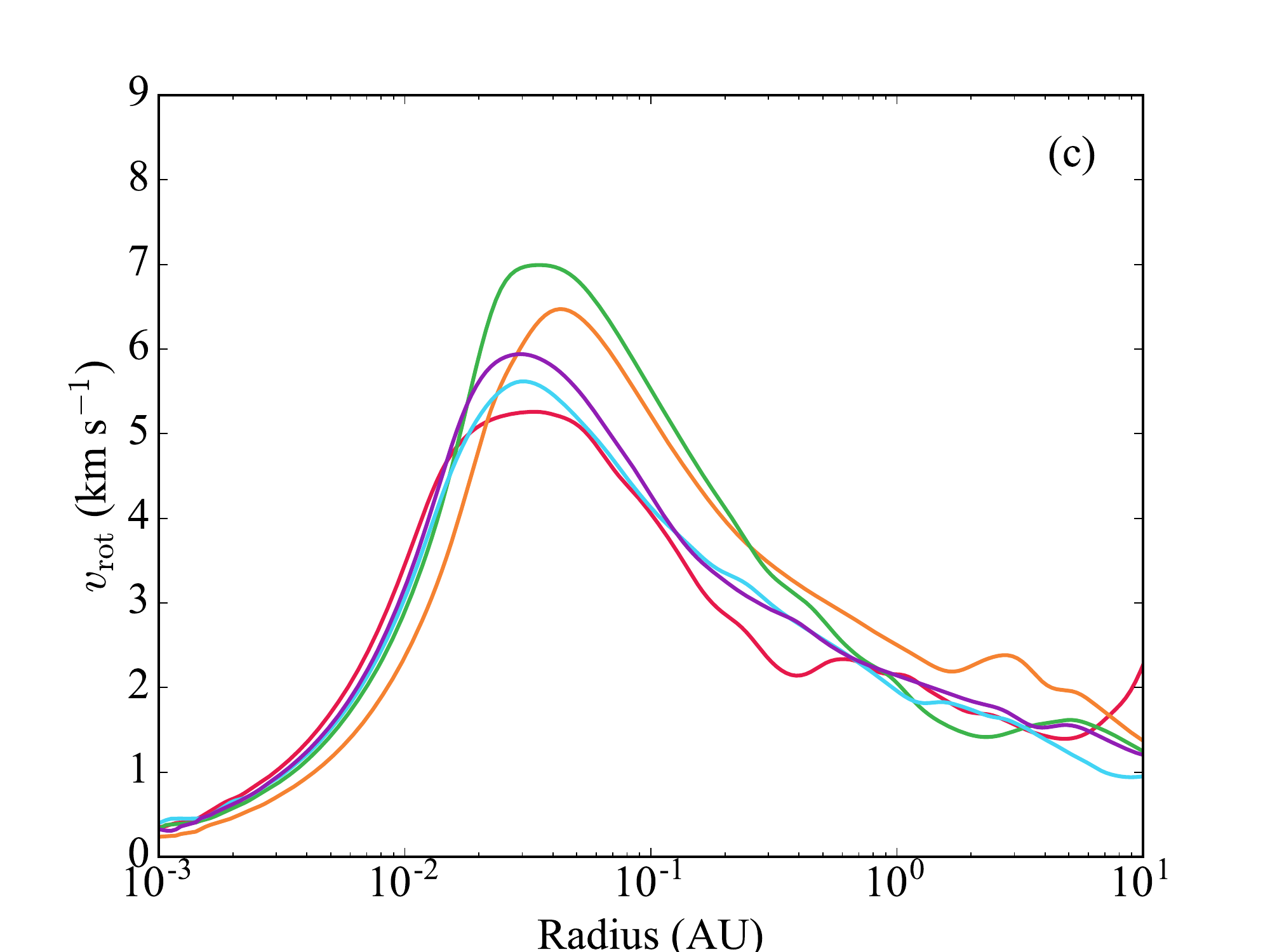}}}
  \subfloat{\resizebox{0.45\hsize}{!}
  {\includegraphics{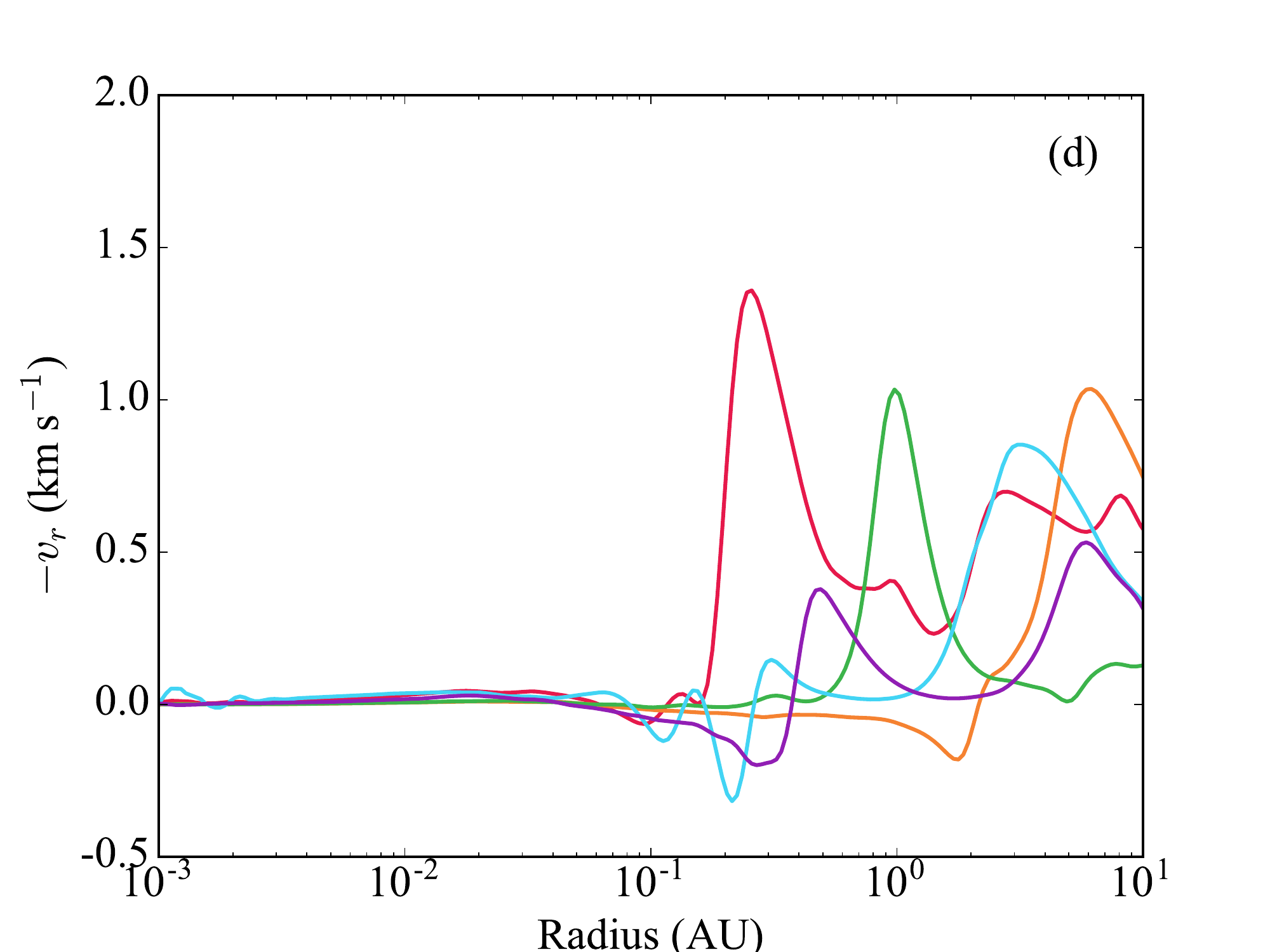}}} \\
  \subfloat{\resizebox{0.45\hsize}{!} 
  {\includegraphics{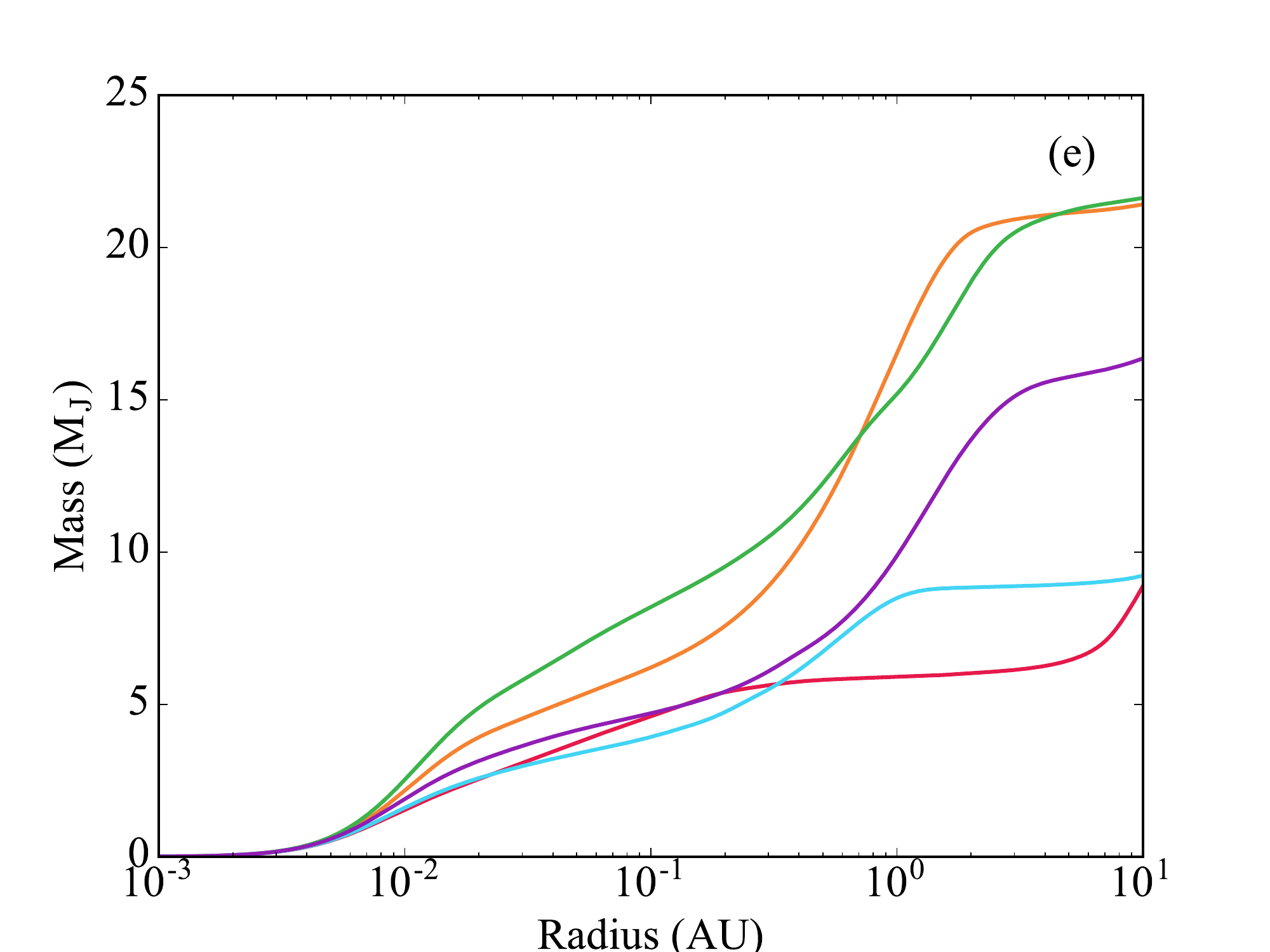}}}
  \subfloat{\resizebox{0.45\hsize}{!}
  {\includegraphics{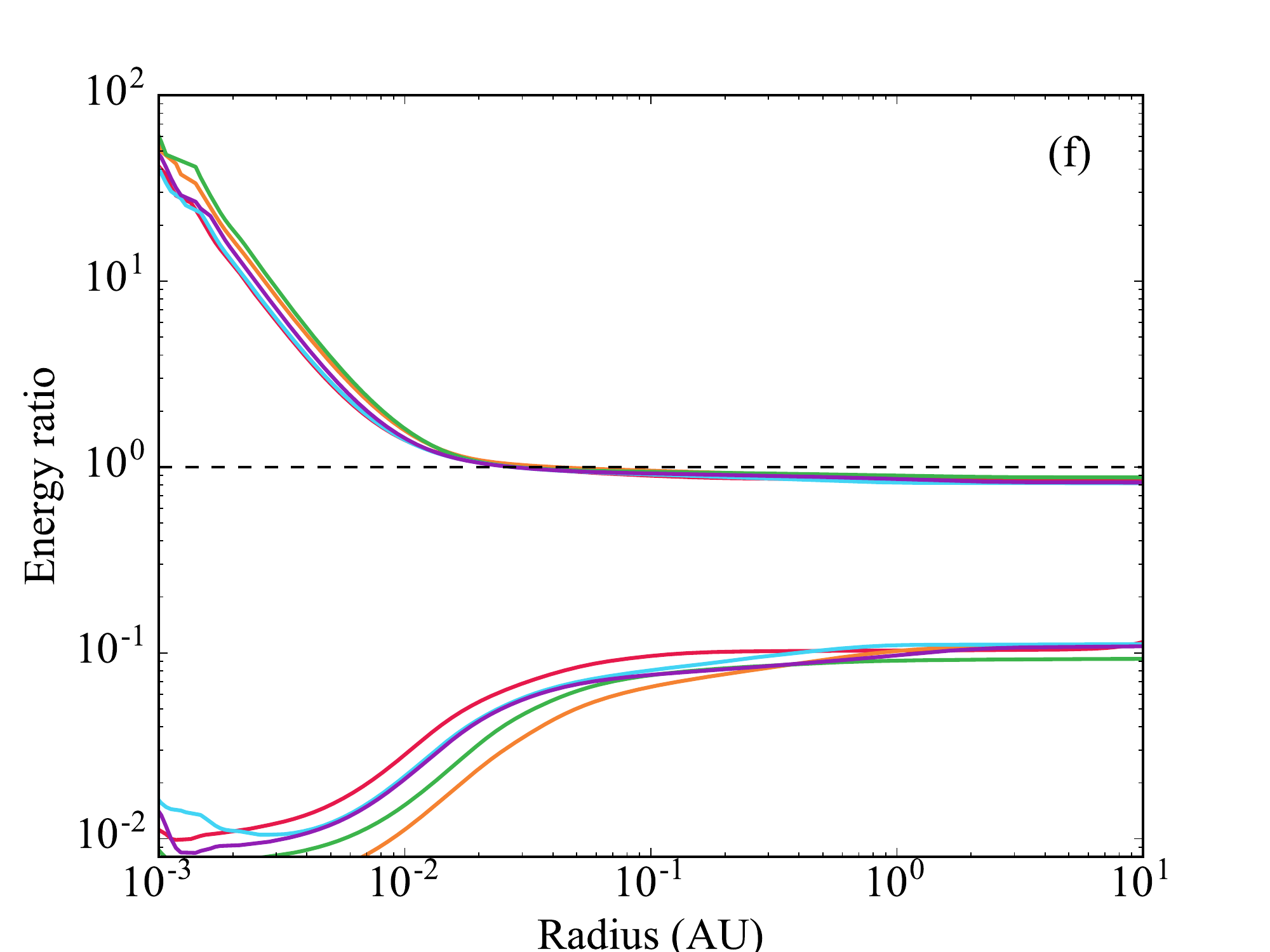}}}
  \caption
  {
    Properties of Type Ib protoplanets formed in the simulations (same as in
    Figure~\ref{fig:typeIa}). The protoplanets presented here are the fragments that have undergone second collapse but  do not show
    any infall velocity signatures indicative of a second core. They are
    structurally similar to the Type Ia protoplanets in Figure~\ref{fig:typeIa}, however the infall velocities here are
    almost zero or slightly negative, that is, indicative of a slowly-expanding
    protoplanet core.
  }
  \label{fig:typeIb}
\end{center}
\end{figure*}

\begin{figure*}
  \begin{center}
  \subfloat{\resizebox{0.45\hsize}{!}
  {\includegraphics{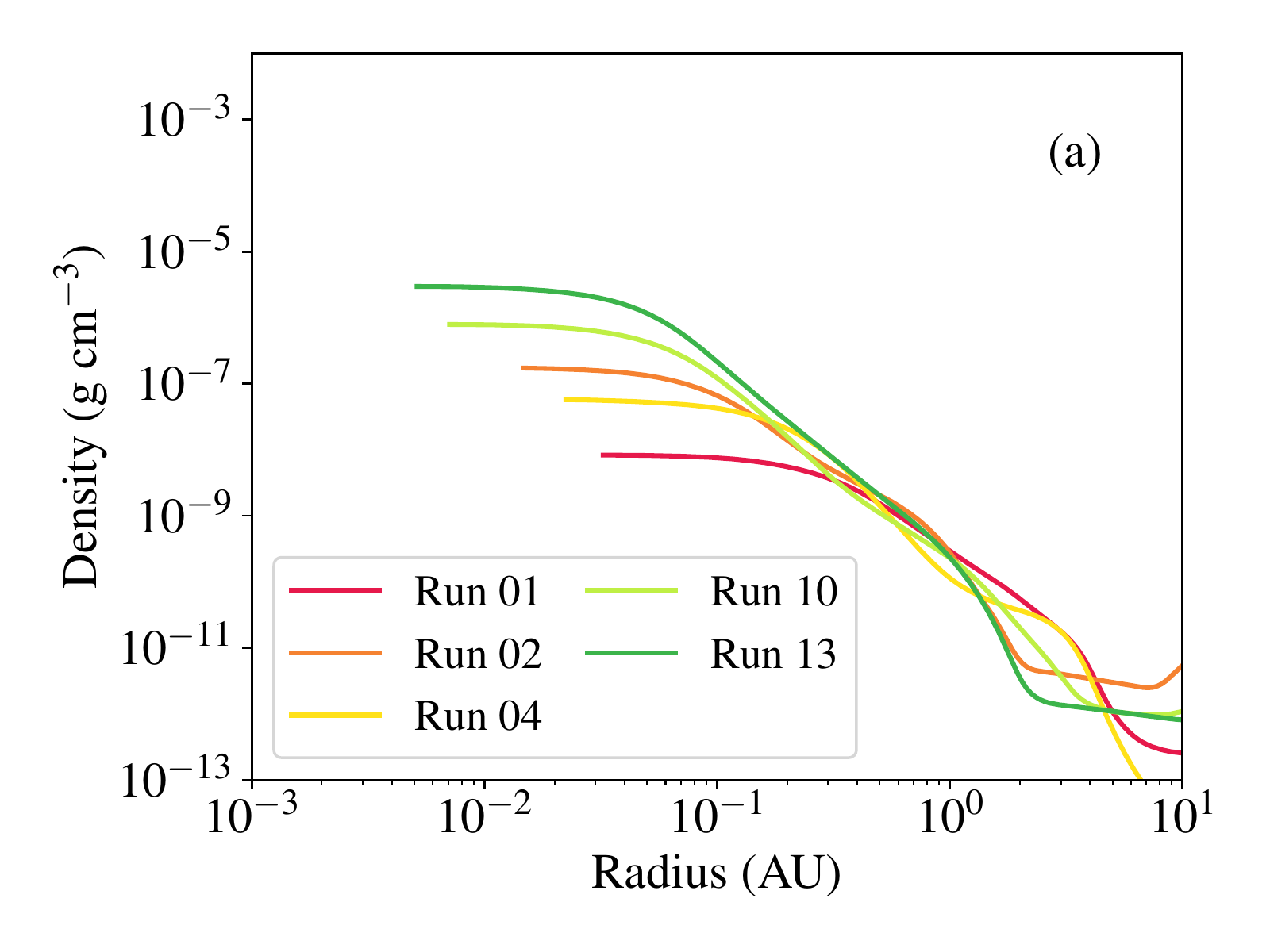}}}
  \subfloat{\resizebox{0.45\hsize}{!}
  {\includegraphics{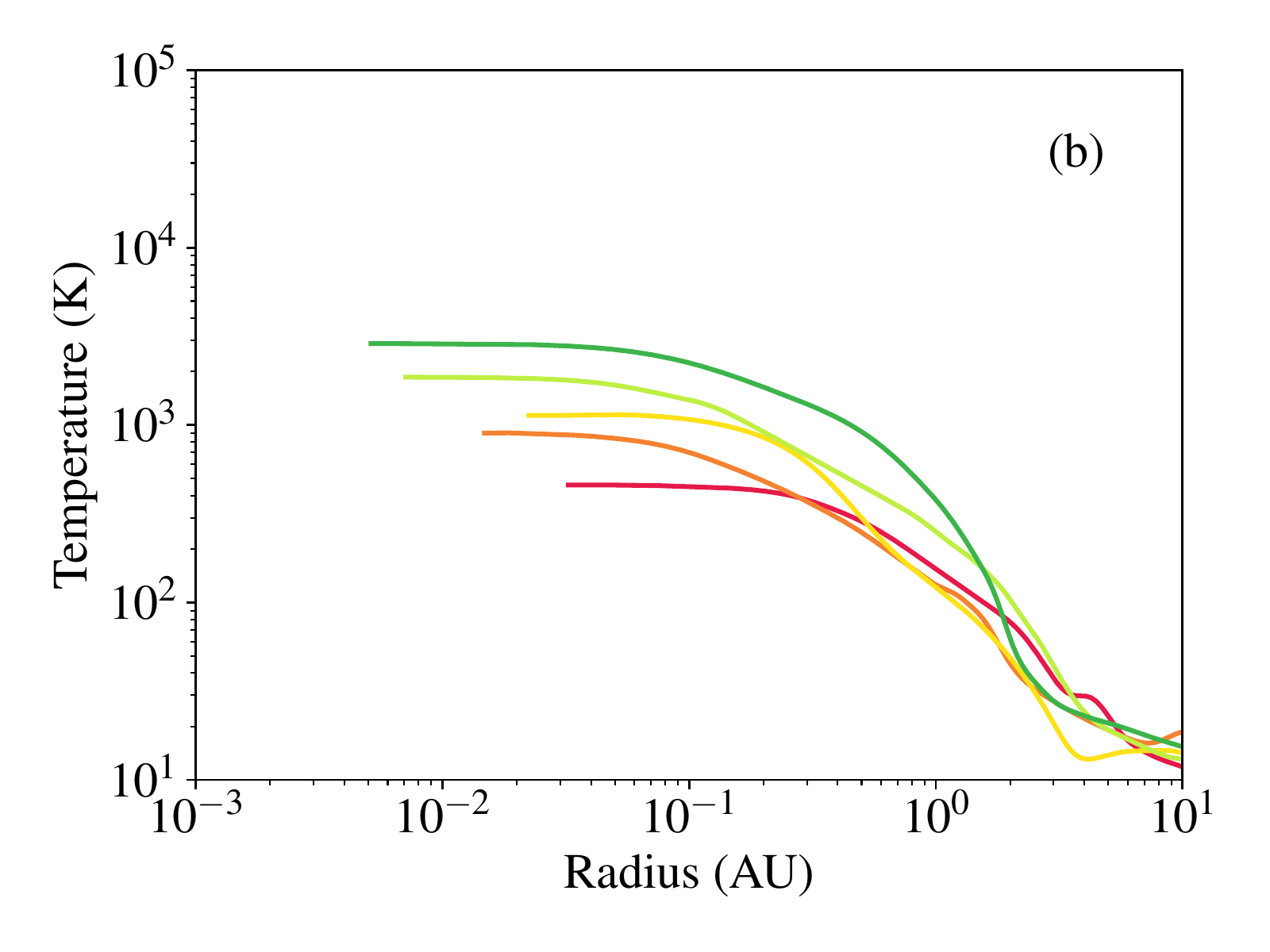}}} \\
  \subfloat{\resizebox{0.45\hsize}{!}
  {\includegraphics{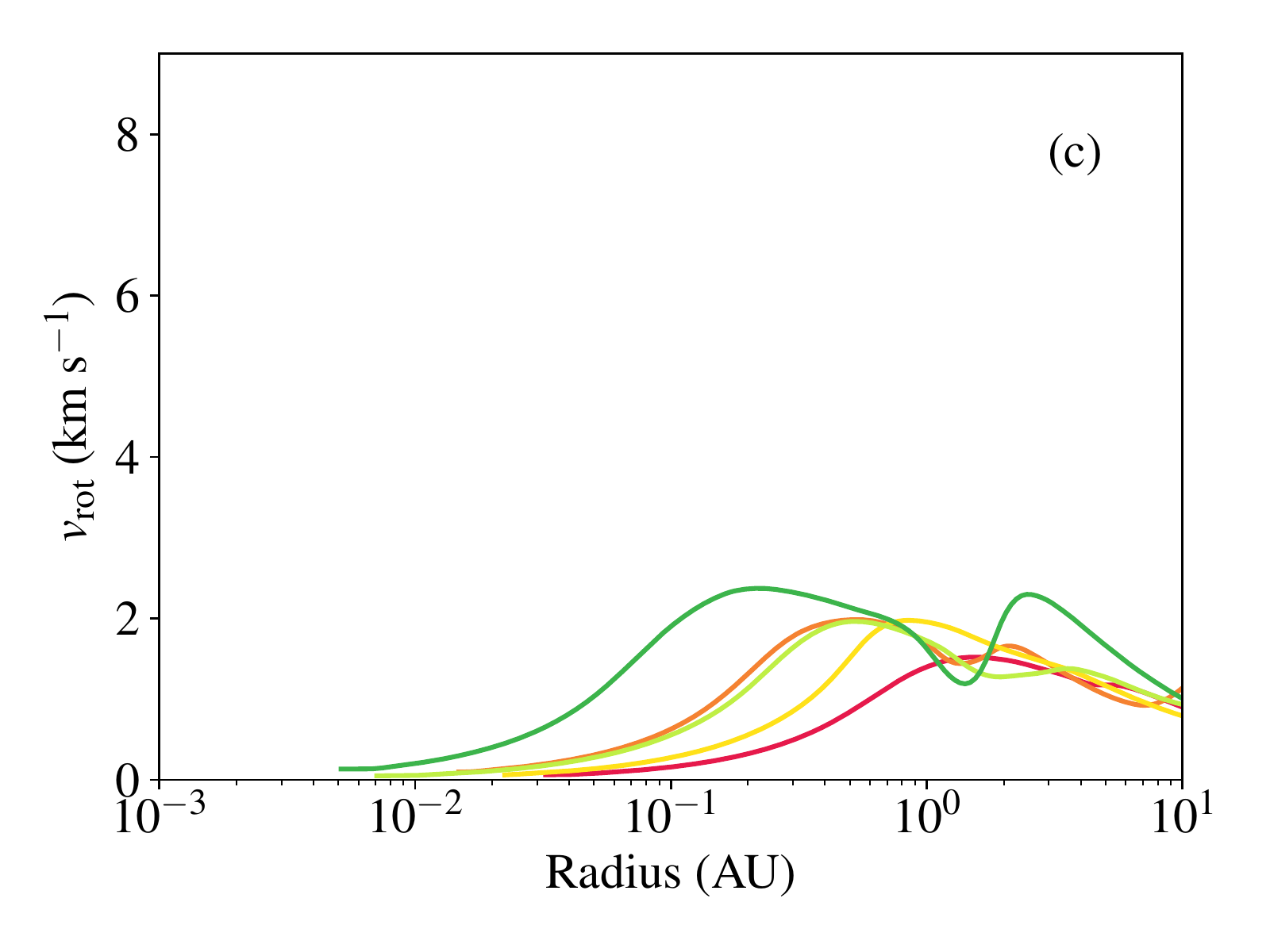}}}
  \subfloat{\resizebox{0.45\hsize}{!}
  {\includegraphics{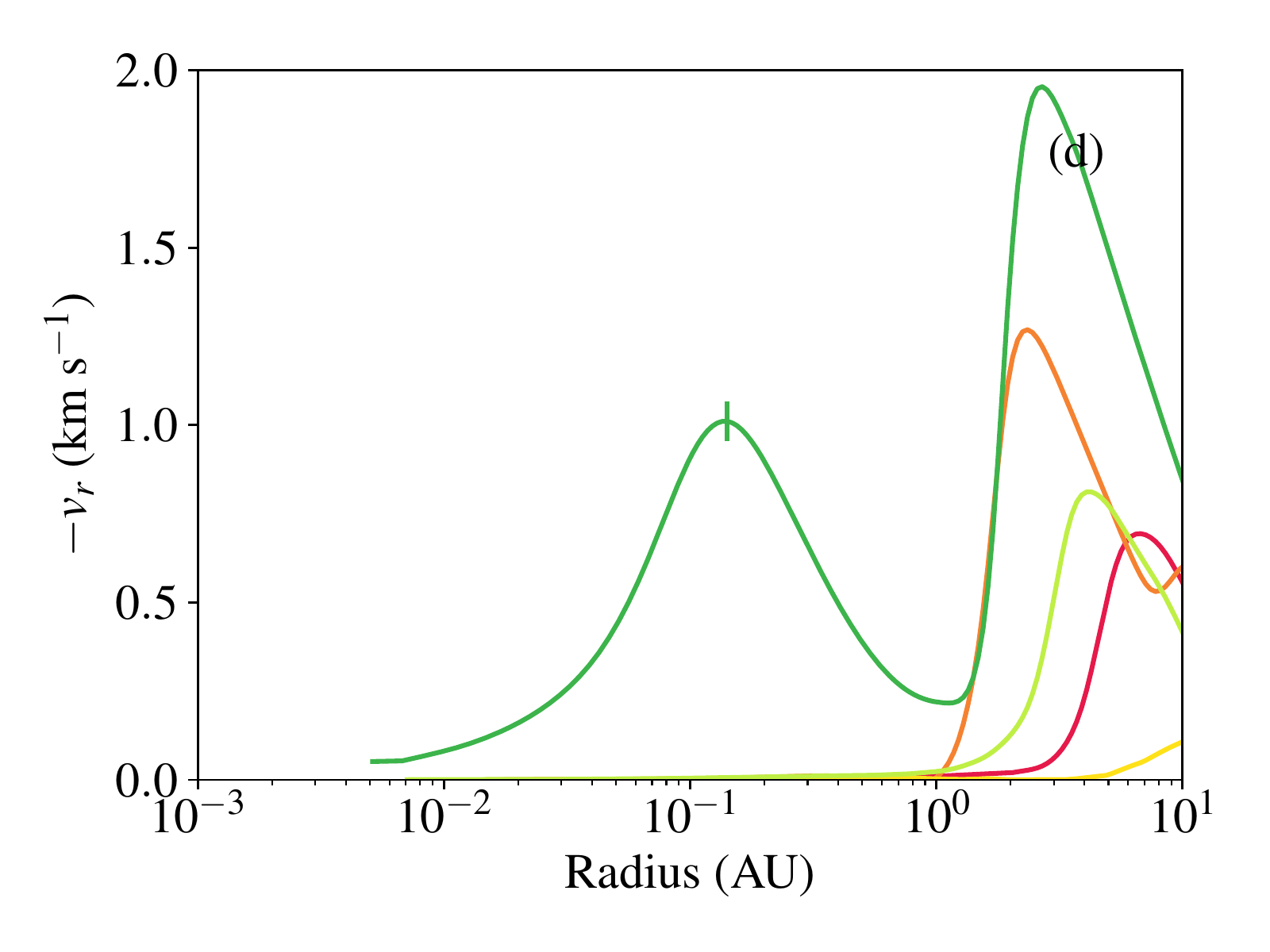}}} \\
  \subfloat{\resizebox{0.45\hsize}{!}
  {\includegraphics{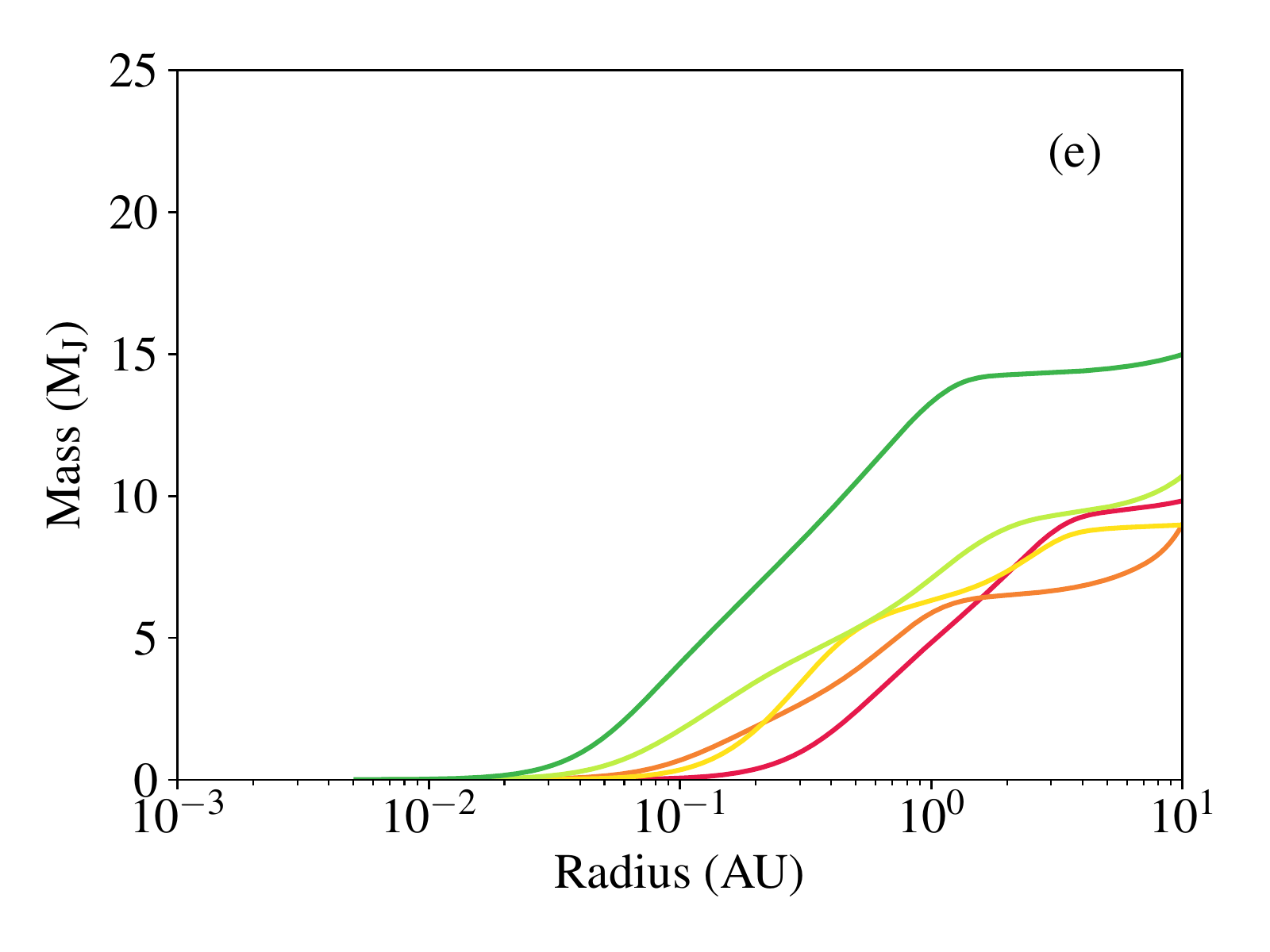}}}
  \subfloat{\resizebox{0.45\hsize}{!}
  {\includegraphics{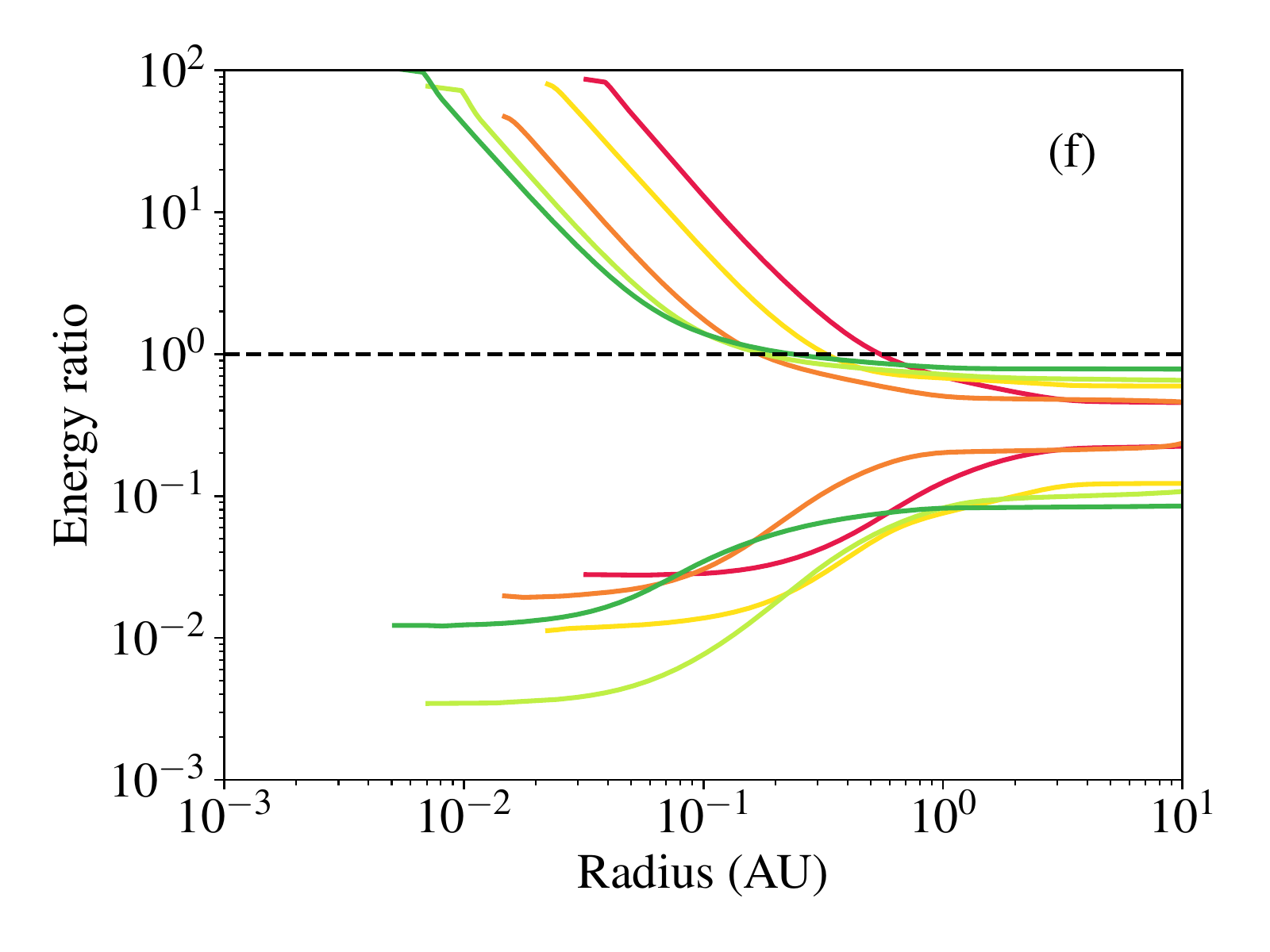}}}
  \caption
  {
    Properties of Type II protoplanets formed in the simulations (same as in
    Figure~\ref{fig:typeIa}, that is, fragments that do not reach a density of  $10^{-3} \textup{ g cm}^{-3}$ at their centres). One of these protoplanets ({Type IIa}; Run 13) undergoes a second collapse and shows evidence of a second core in the radial infall profile. This is depicted by open stars in Figures. However,  most of these protoplanets do not undergo second collapse ({Type IIb}). The short vertical line in (d) indicates the position of the second core for Run 13.
  }
  \label{fig:typeII}
\end{center}
\end{figure*}

Plots of the radial profiles of various properties for all  fragments at central density  $10^{-9}{\rm g\ cm}^{-3}$  (Figures~\ref{fig:fragment_properties}-\ref{fig:noncollapsed_fragment_properties}), and protoplanets (Type I and II, see discussion in Section~\ref{sec:the_properties_of_planets_formed_through_disc_fragmentation})  (Figures~\ref{fig:typeIa}-\ref{fig:typeII}) are presented.
Structurally,  protoplanets are similar to one another, differing only in
mass. The temperature is generally higher within the more massive protoplanets. The rotational velocity is comparable to the infall velocity
but despite this, the ratio between rotational energy and gravitational energy
is generally small throughout, between 0.01 and 0.1. The thermal energy is
comparable to the gravitational energy.

\begin{figure*}
  \begin{center}
    \resizebox{0.7\hsize}{!}{\includegraphics{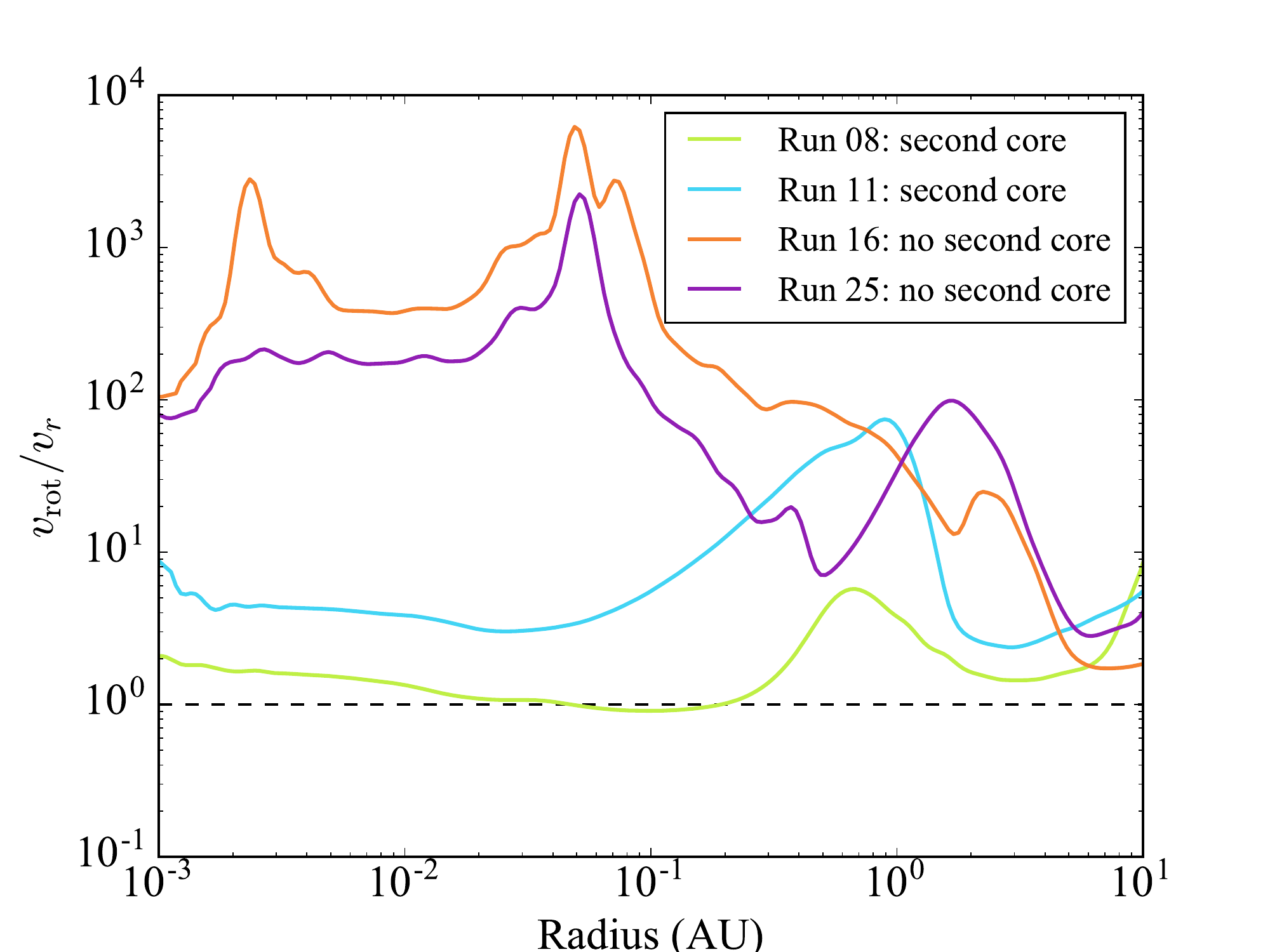}}
    \caption
    {
      Ratio of the azimuthally-averaged rotational-to-infall velocity for a
      set of protoplanets with and without any second core signatures as
      determined from infall velocity peaks. The protoplanets in Runs 8 and 11 (Type Ia protoplanets) show signs of second
      cores in their infall velocities and exhibit values of $v_{\textup{rot}} /
      v_{r} < 10$ in their inner regions. The protoplanets in Runs 16 and 25 (Type Ib protoplanets) do not show second core
      signatures, and their rotational velocity is of the order of 3 magnitudes
      higher than the infall velocity in their inner regions. The significant
      amount of rotation inhibits the formation of an accretion shock around the
      second core.
    }
    \label{fig:v_ratio}
  \end{center}
\end{figure*}

A few of the protoplanets formed in the simulations do not show clear signs of a
second core (Type Ib protoplanets), despite having undergone a second collapse (see Figure~\ref{fig:typeIb} and Figure~\ref{fig:typeII}). These protoplanets have
almost zero infall velocities (or slightly negative in some cases, indicative of
a slow expansion) and they seem to be fast rotating. Figure \ref{fig:v_ratio}
shows azimuthally-averaged radial profiles of the ratio between rotational to
infall velocity. We compare protoplanets which show a clear sign of second core
formation with those that do not. The protoplanets with a second core (Type Ia, IIa; runs 8
and 11, green and blue lines, respectively) have $v_{\textup{rot}} / v_{r} < 10$
in their inner regions, which is relatively low compared to the protoplanets
without second cores (Type Ib, IIb; $v_{\textup{rot}} / v_{r} >10^2$). In these latter cases (runs 16 and 25, orange and purple
lines, respectively), the rotational velocity is a factor of 2 - 4 magnitudes
higher than the infall velocity.

\end{document}